\documentclass[twocolumn,showpacs,preprintnumbers,amsmath,amssymb,superscriptaddress, 
rmp, aps]{revtex4-2}

\usepackage{enumerate}
\usepackage{graphicx}
\usepackage{dcolumn}
\usepackage{bm}
\usepackage[version=3]{mhchem} 
\usepackage{amsmath}
\usepackage{esvect}

\usepackage{array}
\usepackage{graphicx}
\usepackage{graphicx}
\usepackage{dcolumn}
\usepackage{bm}
\usepackage{color}
\usepackage{pifont}
\usepackage{natbib}
\usepackage{hyperref}
\hypersetup{
colorlinks = true,
urlcolor   = black,
linkcolor  = blue,
citecolor  = blue
}
\usepackage{nicefrac}
\usepackage{slashed}

\usepackage{color}
\newcommand{\red}[1]{{\color{black} #1}}

\newcommand{\green}[1]{{\color{black} #1}}

\newcommand{\magenta}[1]{{\color{black} #1}}

\begin{document}

\title{\red{Advances in Honeycomb Layered Oxides:\\
Part II: Theoretical advances in the characterisation of honeycomb layered oxides with optimised lattices of cations}}

\author{Godwill Mbiti Kanyolo}
\email{gmkanyolo@mail.uec.jp}
\email{gm.kanyolo@aist.go.jp}
\affiliation{The University of Electro-Communications, Department of Engineering Science,\\
1-5-1 Chofugaoka, Chofu, Tokyo 182-8585, Japan}
\affiliation{Research Institute of Electrochemical Energy (RIECEN), National Institute of Advanced Industrial Science and Technology (AIST), 1-8-31 Midorigaoka, Ikeda, Osaka 563-8577, Japan}

\author{Titus Masese}
\email{titus.masese@aist.go.jp}
\affiliation{Research Institute of Electrochemical Energy (RIECEN), National Institute of Advanced Industrial Science and Technology (AIST), 1-8-31 Midorigaoka, Ikeda, Osaka 563-8577, Japan}
\affiliation{AIST-Kyoto University Chemical Energy Materials Open Innovation Laboratory (ChEM-OIL), Yoshidahonmachi, Sakyo-ku, Kyoto-shi 606-8501, Japan}


\begin{abstract}
The quest for a successful condensed matter theory that incorporates diffusion of cations, whose trajectories are restricted to a honeycomb/hexagonal pattern prevalent in honeycomb layered materials is ongoing, with the recent progress discussed herein focusing on symmetries, topological aspects and phase transition descriptions of the theory. Such a theory is expected to differ both qualitatively and quantitatively from 2D electron theory on static carbon lattices, by virtue of the dynamical nature of diffusing cations within lattices in honeycomb layered 
\green{materials}. Nonetheless, similarities exist (especially in the case of fermionic cations), whereby quantities such as pseudo-spin and pseudo-magnetic field degrees of freedom are discernible. Herein, we have focused on recent theoretical progress in the characterisation of \magenta {pnictogen- and chalcogen-based} honeycomb layered oxides with emphasis on hexagonal/honeycomb lattices of cations. 
Particularly, we discuss the link between Liouville conformal field theory to expected experimental results characterising the optimal nature of the honeycomb/hexagonal lattices in congruent sphere packing problems. The diffusion and topological aspects are captured by an idealised model, which successfully incorporates the duality between the theory of cations and their vacancies. Moreover, the rather intriguing experimental result that a wide class of silver-based layered materials form stable Ag bilayers, each comprising a pair of triangular sub-lattices, suggests a bifurcation mechanism for the Ag honeycomb lattice into a pair of hexagonal sub-lattices, which ultimately requires conformal symmetry breaking within the context of the idealised model, resulting in a cation monolayer-bilayer phase transition. Other relevant experimental, theoretical and computational techniques applicable to the characterisation of honeycomb layered materials have been availed for completeness. Indeed, this work seeks to demarcate the frontier of this vast field of research, launching new avenues along the way that hold promise in inviting a wider scientific community into this presently divulging field of honeycomb layered materials.
\end{abstract}

\maketitle

\tableofcontents


\section{Introduction}

\red{Theoretical advances in understanding behaviour of condensed matter systems is often steered not only by experimental and computational results, but more so by our ever-growing understanding of continuous symmetries.}\cite{altland2010condensed}
In particular, Noether's theorem guarantees that for every continuous symmetry of the action, there is a corresponding conservation law. 
\red{Indeed, identifying the complete set of continuous symmetries exhibited is rewarding, for it aids in establishing the complete form of the classical action used to not only derive the relevant equations of motion but also quantise the system through the path integral approach to quantum theory}.\cite{zee2010quantum} 
Whilst, not all symmetries of the action are exhibited 
\red{at the quantum regime}, the renormalisation group flow 
determines the relevant energy scales where such symmetries are 
\red{manifest}.\cite{zamolodchikov1986irreversibility} 

One such \red{crucial} 
symmetry is scale invariance, which is 
guaranteed at the fixed points of the renormalisation group flow, 
\red{corresponding to} the critical point of a phase transition where the physics is energy scale independent.\cite{zamolodchikov1986irreversibility, weinberg2000quantum, ginsparg1988applied, poland2019conformal, zamolodchikov1996conformal, cappelli2009ade, guillarmou2020conformal} 
\red{For instance, critical points where two or more phases coexist are vital in elucidating several features in any sought-after theory of phase transitions}.\cite{domb2000phase} 
\red{Under certain rather general conditions, scale invariance almost always implies conformal invariance, suggesting a conformal field theory (CFT) must live at the critical point.\cite{polchinski1988scale, riva2005scale}} Thus, 
\red{CFTs} offer potent theoretical tools for describing condensed matter systems near critical points of phase transitions.\cite{domb2000phase} In two-dimensional (2D) systems, scale invariance implies conformal invariance.\cite{polchinski1988scale} 
\red{In particular, a useful parameter to track is a positive function}, $c(M, T_j)$ which depends on the energy scale, $E$ and couplings $T_j$ labeled by the index $j$, and monotonically decreases under the renormalisation group flow.\cite{komargodski2011renormalization, zamolodchikov1986irreversibility} The so-called $C$-theorem states that such a function must be independent of 
\red{energy} at the fixed points satisfying, $\partial c(E, T_j)/\partial T_j = 0$ with solutions $T_j^{\rm c}$ which 
correspond to the relevant coupling constants for the \red{CFT} 
\red{whilst} the function $c(E, T_j^{\rm c}) = c$ is a constant independent of energy scale known as the central charge.\cite{zamolodchikov1986irreversibility} Thus, 
\red{CFTs}  
\red{are classified not only by their highest weight state, $h$, but also their central charge, $c$ which can be non-vanishing in the quantum regime}.\cite{polchinski1998string2, duff1994twenty} 

\red{In materials science}, the relevant condensed matter systems 
\red{exhibiting phase transitions} are almost always crystalline, 
\red{where} the 
\red{spacial} symmetries are no longer continuous but discrete. Surprisingly, even the time translation symmetry related to energy conservation can be discrete, for instance, in time crystals or completely non-existent if the condensed matter system can be considered 
\red{dissipative}.\cite{sacha2017time, shapere2012classical, wilczek2012quantum} Thus, Noether's theorem applies in the continuum limit, $a_i \rightarrow 0$, where $a_i$ are the lattice spacing of atoms within the crystal along dimension, $i = 1, 2 \cdots d$ with $d$ the number of 
\red{spacial} dimensions of the lattice. 
\red{Thus, exemplars of discrete symmetries are crystalline 
symmetries, which} can be spontaneously broken by non-trivial topologies serving as topological defects introduced by impurities, vacancies, dislocations and/or disclinations.\cite{kanyolo2020idealised, kanyolo2021honeycomb, kanyolo2022cationic, masese2021topological, masese2023honeycomb} In particular, topological defects can be treated as gauge symmetries, which define conserved topological charges or additional spin degrees of freedom.\cite{musevic2006two, mackintosh1991orientational, kamien2002geometry, vitelli2004anomalous, bowick2009two, turner2010vortices, mesarec2016effective} \red{Consequently, the CFTs describing critical phenomena exhibited by such crystalline materials can not only have space and time/space-time symmetries responsible for conservation of energy, momentum and angular momentum in the continuum limit alongside conformal invariance, but also internal symmetries such as U($N$) and SU($N$), with $N \in \mathbb{N}$ due to topological charges and additional spin degrees of freedom.} 
\red{Thus, the topological charges can be related to a geometric theory via the Poincar\'{e}-Hopf theorem\cite{kanyolo2022cationic}, hence introducing invariance under homeomorphisms.}
Moreover, emergent gravity has been previously considered to describe defects in crystals.\cite{kleinert1987gravity, kleinert1988lattice, yajima2016finsler, holz1988geometry, verccin1990metric, kleinert2005emerging} For instance, 
a finite torsion (non-symmetric Christoffel symbols/affine connection, $\Gamma_{\,\,\mu\nu}^{\rho} \neq \Gamma_{\,\,\nu\mu}^{\rho}$) within the context of Einstein-Cartan theory 
\red{is considered to} 
capture various 
\red{features} 
\red{concerning} disclinations and dislocations within crystals.\cite{kleinert1987gravity, kleinert1988lattice, yajima2016finsler, holz1988geometry, verccin1990metric} 
\red{Particularly,} Einstein gravity 
\red{has been proposed to} emerge in a crystal whose kinetic energy terms are restricted to second-order in derivatives\cite{kleinert2005emerging} in accordance with Lovelock's theorem.\cite{lovelock1971einstein} 

\begin{figure*}
\begin{center}
\includegraphics[width=\textwidth,clip=true]{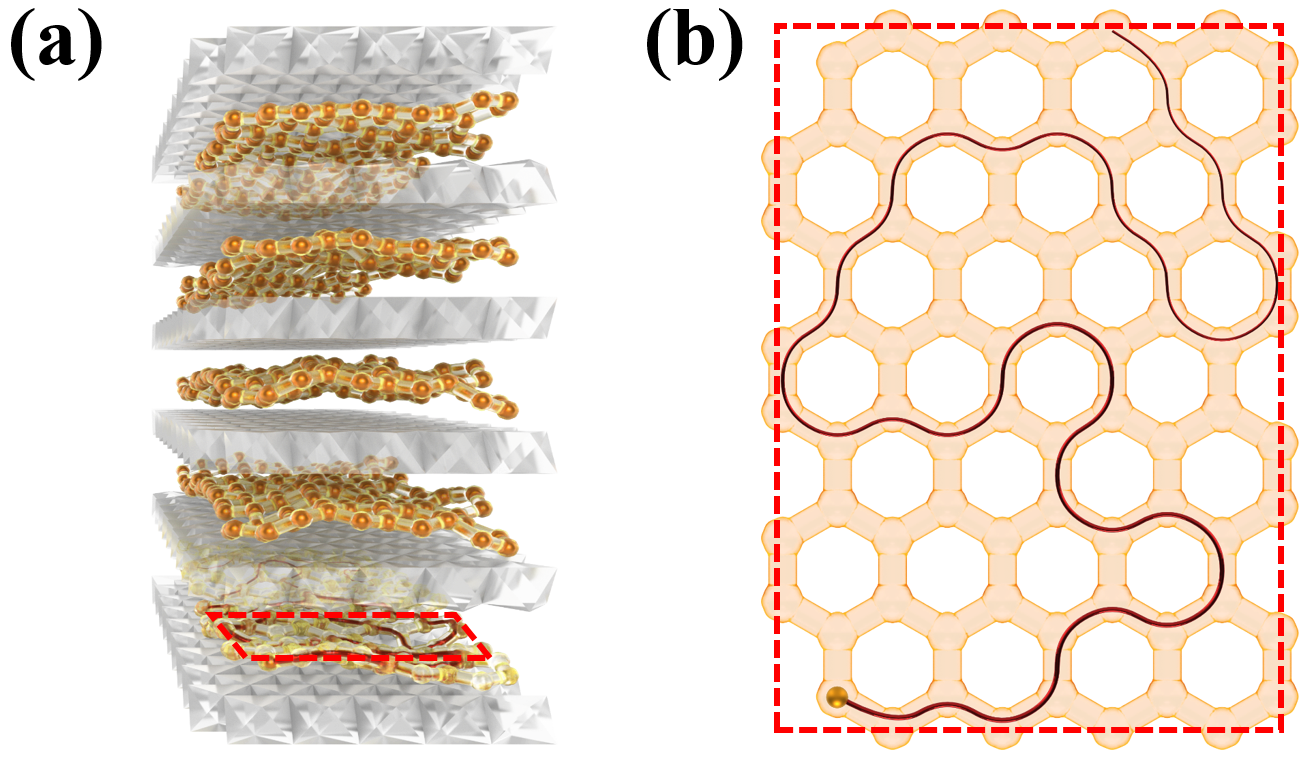}
\caption{
\red{(a) A schematic representation of the structure of exemplar honeycomb layered materials wherein the idealised model applies with the cations represented by reddish yellow (golden) spheres. The red rectangle at the base indicates the approximate location of the schematic in Figure \ref{Fig_2} (b). (b) Schematic depicting honeycomb-shaped diffusion pathways with vacant cationic sites in exemplar honeycomb layered materials such as $A_4MD\rm O_6$, $A_3M_2D\rm O_6$ or $A_2M_2D\rm O_6$ wherein $A$ represents an alkali ion (Li, Na, K, \textit{etc}.) or coinage metal ions such as Ag, whereas $M$ is mainly a transition metal species such as Co, Ni, Cu and Zn, and $D$ depicts a pnictogen or chalcogen metal species such as Sb, Bi and Te. The maroon line shows a possible random diffusion pathway of a single $A$ cation. Figure reproduced with permission.}\cite{kanyolo2022cationic}}
\label{Fig_2}
\end{center}
\end{figure*}

\red{Recent advances have demonstrated that such is the case for a specific class of layered materials commonly referred to in literature as honeycomb layered oxides.}\cite{kanyolo2020idealised, kanyolo2022cationic, kanyolo2021honeycomb, kanyolo2022advances} Their crystalline structure and symmetries can be elucidated by considering the honeycomb layered oxide with the general chemical formula, 
\red {$A_aM_mD_d\rm O_6$ (\magenta {typically with} $0 < a \leq 4$; $0 < m \leq 2$; $0 < d \leq 1$)}, where $A$ is an alkali metal cation 
\red {suchlike K, Na, Li, Rb and Cs}, or coinage metal cations such as 
Ag, Cu and Au, $M$ is a 
\red{transition metal atom (suchlike Ni, Co, Zn, Mn, Cu, $etc.$) or a $s$-block metal atom such as Mg, whilst $D$ is a chalcogen atom (such as Te) or a pnictogen atom (suchlike Bi and Sb).}\cite{kanyolo2021honeycomb, masese2023honeycomb, kanyolo2022advances} For instance, in the case of $A_2\rm Ni_2TeO_6$, where $A =$ Cs, Rb, K, Na, Li, H, Au, Ag, Cu, 
\red{transmission electron microscopy (TEM)} amongst other complementary techniques such as density functional theory (DFT) simulations, X-ray and neutron diffraction (XRD, ND), show the slabs formed by $\rm NiO_6$ octahedra arranged in a honeycomb lattice packing, with the $\rm TeO_6$ octahedra occupying the honeycomb centres.\cite{kanyolo2021honeycomb, matsubara2020magnetism, masese2023honeycomb, tada2022implications, masese2018rechargeable, masese2021Na2Ni2TeO6, masese2021topological, kanyolo2022advances, evstigneeva2011new, karna2017, grundish2019electrochemical, wang2023p2, luke2023rapid, orikasa2014high} 
\red{A hexagonal Bravais lattice or a honeycomb packing of $A$ cations intercalate the spaces between the slabs (interlayers) as shown in Figure \ref{Fig_2}(a), with a single cation sandwiched directly below and above Ni atoms in a prismatic coordination to 6 oxygen atoms, leaving the space directly above and below Te atoms either occupied or vacant respectively.}\cite{kanyolo2022advances} 

Meanwhile, octahedral coordination of cations to 6 oxygen atoms tends to avail a smaller intercalation space between interlayers, which precludes cations 
with a Shannon-Prewitt ionic radius larger than 
\red {$\sim 1.1$ \AA\,} such as $A =$ K, Rb, Cs 
\red{allowing for} $A =$ Li, Na as the only viable alkali metal candidates.\cite{kanyolo2022advances} On the other hand, prismatic coordination precludes Li, which has the smallest Shannon-Prewitt ionic radius of $\sim 0.75$ \AA\, compared to all the other alkali metal atoms.\cite{kanyolo2021honeycomb, kanyolo2022advances} Other cationic coordinations to oxygen atoms such as square planar and tetrahedral have also been theoretically and/or experimentally found, which \red{favour} 
$A =$ Cu and $A =$ Li respectively, due to their small Shannon-Prewitt radii, and a linear/dumbbell coordination expected for $A =$ Cu, Ag, Au coinage metal atoms.\cite{tada2022implications} 
\red{Thus, the Shannon-Prewitt radius of cations,\cite {shannon1976revised} which scales with their atomic number, tends to be directly proportional to the interlayer distance of honeycomb layered oxides and hence can favour or preclude particular cationic coordinations}, suggesting pure electrostatic effects, in addition to valence bond theory, play a 
\red {pivotal} role in determining lattice parameters. 

The search and classification of 
symmetries within 
\red{well-tested and novel} compositions of honeycomb layered oxides is currently the subject of active research, with the focus primarily on 
layered materials that exhibit a 2D hexagonal and/or honeycomb packing of transition metal atoms and/or cations.\cite{kanyolo2022advances, kanyolo2021honeycomb}
\red{For instance, the problem of finding the optimal arrangement of charged atoms (\textit{e.g.} cations) in $d$ dimensions ($d$D) which minimises their electrostatic energy is a congruent sphere packing problem in mathematics} equivalent to the spinless modular bootstrap for 
\red{CFTs} under the algebra U(1)$^c\times$ U(1)$^c$\cite{hales2011revision, cohn2017sphere, viazovska2017sphere, zong2008sphere, cohn2009optimality, cohn2014sphere, afkhami2020high, hartman2019sphere},
\begin{align}\label{sphere_packing_eq}
    \mathcal{Z}_{\Lambda_d}(b) = \sum_{\vec{x}_i, \vec{x}_j \in \Lambda_d}\frac{\exp(-\pi ib\Delta_{\nu}(\vec{x}_i- \vec{x}_j))}{\eta^{2c}(b)},
\end{align}
where $c$ is the central charge, or equivalently the linear programming bound for congruent sphere packing in $d = 2c$ dimensions, $\eta(b) = q^{1/24}(b)\prod_{n = 1}^{\infty}(1 - q^n(b))$ is the Dedekind eta function with $q(b) = \exp(2\pi ib)$ the nome and $b$ a complex-valued variable. It 
\red{was found} that the 2D hexagonal Bravais lattice (and its dual, the non-Bravais bipartite honeycomb lattice)
saturates the linear programming bound in $d = 2,8$ and $24$ dimensions\cite{hartman2019sphere}, 
\red{where $h = \Delta_{\nu}(\vec{x}_i- \vec{x}_j)/2 = -|\vec{x}_i - \vec{x}_j|^2/2\mu = -\nu \in \mathbb{Z}^-$ is the \red{scaling dimension} proportional to the distance between congruent sphere centres (for 2D, this corresponds to the separation distance between vertices in the hexagonal lattice)}, with $\vec{x}_i, \vec{x}_j \in \Lambda_d$ position vectors lying within the $d$-dimensional lattice, $\Lambda_d$ and $\sqrt{\mu} \sim \ell_{\rm P}$ 
\red{of order} lattice constant, $\ell_{\rm P}$.
\red{Meanwhile, the honeycomb lattice is bipartite, 
comprising two hexagonal sub-lattices with partition function, $\mathcal{Z}_{\rm hc}(b, \overline{b}) = \mathcal{Z}_{\Lambda_d}(b)\overline{\mathcal{Z}}_{\Lambda_d}(\overline{b})$, each described by the CFT given by eq. (\ref{sphere_packing_eq})}, which introduces modular invariance in bosonic lattices.\cite{kanyolo2022cationic}

\begin{figure*}
\begin{center}
\includegraphics[width=\textwidth,clip=true]{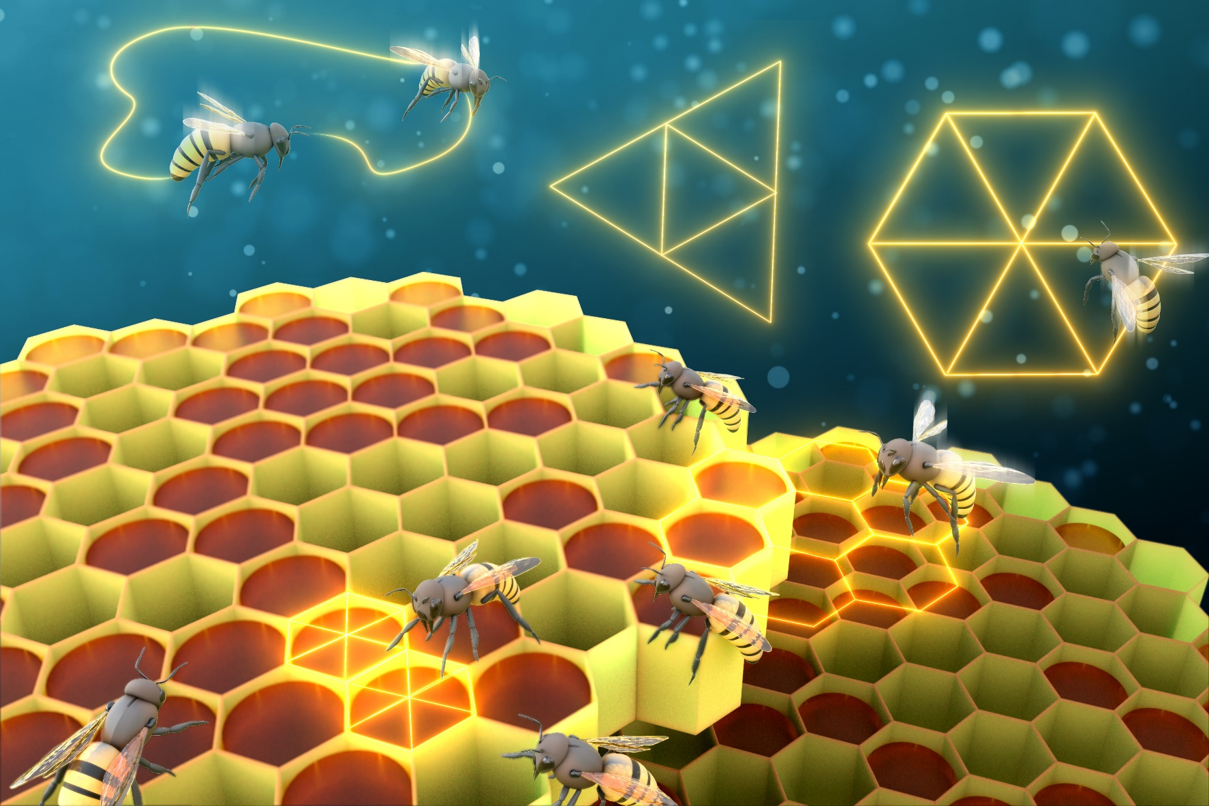}
\caption{
\red{An illustration depicting Hale's conjecture at work in evolution by natural selection.\cite{hales2001honeycomb} Indeed, this conjecture has relevance in biological systems, such as the justification for natural selection favouring the bees that expended the least amount of work to create the largest cross-sectional area for the storage of honey}.\cite{lyon2012mathematical, raz2013application}}
\label{Fig_1}
\end{center}
\end{figure*}

\red{For instance, in the case of $A_aM_mD_d\rm O_6$ with a prismatic coordination of $A$ bosonic atoms to oxygen atoms, defining the valency of $M$ and $D$ respectively as $\mathcal{V}_M$ and $\mathcal{V}_D$, the inequality $m\times \mathcal{V}_M + d\times \mathcal{V}_D > 9$ such as in $A^{1+}_2\rm Ni^{2+}_2Te^{6+}O^{2-}_6$ ($A$ = Na, K) appears to render cationic sites directly below and above the $D$ atoms unfavourable for occupation due to high electrostatic repulsion.}\cite{kanyolo2022advances, kanyolo2021honeycomb} 
\red{This precludes the hexagonal lattice of cations leaving the honeycomb lattice as the better optimised lattice.} Conversely, the hexagonal lattice is the optimised choice \red{when the valency bound is saturated}, $m\times \mathcal{V}_M + d\times \mathcal{V}_D = 9$, such as in $\rm A^{1+}_3Ni^{2+}_2Bi^{5+}O^{2-}_6$ ($A =$ Li, Na, K). Moreover, such higher order interactions of \red{the} cationic lattice with the \red{other} atoms in the slab 
\red{are} expected to 
\red{somewhat} affect the observed lattice pattern, perturbing the stable configurations from the optimised cases that saturate both the linear programming and valency bounds. 
\red{For instance}, in the case of the mixed alkali ${\rm K}_{2 - a}{\rm Na}_a\rm Ni_2^{2+}Te^{6+}O^{2-}_6$ or ${\rm K}_{2 - a}{\rm Na}_a\rm Ni_2^{3+}Te^{4+}O^{2-}_6$ with $m\times \mathcal{V}_{\rm Ni} + d\times \mathcal{V}_{\rm Te} = 10 > 9$ ($0 \leq a < 2$), 
\red{where} the crystal structure 
\red{consists of} 
\red{alternating monolayers} of Na and K lattices, the Na lattice was found to be hexagonal, which differs from the K lattice 
\red{that} retains its \red{expected} honeycomb pattern.\cite{masese2021mixed, berthelot2021stacking} 
\red{The differing lattice of Na from K} is attributed to the appearance of an edge dislocation which exchanges the relative position of Ni and Te along a Burgers vector, 
\red{thus lowering the Te-Te' Coulomb repulsion} across the interlayers containing Na cations. 

Moreover, this mixed alkali state was 
\red{discussed} within the 
\red{context} of the so-called \textit{honeycomb/Hale's conjecture}\cite{hales2001honeycomb}, 
\red{by identifying a correspondence between the area and perimeter of the honeycomb pattern of the lattice to the thermodynamic entropy and free energy respectively}.\cite{masese2021math, masese2021mixed} Indeed, this conjecture has relevance in biological systems, such as the justification for natural selection favouring the bees that expended the least amount of work to create the largest cross-sectional area for the storage of honey\cite{lyon2012mathematical, raz2013application}, as illustrated in Figure \ref{Fig_1}. 
\red{Excluding the mixed alkali and other hybrids, only the hexagonal and honeycomb monolayers of cations have been observed to date
especially for the linear and prismatic coordinations to oxygen atoms}, suggesting some underlying universality of the valence bond and 
\red{conformal field theories governing the formation and stability of the cationic lattices}.\cite{odor2004universality} Moreover, unlike the carbon atoms in a honeycomb lattice of graphene\cite{mecklenburg2011spin, georgi2017tuning}, the cations in honeycomb layered oxides can be mobilised when a relatively low activation energy of $< 1$ eV (Li, Na, K) per cation is available\cite{sau2022insights, matsubara2020magnetism}, suggesting a more elaborate charge transport theory 
\red{for} positive ions, compared to the electron transport in graphene \red{which is} restricted to localised carbon atoms, \textit{albeit} both lattices expected to share particular properties such as pseudo-spin \red{inherited from} 
the honeycomb lattice.\cite{kanyolo2022cationic, mecklenburg2011spin, georgi2017tuning} 

\red{Such universality has motivated the reformulation of the 2D 
molecular dynamics of cations in terms of an idealised model, whereby the (de-)intercalation process of a honeycomb layered oxide cathode is captured by Liouville CFT with $c = 1$, 
corresponding to the aforementioned spinless
modular bootstrap for CFT in the $d = 2c$ sphere packing problem.}\cite{kanyolo2020idealised, kanyolo2022cationic, nakayama2004liouville, polchinski1998string2, zamolodchikov1996conformal, afkhami2020high, hartman2019sphere} In particular, due to charge conservation, each extracted cation 
\red{creates} a neutral vacancy, 
\red{\textit{albeit}} with a pseudo-magnetic moment, at each cationic site 
during the de-intercalation process, whereby the charged cation acquires an Aharanov-Casher phase as it diffuses around the vacancies, along the honeycomb pathways shown in Figure \ref{Fig_2}(b).\cite{aharonov1984topological} Consequently, these vacancies correspond to $\nu$ number of topological defects, where $\nu$ is the first Chern number. As a result, these defects can be treated as topological charges satisfying the Poincar\'{e}-Hopf theorem, where the 
\red{number density of the vacancies (proportional to the charge density of the cations)} 
\red{corresponds} to the Gaussian curvature of an emergent 
2D closed manifold of genus $g = \nu$ with a \red{conformal} metric, 
\begin{align}
    dt^2 = \exp(2\Phi)(dx^2 + dy^2),
\end{align}
where $\Phi$ is a potential satisfying Liouville's equation.\cite{nakayama2004liouville, polchinski1998string2, zamolodchikov1996conformal} Thus, the quantum state with no vacancies ($g = 1$) 
\red{corresponds to} the 2-torus, invariant under the operation of the generators, $S$ and $T$ of the modular group, $\rm PSL_2(\mathbb{Z})$ as expected.\cite{kanyolo2022cationic, cohen2017modular} 

Moreover, the 
\red{classical Liouville CFT} can be recast in terms of 
\red{Einstein's theory of general relativity in ($1+3$)D} by imposing a space-like Killing vector along one of the 
\red{spacial} directions perpendicular to the honeycomb lattice as well as a time-like Killing vector\cite{kanyolo2020idealised}, with the resultant field equations having found applications in quantum black hole information theory.\cite{kanyolo2022local} \red{Thus,} the gravity field equations can be derived from 
\red{the typical Einstein-Hilbert action} with a torsion-free connection\cite{thorne2000gravitation}, $\Gamma_{\,\,\mu\nu}^{\rho} = \Gamma_{\,\,\nu\mu}^{\rho}$, where the topological defects 
\red{are} non-vanishing 
\red{under} a torsion-free manifold 
\red{\textit{albeit} with} the Gauss-Bonnet term 
also present 
\red{in the action}.\cite{lovelock1971einstein} 
Thus, cationic diffusion in honeycomb layered oxides effectively serve as vital testing grounds for theories of $d = 2, 3$, $(d + 1)D$ 
\red{emergent} gravity.\cite{gross1991two, holz1988geometry}

\red{Fairly recently, a diverse class of layered materials exhibiting bilayered arrangements of cations have been identified, with Ag-based oxides and halides} such as 
\red {$\rm Ag_3O$ (or equivalently as $\rm Ag_6O_2$), ${\rm Ag_2}M\rm O_2$ ($M$ = Rh, Mn, Fe, Cu, Ni, Cr, Co), $\rm Ag_2F$ and more recently ${\rm Ag_2}M_2\rm TeO_6$ (where $M$ = Zn, Cu, Co, Mg, Ni, Mg) serving as exemplars,}
suggesting a bifurcation mechanism for the bipartite honeycomb lattice into its two hexagonal sub-lattices.\cite{allen2011electronic,schreyer2002synthesis,matsuda2012partially, ji2010orbital,yoshida2020static, yoshida2011novel, yoshida2008unique, yoshida2006spin, masese2023honeycomb, argay1966redetermination, beesk1981x, taniguchi2020butterfly} 
The bifurcation mechanism appears to require pseudo-spin and pseudo-magnetic degrees of freedom analogous to graphene, whereby rather universal 
\red{properties} are thought to lead to the hexagonal bilayers.\cite{masese2023honeycomb} 
\red{These properties include the existence of subvalent states of Ag, in addition to a bifurcated honeycomb lattice stabilised by unconventional Ag-Ag' weak bonding at conspicuously short distances, instigated by $sd$ hybridisation of 4$d$ and 5$s$ orbitals of monovalent Ag atoms. This is akin to elemental Ag metal bonding due to the so-called argentophilic interactions.}\cite{jansen1980silberteilstrukturen} 
\red{In the case of} graphene-based systems\cite{allen2010honeycomb, mecklenburg2011spin, georgi2017tuning, kvashnin2014phase}, 
\red {2{\it sp}$^2$} hybridisation in carbon with valency 4+ leads to three $\sigma$-bonds and a leftover $p_z$ orbital electron which can form a $\pi$-bond with an adjacent carbon atom, leading to a trigonal planar geometry. This 
\red{leftover} $p_z^1$ orbital electron is responsible for the rather differing properties of graphene and graphite compared to diamond, whose hybridisation instead is 2{\it sp}$^3$.\cite{kvashnin2014phase} 

Of particular interest is the excellent conduction of carbon atoms in graphene, facilitated by the 
\red{itinerant} $p_z^1$ orbital electron moving at the Fermi velocity - speed of light in graphene layers - with two helicity states 
\red{that can be associated with} the pseudo-spin degrees of freedom at the Dirac point.\cite{mecklenburg2011spin, allain2011klein, yb2005experimental} Essentially, the bipartite nature of the honeycomb lattice in graphene requires the wavefunction of the conduction electron at the Dirac points to be described by a 2D mass-less Dirac spinor, with each component representing the helicity states known as pseudo-spins. An analogous situation can be 
\red{conceived} for honeycomb lattices of Cu, Ag and Au ($n =$ 3, 4 and 5 respectively) fermionic cations with closed $nd^{10}$ and half-filled $(n + 1)s^1$ orbitals. 
\red{Like in graphene\cite{balandin2011thermal, stankovich2006graphene}, electrical and thermal conductivity in the honeycomb lattices of coinage metal atoms is expected to be excellent - of the order comparable to their elemental values, whereby the $(n + 1)s^1$ valence electrons are itinerant.} Due to electrostatic screening of the 
\red{electric charge of the nucleus} and other factors\cite{schwarz2010full}, the $(n + 1)s^1$ orbital energy level is located 
at close proximity to the degenerate $nd^{10}$ orbitals, which encourages $sd_{z^2}$ hybridisation 
\red{resulting in} two degenerate states, ($(n + 1)s^2$, $nd_{z^2}^1$) in addition to ($(n + 1)s^1$, $nd_{z^2}^2$ state). 
\red{It is the $nd_{z^2}^1$ orbital that is analogous to the $p_z^1$ orbital of carbon in graphene whose helicity state is associated with the pseudo-spin in coinage metal atoms.} 

In addition, exchange interactions of two pseudo-spins adjacent to each other in a unit cell 
\red{have been proposed to be} mediated by conduction electrons\cite{masese2023honeycomb, kanyolo2022advances}, whereby the two cations are treated as magnetic impurities within the context of Ruderman–Kittel–Kasuya–Yosida (RKKY) interaction.\cite{ruderman1954indirect, kasuya1956prog, yosida1957magnetic} Moreover, 
\red{analogous to folded, stretched or strained graphene, a finite Gaussian curvature can act as a pseudo-magnetic field coupling to} the pseudo-spin degree of freedom via the topological charge/Euler characteristic, 
\red{contributing to the $sd$ hybridisation. This interaction serves as a metallophilic interaction between elements in group 11 (numismophilicity), modelled by the 1D Ising Hamiltonian density}\cite{kanyolo2022cationic},
\begin{align}
    \mathcal{H}_{\rm Ising} = -J_{\rm RKKY}^AS_1\cdot S_2 - B(S_1 + S_2),
\end{align}
where $J_{\rm RKKY}^A$ is the RKKY Heisenberg coupling ($|J_{\rm RKKY}^{\rm Cu}| < |J_{\rm RKKY}^{\rm Ag}| < |J_{\rm RKKY}^{\rm Au}|$) and $B$ is the pseudo-magnetic field term proportional to the Gaussian curvature in 2D systems. Surprisingly, the 
pseudo-spin states can also be \red{treated as the chiral states of the cations and} distinguished by 
the Gell-Mann–Nishijima formula\cite{zee2010quantum}, 
\begin{align}\label{Gell-Mann–Nishijima_eq}
    Q = 2I + Y,
\end{align}
where $Y$ is the U($1$) electric charge (playing the role of hypercharge), $I$ is the $z$-component of SU($2$) isospin, $Q$ is the effective charge and eq. (\ref{Gell-Mann–Nishijima_eq}) originates from $\rm SU(2)\times U(1)$ symmetry (breaking) in 2D.\cite{masese2023honeycomb}

\red{Electronically, there are three coinage metal atom states, depending on the occupancy of the $nd$ and $(n + 1)s$ orbitals. Due to the odd number of electrons, the neutral atom is a fermion (as expected) with its spin state inherited from the spin of the valence electron. Thus, due to $sd_{z^2}$ hybridisation, \red{a single spin up or down electron} can either be in the $nd_{z^2}^1$ or $(n + 1)s^1$ orbital with all the remaining lower energy orbitals fully occupied. Nonetheless, the valency corresponds to the number of electrons in the $ns$ orbital ($ns^2$ or $ns^1$). This results in two valence states, $A^{2+}$ and $A^{1+}$ ($A = \rm Cu, Ag, Au$). Moreover, in order to become closed shell in chemical reactions, the coinage metal atom can either be an electron donor with valency $2+, 1+$ or a receptor/anion with valency $1-$, whereby the receptor $A^{1-}$ achieves closed shell $nd_{z^2}^2$ and $(n + 1)s^2$ orbitals forming stable bonds. Indeed, this anion state has been observed in coinage metal cluster ions as $\rm Ag_N^{1-} (N \in \mathbb{Z}^+)$\cite{minamikawa2022electron, ho1990photoelectron, dixon1996photoelectron, schneider2005unusual}, whereas the isolated anion state ($N = 1$) is readily observed in compounds such as $\rm CsAu\cdot NH_3$ due to enhanced relativistic effects of Au.\cite{jansen2008chemistry} Thus, the $A^{1+}$ and $A^{1-}$ valence states are related by isospin rotation (SU($2$)) with the isospin given by $I = \mathcal{V}_A/2$ where $\mathcal{V}_A = 1+, 1-$ are the valence states, and $Y = 0$ is the electric charge of the neutral atom. Meanwhile, the $A^{2+}$ state is an isospin singlet with electric charge, $Y = \mathcal{V}_A = 2+$. Nonetheless, these three cation states $A^{2+}, A^{1-}$ and $A^{1+}$ must have an effective charge, $Q = +2, -1$ and $Q = +1$ respectively, obtained by the Gell-Mann–Nishijima formula and \red{are} treated as independent ions related to each other by $\rm SU(2)\times U(1)$, forming the basis for fractional valent (subvalent) states. Due to $sd$ hybridisation, all these three states are degenerate on the honeycomb lattice. The degeneracy between $\rm Ag^{2+}$ and $\rm Ag^{1-}$ corresponds to right-handed and left-handed chirality of $\rm Ag$ fermions on the honeycomb lattice, treated as the pseudo-spin.\cite{masese2023honeycomb}}

For illustration purposes, the simple case of the bilayered $\rm Ag_2NiO_2$, which requires the existence of the subvalent state $\rm Ag^{1/2+}$ to be electronically neutral is replaced by $\rm Ag_2NiO_2 = Ag^{2+}Ag^{1-}Ni^{3+}O_2^{2-}$ instead, which already implies bifurcation of the honeycomb lattice. Hybrids with a stable honeycomb monolayer and hexagonal bilayer arranged along the [001] plane in an alternating fashion can also be explained, 
$\rm Ag_3Ni_2O_4 = \frac{1}{2}(Ag_2^{1+}Ni_2^{3+}O_4^{2-})(Ag_2^{2+}Ag_2^{1-}Ni_2^{3+}O_4^{2-})$ with a subvalent state, $\rm Ag^{2/3+}$. \red{Thus, summarising the possible fractional subvalent states of Ag in these materials is a matter of considering the various ratios of coinage metal atoms in the possible lattices. In this case, the lattice with $1:1$ left-right chiral ($A^{2+}, A^{1-}$) is bilayered (bifurcated honeycomb) with sub-valency $1/2+$, whilst the lattice with $1:1$ left chiral ($A^{1+}, A^{1+}$) is hexagonal with valency $1+$.}
Note that, $sd_{z^2}$ hybridisation tends to occur efficiently whenever the $d_z^2$ orbital is isolated from the rest of the $d^{10}$ orbitals by crystal field splitting. Thus, the bifurcation mechanism is favoured in layered crystal structures whose Ag atoms at the monolayer-bilayer critical point ($T = T_{\rm c}$, with $T_{\rm c} = m$ the critical temperature/effective cationic mass on the honeycomb lattice) of the lattice exhibit prismatic or linear coordinations to O atoms, since these systems would have an isolated $nd_{z^2}$ orbital according to crystal field splitting theory.\cite{burns1993mineralogical, ballhausen1963introduction, jager1970crystal, de19902} 
\red{T}he bifurcation of the honeycomb lattice is 
\red{analogous to} Peierls distortion which, \textit{e.g.} in the dimerisation of 
polyacetylene\cite{garcia1992dimerization, peierls1979surprises, peierls1955quantum}, results in a metal-insulator phase transition.\cite{stewart2012evidence} Finally, more complicated structures may have different ratios and combinations leading to sub-valence states, $+1/3$ ($\rm 2Ag^{1+}, Ag^{1-}$) or $+4/5$ ($\rm Ag^{2+}, Ag^{1-}, 3Ag^{1+}$), provided $sd_{z^2}$ hybridisation is guaranteed.\cite{pettifor1978theory, lacroix1981density, manh1987electronic, gallagher1983positive, horn1979adsorbate}
In principle, subvalent Ag cations have also been reported in Ag-rich oxide compositions such as 
\red {$\rm Ag_{16}B_4O_{10}$, $\rm Ag_3O$, $\rm Ag_{13}OsO_6$, $\rm Ag_5Pb_2O_6$, $\rm Ag_5GeO_4$, $\rm Ag_5SiO_4$,} the halides such as $\rm Ag_2F$, and the theoretically predicted $\rm Ag_6Cl_4$.\cite{derzsi2021ag, kovalevskiy2020uncommon, ahlert2003ag13oso6, jansen1992ag5geo4, jansen1990ag5pb2o6, argay1966redetermination, beesk1981x, bystrom1950crystal}

\green{Finally}, computational modelling methods 
\red{suchlike} molecular dynamics (MD) and first-principles density functional theory (DFT) have been utilised not only to predict various 
\red{mesoscopic and nanoscopic} properties of 
\red{honeycomb layered oxides} ({\it e.g.} phase stability, operating voltage, cation migration barriers, defect formation, band structure, {\it etc.}), \green{but also successfully propel a vast array of theoretical strides within experimental reach}.\cite {sau2015role, sau2015ion, sau2016influence, sau2016ion, sau2014molecular, sau2022insights, huang2020, sau2022ring, bianchini2019nonhexagonal} \green{Particularly, c}omputational modelling techniques avail exclusive insights into the mechanisms dictating the physicochemical properties of materials, particularly honeycomb layered oxides at the atomic level, and are thus invaluable tools in the design of \green{honeycomb layered} materials. \green{Consequently, b}ased on all the aforementioned theoretical advances in layered materials with honeycomb lattices, a \red{fairly} complete 
\red{treatise} that tackles their characterisation is warranted. Thus, this treatise seeks to 
\red{elucidate the theoretical frameworks that successfully tackle several novel phenomena within these condensed matter systems, redefining} the frontier of their research and applications. Other relevant experimental, theoretical and computational techniques applicable to the characterisation of honeycomb layered 
\green{materials} have been availed for completeness. We conclude by 
\red {envisaging} future research directions where interesting physicochemical, 
\red {topological and electromagnetic} properties could be lurking, particularly 
as testing grounds for ideas in emergent conformal field and 2D quantum gravity theories. \green{Indeed, this work seeks to demarcate the frontier of this vast field of research, launching new avenues along the way that hold promise in inviting a wider scientific community into this presently divulging field of honeycomb layered materials.}  

Hereafter, we shall set Planck's constant, the speed of electromagnetic waves in the material, $\tilde{c}$, Boltzmann's constant, $k_{\rm B}$ and the elementary charge of the cations, $q_{\rm e}$ to unity, $\hbar = \tilde{c} = k_{\rm B} = q_{\rm e} = 1$, and employ Einstein summation convention unless explicitly stated otherwise. 
\red{Throughout}, valence states, $x\pm$ will be distinguished from charged states, $\pm x$ where $x$ is a number or fraction. 

\section{Theoretical models}

\subsection{Molecular Dynamics}

\red{Molecular Dynamics (MD) is a potent tool used to predict behaviour of molecules and atoms interacting by a force field (a.k.a interatomic potential) in regimes that go beyond analytic methods.}\cite{tuckerman2010statistical} The dynamics of molecules is primarily governed by the form of the interatomic potential employed in the simulation. A relevant example of an interatomic potential previously employed to accurately predict the dynamics of cations within honeycomb layered oxides (\textit{e.g.} $A_2\rm Ni_2TeO_6$, $A =$ Li, Na, K or ${\rm Na_2}M_2\rm TeO_6$ 
\red {($M =$ Zn, Co, Ni and Mg))} is the Vashishta-Rahman 
interatomic potential,\cite{sau2016influence, sau2015ion, sau2015role} 
\begin{align}\label{Vashishta-Rahman_eq}
    U_{ij} (r_{ij}) = \frac{q_{\rm e}^{i}q_{\rm e}^{j}}{r_{ij}} + \frac{A_{ij}(\tilde{\sigma}_{i} + \tilde{\sigma}_{j})^{n_{ij}}}{r_{ij}^{n_{ij}}} - \frac{P_{ij}}{r_{ij}^4}- \frac{C_{ij}}{r_{ij}^6},
\end{align}
where $\tilde{\sigma}_i$, $\tilde{\sigma}_j$ is the ionic radius of the $i$-th, $j$-th cation, $q_{\rm e}^i, q_{\rm e}^j$ is the cationic charge, the parameters, $A_{ij}$, $P_{ij}$ and $C_{ij}$ 
\red{arise from the} 
\red{repulsive energy due to electron orbital overlap},
\red{averaged charge dipole interactions} and 
\red{the constant of dispersion between cationic pairs $i$ and $j$}, respectively. Other potentials of interest include 
\red{the popular} 
\red {Lennard-Jones and Born-Mayer (Buckingham) potentials}, which follow 
\red{similar} treatments.\cite{lennard1931cohesion, buckingham1938classical} 

\red{These parameters can be desirably determined experimentally or using empirical fitting to obtain the experimentally reported bond lengths when the values are not readily available in literature.} For instance, the bond lengths can be determined experimentally from lattice parameters 
\red{obtained} from determining the radial distribution function, $g(r)$, between the framework ion pairs, $M$\textendash O, O\textendash O, and Te\textendash O using techniques such as neutron diffraction (ND) and X-ray diffraction (XRD).\cite{grundish2019electrochemical, masese2018rechargeable, evstigneeva2011new, matsubara2020magnetism} 
\red{Thus, the rather considerable number of free parameters to be fixed experimentally necessarily limits the predictive power of MD simulations.} Nonetheless, the simulation results can be predictive when the theory allows for a small subset of stable structures, such as predicting cationic hopping activation energies in alkali-based honeycomb layered oxides after some fine-tuning of structural parameters to obtain stable structures.\cite{sau2015ion, sau2015role, sau2016influence, sau2022insights} 
\red{In the MD simulations, it is essential to track the number of cations in moles (N) and total energy (E), volume (V) and pressure (P) of the thermodynamic system, giving} rise to three particularly useful simulation methods, the micro-canonical ensemble (NVE), canonical ensemble (NVT) and isothermal-isobaric ensemble (NPT). The naming appropriately defines the elements kept constant in each simulation method. For instance, in situations where the system with a fixed particle number 
\red{is adiabatic}, energy is conserved and the volume is fixed, rendering the NVE simulations suitable, whereas thermodynamic systems at equilibrium at constant temperature or pressure are well-suited for NVT or NPT simulations respectively. Microscopically, the classical dynamic theory of cations obeys Newtonian mechanics governed by the interatomic potential. This implies that NVE is 
\red{the most} suited for such simulations. 

Nonetheless, previous MD simulations of cationic behaviour in honeycomb layered oxides using 
either NVE and NPT/NVT have yielded consistent results, suggesting crystalline stability is achieved at constant temperature and pressure.\cite{sau2022insights} In particular, employing a particular form of the Vashishta-Rahman potential and the Parrinello-Rahman isobaric-isothermal (NPT) MD method,\cite{parrinello1981polymorphic} which allows for changes in the simulation box sizes whilst keeping angles fixed, a series of MD simulations can be 
\red {performed} at constant atmospheric pressure and various temperature ranges where the honeycomb layered oxide is known to form a stable structure using barostatting and thermostatting techniques, 
\red{which involve coupling some dynamical variables to the simulation box and the velocities of the cations respectively}.\cite{nose1984molecular}
\red{The LAMMPS software package is typically used to carry out such simulations, which includes provisions to apply periodic boundary conditions and the Ewald summation technique for the convergence of long-range Coulombic interactions whenever necessary}.\cite{plimpton1995fast, thompson2022} 

\red{In succeeding subsections, we introduce the theoretical treatments and considerations employed to define the self-diffusion coefficient and Haven's ratio especially applicable to alkali cations in honeycomb layered oxides.}

\subsubsection{Diffusion coefficient}

\red{We shall begin from the Green-Kubo relation, which represents the exact expression relating transport coefficient such as conductance and time correlation functions}.\cite{green1954markoff, kubo1957statistical} 
\red{In the case of $\nu$ cations, the conductance $G_i$ experienced by the $i$-th particle is averaged in two dimensions and is given by,}
\begin{subequations}\label{Green_Kubo_eq}
\begin{align}\label{Green_Kubo_eq1}
    G_{i} = \frac{1}{2}\sum_{j = 1}^{\nu}G_{ij} = \frac{\beta}{2}\int_{0}^{\infty} ds\sum_{j = 1}^{2}\langle\dot{\mathcal{Q}}_i(0)\dot{\mathcal{Q}}_j(s) \rangle,
\end{align}
where $\beta = 1/k_{\rm B}T$ is the inverse temperature, $\mathcal{Q}_i(s), \mathcal{Q}_j(s)$ are charges and $\dot{\mathcal{Q}}_i(s) = d\mathcal{Q}_i(s)/ds = \mathcal{I}_i(s), \dot{\mathcal{Q}}_j(s) = d\mathcal{Q}_j(s)/ds = \mathcal{I}_j(s)$ are the current of $i,j$-th particle respectively evaluated at time, $s$. 
\red{We employ indices $i,j,k$ for the particle species and $a,b,c$ for the spacial coordinates.} 

\red{However, a more comprehensive expression using the conductivity tensor, $\sigma_{abij}(\Vec{r})$ is warranted,}
where the average conductance tensor 
\red{experienced} by the $i$-th particle is given by,
\begin{multline}\label{Green_Kubo_eq2}
    \sigma_{abi} = -\frac{1}{2}\beta \sum_{j = 1}^{\nu}\int_{0}^{\infty} ds \int dV' \langle J_{ai}(s,\vec{r}\,')J_{bj}(0,\Vec{r}) \rangle\\
    \equiv \frac{1}{2}\sum_{j = 1}^{\nu}\sigma_{abij}(\Vec{r}),
\end{multline}
\end{subequations}
$dV'$ is the volume element, $J_{ai}(t,\Vec{r}) = \rho_i(\Vec{r})v_{ai}(t)$ is the current density, $\rho_i(\Vec{r})$ is the charge density satisfying the equilibrium condition $\partial \rho_i(\Vec{r})/\partial t = 0$ (
\red{only applicable in the case of NPT- and NVT-MD simulations}) and $v_{ai}(s) \equiv \Vec{v}_i(s) = d\Vec{r}_i(s)/ds$ 
\red{corresponds to the velocity vector, which satisfies the Langevin equation}\cite{lemons1997paul},
\begin{align}\label{Langevin_eq2}
    m_i\frac{d}{ds}v_{ai}(s) = -\sum_{bj}(\mu_{abij})^{-1}v_{bj}(s) + q_{\rm e}^iE_{ai}(s). 
\end{align}
\red{Here, $E_{ai}(s)$ the electric field acting on the $i$-th species, $m_i$ is the mass, and $\mu_{abij}$ the mobility tensor.} 
\red{Substituting $J_{ai}(t,\Vec{r}) = \rho_i(\Vec{r})v_{ai}(t)$} into (\ref{Green_Kubo_eq2}) yields,
\begin{multline}\label{calculation_eq1}
    \sigma_{abij}(\vec{r}) = -\frac{1}{2}\beta\rho_i(\Vec{r})\int dV'\rho_j(\vec{r}\,') \int_{0}^{\infty} ds \left \langle \dot{r}_{ai}(s)\dot{r}_{bj}(0) \right \rangle\\
    = \frac{q_j\beta\rho_i(\Vec{r})}{2}\int_{0}^{\infty} ds \left \langle \dot{r}_{ai}(s)\dot{r}_{bj}(0) \right \rangle,
\end{multline}
where $q_{\rm e}^i = \int dV'\rho_i(\vec{r}\,')$ is the charge of the $i$-th cation. \red{At thermal equilibrium},  
$d\vec{v}_i/ds = 0$ in eq. (\ref{Langevin_eq2}), 
resulting in $J_{ai}(\vec{r}) = \rho_i(\vec{r})v_a(0) = q\rho_i(\vec{r})\sum_{bj}\mu_{abij}E_{bj}(0) = \sum_{bj}\sigma_{abij}E_{bj}(0)$, where,
\begin{align}\label{conductivity_eq}
   \sigma_{abij}(\vec{r}) = q_{\rm e}^i\rho_i(\vec{r})\mu_{abij}.
\end{align}

\red{At equilibrium, the velocity vector also satisfies the diffusion equation,}
\begin{multline}
    0 = \frac{\partial \rho_i(\vec{r})}{\partial s} = \sum_a\frac{\partial}{\partial r_a}\left (\rho_i(\vec{r})v_{ai} \right )\\
    + \sum_{abj}\frac{\partial}{\partial r_a}\left (D_{abij}\frac{\partial}{\partial r_b} \rho_j(\vec{r})\right ),
\end{multline}
where $\rho(\vec{r}) \propto \exp(-\beta U_i(\vec{r}))$, $D_{abij}$ is the diffusion coefficient (tensor),
\begin{align}\label{eq. 4}
    U_i(\vec{r}) = \frac{1}{2}\sum_{j = 1}^{\nu}U_{ij}(\vec{r}),
\end{align}
is the electromagnetic potential 
\red{experienced} by an individual cation, $U_{ij}$ is the potential used in the simulation (\textit{e.g.} the Vashishta-Rahman interatomic potential) and $-\partial U_i(\vec{r})/\partial r_a = q_{i}^{-1}E_{ai}(\vec{r})$ is the electric field, one can check that the Einstein-Smoluchowski equation,
\begin{align}\label{Einstein_Smoluchowski_eq}
    \mu_{abij} = \beta D_{abij},
\end{align}
imposes the equilibrium condition on eq. (\ref{Langevin_eq2}).

\red{Consequently, following} 
eq. (\ref{conductivity_eq}) and eq. (\ref{Einstein_Smoluchowski_eq}), 
eq. (\ref{calculation_eq1}) can be re-written into 
\red{a formula} for the diffusion coefficient,
\begin{align}
    D_{abij} = -\int_{0}^{\infty} ds \left \langle \dot{r}_{ai}(s)\dot{r}_{bj}(0) \right \rangle. 
\end{align}
Integrating by parts and 
\red{neglecting} the boundary term, 
\red{we find},
\begin{align}\label{calculation_eq2}
    D_{abij} = \int_{0}^{\infty} ds \left \langle \ddot{r}_{ai}(s)r_{bj}(0) \right \rangle. 
\end{align}
Using the Langevin equation 
\red{given in eq. (\ref{Langevin_eq2})}, where $\ddot{r}_{ai}(s) = \dot{v}_{ai}(s)$ and $\left \langle E_{ai}(s)r_{bj}(0) \right \rangle = 0$, eq. (\ref{calculation_eq2}) 
\red{becomes},
\begin{align}\label{calculation_eq3}
    D_{abij} = -\frac{1}{m_i}\sum_{ck}\mu_{acik}^{-1}\int_{0}^{\infty} ds \left \langle \dot{r}_{ck}(s)r_{bj}(0) \right \rangle. 
\end{align}
\red{Imposing the spacial isotropic condition, we can write} 
$D_{abij} = D_{ij}\delta_{ab}$. \red{Moreover,} by the Einstein-Smoluchowski equation given by eq. (\ref{Einstein_Smoluchowski_eq}), $\mu_{abij} = \mu_{ij}\delta_{ab}$, where $\delta_{ab}$ is the Kronecker delta, $D_{ij}$ is the diffusion coefficient tensor and $\mu_{ij}$ the mobility tensor. Moreover, 
the central limit theorem 
\red{guarantees that} the distribution function for varied 
\red{spacial} directions in the correlation is Gaussian, and hence the off-diagonal correlation functions vanish, or at least are negligible compared to the diagonal elements. This corresponds to $D_{ij} = D_i\delta_{ij}$, where $D_i$ is the diffusion coefficient as seen by particle $i$, and $\mu_{ij} = \mu_i\delta_{ij}$, where $\mu_i$ is the mobility. 

To make further progress, additional assumptions can be applied. 
\red{For instance, within the} Drude model, 
mobility is related to the average time between collisions (mean free time), $t$ by $\mu_j = t/m_j$ 
\red{which serves as the integration cut-off scale in the Green-Kubo relation}. \red{Consequently,} 
\red{applying these conditions to} eq. (\ref{calculation_eq3}), 
\red{the diffusion coefficient experienced by the $j$-th cation is given by},
\begin{multline}\label{diffusion_const_eq1}
    D_j = D_j\frac{1}{2}\sum_{ab}\delta_{ab} 
    = \frac{1}{2}\sum_{ai}D_i\delta_{ij}\delta_{ab} 
    = \frac{1}{2}\sum_{abi}D_{abij}\\ 
    = -\lim_{t \rightarrow \infty}\frac{1}{2t}\sum_{abi}\int_{0}^{t} ds \frac{d}{ds}\left \langle r_{ai}(s)r_{bj}(0) \right \rangle,
\end{multline}
where we have used $\sum_{ab}\delta_{ab} = 2$ for two dimensions. 

\subsubsection{Haven's ratio} 

\red{The self/tracer-diffusion coefficient, $D_{\rm T}$ which describes only the correlations of the same particle lacks any inter-particle correlation information, and hence differs from the physical diffusion coefficient, $D_{\sigma}$ by the Haven's ratio, $D_{\rm T}/D_{\sigma} \equiv H_{\rm R} \leq 1$, less than unity}.\cite{vargas2020dynamic} Meanwhile, the physical diffusion coefficient is calculated from eq. (\ref{diffusion_const_eq1}) as the average, 
\begin{align}\label{Average_D_eq}
    D_{\sigma} = \frac{1}{\nu}\sum_{j = 1}^{\nu}D_j.
\end{align}
\red{Thus, Haven's ratio ($H_{\rm R}$) corresponds to} the quotient of the mean-square displacement without cross-terms (MSD) and mean-sqare displacement including cross-terms (MSD$^*$)
\begin{subequations}\label{Havens_eq}
\begin{multline}
    {\rm MSD^*} = \frac{1}{\nu}\left \langle \sum_{i = 1}^{\nu}\Delta \vec{r}_{i}(t)\cdot \sum_{j = 1}^{\nu}\Delta \vec{r}_{j}(t) \right \rangle\\
   =\frac{1}{\nu} \left \langle \sum_{i = 1}^{\nu} \Delta \vec{r}_{i}(t)^2+\sum_{i = 1}^{\nu}\sum_{j = 1, i \neq j}^{\nu}\Delta \vec{r}_{i}(t)\cdot\Delta \vec{r}_{j}(t) \right \rangle\\
   =\frac{1}{\nu}\left \langle \sum_{i = 1}^{\nu}\Delta \vec{r}_{i}(t)^2\right \rangle\
    +\frac{1}{\nu}\left \langle \sum_{i = 1}^{\nu}\sum_{j = 1, i \neq j}^{\nu} \Delta \vec{r}_{i}(t)\cdot \Delta \vec{r}_{j}(t) \right \rangle\\
=\underbrace{\frac{1}{\nu}\left \langle \sum_{i = 1}^{\nu} \Delta \vec{r}_i^2\right \rangle}_{\rm self}
    +\underbrace{
    {\frac{1}{\nu}\left \langle \sum_{i = 1}^{\nu}\sum_{j = 1, i \neq j}^{\nu} \Delta \vec{r}_i\cdot\Delta \vec{r}_j \right \rangle}}_{\rm cross-terms}
=\frac{\rm MSD}{H_R}.
\end{multline}
where $\Delta \vec{r}_j = \vec{r}_j(t) - \vec{r}_j(0)$ and, 
\begin{align}
    {\rm MSD} = \left \langle \frac{1}{\nu}\sum_{j = 1}^{\nu}(\vec{r}_j(t) - \vec{r}_j(0))^2\right \rangle,
\end{align}
\red{Haven's ratio can be expressed by} the centre of mass coordinate, $\vec{r}$ of $\nu$ particles\cite{haarmann2021ionic, deng2017enhancing, marrocchelli2013effects},
\begin{multline}\label{cm_eq}
    {\rm MSD^*} = \nu\left \langle \frac{\sum_{i = 1}^{\nu}\Delta \vec{r}_{i}(t)}{\nu}\cdot \frac{\sum_{j = 1}^{\nu}\Delta \vec{r}_{j}(t)}{\nu} \right \rangle\\
    =\nu\left \langle (\vec{r}(t) - \vec{r}(0))\cdot (\vec{r}(t) - \vec{r}(0)) \right \rangle\\
    = \nu\left \langle (\vec{r}(t) - \vec{r}(0))^2\right \rangle =\frac{\rm MSD}{H_R}.
\end{multline}
\end{subequations}
\red{Consequently}, using,
\begin{multline}
    \sum_{abij}\left \langle r_{ai}(t)r_{bj}(0) \right \rangle
    = \frac{\sum_{abij}\delta_{ab}\delta_{ij}\left \langle r_{ai}(t)r_{ai}(0) \right \rangle}{H_{\rm R}},
\end{multline}
with Haven's ratio, $H_{\rm R}$ defined in eq. (\ref{Havens_eq}) and $\left \langle r_{ai}(t)r_{ai}(t) \right \rangle = \left \langle r_{ai}(0)r_{ai}(0) \right \rangle$ which assumes time translation symmetry, eq. (\ref{Average_D_eq}) becomes, 
\begin{multline}\label{Final_eq}
    D_{\sigma} = \frac{1}{\nu}\sum_{j = 1}^{\nu}D_j\\
    = \lim_{t \rightarrow \infty}\frac{H_{\rm R}^{-1}}{4t}\left \langle \frac{1}{\nu}\sum_{j = 1}^{\nu}(\vec{r}_j(t) - \vec{r}_j(0))^2\right \rangle = H_{\rm R}^{-1}D_{\rm T},
\end{multline}
where $D_{\rm T} = \lim_{t \rightarrow \infty}{\rm MSD}/4t$, $\rm MSD$ is the self-mean square displacement, as defined in eq. (\ref{Havens_eq}) and $D_{\rm T}$ is the self/tracer-diffusion coefficient. Thus, the equivalence of eq. (\ref{Final_eq}) to the Green-Kubo relations given in eq. (\ref{Green_Kubo_eq})) is only valid \red{for long mean free times}
corresponding to the limit, $t \rightarrow \infty$. 
\red{Nonetheless, a finite cut-off avails eq. (\ref{Final_eq}) the advantage over the Green-Kubo formula, which suffers from large fluctuation contributions proportional to simulation time scales.}\cite{bhargava2005dynamics, hansen1975statistical, dommert2008comparative}

\subsubsection{Activation energy and total potential energy}

\red{The self/tracer-diffusion coefficient, $D_{\rm T}$ depends on temperature ($T$) following Arrhenius equation},  
\begin{align}\label{Arrhenius_eq}
    D_{\rm T} = D_0\exp\left (\frac{-E_{\rm a}}{k_{\rm B}T}  \right ),
\end{align}
where $D_0$ is the pre-exponential factor, $E_{\rm a}$ represents the activation energy of ion hopping, and $k_{\rm B}$ is the Boltzmann constant. The total potential energy 
is calculated as,
\begin{align}
    U_{\rm T} = \frac{1}{2}\sum_{i,j = 1}^{\nu}U_{ij} = \sum_{j = 1}^{\nu}U_j,
\end{align}
where $U_{ij}$ is the interatomic 
\red{potential} such as the Vashishta-Rahman potential given in eq. (\ref{Vashishta-Rahman_eq}) and $U_j$ is the potential energy of individual cations given in eq. (\ref{eq. 4}). Simulations typically seek to 
\red{predict or confirm} the activation energy $E_{\rm a}$ of $A$ cations within a 
\red{particular} class of honeycomb layered oxides, which is obtained from eq. (\ref{Arrhenius_eq}).\cite{sau2015ion, sau2015role, sau2016influence, sau2022insights} 

\subsubsection{Cation population density}

\red{Moreover, since the population density for a particular cation, $j$ is given by},
\begin{align}\label{population_eq}
    P_j = P^j_0\exp(-\beta U_j),
\end{align}
finding all the independent parameters of $U_j$ experimentally or using empirical fitting to attain the experimentally reported bond lengths completely yields the population profile of the cations, where $P^j_0$ are constants fixed by normalisation, $\int d^{\,2}r\, P_j(\vec{r}) = 1$. 
\red{However, the thermodynamical system minimises the free energy instead of the interaction potential energy, which means it does not sit at the theoretical minimum of the potential due to the contribution from entropy.}
\red{Consequently}, a careful consideration of the entropy term must be carried out to correctly define the population density, which necessarily deviates from eq. (\ref{population_eq}). 
\magenta{Dropping the index, $j$ for brevity, it is prudent to} define the relative free energy experienced by 
individual cations, $\Delta F$ defined relative 
\red{the} maximum population density of cations, $P_{\rm max.}$ as, 
\begin{align}\label{change_free_energy_eq}
    \Delta F = \Delta U - T\Delta S,
\end{align}
where $\Delta F = -k_{\rm B}T\ln(P/P_{\rm max.})$, $\Delta U = U - U_{\rm max.}$ is the relative potential energy and $\Delta S = S - S_{\rm max.}$ is the relative entropy. 

\red{Whilst} eq. (\ref{change_free_energy_eq}) 
\red{is intuitive}, it is prudent to note that since we are dealing with diffusion, it is plausible that all quantities are diffusion-path dependent. Therefore, \red{there is need to avail} 
a rigorous derivation of eq. (\ref{change_free_energy_eq}). In particular, consider a thermodynamic system of two variables $a, b$ (\textit{e.g.} the coordinates in 2D) such that the probability of finding the system in a given configuration is given by the Boltzmann formula,\cite{reif2009fundamentals}
\begin{align}\label{probability_eq}
    \mathcal{P}_{ab} = \frac{1}{Z}\exp(-\beta H_{ab}),
\end{align}
where $\beta = 1/k_{\rm B}T$ is the inverse temperature, $H_{ab}$ is the Hamiltonian/energy and $Z$ is the partition function of the system defined by the total probability, $\sum_{ab} \mathcal{P}_{ab} = 1$, \textit{i.e.},
\begin{align}\label{partition_eq}
    Z = \sum_{ab}\exp(-\beta H_{ab}).
\end{align}
\red{We can} transform eq. (\ref{probability_eq}) into a statement about averages since it is the averages of quantities that are explicitly measured in experiments. Taking the natural logarithm of eq. (\ref{probability_eq}) yields,
\begin{subequations}
\begin{align}\label{log_eq}
    -\ln (Z) = \ln P_{ab} +  \beta H_{ab}.
\end{align}
\red{The} average of eq. (\ref{log_eq}) \red{can be performed} by multiplying by the probability, $\mathcal{P}_{ab}$ and 
\red{taking} the sum over the indices $a,b$ to 
\red{obtain},
\begin{align}\label{log_eq2}
    -\sum_{ab}\mathcal{P}_{ab}\ln Z = \beta\sum_{ab}\mathcal{P}_{ab}H_{ab} + \sum_{ab}\mathcal{P}_{ab}\ln \mathcal{P}_{ab}. 
\end{align}
\end{subequations}
Recalling that $\sum_{ab} \mathcal{P}_{ab} = 1$, eq. (\ref{log_eq2}) transforms into, 
\begin{align}\label{free_energy_eq}
    F = E - TS,
\end{align}
where,
\begin{subequations}
\begin{align}
    E = \sum_{ab}\mathcal{P}_{ab}H_{ab},\\
    F = -\beta^{-1}\sum_{ab}\mathcal{P}_{ab}\ln Z = -\beta^{-1}\ln Z,\\
    S = -k_{\rm B}\sum_{ab}\mathcal{P}_{ab}\ln \mathcal{P}_{ab},
\end{align}
\end{subequations}
are the average energy, 
free energy and 
entropy, respectively with $T = 1/k_{\rm B}\beta$.

\red{Whilst the lattice can be discrete, in MD simulations, we explicitly deal in the continuum limit} where the probability $\mathcal{P}_{ab}$ transforms into probability density, $P(x,y)$, 
\begin{subequations}
\begin{align}
\label{probability_eq2}
    \mathcal{P}_{ab} \rightarrow \mathcal{P}(x,y) \equiv \frac{P(x,y)}{\Omega(x,y)},\\
    \sum_{ab} \rightarrow \int dxdy\,\Omega(x,y),
\end{align}
where $dxdy$ is the area element of the manifold, $\Omega$ is interpreted as a `density of states' for the system with dimensions of inverse area, \textit{i.e.} $[\Omega] = \rm 1/(length)^2$ and can be taken to be fairly constant over the plane, $x,y$ and $P(x,y) = \Omega \mathcal{P}(x,y)$ is the probability density we seek. This transformation is defined such that it preserves the dimensionality of the quantities. 
\red{In this paradigm, the statement, $\sum_{ab}\mathcal{P}_{ab} = 1$ transforms into,}
\begin{align}
    \int dxdy\, \Omega(x,y) \mathcal{P}(x,y) = \int dxdy\,P(x,y) = 1.
\end{align}
\end{subequations}
Finally, we can now rewrite all quantities in eq. (\ref{free_energy_eq}) in terms of the probability density $p(x,y)$ as, 
\begin{subequations}\label{thermodynamic_eq}
\begin{align}
    E = \sum_{ab}\mathcal{P}_{ab}H_{ab} \rightarrow \int dxdy\, P(x,y)H(x,y),\\
    Z = \sum_{ab}\exp(-\beta H_{ab}) \rightarrow \int dxdy\,\Omega\exp(-\beta H(x,y)),
\end{align}
and, 
\begin{multline}\label{entropy_eq}
    S = -k_{\rm B}\sum_{ab}\mathcal{P}_{ab}\ln \mathcal{P}_{ab}\\
    = - k_{\rm B}\int dxdy\,P(x,y) \ln \left ( \frac{P(x,y)}{\Omega} \right ),
\end{multline}
\end{subequations}
where the free energy 
\red{is maintained as} $F = -\beta^{-1}\ln Z$. 

\red{The coordinates $x(s), y(s)$ can be parameterised by a single variable $s$ representing the path between points on the $ab$ plane.} Thus, for a single thermodynamic system within an ensemble of $n$ such paths at thermal equilibrium with each other (at temperature $T$), 
the energy and free energy will be given by $H(s_1)$ and $F(s_1) = -\beta^{-1}\ln(Z(s_1))$ respectively, where 
\begin{subequations}
\begin{multline}
    F(s_1\cdots s_n) = -\beta^{-1}\ln Z(s_1\cdots s_n)\\
    = -\beta^{-1}\sum_{s = s_1}^{s = s_n} \ln(Z(s)) = \sum_{s = s_1}^{s = s_n}F(s),
\end{multline}
is the free energy of the ensemble, whilst $Z(s_1\cdots s_n)$ and $Z(s)$ respectively are given by,
\begin{align}
    Z(s_1\cdots s_n) = \prod_{s = s_1}^{s = s_n}Z(s),\\
    Z(s) = \exp(-\beta F(s)).
\end{align}
Thus, the probability density at $s$ 
\red{is given by},
\begin{multline}\label{ps_eq}
    P(s) = \frac{1}{\Omega Z(s_1\cdots s_n)}\exp(-\beta F(s))\\
    = \frac{Z(s)}{\Omega Z(s_1\cdots s_n)}.
\end{multline}
\end{subequations}
\red{Proceeding, we can define} the difference in free energy (relative free energy) between any path, \textit{e.g.} $s = s_1$ and 
\red{another} arbitrary path $s$ by,
\begin{multline}
    \Delta F = F(s) - F(s_1)\\
    = -\beta^{-1}\ln\left ( \frac{Z(s)}{Z(s_1)} \right ) = -\beta^{-1}\ln\left (\frac{P(s)}{P(s_1)} \right ). 
\end{multline}
For convenience, we can set $s_1 = s_{\rm max.}$, where $s_{\rm max.}$ is the path that 
maximises the probability density, $P(s)$ in eq. (\ref{ps_eq}), thus 
arriving at eq. (\ref{change_free_energy_eq}).

\red{\subsection{Idealised model of cationic diffusion}}

\red{\subsubsection{Candidate materials}}

We shall 
\red{introduce} the idealised model of cationic diffusion\cite{kanyolo2020idealised, kanyolo2021honeycomb},
\red{which lays the groundwork for our formalism thereafter}. The idealised model 
\red{applies to a \red{wide} class of} layered materials, where 
\red {mobile cations (positively charged ions)} are sandwiched between the layers of immobile ions forming adjacent series of slabs within a stable crystalline structure, as shown in Figure \ref{Fig_2}(a). 
\red{In a majority of these exemplars, the mobility of the cations can be traced to extremely weak chemical bonds whose strength is correlated with the strength of emergent forces such as Van der Waals interactions and the inter-layer distance between the slabs,\cite{sun2019adverse, delmas2021, dresselhaus1981,whittingham2004,kanyolo2021honeycomb}} \red{\textit{viz.}},
\begin{enumerate}[(i)]
\red{
    \item Layered polyanion-based compounds consisting of pyrophosphates suchlike \red{${\rm K_2}M\rm P_2O_7$ ($M$ = Cu, Ni, Co) and $\rm Na_2CoP_2O_7$, pyrovanadates such as $\rm Rb_2MnV_2O_7$ and $\rm K_2MnV_2O_7$}, oxyphosphates such as $\rm LiVOPO_4$ and $\rm NaVOPO_4$, layered $\rm KVOPO_4$, diphosphates such as Na$_3\rm V(PO_4)_2$, fluorophosphates such as Na$_2\rm FePO_4F$, oxysilicates such as Li$_2\rm VOSiO_4$ and hydroxysulphates such as Li$\rm FeSO_4OH$;\cite{barpanda2012high, prakash2006, liu2018novel, niu2019review, barpanda2018polyanionic, jin2020polyanion, masquelier2013polyanionic, yahia2007crystal, liao2019KVOPO4}
    \item Graphite intercalation compounds such as $\rm CsC_8$, $\rm RbC_8$, $\rm LiC_6$ and $\rm KC_8$, including their intermediate compositions, \red{for instance}, ${\rm LiC}_{9n}$, ${\rm LiC}_{6n}$ ($n > 1$) and ${\rm KC}_{12n}$ ($n > 1$) ($n \geq 2$);\cite{dresselhaus1981, hosaka2020, jian2015carbon, guerard1975, dresselhaus2002intercalation}
    \item Layered metal \red{(di)}chalcogenides (suchlike $\rm LiVS_2$, $\rm Li_2FeS_2$, $\rm KCrTe_2$, $\rm KNi_2Se_2$, $\rm KMoS_2$, $\rm LiMoS_2$, $\rm K_{\it x}MoSe_2$, $\rm K_{\it x}WS_2$, $A_x\rm CrS_2$ and $A_x\rm TiS_2$ (where $A = \rm Ag, K, Na, Li, {\it etc.}$) and $0 < x < 1$) and trichalcogenides such as $\rm Na_2TiS_3$ and $\rm Na_2TiSe_3$;\cite{chia2015, johnson1982lithium, whittingham1978chemistry, leube2022layered, freitas2015, shang2018, fang2018, yu2018, fang2019, neilson2012, murphy1977, brec1980, bae2021}
    \item Layered metal carbides (suchlike MXenes) and layered metal nitrides such as $\rm MgMoN_2$, $\rm LiMoN_2$, $\rm NaTaN_2$;\cite{verrelli2017, jiang20222d, rauch1992, elder1992, bae2021}
    \item Layered transition metal oxides such as $A_xM\rm O_2$ (where $A = \rm Ag, K, Na, Li, {\it etc}.$, $M$ is a transition metal \red{or a combination of multiple transition metals} and $0 < x < 1$), $A_xM\rm O_3$, $D_x\rm V_2O_5$ (where $D = \rm Mg, Ca, Al, Ag$), $A_y\rm V_2O_5$ (where $0 < y < 2$), $\rm Mg_2Mo_3O_8$, $\rm Ca_3Co_4O_9$,  potassium polytitanates entailing the compositions of $\rm K_2Ti_{\it n}O_{2{\it n}+1}$ ($n = 2,3,4,5,6,7,8$) and other layered alkali titanates such as $\rm Na_2Ti_3O_7$ and $\rm Cs_2Ti_5O_{11}$ and $\rm Ca_3Co_4O_9$;\cite{whittingham2004, goodenough2013, delmas2021, xu2017, masset2000Ca3Co4O9, galy1992vanadium, shannon1971chemistry1, shannon1971chemistry2, shannon1971chemistry3,shirpour2014lepidocrocite, gautam2016, verrelli2017, cid1962K2Ti6O13, izawa1982ion, marchand1980tio2, vitoux2020, smirnova2005crystal}
    \item Honeycomb layered oxides consisting mainly of compositions such as $A_4MD\rm O_6$, $A_3M_2D\rm O_6$ or $A_2M_2D\rm O_6$ wherein $A$ represents coinage metal ions suchlike Ag or an alkali-ion (K, Na, Li,\textit{etc}.), whereas $D$ depicts a chalcogen or pnictogen metal species such as Te, Bi, Sb, amongst others or transition metal atoms such as Ru and $M$ is mostly a transition metal species suchlike Co, Zn, Cu, Ni, \textit{etc}. or Mg (and/or a combination of multiple transition metals).\cite{kumar2012novel, grundish2019electrochemical, nalbandyan2013crystal, skakle1997synthesis, smirnova2005subsolidus, politaev2010mixed, berthelot2012new, zvereva2012monoclinic, seibel2013structure, nagarajan2002new, zvereva2016orbitally, stratan2019synthesis, brown2019synthesis, uma2016synthesis, yadav2019new, zvereva2013new, roudebush2013structure, derakhshan2007electronic, viciu2007structure, evstigneeva2011new,yadav2022,yadav2022influence,bera2022magnetism, haraguchi2021, li2020superlattice, voronina2021, jia2019, liao2022, feng2022} It also encompasses compositions such as $AM_2D\rm O_6$ (where $A$ represents alkaline-earth metal atoms suchlike Mg, Ca, Ba, \textit{etc}).\cite {song2022influence}}
\end{enumerate}
\red {The aforementioned materials} are poised to exhibit emergent quantum geometries commensurate with 
\red{2D Liouville theory}.\cite{kanyolo2020idealised, kanyolo2022cationic, zamolodchikov1996conformal}

\subsubsection{Cationic vacancies as topological defects}

\red{In particular}, the radial distribution function (pair correlation function) for the cations, $g(\vec{r})$ 
\red{is the conditional probability density that a cation will be found at position $\vec{r}$ at each inter-layer, relative to another cation within the same inter-layer}. Equivalently, it is the average density of 
\red{a cation} at $\vec{x}$ relative to a tagged particle.\cite{chandler1987introduction} 
This 
\red{requires} that the 2D number density given by, 
\begin{subequations}\label{radial_norm_eq}
\begin{align}
    \rho_{\rm 2D}(\vec{r}) = \rho_0g(\vec{r}),
\end{align}
is normalised as\cite{tuckerman2010statistical}, 
\begin{align}\label{normalisation_eq}
    \int_{\mathcal{A}} d^{\,2}r\,\rho_{\rm 2D}(\vec{r}) = \nu - 1,
\end{align}
\end{subequations}
where 
\red{$\nu$ is the number of cations within the inter-layer, $\rho_0$ is the bulk number density and the integration is performed over some emergent 2D Euclidean manifold, $\mathcal{A}$ at each inter-layer.} Moreover, the 
normalisation \red{given by}, 
$\nu - 1$ instead of 
\red{simply} $\nu$, 
\red{is understood} to arise from excluding the contribution of the reference cation, \red{as per the standard normalisation of the pair correlation function (given above)}.\cite{tuckerman2010statistical}

\begin{figure*}
\begin{center}
\includegraphics[width=\textwidth,clip=true]{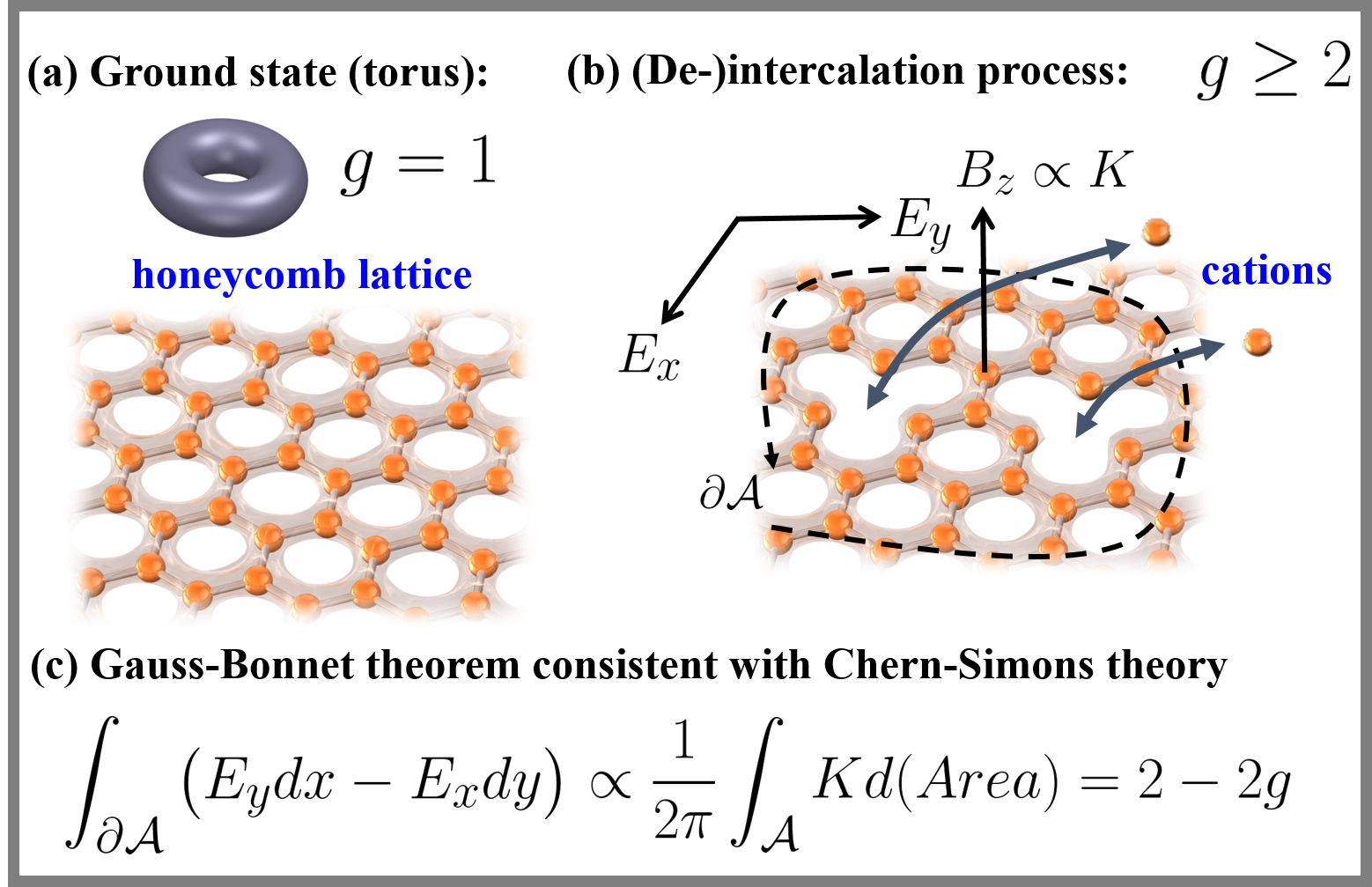}
\caption{
\red{Quasi-stable configurations of the cation honeycomb lattice in the ground and excited states ((de)-intercalation process) and the directions of the electric and pseudo-magnetic fields.} (a) The cations arranged in a honeycomb fashion with a single reference cationic vacancy (not shown, $g = 1$) topologically equivalent to the torus. This represents the ground state of the system consistent with eq. (\ref{radial_norm_eq}). (b) The extraction process of $g \geq 2$ cations from the honeycomb surface by applied electric fields $E_{x}$ and $E_{y}$ in the $x$ and $y$ directions respectively. The extraction process introduces Gaussian curvature, $K$ which can be interpreted as a pseudo-magnetic field, $B_z$ in the $z$ direction, which vanishes in the ground state ($g = 1$). Thus, the vacancies created by this extraction process can be counted by tracking the change in the electric fields at the boundary, $\partial \mathcal{A}$ of some emergent manifold, $\mathcal{A}$ representing the honeycomb lattice with vacancies. (C) Stokes' theorem on the emergent manifold consistent with Chern-Simons theory. Applying Stokes's theorem yields the Gauss-Bonnet theorem, where $g$ is the genus of $\mathcal{A}$.}
\label{Fig_3}
\end{center}
\end{figure*}

\red{Meanwhile}, the 
coordinate, $\vec{x} = (\vec{r}, z)$ 
\red{tracks the average diffusion path} 
of 
\red{the centre of mass of the cations}
and hence 
\red{is taken to} obey the Langevin equation\cite{lemons1997paul}, 
\begin{align}\label{Langevin_eq}
    \vec{n}\times\frac{d^2\vec{x}}{dt^2} = -\vec{p}(t) - \vec{\eta}(t),
\end{align}
where $t$ is the proper length on the 2D manifold, $\mathcal{A}$ with the metric,
\begin{align}\label{2D_metric_eq}
    dt^2 = g_{ab}dr^adr^b,
\end{align}
$\vec{\eta}(t)$ is the acceleration and $\vec{p} = 2md\vec{x}(t)/dt$ is the centre of mass momentum acting as the friction component with $m$ the average effective mass of the cations and $1/2m$ 
\red{plays} the role of a mean time between collisions, assumed to be equivalent in all slabs due to translation invariance along the unit vector, $\vec{n} = (0, 0, 1)$ normal to the slabs (where the unit vector points in the $z$ direction). 
\red{This guarantees that in the continuum limit, the 3D crystal admits \red{not only} a time-like Killing vector leading to energy conservation, \red{but also} a $z$-like Killing vector, which leads to momentum conservation in the $z$ direction.}\cite{kanyolo2020idealised} In fact, the additional constraint $mdz/dt = p_z = 0$ restricts cationic motion within the $x-y$ plane, which in turn allows one to effectively set $z = 0$ without loss of generality. We 
have, $\vec{x} = (\vec{r}, 0)$ where $\vec{r} = (x, y)$. Thus, it turns out that eq. (\ref{Langevin_eq}) is the analogue of the Gauss-Bonnet theorem\cite{wu2008historical} on the manifold, $\mathcal{A}$, 
\begin{align}\label{Gauss_Bonnet_eq}
    \chi(\mathcal{A}) = \frac{1}{2\pi}\int_{\mathcal{A}}d^{\,2}r\sqrt{\det(g_{ab})}K + \frac{1}{2\pi}\int_{\partial \mathcal{A}} k_{\rm g}(t)dt, 
\end{align}
as illustrated in Figure \ref{Fig_3}, where $\chi(\mathcal{A})$ is the Euler characteristic (or sometimes referred to as Euler-Poincar\'{e} characteristic) and,
\begin{subequations}
\begin{align}
    \frac{1}{2\pi}\int_{\partial \mathcal{A}} \vec{p}\cdot d\vec{x} = -\chi \in \mathbb{Z},
\end{align}
interpreted as the old quantum condition\cite{pauling2012introduction, ishiwara2017universal}, whilst the 
\red{geodesic and Gaussian curvatures} respectively are given by,
\begin{align}
    k_{\rm g}(t) = \frac{d\vec{x}(t)}{dt}\cdot\left(\vec{n}\times\frac{d^2\vec{x}(t)}{dt^2}\right),\\
    K(\vec{x}) = \frac{\vec{n}\cdot\vec{\nabla}\times\vec{\eta}}{\sqrt{\det(g_{ab})}}.
\end{align}
\end{subequations}

\red{At thermal equilibrium}, the geodesic curvature 
\red{vanishes}, $k_{\rm g} = 0$ leading to 
\red{momentum conservation} in the $\vec{x}$ direction, $d^2\vec{x}/dt^2 = 0$ which implies the Gauss-Bonnet theorem expression 
\red{is devoid of} boundary terms and cations diffuse along geodesics, hence 
\red{simplifies to},
\begin{align}\label{vacancy_genus_eq}
    \chi(\mathcal{A}) = 2 - 2g = \frac{1}{2\pi}\int_{\mathcal{A}}d^{\,2}r\sqrt{\det(g_{ab})}K,
\end{align}
where $g$ is the genus of the emergent 2D manifold, $\mathcal{A}$. To consistently introduce the electric field, we shall require the acceleration and the friction terms respectively, in the Langevin equation, to take the forms,
\begin{subequations}\label{forms_eq}
\begin{align}
    \vec{\eta} = \frac{2\pi}{m}(\vec{n}\times\vec{E}),\\
    \vec{p} = -\vec{\nabla}\Phi_{\rm AC}(\vec{x}),
\end{align}
\end{subequations}
where $\vec{E} = (E_x, E_y, 0)$ is the analogue electric field on $\mathcal{A}$ \red{responsible for} the cation extraction (de-intercalation) and re-insertion (intercalation) processes in an electrode-electrolyte \red{setup} forming a cell or battery\cite{goodenough2013} and $\Phi_{\rm AC}$ plays the role of the particle action. 
\red{Intuitively}, $\Phi_{\rm AC}$ \red{is analogous to} the Aharonov-Casher phase\cite{aharonov1984topological}, where the fictitious magnetic moment corresponds to, $\vec{\mu} = 2\pi\vec{n}/m$.\cite{kanyolo2019berry} Thus, 
\red{it is intuitive to view the cations (positively charged ions) diffusing along paths, $\vec{x}(t)$ around neutral vacancies} with a magnetic moment, $\vec{\mu}$.

\red{Moreover}, taking the analogue of Gauss' law (of electromagnetism) to be given by, 
\begin{align}\label{Gauss_eq}
    \vec{\nabla}\cdot\vec{E} = 2m\rho_{\rm 2D}(\vec{x}),
\end{align}
the Gauss-Bonnet theorem 
becomes,
\begin{subequations}\label{Euler_char_eq}
\begin{multline}
    \chi(\mathcal{A}) = -\frac{1}{2\pi}\int_{\partial \mathcal{A}}\vec{p}\cdot d\vec{x} = \frac{1}{2\pi}\int_{\partial \mathcal{A}}\vec{\eta}\cdot d\vec{x}\\
    = \frac{1}{m}\int_{\partial \mathcal{A}} d\vec{x}\,\cdot(\vec{n}\times\vec{E})
    = -\frac{1}{m}\int_{\mathcal{A}} d^{\,2}r\,\vec{\nabla}\cdot\vec{E}\\
    = -2\int_{\mathcal{A}} d^{\,2}r\,\rho_{\rm 2D} = 2 - 2\nu,
\end{multline}
where we have used eq. (\ref{radial_norm_eq}) 
\red{in the last line}. This implies that the Gaussian curvature of the 2D manifold ought to be proportional to the \red{effective} 
2D charge density,
\begin{align}\label{K_eq}
    K(\vec{x}) + \frac{4\pi\rho_{\rm 2D}(\vec{x})}{\sqrt{\det(g_{ab})}} = 0.
\end{align}
\end{subequations}
Thus, 
\red{the genus of the manifold can be interpreted as the number of cationic vacancies, $g = \nu$ as required}. Moreover, 
\red{when the material lacks the activation energy needed to dislodge cations from the lattice introducing vacancies},
$\rho_{\rm 2D} = 0$, which requires the Euler characteristic to identically vanish, $K = 0$. Consequently, the Gauss-Bonnet theorem in eq. (\ref{Gauss_Bonnet_eq}) already requires an emergent 2D manifold ($\mathcal{A}$) description for the 
diffusion processes \red{in 2D}. 
\red{Moreover}, since the energy, $E$ of the cations/vacancies is proportional to their number, $\beta E = -\chi$, the free energy equation, $F = E - \mathcal{S}/k_{\rm B}\beta$ (eq. (\ref{free_energy_eq})) suggests that the other terms in the theorem correspond to the entropy $-\mathcal{S}/k_{\rm B} = \frac{1}{2\pi}\int_{\mathcal{A}}Kd(Area)$ and free energy $\beta F = -\int_{\partial \mathcal{A}} k_{\rm g}dt$ terms, where $\beta$ is the inverse temperature. 
\red{Minimising the free energy and maximising the entropy of the system corresponds to maximising area of the manifold $\mathcal{A}$ and minimising its perimeter $\partial\mathcal{A}$. Since the manifold describes a lattice of cations, by Hale's conjecture\cite{hales2001honeycomb}, this geometric description describes the honeycomb lattice as the stable thermodynamic configuration.\cite{masese2021math, masese2021mixed}} 

\red{Finally}, the Langevin equation in eq. (\ref{Langevin_eq}) motivates the equilibrium Fokker-Planck equation\cite{risken1996fokker} of the form,
\begin{subequations}\label{Fokker_Planck_eq}
\begin{align}
    \rho_{\rm 2D}(\vec{x}) = \rho_0\exp\left(-\frac{\nu}{2}\Phi_{\rm AC}(\vec{x})\right ),\\
    0 = \frac{\partial \rho_{\rm 2D}}{\partial t} = - \vec{\nabla}\cdot(\rho_{\rm 2D}\vec{v}) + D\vec{\nabla}^2\rho_{\rm 2D},
\end{align}
where $\vec{v} = d\vec{x}/dt$ is the centre of mass velocity, $D = 1/M$ plays the role of diffusion coefficient with $M = m\nu$ the total effective mass of the cations and,
\begin{align}
    g(\vec{x}) = \exp\left(-\frac{\nu}{2}\Phi_{\rm AC}(\vec{x})\right ),
\end{align}
\end{subequations}
is the \red{pair correlation function} 
appearing in eq. (\ref{radial_norm_eq}). Thus, imposing a Boltzmann distribution (which 
\red{corresponds to} the number density of a system of particles of total effective mass, $M$ with dynamics governed by a gravitational 2D potential, $\Phi_{\rm AC}$) at equilibrium given by,
\begin{subequations}\label{Boltzmann_eq}
\begin{align}
    \rho_{\rm 2D} \propto \exp\left(-\frac{\beta M}{2}\Phi_{\rm AC}(\vec{x})\right ),
\end{align}
where $\beta$ is the inverse temperature implies the conditions,
\begin{align}\label{mass_beta_eq}
    \beta M = \nu,\\
    \beta = 1/m,
\end{align}
\end{subequations}
must be satisfied. 

\red{Moreover}, the 
\red{peculiar} relation, $M = 1/D$ can be better understood 
\red{by applying} the virial theorem\cite{marc1985virial},
\begin{subequations}\label{virial_eq}
\begin{align}
     \nu/\beta = \left \langle \sum_{j = 1}^{\nu} \frac{\vec{p}_j\cdot\vec{p}_j}{2m} \right \rangle
     = \frac{1}{2}\left \langle \sum_{j = 1}^{\nu}\vec{r}_j\cdot\frac{\partial V(\vec{r}_j)}{\partial \vec{r}_j } \right \rangle,
\end{align}
where 
\red{the averages are evaluated at equilibrium using},
\begin{align}
    \langle \cdots \rangle = \frac{\int (\cdots) \exp\left (-\beta H(\vec{p}_j,\vec{r}_j)  \right )\prod_{j = 1}^{\nu}d^2p_jd^2r_j}{\int \exp\left (-\beta H(\vec{p}_k,\vec{r}_k)  \right )\prod_{k = 1}^{\nu}d^2p_kd^2r_k},
\end{align}
\end{subequations}
and we have introduced individual particle coordinates, $\vec{x}_j = (\vec{r}_j, 0)$ with $\vec{r}_j = (x_j, y_j)$ the 2D coordinates. 
\red{Proceeding}, we shall consider the 
\red{particular} Hamiltonian 
\red{for} the cations,
\begin{subequations}
\begin{align}
   H(\vec{p}_j,\vec{r}_j) = \frac{\vec{p}_j\cdot\vec{p}_j}{2m} + V(\vec{r}_j),
\end{align}
\red{with momenta, $\vec{p}_j$, displacement vectors, $\vec{r}_j$ and $m = 1/\beta$ a mass per cation parameter defined as the inverse of the mean time/path between collisions},
\begin{multline}\label{V_eq}
    V(\vec{r}_{j}) = \frac{m}{4\mu}\sum_{k = 1}^{\nu}\vec{r}_k\cdot\vec{r}_j + \cdots\\
    = U(\vec{r}_j) + \frac{m}{4\mu}\sum_{k \neq j}^{\nu}\vec{r}_k\cdot\vec{r}_j + \cdots,
\end{multline}
\end{subequations}
\red{displaying the leading interaction term in the potential energy defined inversely proportional to $\mu$, the self-mobility of the cations} and,
\begin{align}\label{quadratic_eq}
    U(\vec{r}_j) = \frac{m}{4\mu}\vec{r}_j\cdot\vec{r}_j.
\end{align}

\red{Typically, other terms such as the Vashishta-Rahman potential\cite{Vashishta1978}, which capture interactions of the cations with the slabs atoms especially oxygen, contribute higher order terms (represented by $\cdots$) potentially neglected herein.} 
\red{This requires that} the diffusion coefficient, including cation-cation correlation terms\cite{vargas2020dynamic}, 
\red{satisfy},
\begin{subequations}
\begin{multline}\label{D_eq}
   D_{\sigma} = \frac{1}{4\beta}\left \langle \frac{1}{\nu}\sum_{j = 1}^{\nu}\sum_{k = 1}^{\nu} \vec{r}_j\cdot\vec{r}_k \right \rangle \simeq \frac{\mu}{\nu}\left \langle \sum_{j = 1}^{\nu} V(r_j) \right \rangle\\
   = \frac{\mu}{\nu}\left \langle \sum_{j = 1}^{\nu}\left (\vec{r}_j\cdot\frac{\partial V(\vec{r}_j)}{\partial \vec{r}_j } - U(\vec{r}_j)\right )\right \rangle + \cdots\\
   = \frac{2\mu}{\nu}\left \langle\sum_{j = 1}^{\nu}\frac{\vec{p}_j\cdot\vec{p}_j}{2m} \right \rangle - \frac{\mu}{\nu}\left \langle \sum_{j = 1}^{\nu} U(\vec{r}_j)\right \rangle + \cdots\\
   = 2\mu/\beta - \frac{\mu}{\nu}\left \langle\sum_{j = 1}^{\nu} U(\vec{r}_j)\right \rangle + \cdots,
\end{multline}
\red{as} $\beta \rightarrow \infty$, where we have used the result in eq. (\ref{V_eq}) and the virial theorem in eq. (\ref{virial_eq}). 
\red{Observe} that, when cation-cation correlation ($j \neq k$) terms 
\red{vanish}, $V(\vec{r}_j) \simeq U(\vec{r}_j)$, 
\red{whereas} $1/\sqrt{2\mu}$ 
\red{plays the role of} the frequency of $\nu$ harmonic oscillators and the diffusion coefficient 
\red{becomes} the self/tracer-diffusion coefficient, 
\begin{multline}\label{ES_eq}
    D_{\sigma} \simeq D_{\rm T} = \frac{1}{4\beta}\left \langle \frac{1}{\nu}\sum_{j = 1}^{\nu} \vec{r}_j\cdot\vec{r}_j \right \rangle = \frac{\mu}{\nu}\left \langle \sum_{j = 1}^{\nu} U(r_j) \right \rangle\\
    = \frac{\mu}{2\nu}\left \langle \sum_{j = 1}^{\nu} \vec{r}_j\cdot\frac{\partial U(\vec{r}_j)}{\partial \vec{r}_j } \right \rangle = \mu/\beta,
\end{multline}
\end{subequations}
which 
\red{is} the Einstein-Smoluchowski 
\red{relation}, consistent with eq. (\ref{D_eq}). Thus, defining the total mass as the average potential energy, 
\begin{subequations}
\begin{align}
    \left \langle \sum_{j = 1}^{\nu} U(\vec{r}_j) \right \rangle \equiv M,
\end{align}
$D = 1/M$ requires the Einstein-Smoluchowski 
\red{relation} be equivalent to $\mu = \beta/M$. 
\red{Consequently}, since the mobility is considered a constant, we 
\red{can} re-define it as $\mu = 8\pi G$, where $G \sim \ell_{\rm P}^2$ and $\ell_{\rm P}$ is taken to be the lattice constant with dimensions of $\rm length$. Finally, we 
\red{obtain},
\begin{align}\label{S_eq}
    \beta = \mu/D = \mu M = 8\pi GM,\\
    \nu/2 = M/2m = \beta M/2 = 4\pi GM^2 = A/4G \sim \mathcal{S},
\end{align}
\end{subequations}
where $G$ is a `gravitational' constant, in obvious comparison with 
\red{Schwarzschild black hole thermodynamics} with $A$ the black hole area and $\mathcal{S}$ the entropy.\cite{hawking1976black}

\red{\subsubsection{A gravitation description}}

\red{A} crucial observation is that the analogue electric field, $\vec{E}$ in eq. (\ref{Gauss_eq}) is sourced by $2m\rho_{\rm 2D}$, 
\red{with} dimensions of mass density (instead of charge density). 
Introducing a dual potential\cite{kanyolo2020idealised, kanyolo2022cationic},
\begin{subequations}\label{def_Phi_eq}
\begin{align}
    \Phi = \int d\vec{x}\cdot\vec{n}\times\vec{\nabla}\Phi_{\rm AC},
\end{align}
\red{Gauss law in eq. (\ref{Gauss_eq}) can be transformed using},
\begin{align}
    \vec{\nabla}\Phi = \frac{2\pi}{m}\vec{E},
\end{align}
\end{subequations}
and eq. (\ref{2D_metric_eq}) 
\red{into},
\begin{subequations}\label{C_metric_eq}
\begin{align}
    \nabla^2\Phi(\vec{x}) = 4\pi\rho_{\rm 2D}(\vec{x}) = -K(\vec{x})\sqrt{\det(g_{ab})},
\end{align}
\red{thus fixing} the 2D metric in eq. (\ref{2D_metric_eq}) to 
\red{be conformal},
\begin{align}\label{Liouville_metric_eq}
    dt^2 = \exp(2\Phi)(dx^2 + dy^2) = g_{ab}dr^adr^b.
\end{align}
\end{subequations}
\red{Consequently, a 2D gravitational description for the idealised dynamics of the cations exists, where}
eq. (\ref{vacancy_genus_eq}) and eq. (\ref{Euler_char_eq}) reveal that the energy needed to create cationic vacancies in the vacuum must always balance the energy due to motion, since $\vec{p} \neq 0$ when $g \neq 1$. 
\red{As per eq. (\ref{normalisation_eq}), the vacuum appears to correspond to $g = \nu = 1$, when there is a reference particle. Since we have the freedom to include or exclude the reference particle, we shall introduce the convention, $k = \nu + 1$, whereby $\nu = 0$ is the ground state with no vacancies, and $k = g$ is the genus when the reference particle is included}. The ground and excited states (achieved by de-intercalation of cations) in the honeycomb lattice are illustrated in Figure \ref{Fig_3}. 
\red{Consequently, the vacuum, $g = 1$ ($\int K d(Area) = 0$) is two-fold degenerate, with $K \neq 0$ and $K = 0$ corresponding to the Gaussian curvature of the 2-torus and flat-torus respectively.}

It 
\red{has been} shown 
that the 
\red{aforementioned} idealised model 
\red{is consistent with} the following tensor field equations in $d = 1 + 3$ dimensional space-time\cite{kanyolo2020idealised, kanyolo2021reproducing, kanyolo2022local}, 
\begin{align}\label{CFE_eq}
    \nabla_{\mu}K^{\mu}_{\,\,\nu} = \beta\Psi^*\partial_{\nu}\Psi,
\end{align}
where $K_{\mu\nu} = R_{\mu\nu} + iF_{\mu\nu}$ is a complex-Hermitian tensor, $R_{\mu\nu} = R^{\rho}_{\,\,\mu\rho\nu}$ is the Ricci tensor, $R^{\rho}_{\,\,\mu\sigma\nu}$ the Riemann tensor, $F_{\mu\nu} = \partial_{\mu}A_{\nu} - \partial_{\nu}A_{\mu}$ is the (analogue) electromagnetic tensor, $A_{\mu}$ is the analogue gauge potential, $\Psi = \sqrt{\rho}\exp(im\int ds)$ is a complex-valued function playing the role of the quantum mechanical wave-function of the cations and,
\begin{align}\label{rho_eq}
    \rho(\vec{x}) = -2\rho_{\rm 2D}(\vec{x})\xi^{\mu}u_{\mu}/\beta,
\end{align}
is the 3D number density, $x^{\mu} = (t, \vec{r}, z) = (t, x, y, z)$ are the coordinates with $t$ and $z$ assumed 
\red{to have the same integration cut-off}, $\beta = 1/m$, and $u^{\mu} = dx^{\mu}/d\tau$ the four-velocity 
\red{satisfying the space-time metric}, 
\begin{align}
    d\tau^2 = -g_{\mu\nu}dx^{\mu}dx^{\nu},
\end{align}
with $\tau$ the proper time. To see this, we require translation 
\red{invariance} along the $t$ and $z$ coordinates where, 
\begin{subequations}\label{Killing_eq}
\begin{align}
    \xi^{\mu} = (1, \vec{0}),
\end{align}
is a time-like Killing vector 
\red{requiring} energy conservation, and,
\begin{align}
    n^{\mu} = (0, \vec{n}),
\end{align}
\end{subequations}
is the $z$-like Killing vector, where $\vec{n} = (0, 0, 1)$ is the unit normal vector to the 2D manifold given by $\mathcal{A}$.

\red{In fact, using $F_{0i} = \vec{E} = (E_x, E_y, 0)$ and $\frac{1}{2}\varepsilon_{ijk}F_{jk} = \vec{B} = 0$ with $\varepsilon_{ijk}$ the 3D Levi-Civita symbol normalised as $\varepsilon_{123} = 1$, eq. (\ref{def_Phi_eq}) follows from the phase equations of motion}\cite{kanyolo2020renormalization, kanyolo2020rescaling},
\begin{subequations}\label{phase_eq}
\begin{align}
    \partial_{\mu}\Phi =  \beta\xi^{\nu}\eta_{\sigma\mu}\eta_{\rho\nu}F^{\sigma\rho},\\
    \partial_{\mu}\Phi_{\rm AC} = \beta n^{\nu}\,\eta_{\sigma\mu}\eta_{\rho\nu}^*F^{\sigma\rho},
\end{align}
\end{subequations}
where $\eta_{\mu\nu}$ is the Minkowski metric tensor, 
\begin{align}\label{Bianchi_eq}
    \partial_{\mu}\!^*F^{\mu\nu} = 0,
\end{align}
is the U($1$) Bianchi identity and $^*F_{\mu\nu} = \frac{1}{2}\varepsilon_{\mu\nu\sigma\rho}F^{\sigma\rho}$ is the dual field strength with $\varepsilon_{\mu\nu\sigma\rho}$ the 4D Levi-Civita symbol normalised as $\varepsilon_{1234} = 1$. 
\red{Proceeding, the real and imaginary parts respectively of eq. (\ref{CFE_eq}) correspond to},
\begin{subequations}\label{Real_Imaginary_eq}
\begin{align}
\partial_{\mu}R = \beta\partial_{\mu}\rho,\\
\nabla_{\mu}F^{\mu\nu} = J^{\nu} = -\rho u^{\nu},
\end{align}
\end{subequations}
where we have used the Bianchi identity, $\nabla^{\mu}R_{\mu\nu} = \frac{1}{2}\partial_{\nu}R$. 
\red{Since the Lie derivative of the Ricci scalar along the direction of a Killing vector must vanish, $\xi^{\mu}\partial_{\mu}R = n^{\mu}\partial_{\mu}R = 0$, we must have $\partial \rho(\vec{x})/\partial t = \partial \rho(\vec{x})/\partial z = 0$ as expected.} 

Moreover, we shall introduce the Newtonian potential, $\Phi$, satisfying,
\begin{align}\label{u_eq}
    u^{\mu} = \exp(\Phi)\xi^{\mu} = (\exp(\Phi), \vec{0}),
\end{align}
which, for a diagonalised metric tensor, implies that $g_{00} = \xi^{\mu}\xi_{\mu} = u^{\mu}u_{\mu}\exp(-2\Phi) = -\exp(2\Phi)$. In the 
\red{Newtonian} limit,
\begin{align}
g_{\mu\nu} \simeq \begin{pmatrix}
-\exp(-2\Phi) & 0 & 0 & 0 \\
0 & 1 & 0 & 0 \\
0 & 0 & 1 & 0\\
0 & 0 & 0 & 1
\end{pmatrix},
\end{align}
and using eq. (\ref{Euler_char_eq}) and eq. (\ref{rho_eq}), we 
\red{obtain the energy at equilibrium},
\begin{multline}\label{dimensionless_eq}
    \mathcal{E} = \int_{\mathcal{V}}d^{\,3}x\sqrt{-\det(g_{\mu\nu})}\,T^{00}\\
    = \int_{\mathcal{V}}d^{\,3}x\sqrt{-\det(g_{\mu\nu})}\,\rho u^{0}u^0\\
    = \int_{\mathcal{V}}d^{\,3}x\,\rho u^{0} = 2\int_{\mathcal{A}}d^{\,2}r\,\rho_{\rm 2D}\\
    = -\frac{1}{2\pi}\int_{\mathcal{A}}d^{\,2}r\,\sqrt{\det(g_{ab})}K = -\chi(\mathcal{A}),
\end{multline}
where $\mathcal{V} = \ell\times\mathcal{A}$ is a 3D manifold built up by slices of the emergent 2D manifold, $\mathcal{A}$ stacked along the $z$ coordinate (1D manifold, $\ell$), with a cut-off distance 
\red{along} $z$ given by $(\Delta z)_{\rm cut-off} = \beta = 1/m$, and we have introduced the tensor, $T^{\mu\nu} = \rho u^{\mu}u^{\nu}$. Thus, the imaginary part in eq. (\ref{Real_Imaginary_eq}) corresponds to eq. (\ref{Gauss_eq}) with $\vec{E} = (F_{01}, F_{02}, F_{03})$. 
\red{Meanwhile}, since the space-time metric is given by,
\begin{multline}\label{metric_ST_eq}
    d\tau^2 \simeq -g_{\mu\nu}dx^{\mu}dx^{\nu}\\
    = \exp(-2\Phi)dt^2 - dx^2 - dy^2 - dz^2,
\end{multline}
we can 
\red{recover} the conformal metric in eq. (\ref{C_metric_eq}) by considering trajectories of mass-less particles, $d\tau  = 0$ restricted to the plane perpendicular to the $z$ direction, $dz = 0$ with unit normal vector $\vec{n} = (0, 0, 1)$. For 
\red{trajectories of massive particles}, $d\tau \neq 0$, and we have $-1 = dz^2/d\tau^2 = (dz^2/dt^2)(dt^2/d\tau^2) = v_z^2\exp(2\Phi)$, where $v_z = dz/dt$ is the velocity along the $z$ direction. Thus, since 
\red{cationic motion is restricted to 2D},
we must either have, $v_z = \sqrt{g_{00}} \rightarrow 0$, which implies the limit,
\begin{align}\label{vz_eq}
    \lim_{v_z \rightarrow 0} g_{00}(v_z) = 0,
\end{align}
\red{and} corresponds to $\Phi \rightarrow \infty$, or the mass-less condition ($d\tau = dz = 0$) on the honeycomb lattice for arbitrary $\Phi$. 

In addition, 
\red{introducing} the bra-ket notation, 
\begin{subequations}
\begin{align}
    \langle \Psi|\mathcal{O}|\Psi\rangle = \int_{\mathcal{V}}d^{\,3}x\,\Psi^*\mathcal{O}\Psi,
\end{align}
where $\mathcal{O}$ 
\red{is an arbitrary} Hermitian operator acting on $\Psi$, 
\red{the Euler characteristic can be re-written} using eq. (\ref{CFE_eq}) as, 
\begin{multline}\label{energy_eq}
    \chi(\mathcal{A}) = \int_{\mathcal{V}}d^{\,3}x\sqrt{-\det(g_{\mu\nu})}\,\rho u^{0}u^0
    = \int_{\mathcal{V}}d^{\,3}x\,\rho u^{0}\\
    = \int_0^{-i\beta}dt\langle \Psi(\vec{x})|g^{0\nu}\frac{\partial}{\partial x^{\nu}}|\Psi(\vec{x}) \rangle\\
    = -\int_0^{-i\beta}dt\,\langle \Psi(\vec{x})|\exp(\Phi(\vec{x}))\frac{\partial}{\partial t}\exp(\Phi(\vec{x}))|\Psi(\vec{x}) \rangle\\
    = -\int_0^{-i\beta}dt\langle \Psi'(\vec{x}(t))|\frac{\partial}{\partial t}|\Psi'(\vec{x}(t)) \rangle\\
    = \frac{1}{2\pi}\int_{\partial\mathcal{A}}\vec{\eta}\cdot d\vec{x} = \frac{1}{2\pi}\int_{\mathcal{A}} d^{\,2}r\,\vec{n}\cdot(\vec{\nabla}\times\vec{\eta}),
\end{multline}
where we have used $\beta = 1/m$, the time-time component, $g^{00} = -\exp(2\Phi) = 1/g_{00}$ of the inverse metric tensor, $g^{\mu\nu}$ which satisfies $g^{\mu\sigma}g_{\nu\sigma} = \delta^{\mu}_{\,\,\nu}$ and we have set $|\Psi'(\vec{x})\rangle = \exp(\Phi(\vec{x}))|\Psi(\vec{x}) \rangle$. Moreover, 
\red{the particle trajectories have been treated as periodic in $-i\beta$}, in order for,
\begin{align}
    \vec{\eta} = \langle \Psi'(\vec{x})|\vec{\nabla}|\Psi'(\vec{x})\rangle = \frac{2\pi}{m}(\vec{n}\times \vec{E}),
\end{align}
\end{subequations}
given in eq. (\ref{forms_eq}) 
to correspond to the Berry connection, \red{interpreted as a Berry connection, $2\rho_{\rm 2D}$ as the Berry curvature\cite{berry1984quantal, cohen2019geometric} and the Euler characteristic of the 2D manifold, $\mathcal{A}$ given by $\chi(\mathcal{A})$, related to the first Chern-number (number of topological charges/vacancies) by the Poincar\'{e}-Hopf theorem.\cite{chern1946characteristic, kanyolo2022cationic}} 

\begin{table*}
    \caption{
    \red{Non-exhaustive list} of gauge (diffusion)/gravity (geometry) duality 
    \red{exhibited by} the idealised model in layered materials.}\label{Table_1}
    \centering
    \resizebox{0.5\textwidth}{!}{
    \begin{tabular}{|l|l|}
    \hline
    \textbf{gauge (diffusion)} & \textbf{gravity (geometry)}\\
    \hline \hline
    cation & vacancy\\
    lattice, $\Lambda$ & manifold, $\mathcal{A}$\\
    lattice constant, $\sqrt{\mu}$ & (reduced) Planck length, $\ell_{\rm P}$\\
    mobility, $\mu$ & Planck area, $\ell_{\rm P}^2$\\
    entropy, $\mathcal{S}$ & area, $A$\\
    free energy/work done & perimeter\\
    sphere packing (2D) & Hale's conjecture\\
    pseudo-magnetic field, $B_z$ & Gaussian curvature, $K$\\
    (acceleration $\times$ velocity)$_z$ & geodesic curvature, $k_{\rm g}$\\
    Chern number, $\nu$ & genus, $g$ 
    \\
    U(1) gauge field, $A_{\mu}$ & (time-like) Killing vector, $\xi_{\mu}$\\
    time, $t$ & line element, $t$\\
    number density, $\rho$ & Ricci scalar, $R$\\
    temperature, $1/\beta$ & effective mass, $m$\\
    Diffusion coefficient, $D = \mu/\beta$ & inverse mass, $1/M = m\ell_{\rm P}^2$\\
    (number) fluctuation, $\partial\rho/\partial t\neq 0$ & dissipation, $\nabla_{\mu}\xi_{\nu} + \nabla_{\nu}\xi_{\mu} \neq 0$\\
    Euler-Poincar\'{e} formula: & Gauss-Bonnet theorem:\\
    Faces - Edges + Vertices $= \chi(\Lambda)$ & $\int_{\mathcal{A}}KdA + \int_{\partial \mathcal{A}}k_{\rm g}dt = 2\pi\chi(\mathcal{A})$\\
    \hline
    \end{tabular}
    }
\end{table*}

\red{In addition}, the real part of eq. (\ref{CFE_eq}) given in eq. (\ref{Real_Imaginary_eq}) can be 
\red{integrated} to yield, 
\begin{subequations}
\begin{align}\label{real_eq}
    R = \beta\rho + 4\Lambda,
\end{align}
where we have introduced the integration constant, $4\Lambda = 0$ that can be considered to vanish. Thus, eq. (\ref{real_eq}) should reduce to eq. (\ref{K_eq}) in 2D after imposing the Killing vectors. \red{Specifically, eq. (\ref{real_eq})}
\red{corresponds to} the trace of Einstein Field Equations,
\begin{align}\label{EFE_eq}
    R^{\mu\nu} = \beta\left (T^{\mu\nu} - \frac{1}{2}(T^{\alpha\beta}g_{\alpha\beta})g^{\mu\nu} \right ), 
\end{align}
\end{subequations}
where we have used $T^{\mu\nu} = \rho u^{\mu}u^{\nu}$, $u^{\mu}u_{\mu} = -1$ and $\beta = 8\pi GM$ is the coupling.\cite{kanyolo2022local, kanyolo2020idealised} 
\red{It is worth noting that, considering emergent gravity within crystals to describe defects is not entirely a novel idea, since it has been considered in great detail for disclinations and dislocations within the context of classical geometries with torsion.}\cite{kleinert1987gravity, kleinert1988lattice, yajima2016finsler, holz1988geometry, verccin1990metric, kleinert2005emerging} 

\red{A potential challenge with the idealised model} is that the cations are charged whilst 
the emergent gravitational field, $\Phi$ is not. However, this poses no problem since 
the gravity description 
\red{arises from} the 
neutral cationic vacancies and not necessarily the cations themselves. 
\red{Nonetheless}, since both descriptions 
\red{are equivalent when the cations are considered neutral},
the electromagnetic potential $A_{\mu}$, which couples to the charged cations ought to have 
\red{an analogue} that couples to the neutral cationic vacancies. It suffices to consider $u_{\mu} = \beta A_{\mu}$ \red{as a Killing vector,} $\nabla_{\mu}u_{\nu} + \nabla_{\nu}u_{\mu}$, 
in order for,
\begin{align}\label{Killing_eq2}
    \nabla_{\mu}\nabla^{\mu}u^{\nu} = R^{\mu\nu}u_{\mu}.
\end{align}
Thus, 
\red{plugging in} eq. (\ref{EFE_eq}) into eq. (\ref{Killing_eq2}) yields, 
\begin{subequations}\label{killing_EFE_eq}
\begin{align}
    \nabla_{\mu}\nabla^{\mu}u^{\nu} = \beta\left (T^{\mu\nu} - \frac{1}{2}(T^{\alpha\beta}g_{\alpha\beta})g^{\mu\nu} \right )u_{\mu}. 
\end{align}
\end{subequations}
\red{Thus, using $T^{\mu\nu} = \rho u^{\mu}u^{\nu}$ we find},
\begin{align}
    \nabla_{\mu}\nabla^{\mu}u^{\nu} = -\frac{1}{2}\beta\rho u^{\nu},
\end{align}
which corresponds to the imaginary part of eq. (\ref{CFE_eq}) given in eq. (\ref{Real_Imaginary_eq}). 
\red{Indeed, this corresponds to a glimpse of the so-called gauge/gravity duality (with the most famous example, the Anti-de Sitter/conformal field theory (AdS/CFT) duality\cite{maldacena1999large, hubeny2015ads, ryu2006holographic, susskind1995world, bousso2002holographic}) whereby concepts in gauge 
theory have equivalent concepts on the gravity 
side, as summarised} in Table \ref{Table_1}.

\red{\subsubsection{Weighted pair correlation function}}

The pair correlation function 
\red{is given by} the Boltzmann factor 
in eq. (\ref{Boltzmann_eq}), 
equivalent to,
\begin{subequations}
\begin{multline}
    g(r^a) = \exp\left (M\int dt \right )\\
    = \exp\left (\frac{\nu}{4}\int d^{\,2}r\sqrt{\det(g_{ab})}R_{\rm 2D} \right ),
\end{multline}
where we have used eq. (\ref{2D_metric_eq}), $\nu = \beta M = M/m$,  the 2D momentum $p_a = 2mv_a = 2mg_{ab}dx^b/dt$
\begin{multline}
    \Phi_{\rm AC} = -M\int dt = -\frac{\nu}{2}\int p_adr^a\\
    = \frac{\nu}{4}\int d^{\,2}r\sqrt{\det(g_{ab})}R_{\rm 2D},
\end{multline}
\end{subequations}
and 
\red{the 2D Ricci scalar, $R_{\rm 2D} = R_{bd}g^{bd} = 2K$ with $R_{abcd} = K(g_{ac}g_{bd} - g_{ad}g_{bc})$ the 2D Riemann tensor and $R_{bd} = R_{abcd}g^{ac} = Kg_{bd}$ the 2D Ricci tensor}. 

Moreover, 
we expect the pair correlations to be calculated for varied vacancies as cations are 
\red{intercalated/de-intercalated} 
\red{in the layered material}. Thus, using eq. (\ref{S_eq}) and introducing the number of microstates of the system, $\mathcal{N}$ and entropy $\mathcal{S}$ respectively by,
\begin{subequations}\label{averages_eq}
\begin{align}
    \mathcal{N} \equiv \exp(\pi\nu),\\
    \mathcal{S} = \ln \mathcal{N} = \pi\nu,
\end{align}
\end{subequations}
the weighted sum over the vacancy numbers corresponding to distinct topologies of $\mathcal{A}$ yields, 
\begin{align}\label{Large_N_cations_eq}
    \langle \left.g(r^a)\right\vert_{\mathcal{A}}\rangle
    = \sum_{\nu = 0}^{\infty}f_{\nu}(\lambda)\left.g(r^a)\right\vert_{\mathcal{A}} = \sum_{\nu = 0}^{\infty}f_{\nu}(\lambda)\mathcal{N}^{\chi(\mathcal{A})},
\end{align}
where $f_{\nu}(\lambda)$ 
\red{in specific cases discussed later corresponds to the number of vectors of norm $2\nu$} within a given lattice, which is assumed to depend on 
\red{other} \red{lattice-dependent} variables 
indicated as $\lambda$, to be 
defined \red{where, 
\begin{align}\label{def_f_eq}
    f_0(\lambda) = 1,
\end{align}
}
and we have used $\chi(\mathcal{A}) = 2 - 2g = -2\nu$, 
\begin{subequations}\label{average_g_eq}
\begin{align}
    \left.g(r^a)\right\vert_{\mathcal{A}} = \exp\left (\frac{\nu}{4}\int_{\mathcal{A}} d^{\,2}r\sqrt{\det(g_{ab})}R_{\rm 2D} \right ),
\end{align}
and,
\begin{align}
    \frac{1}{4\pi}\int_{\mathcal{A}} d^{\,2}r\sqrt{\det(g_{ab})}R_{\rm 2D} = \chi(\mathcal{A}).
\end{align}
\end{subequations} 
Thus, it is instructive to take the weighted sum of the \red{pair correlation function} 
for different topologies given in eq. (\ref{Large_N_cations_eq}) to correspond to the partition function, $\mathcal{Z} = \langle \left.g(r^a)\right\vert_{\mathcal{A}}\rangle$. \red{Following} 
this argument, 
eq. (\ref{average_g_eq}) 
\red{is written as},
\begin{subequations}
\begin{align}\label{QG_eq1}
    \mathcal{Z} = \ln Z = \sum_{\nu = 0}^{\infty}f_{\nu}(\lambda)\mathcal{N}^{\chi(\nu)},
\end{align}
where,
\begin{align}
    Z = \int D[\varphi]\exp\left (i\frac{1}{g_{\rm c}^2}{\rm Tr}(\mathcal{L}(\varphi)) \right ),
\end{align}
\end{subequations}
$Z$ is the partition function of an unidentified Hermitian matrix field theory, $\varphi^{\dagger} = \varphi$, where $\lambda/\mathcal{N} = g_{\rm c}^2$ is the coupling of the theory and $\mathcal{L}(\varphi)$ is a function of $\varphi$ invariant under U($\mathcal{N}$).\cite{aharony2000large, t1993planar} 

Finally, making the 
\red{identification},
\begin{align}\label{identification_eq}
    \sum_{\mathcal{A} \in \nu} f_{\nu}(\lambda) \leftrightarrow \int \mathcal{D}[g_{ab}(\mathcal{A})],
\end{align}
eq. (\ref{QG_eq1}) becomes the partition function of 2D quantum gravity in Euclidean signature\cite{gross1991two},
\begin{align}\label{QG_eq2}
    \mathcal{Z} = \int \mathcal{D}[g_{ab}]\exp\left (\frac{1}{2\kappa}\int_{\mathcal{A}} d^{\,2}r\sqrt{\det(g_{ab})}\,R_{\rm 2D}\right ),
\end{align}
where the dimension-less coupling constant 
\red{corresponds to} $\kappa = 2/\nu$. This observation will be further explored in subsequent sections.

\subsection{Conformal Field Theory}

Whether a 
physical theory is particularly soluble or not is 
predicated on the number of 
\red{conservation laws} exhibited by the system.\cite{neuenschwander2017emmy, kitaev2006anyons} 
\red{In particular, since Noether's theorem requires 
every continuous symmetry to guarantee a corresponding conservation law}\cite{neuenschwander2017emmy}, 
\red{highly-symmetric theories tend to be} 
\red{the most soluble}. 
Consequently, 
\red{physicists} can often exploit 
continuous symmetries of the physical system to completely find 
the solutions of 
the theory, in what can be dubbed as `bootstrapping' the theory in question.\cite{poland2019conformal} 

At the heart of all physical quantum theories is 
group theory, which requires that each continuous internal symmetry (here labeled by a positive integer, $n \in \mathbb{N}$) be 
\red{generated by a Hermitian quantum operator}, $Q_n(t) = \exp(i\mathcal{H}t)Q_n(0)\exp(-i\mathcal{H}t) \equiv Q_n = Q_n^{\dagger}$ which must commute with 
\red{the Hamiltonian, $\mathcal{H} = \mathcal{H}^{\dagger}$ of the system (the generator of time translations)} to be considered a symmetry of the system, $i\partial_t Q_n(t) = [Q_n(t), \mathcal{H}] = 0$, thus satisfying Noether's theorem. Consequently, conservation laws require $Q_n$ and $\mathcal{H}$ 
\red{to be simultaneously diagonalisable} (\textit{i.e.} operators that commute share common eigenstates)\red{, which in turn restricts the size of the irreducible representation of the Lie algebra of the group (Hilbert/Fock space \textit{etc}) to} 
a subset 
\red{spanned only by the eigenvectors} $|Q_n\cdots Q_n'\rangle_{n \neq n'}$ and their linear combinations,
\red{since only these states correspond to the eigenstates of the Hamiltonian}. For instance, this implies that, physical systems that admit a Lie group of size, $N$ with a large number of generators 
often admit 
\red{known} solutions. Indeed, this is the philosophy exploited when embarking on solving matrix field theories in the large $N$ limit.\cite{t1993planar, aharony2000large}

On the other hand, quantum field theories (QFTs) in $d$-dimensional Minkowski space-time, $\eta_{\mu\nu}$ have space-time symmetries 
generated by 
$d + d(d - 1)/2 = d(d + 1)/2$ \red{number of Poincar\'{e} group generators}\cite{ohlsson2011relativistic, francesco2012conformal}, corresponding to $d$ translations, $p_{\mu} = -i\partial_{\mu}$ and $d(d - 1)/2$ rotations and boosts, $M^{\mu}_{\,\,\nu} = i(x^{\mu}\partial_{\nu} - x_{\nu}\partial^{\mu})$, alongside the set of all
internal symmetries, $\{ Q_n \}$ which further constrain the representation space. For instance, the Coleman-Mandula theorem requires 
\red{such QFTs with a mass (inverse-length) scale to have no} 
internal symmetries with space-time indices\cite{mandula2015coleman}, hence 
\red{guaranteeing} internal symmetries 
\red{are} strictly 
generated by space-time scalars. \red{Nonetheless}, 
\red{relativistic QFTs with no mass scale offer a recourse to circumventing this no-go theorem}.\cite{mandula2015coleman, weinberg2000quantum} 
As a result, 
\red{the emergent conformal symmetry} near critical points of physical systems where phase transitions occur\cite{domb2000phase}, 
\red{further augments the constraints} on such 
theories known as conformal field theories (CFTs)\cite{ginsparg1988applied}, by introducing additional conformal symmetries (scale transformations, $D = ix^{\mu}\partial_{\mu}$ and special conformal transformations, $C_{\mu} = i(x^{\alpha}x_{\alpha}\partial_{\mu} - 2x_{\mu}x^{\alpha}\partial_{\alpha})$).\cite{weinberg2000quantum, ginsparg1988applied} Provided conformal symmetries are \red{adhered to} 
in calculations of quantum correlation functions, CFTs can be completely solved by conformal bootstrap.\cite{poland2019conformal} 

The most 
\red{well-known} examples are Liouville conformal field theory\cite{zamolodchikov1996conformal} and 
2D CFTs known as Virasoro minimal models\cite{ginsparg1988applied}, which have successfully been classified and completely solved.\cite{cappelli2009ade, guillarmou2020conformal} \red{The Virasoro algebra for the 2D CFTs is given by,\cite{polchinski1998string2, ginsparg1988applied}}
\begin{subequations}\label{Virasoro_eq}
\begin{align}
    [L_n, L_m] = (n - m)L_{n + m} + \frac{c}{12}(n^3 - n)\delta_{(n + m), 0},\\
    [\overline{L}_n, \overline{L}_m] = (n - m)\overline{L}_{n + m} + \frac{\overline{c}}{12}(n^3 - n)\delta_{(n + m), 0},
\end{align}
\end{subequations}
spanned by two copies of commuting generators, $L_n = L_{-n}^{\dagger}$ and $\overline{L}_n = \overline{L}_{-n}^{\dagger}$ \red{for all integers} $n \in \mathbb{Z}$, where $[\overline{L}_n, L_m] = 0$, 
$c, \overline{c}$ is a real-valued constant 
\red{(the central charge)},
\red{satisfying} $[L_n, c] = [L_n, \overline{c}] = [\overline{L}_n, c] = [\overline{L}_n, \overline{c}] = 0$ and $\delta_{m,n}$ is the Kronecker delta. \red{Meanwhile, the representations are characterised by}
a highest weight primary state, $|h\rangle, |\overline{h}\rangle$ satisfying, $L_0|h \rangle = h|L \rangle$, $\langle h|L_0 = \langle h|h$ or $\overline{L}_0|\overline{h} \rangle = \overline{h}|\overline{h} \rangle$, $\langle \overline{h}|\overline{h}_0 = \langle \overline{h}|\overline{h}$, $L_{|n| \neq 0}|h \rangle = 0$, $\langle h|L_{-|n| \neq 0} = 0$ and $\overline{L}_{|n| \neq 0}|\overline{h} \rangle = 0$, $\langle \overline{h}|\overline{L}_{-|n| \neq 0} = 0$. The rest, $L_{-|n| \neq 0}|h \rangle \neq 0$, $\overline{L}_{-|n| \neq 0}|\overline{h} \rangle \neq 0$ and $\langle h|L_{|n| \neq 0} \neq 0$, $\langle \overline{h}|\overline{L}_{|n| \neq 0} = 0$ can be computed by 
\red{applying} the Virasoro algebra in eq. (\ref{Virasoro_eq}). The basis vectors of the representation form a Verma module,
\begin{align*}
    N = 0: |h\rangle
\end{align*}
\begin{align*}
    N = 1: L_{-1}|h\rangle
\end{align*}
\begin{align*}
    N = 2: L_{-2}|h\rangle,\,\, L_{-1}^2|h\rangle
\end{align*}
\begin{align*}
    N = 3: L_{-3}|h\rangle,\,\, L_{-2}L_{-1}|h\rangle,\,\, L_{-1}L_{-2}|h\rangle,\,\, L_{-1}^3|h\rangle
\end{align*}
\begin{align*}
    \vdots 
\end{align*}
\begin{align*}
    N \in \mathbb{N}: \mathcal{V}_{N}
\end{align*}
where $\mathcal{V}_{N}$ contains $p(N)$ number of elements equal to the partition function of the positive number, $N$.\cite{polchinski1998string2} 

\red{However, the Verma module, including the linear combination of its elements is not irreducible, since it includes elements with vanishing norms called null states. Nonetheless, the quotient of the Verma module by the null states is irreducible.} Finally, ensuring all elements of the quotient are positive definite (unitarity) determines the allowed values of $h$, $\overline{h}$ and $c, \overline{c}$ in the models, which completely fixes the critical exponents near the point of occurrence of phase transitions known as the critical point.\cite{domb2000phase} 
\red{On the other hand}, a finite central charge ($c \neq 0, \overline{c} \neq 0$) in eq. (\ref{Virasoro_eq}) governs the commutation relations of local conformal transformations, which requires that the entire infinite dimensional conformal symmetry 
\red{to be} broken. Nonetheless, the global conformal transformations of the ground state are generated by a linear combination of the generators in eq. (\ref{Virasoro_eq}) which satisfy, $[L_n, L_{-n}] = 2nL_{0}$, $[\overline{L}_n, \overline{L}_{-n}] = 2n\overline{L}_{0}$ with $\overline{c}, c \neq 0$. This singles out the condition, $n^3 - n = 0$, which requires that $n = 0, \pm 1$, corresponding to the unbroken global conformal transformations.

\red{\subsubsection{Liouville's equation}}

\begin{figure*}
\begin{center}
\includegraphics[width=\textwidth,clip=true]{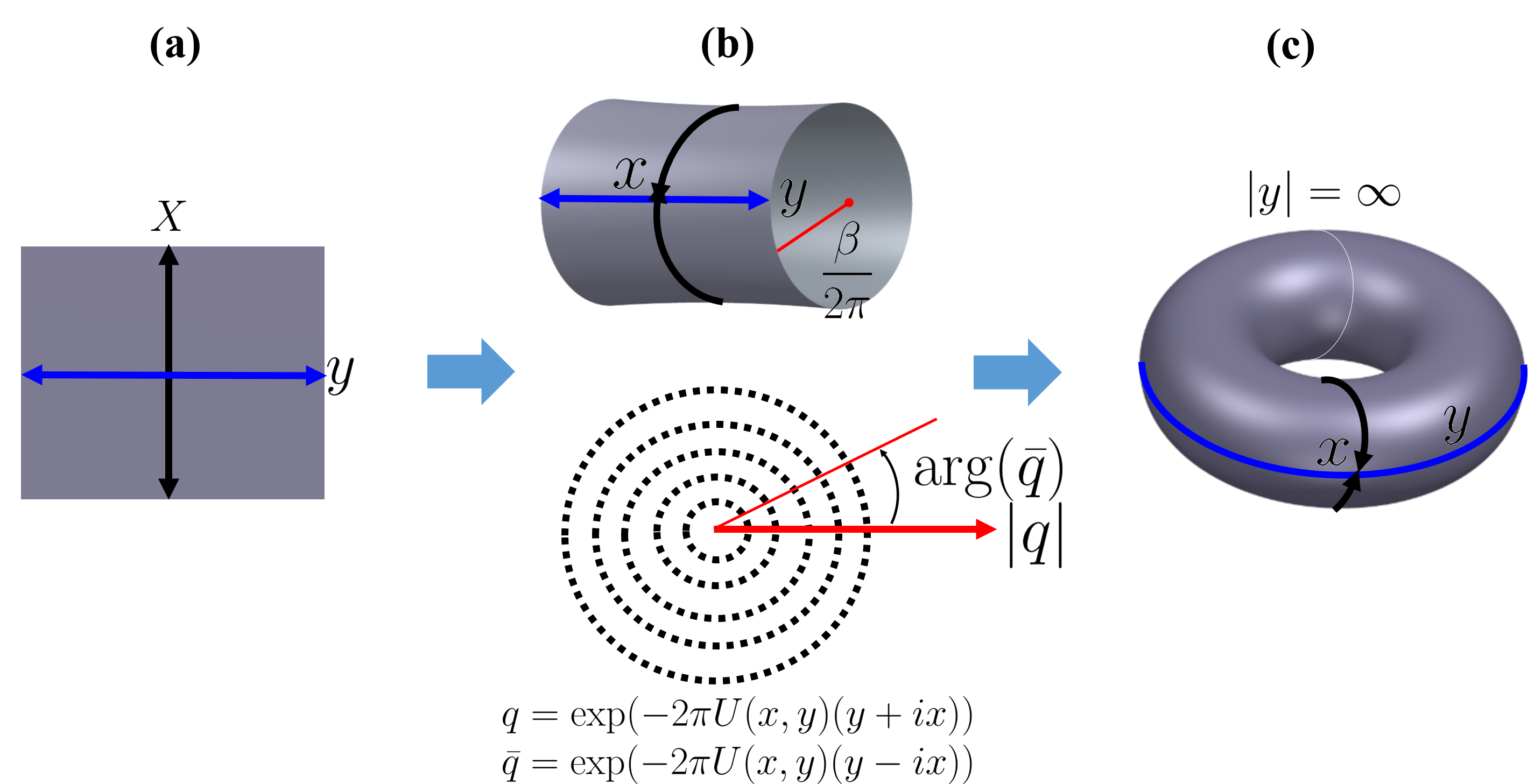}
\caption{Transformation of the flat plane to a cylinder and 2-torus by modular transformations.
(a) The flat plane with coordinates $X, y$ and Minkowski signature. (b) A cylinder obtained by a Wick rotation, $X = ix$ where $x \rightarrow x + \beta$ is periodic, which obtains the radius of the cylinder to be $\beta/2\pi$. The cylinder can be represented by new polar coordinates, $q = \exp(-U(x,y)(y + ix))$ and $\overline{q} = \exp(-U(x,y)(y - ix))$ on the complex plane where $U(x,y)$ is a potential, represented by concentric dashed circles with a radial direction $|q| = |\overline{q}|$ and argument ${\rm arg}(\overline{q}) = -{\rm arg}(q) = 2\pi U(x,y)x$. 
Thus, the winding number around the cylinder corresponds to $\beta U(x,y) = \nu$ which maps the red arrow in the complex plane back to itself. (c) 2-torus of genus 1 ($g = 1$) obtained by mapping the infinity of the complex plane, $q, \overline{q} = +\infty$ to the origin, $q, \overline{q} = 0$ corresponding to $y = -\infty$ and $y = +\infty$ respectively where the potential is taken to be positive definite, $U(x,y) > 0$.}
\label{Fig_4}
\end{center}
\end{figure*}

Given a 2D flat pseudo-Riemannian manifold with the metric, 
\begin{subequations}\label{metric_eq}
\begin{align}
    dt^2 = dy^2 - dX^2,
\end{align}
as depicted in Figure \ref{Fig_4} (a), where $y, X$ are the coordinates and $dt$ is the proper distance, a Wick rotation $X = ix$ transforms the flat pseudo-Riemannian manifold into a cylinder with the Riemannian metric,
\begin{align}\label{metric_Wick_eq}
    dt^2 = dy^2 + dx^2,
\end{align}
\end{subequations}
since, $x' = x + \beta$ is periodic\cite{zee2010quantum} in the inverse temperature, $\beta$ as depicted in Figure \ref{Fig_4} (b). This means that, under translations \red{along the circumference of the cylinder}, $x \rightarrow x + \beta$, the new coordinates, $q = \exp(-2\pi bU(b, \overline{b}))$, $\overline{q} = \exp(-2\pi\overline{b}U(b, \overline{b}))$ (where $b = y + X = y + ix$, $\overline{b} = y - X = y - ix$ and $U(b, \overline{b})$ is a (potential energy) function of $b, \overline{b}$) transform as $q' = \exp(-2\pi b'U(b', \overline{b})) = q$ and $\overline{q}' = \exp(-2\pi\overline{b}'U(b, \overline{b}')) = \overline{q}$ when, 
\begin{align}\label{nu_eq}
    \beta U(b, \overline{b}) = \nu \in \mathbb{N},
\end{align}
and,
\begin{align}\label{modular_eq1}
    U(b, \overline{b}) = U(b', \overline{b}) = U(b, \overline{b}') = U(b', \overline{b}'),
\end{align}
where $\nu \in \mathbb{N}$ is the winding number. 

Moreover, the cylinder can be 
\red{transformed into} a 2-torus, $T^2$ as shown in Figure \ref{Fig_4} (c), by making the identification, $|q| = 1/|q|$ at $|q| = 0, \infty$ corresponding to $|y| = \infty$. Thus, 
\red{as} $|y| \rightarrow \infty$ corresponding to the edges of the cylinder, we are allowed to choose new coordinates, $q \rightarrow 1/q$ and $\overline{q} \rightarrow 1/\overline{q}$. 
\red{This can be implemented by the transformations}, $b \rightarrow 1/b$ and $\overline{b} \rightarrow 1/\overline{b}$, 
whilst ensuring the potential energy satisfies, 
\begin{align}\label{modular_eq2}
     U(b, \overline{b}) = -b^{-2}U(1/b, \overline{b}) =  -\overline{b}^{-2}U(b, 1/\overline{b}).
\end{align}
\red{Consequently}, on the 2-torus, the potential energy of the form $U(\overline{b}, b) = U(k, \overline{k})$, with $k = ib$, $\overline{k} = -i\overline{b}$, 
\red{can be considered} a modular form of weight 2\cite{cohen2017modular},
\begin{subequations}\label{modular_eq}
\begin{align}
    U(Q\cdot k, \overline{k}) = (\gamma k + \delta)^2U(k, \overline{k}),\\
    U(k, Q\cdot\overline{k}) = (\gamma \overline{k} + \delta)^2U(k, \overline{k}),\\
Q \cdot k = \begin{pmatrix}
 \alpha & \beta \\
 \gamma & \delta \\
\end{pmatrix} \cdot k = \frac{\alpha k + \beta}{\gamma k + \delta},\\
Q \cdot \overline{k} = \begin{pmatrix}
 \alpha & \beta \\
 \gamma & \delta \\
\end{pmatrix} \cdot \overline{k} = \frac{\alpha \overline{k} + \beta}{\gamma \overline{k} + \delta},
\end{align}
\end{subequations}
where $Q \in \rm SL_2(\mathbb{Z})/Z_2 \equiv PSL_2(\mathbb{Z})$ is an element of the quotient of the special linear group by the sign group, $Z_2$. These transformations have been illustrated in Figure \ref{Fig_4}. \red{This requires that the translations along $x$ displayed in eq. (\ref{modular_eq1}) be generated by},
\begin{subequations}
\begin{align}
    Q_1 = \begin{pmatrix}
 1 & 1 \\
 0 & 1 \\
\end{pmatrix} \in Q,
\end{align}
\red{in order for},
\begin{align}\label{T_beta_eq}
    Q_1^{\beta} = \begin{pmatrix}
 1 & 1 \\
 0 & 1 \\
\end{pmatrix}^{\beta} = \begin{pmatrix}
 1 & \beta \\
 0 & 1 \\
\end{pmatrix} \not\in Q, 
\end{align}
\end{subequations}
\red{to correspond} to a complete winding along $x$. 
\red{Meanwhile}, 
\begin{align}\label{Q_2_eq}
    Q_2 = \begin{pmatrix}
 0 & -1 \\
 1 & 0 \\
\end{pmatrix} \in Q,
\end{align}
corresponds to translations along $y$ given in eq. (\ref{modular_eq2}). 

\red{However, since $\beta$ is not necessarily an integer, $Q_1^{\beta} \not\in Q$ is not an element of $\rm PSL_2(\mathbb{Z})$.} 
\red{This can be corrected by re-scaling} the coordinates, $(x,y) \rightarrow U^{-1}(b, \overline{b})(x,y)$, in order for the circumference of cylinder to be periodic in $\beta U(b, \overline{b}) = \nu \in \mathbb{Z}$ instead, where,
\begin{align}\label{T_nu_eq}
Q_1^{\nu} = \begin{pmatrix}
 1 & 1 \\
 0 & 1 \\
\end{pmatrix}^{\nu} = \begin{pmatrix}
 1 & \nu \\
 0 & 1 \\
\end{pmatrix} \in Q.
\end{align}
This also redefines the aforementioned coordinates, $\overline{q}, q$ to,
\begin{align}\label{q_eq}
    q(k) = \exp(i2\pi k),\,\,\overline{q}(\overline{k}) = \exp(-i2\pi \overline{k}),
\end{align}
where winding around the circumference of the cylinder is given by $k' = k + \nu$ and $\overline{k}' = \overline{k} - \nu$. Moreover, setting $\mathcal{G} = \beta^{-1}U^{-1}$, the re-scaling 
\red{requires that}, $dx' = d(\mathcal{G}x) = xd\mathcal{G} + \mathcal{G}dx$ and $dy' = d(\mathcal{G}y) = yd\mathcal{G} + \mathcal{G}dy$. Thus, we 
\red{obtain},
\begin{multline}\label{transf_eq}
    dt'^2 = dx'^2 + dy'^2 = \mathcal{G}^2(dx^2 + dy^2)\\
    + (x^2 + y^2)d\mathcal{G}^2 + \frac{1}{2}(d(x^2) + d(y^2))d(\mathcal{G}^2).
\end{multline}
\red{Indeed, since modular group on a torus implies the conformal group in 2D, these transformations require the metric in eq. (\ref{metric_eq}) be defined up to a conformal factor,}
\begin{subequations}\label{Liouville_eq}
\begin{align}
    dt^2 = g_{ab}dr^adr^b = \exp(2\Phi(x,y))(dx^2 + dy^2),
\end{align}
corresponding to eq. (\ref{Liouville_metric_eq}), which satisfies Liouville's equation, 
\begin{align}
    \nabla_{\rm 2D}^2\Phi(x,y) = -K(x,y)\exp(2\Phi(x,y)),
\end{align}
\end{subequations}
given in eq. (\ref{C_metric_eq}), requiring the last line in eq. (\ref{transf_eq}) to vanish, 
\begin{subequations}\label{vanish_eq}
\begin{align}
    (x^2 + y^2)d\mathcal{G}^2 + \frac{1}{2}(d(x)^2 + d(y)^2)d(\mathcal{G})^2 = 0.
\end{align}
\red{Here, $g_{ab}$ is the metric tensor, $r^a = (x, y)$ are the 2D coordinates, $\nabla_{\rm 2D}^2 = \partial^2/\partial x^2 + \partial^2/\partial y^2$ is the 2D Euclidean Laplace operator, $K(x,y)$ is the Gaussian curvature and,} 
\begin{align}\label{C_factor_eq}
    \mathcal{G}(x,y) = \exp(2\Phi(x,y)),
\end{align}
is the conformal factor. 

\red{We can solve eq. (\ref{vanish_eq}) by first rearranging it to yield},
\begin{multline}
    d\ln(\mathcal{G}(x,y)) = \frac{\partial \ln \mathcal{G}(x,y)}{\partial x}dx + \frac{\partial \ln \mathcal{G}(x,y)}{\partial y}dy\\
    = -2\left (\frac{xdx}{x^2 + y^2} + \frac{ydy}{x^2 + y^2} \right ). 
\end{multline}
\end{subequations}
Thus, using eq. (\ref{C_factor_eq}), $\partial\Phi(x,y)/\partial x = -x/(x^2 + y^2)$ and $\partial\Phi(x,y)/\partial y = -y/(x^2 + y^2)$, \red{we find},
\begin{subequations}\label{Phi_eq}
\begin{align}\label{Phi_eq2}
    \Phi(\vec{r}) = -\frac{1}{2}\ln\left (\frac{K_0}{2}\vec{r}\cdot\vec{r}\right ), 
\end{align}
where $K_0$ is a constant with dimensions of Gaussian curvature. Since $\overline{b}b = \overline{k}k = \vec{r}\cdot\vec{r} = x^2 + y^2$, we obtain,
\begin{align}\label{U_eq}
   \mathcal{G} = \exp(2\Phi) = \frac{4\mu}{\overline{b}b} = \frac{4\mu}{\overline{k}k} = \frac{4\mu}{\vec{r}\cdot\vec{r}}, 
\end{align}
\end{subequations}
with $\mu = 8\pi G = 1/2K_0$. Since $\mathcal{G}^{-1} = \beta U = \nu \in \mathbb{Z}$, the allowed areas, $A(\vec{r}) = \vec{r}\cdot\vec{r}$ on the manifold $\mathcal{A}$ must take on integer values, $\nu \in \mathbb{Z}$ proportional to $G$, 
\begin{align}\label{AG_eq}
    \nu = \frac{A(\vec{r})}{2\mu} = \beta U(\vec{r}).
\end{align}
Thus, for consistency with eq. (\ref{S_eq}), we ought to consider the average $\nu \rightarrow \langle \nu \rangle$ instead, where $\vec{r}$ is the centre of mass coordinate with,
\begin{subequations}
\begin{align}
    \langle U(\vec{r}) \rangle \equiv M,\\
    U(\vec{r}) = \frac{m}{4\mu}\vec{r}\cdot\vec{r},
\end{align}
\end{subequations}
and $\beta = 1/m$. 
\red{Equivalently, the self/tracer-diffusion coefficient can also be calculated from the centre of mass coordinate as,}
\begin{align}
    D_{\sigma} \simeq D_{\rm T} = \frac{1}{4\beta}\left \langle \frac{1}{\nu}(\vec{r}\cdot\vec{r}) \right \rangle = \frac{\mu}{\beta M}\langle U(r) \rangle = \mu/\beta,
\end{align}
\red{which is consistent with} the centre of mass virial theorem,
\begin{subequations}
\begin{align}
    \nu/\beta = \left \langle \frac{\vec{p}\cdot\vec{p}}{2m} \right \rangle
     = \frac{1}{2}\left \langle \vec{r}\cdot\frac{\partial U(\vec{r})}{\partial \vec{r}} \right \rangle = \left \langle U(\vec{r}) \right \rangle = M,
\end{align}
where,
\begin{align}
    \langle \cdots \rangle = \frac{\int (\cdots) \exp\left (-\beta H(\vec{p},\vec{r}) \right )d^2pd^2r}{\int \exp\left (-\beta H(\vec{p},\vec{r}) \right )d^2pd^2r},
\end{align}
and,
\begin{align}
    H(\vec{p},\vec{r}) \simeq \frac{1}{\nu}\left (\frac{\vec{p}\cdot\vec{p}}{2m} + U(\vec{r}) \right ),
\end{align}
\end{subequations}
is the Hamiltonian.

\red{Due to the constraint in eq. (\ref{Phi_eq}), we can associate the Gaussian curvature for point particles with the form,}
\begin{align}\label{K_curvature_eq}
    K(\vec{r}) = -2\pi K_0|\vec{r}\,|^2\sum_{j = 0}^g f_j\delta^2(\vec{r} - \vec{r}_j),
\end{align}
where $\delta^2(\vec{r}) = \delta(x)\delta(y)$ is the 2D Dirac delta function and $f_{j = 0} = -1$ and $f_{j \geq 1} = 1$ is defined to account for the reference particle at $\vec{r}_{j = 0}$ in the radial distribution function. Evidently, using eq. (\ref{K_eq}), eq. (\ref{rho_eq}) and eq. (\ref{Phi_eq}), eq. (\ref{K_curvature_eq}) corresponds to eq. (\ref{density_eq}) with $d = 2$. Moreover, 
we can choose the constant to be as large as possible, $K_0 = 1/2\mu \sim 1/\ell_{\rm P}^2 \rightarrow \infty$, in order for $\beta U(\vec{r}) = A(\vec{r})/2\mu = \nu \rightarrow \infty$ (in eq. (\ref{AG_eq})) which corresponds to the continuum limit of the lattice, or alternatively $\beta \rightarrow \infty$ at 
\red{fixed} values of $U(\vec{r})$, which corresponds to the \red{theory} 
at zero temperature. 
Consequently, we can consider the Gaussian definition for the 2D delta function,
\begin{multline}
    \delta^2(\vec{r}) = (K_0/\pi)\exp\left (-\frac{K_0}{2}(\vec{r}\cdot\vec{r}) \right )\\ = (K_0/\pi)\exp(-\beta U(\vec{r})),
\end{multline}
which renders the Gaussian curvature, $K(\vec{r})$ a function of $\mathcal{G}(\vec{r})$ and hence only depends on $U(\vec{r})$.

Finally, we recognise that the metric, 
\begin{subequations}
\begin{align}
    dt'^2 = dx^2 - dY^2,
\end{align}
is related to the metric in eq. (\ref{metric_Wick_eq}) by the the Wick rotation, $Y = iy$, which yields,
\begin{align}
    dt'^2 = dx^2 + dy^2 = dt^2.
\end{align}
\end{subequations}
This corresponds to a dual cylinder with coordinates $k = x + Y$, $\overline{k} = x - Y$, $k = ib$ and $\overline{k} = -i\overline{b}$, where the right-left edges of a flat plane are connected instead of the top-bottom ones. 
\red{This corresponds to the following replacements},
\begin{subequations}\label{replacements_eq}
\begin{align}
    k, \overline{k} \rightarrow \overline{b}, b,\\
    b, \overline{b} \rightarrow \overline{k}, k,\\
    \nu \rightarrow \nu,
\end{align}
\end{subequations}
\red{This has great utility since it renders the difference between the coordinates $b$ and $k$ in the theory a matter of convention}.

\red{\subsubsection{Partition function}}

To find the appropriate partition function, we shall consider the Liouville action\cite{alvarez2013random}, 
\begin{multline}\label{Liouville_action_eq}
   S_{b} = \int \frac{d^{\,2}X}{b}\sqrt{\det(\tilde{g}_{ab})}\left (\tilde{g}^{ab}\frac{\partial\phi}{\partial X^{a}}\frac{\partial\phi}{\partial X^{b}} + K\exp(2b\phi) \right )\\
   + \int \frac{d^{\,2}X}{b}\sqrt{\det(\tilde{g}_{ab})}\left (Q(b)\tilde{R}\phi \right ),
\end{multline}
where 
$\tilde{g}_{ab} = \exp(-2b\phi)g_{ab}$ is the 2D metric tensor with $g_{ab}$ given in eq. (\ref{Liouville_eq}), $K$ is the Gaussian curvature associated with $g_{ab}$ and $\tilde{R}$ is the Ricci scalar associated with $\tilde{g}_{ab}$, $Q(b)$ is a parameter dependent on $k$ and genus $h$. 
\red{Setting $\Phi = b\phi$, $r^a = b X^a$ and the metric in eq. (\ref{Liouville_action_eq}) to the 2D identity matrix, $\tilde{g}_{ab} = \delta_{ab} = \exp(-2\Phi)g_{ab}$, we can get rid of the last term since, $\tilde{R} = 0$ even when $Q(b) \neq 0$}.

\red{Thus, t}he Liouville action reduces to,
\begin{align}\label{Liouville_action_eq2}
   S_{b} = b\int d^{\,2}r\left (\vec{\nabla}\Phi\cdot\vec{\nabla}\Phi + K\sqrt{\det(g_{ab})} \right ), 
\end{align}
where $\sqrt{\det(g_{ab})} = \exp(2\Phi)$. Thus, for arbitrary $Q(b)$, eq. (\ref{Liouville_action_eq2}) can be varied with respect to $\Phi(\vec{x})$ to yield eq. (\ref{Liouville_eq}). Moreover, we define the path integral as, 
\begin{subequations}\label{path_int_eq}
\begin{align}
    \mathcal{Z} = \int\mathcal{D}[g_{ab},\Phi]\sum_j\exp\left (iS_{b_j}(\Phi, g_{ab}) \right ),
\end{align}
\red{which, after summation over $j$ and functional integration over $\Phi$ yields,} 
\begin{align}\label{Z_cosh_eq}
    \mathcal{Z} = \int \mathcal{D}[g_{ab}]2\cosh(\pi k(\chi + \theta)),
\end{align}
\end{subequations}
where $b_j = (b, \overline{b})$, $\overline{b}_j = (\overline{b}, b)$, $b = ik$ and $\overline{b} = -ik$, $2\pi\chi = \int d^{\,2}r\sqrt{\det(g_{ab})}K$ is the Gauss-Bonnet theorem, $\chi$ is the Euler characteristic \red {(or sometimes referred to as Euler-Poincar\'{e} characteristic or Euler number)}, 
\begin{subequations}
\begin{align}
    \pi\theta = \frac{Area}{2}\int\frac{d^{\,2}p}{(2\pi)^2} \ln p^2 \rightarrow \pi\sum_p\ln p + C,
\end{align}
\red{is the divergent vacuum energy of $\Phi$ with $g_{ab}$ and $\Phi$ approximated as separate non-interacting fields, $C$ is a constant that will be set to vanish by regularisation}, $Area = \frac{1}{\pi}\int d^{\,2}r$ is the area element and $Area\int d^{\,2}p/(2\pi)^2 \rightarrow \sum_p$, $\vec{p}$ are the allowed momenta/energies of the bosonic field, $\Phi$. 
\red{Thus, Riemann zeta function (or $p$-series) regularisation requires}\cite{hawking1977zeta}, 
\begin{align}\label{regularisation_eq}
    \theta(s) = \sum_p p^{-s}\ln p + C = \sum_p\frac{1}{p^{s - 1}},\\
    C = \sum_p\frac{1}{p^{s - 1}} - \sum_p p^{-s}\ln p,
\end{align}
\end{subequations}
where $\theta = \theta(s = 0)$. 
\red{Thus, taking the limit for a large number of cations, $\theta \ll k\chi(g) \rightarrow \infty$ and using the identification in eq. (\ref{identification_eq}), eq. (\ref{Z_cosh_eq}) corresponds to eq. (\ref{Large_N_cations_eq}) provided that $k \simeq \nu = g$.} 

\red{In Liouville CFT}, it is known that the field $V(\alpha) = \exp(2\alpha\phi)$ is primary when the scaling dimension is given by\cite{zamolodchikov1996conformal, nakayama2004liouville}, 
\begin{subequations}
\begin{align}
    h(k, \nu) = \alpha(k, \nu)(Q(b(k)) - \alpha(k, \nu)),\\
    \overline{h}(k, \nu) = -\alpha(k, \nu)(\overline{Q}(\overline{b}(k)) + \alpha(k, \nu)),
\end{align}
\end{subequations}
\red{whilst the marginal condition for the primary field that guarantees conformal invariance of the theory is $h = \overline{h} = 1$ with $\alpha = b = -\overline{b}$ which yields}, 
\begin{align}\label{Q_eq}
    Q(b) = b(k) + 1/b(k) = i(k - 1/k) = -\overline{Q}(\overline{b}),
\end{align}
with $b(k) = ik, \overline{b}(k) = -ik$ and $\overline{Q}(\overline{b}) = \overline{b} + 1/\overline{b}$. \red{The central charge is given by,} 
\begin{subequations}
\begin{align}
    c(b) = 1 + 6Q^2(b) = 1 + 6\overline{Q}^2(\overline{b}) = \overline{c}(b),
\end{align}
\red{which can be written as,} 
\begin{align}
    c(E, T) = 1 - 6\left (\frac{E^2 - T^2}{ET} \right )^2, 
\end{align}
\end{subequations}
where $T = 1/\beta$ is the temperature and 
\red{we have defined the energy, $E(k) = m \pm M = mk$,}
\begin{align}\label{k_eq}
    k = E/T = \beta (m \pm M) = 1 \pm \nu.
\end{align}
This identification of $k$ with $\nu$ 
\red{is somewhat peculiar} since $k$ originates from the coordinates $x, y$ whereas $\nu$ is the winding number corresponding to the number of cations. 
\red{Nonetheless, this concern can be alleviated by recognising that, the unit lengths of the honeycomb lattice unit cell satisfy,} $k/Y = (Y \pm x)/Y = 1 \pm x/Y = 1 \pm \nu$, where $\nu = g$ is the number of pairs of cations within the unit cell which corresponds to the genus of $\mathcal{A}$ given by eq. (\ref{Euler_char_eq}), suggesting we set $Y = 1$. Physically, $E = m \pm M$ is the 
\red{mass of the reference cation including or excluding the rest of the cations.} The values $k, \nu$ can also be interpreted as the eigenvalues of bosonic or fermionic operators $aa^{\dagger}, a^{\dagger}a$ or $dd^{\dagger}, d^{\dagger}d$ respectively, whereby $a^{\dagger}, a$ or $d^{\dagger}, d$ are the creation, annihilation operators satisfying the commutation or anti-commutation relations,
\begin{align}\label{commutation_eq}
    [a, a^{\dagger}]_- = 1,\,\, [d, d^{\dagger}]_+ = 1.
\end{align}
For instance, for integer values of $h = \alpha(Q - \alpha)$ in the case for bosonic operators, we can 
\red{set, $\alpha = Q/2 + iP/2$ with $[P, Q] = 2i$ which corresponds to the anti-Hermitian operators, $Q = a^{\dagger} - a$ and $P = i(a^{\dagger} + a)$, and hence,}
\begin{align}
    h = -Q^2/4 - P^2/4 + i[P, Q]/4 = -\nu,
\end{align}
where $\nu$ is the energy eigenstate of a single harmonic oscillator. Thus, the Euler characteristic (with $h = \overline{h}$) corresponds to $\chi(k) = 2h = 2\overline{h} = -2\nu = 2 - 2k$, 
\red{which is consistent with} $k = \nu + 1$.  

\red{Taking a different approach}, the C-theorem suggests the appropriate value of $c$ for the conformal field theory 
is obtained at the fixed point\cite{zamolodchikov1986irreversibility}, 
\begin{align}
    \frac{\partial c(E, T)}{\partial T} = 12\frac{(E^4 - T^4)}{E^2T^3} = 0,
\end{align}
which yields $T = E$ for real values of $E \geq 0$, which is equivalent to $M = 0$ for $m \neq 0$. This solution corresponds to 
\red{the torus, $k = \nu + 1 = 1$ with $c = 1$.} Meanwhile, the scaling dimension and spin 
\red{are given by},  
\begin{subequations}\label{delta_eq}
\begin{align}
    \Delta = h + \overline{h} = 2\alpha(Q(b) - \alpha) = -2\alpha(\overline{Q}(\overline{b}) + \alpha),\\
    \sigma = i(L - \overline{L}) = 2\alpha i (Q(b) + \overline{Q}(\overline{b})) = 0,
\end{align}
\end{subequations}
respectively. 

\red{Now, considering the partition function,} 
\begin{multline}\label{partition_CFT_eq}
    \mathcal{Z} = \sum_{j}\sum_{\nu = 0}^{\infty}f_{\nu}\langle h |\left (q^{-(L_0 - c/24)}(b_j) \right )|h \rangle\\
    =  \sum_{j}\sum_{\nu = 0}^{\infty}\bar{f}_{\nu}\langle \overline{h}|\left (\overline{q}^{-(\overline{L}_0 - \overline{c}/24)}(\overline{b}_j) \right )|\overline{h}\rangle\\
    = 2\sum_{\nu = 0}^{\infty}f_{\nu}\cosh(\pi\beta E_{\rm CFT}(k, \nu)),
\end{multline}
where $f_{\nu} = \overline{f}_{\nu}$ is real-valued, $q(b_j) = \exp(i2\pi b_j)$ and $\overline{q}(\overline{b}_j) = \exp(-i2\pi\overline{b}_j)$ are the nome, and $b_j = (b, \overline{b})$, $\overline{b}_j = (\overline{b}, b)$, arriving at the energy,
\begin{subequations}\label{E_CFT_eq}
\begin{multline}
    E_{\rm CFT}(h, c) = 2E(h - c/24)\\
    = 2E(\overline{h} - \overline{c}/24) = E_{\rm CFT}(\overline{h}, \overline{c}),
\end{multline}
with $k = \beta E$ from eq. (\ref{k_eq}). Consequently, \red{defining momentum of the primary field, $V(\alpha) = \exp(2\alpha\phi)$ as,
\begin{align}\label{alpha_eq}
    \alpha(k, \nu) = Q(k)/2 \pm i\sqrt{\chi(\nu)/2}, 
\end{align}  
with $\chi(\nu) = 2 - 2\nu$ the modified Euler characteristic due to the reference cation, and using eq. (\ref{delta_eq}), we find,}
\begin{multline}
    E_{\rm CFT}(k, \nu) = 2E(k)\alpha(k, \nu)(Q(k) - \alpha(k, \nu))\\
    - E(k)Q^2(k)/2 - E(k)/12 = E(k)\left (\chi(\nu) - \frac{1}{12} \right ),
\end{multline}
\end{subequations}
Thus, to reconcile with eq. (\ref{path_int_eq}), 
\red{we shall apply} eq. (\ref{identification_eq}) whereby, the extra factor of $-1/12$ must correspond to the vacuum energy, $\theta = \theta(s = 0)$ after regularisation. This requires the momenta $p \in \mathbb{N}$ be positive integers, which yields the expression $\theta = 1 + 2 + 3 + \cdots = -1/12$ by regularisation. 
\red{Consequently, by assuming the frequencies of the bosonic field, $\Phi$ take on integer values inadvertently introduces the bosonic string into the formalism.\cite{polchinski1998string1, polchinski1998string2}}

\subsubsection{Conformal invariance}

\red{Recall that the marginal condition is guaranteed by $h = \overline{h} = 1$, which now translates to,} 
\begin{subequations}
\begin{align}\label{Invariance_eq}
   c = \overline{c} = 1 + 6Q^2(k) = 25 - 12\chi(\nu)
   .
\end{align}
\red{Consequently, the Euler characteristic $\chi(\nu) = 2 - 2\nu$ yields},
\begin{align}\label{primes_c_eq}
    c = \overline{c} = 1 + 24\nu,
\end{align}
\end{subequations}
\red{which corresponds to $Q^2(k) = -(k - 1/k)^2 = 4\nu$}. In this case, unitarity is only achieved for the two sphere, $\nu = 0$ (which corresponds to the 2-torus satisfying $k = \nu + 1$, as expected). 
\red{The marginal condition is satisfied only for $k = 1$, requiring $k > 1$ Riemann surfaces to break conformal invariance.}
\red{Indeed, this is consistent with $\nu$ as the number of cationic vacancies in the lattice, with $g = k = \nu + 1$ the genus given by the Gauss-Bonnet theorem discussed earlier.} For the honeycomb lattice considered, we indeed have a conformal invariant theory with $\nu = g - 1 = 0$ in the ground state. 

\red{In the subsequent sections, i}t will be instructive 
to explore whether such symmetry breaking is responsible for any particular condensed matter mechanisms and/or behaviour such as phase transitions in honeycomb layered oxides.

\section{Relevant symmetries of the honeycomb/hexagonal lattice of bosonic cations}

\subsection{Modular invariance}

\subsubsection{Honeycomb lattice}

\begin{figure*}
\begin{center}
\includegraphics[width=\textwidth,clip=true]{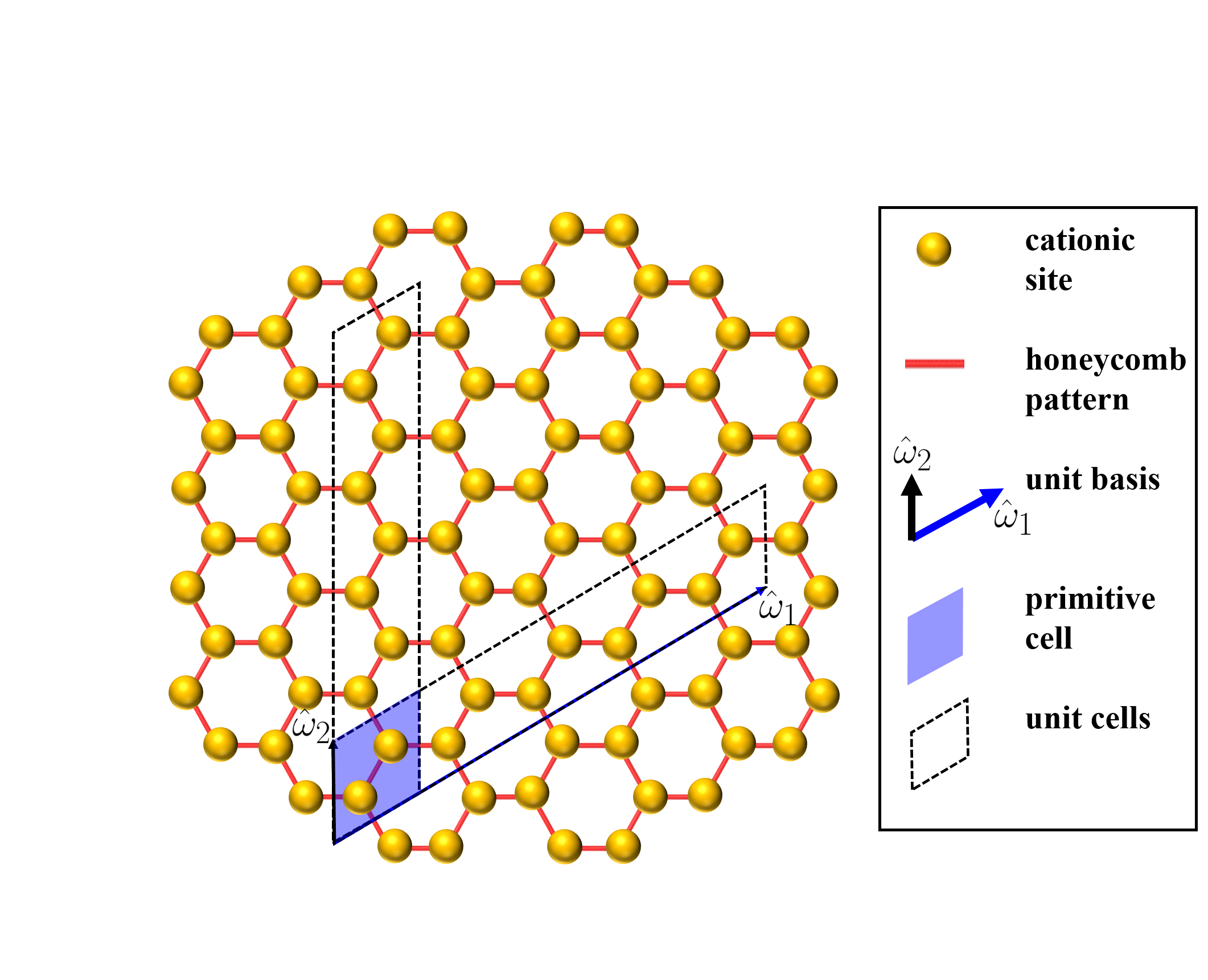}
\caption{
\red{Honeycomb lattice of cations, viewed in the [001] direction, depicting two unit cells as dashed parallelograms. The unit basis $\hat{\omega}_1, \hat{\omega}_2$ for one of the unit cells is shown where the blue rhombus corresponds to a primitive cell of the honeycomb lattice. The cations/cationic sites are depicted by grey spheres whereas the honeycomb lattice is shown in red.}
}
\label{Fig_5}
\end{center}
\end{figure*}

The search and classification of 
symmetries within 
\red{well-tested} and novel compositions of honeycomb layered oxides is currently the subject of active research, with the focus primarily on such layered materials that exhibit a 2D hexagonal and/or honeycomb 
packing of transition metal atoms and/or cations.\cite{kanyolo2021honeycomb, kanyolo2022cationic, kanyolo2022advances} 
\red{The honeycomb lattice within crystalline geometries has been the centre of frontier research ranging from graphene to other layered materials with Kitaev physics and energy storage applications}.\cite{kalantar2016two, kubota2020electrochemistry, kanyolo2021honeycomb, liu2019recent, he2012layered,schnelle2021magnetic, mcclelland2020muon, yao2023magnetic, liu2023non, fu2023suppression, guang2023thermal, mukherjee2023linear, xiang2023magnetic, vavilova2023magnetic} 
\red{Of particular interest is the role the symmetries of the lattice play in introducing conservation laws via Noether's theorem}, which offers vital clues not only towards establishing the explicit form of the relevant 
\red{partition function} but also considering valid idealised models which capture the characteristic properties of the condensed matter system. In particular, the honeycomb lattice is spanned by the 
basis, $\omega_1$ and $\omega_2$ defining a rhombus (primitive cell) enclosing a pair of cation sites as shown in Figure \ref{Fig_5}. The primitive cell corresponds to the shaded rhombus, where $\omega_1/\omega_2 = k = 1$. On the other hand, the unit cell is defined as $\omega_1/\omega_2 = k$ 
\red{indicated by the dashed lines} where $k = 1$ is a special case, hence making a necessary distinction between the primitive cell $k = 1$ and the unit cell, $k \in \mathbb{N}$. Consequently, each unit cell, $k$ on the honeycomb lattice is related to another unit cell $k'$ by $k' = k + \nu$, where $\nu \in \mathbb{N}$ is a positive integer.

\red{Proceeding, since the 
basis}, 
\begin{subequations}\label{V_plus_eq}
\begin{align}
\hat{\omega}_2 + \hat{\omega}_1 = 
\begin{pmatrix}
\omega_1 \\ \omega_2
\end{pmatrix},
\end{align}
under re-scaling by $\omega_2$ becomes, 
\begin{align}
\hat{V}_+(k) = \hat{1} + \frac{\hat{\omega}_1}{\omega_2} = 
\begin{pmatrix}
\omega_1/\omega_2 \\ 1
\end{pmatrix} = 
\begin{pmatrix}
k \\ 1
\end{pmatrix},
\end{align}
\end{subequations}
we can check that the modular transformation $Q_1^{\nu} \in Q$, given in eq. (\ref{T_nu_eq}), relates the unit vectors to each other by, 
\begin{align}
    [Q^{\nu}]^{-1}\hat{V}_+(k) = \hat{V}'_+(k) = Q^{-\nu}
\begin{pmatrix}
k \\ 1
\end{pmatrix} = 
\begin{pmatrix}
k - \nu \\ 1
\end{pmatrix},
\end{align}
which yields the primitive vector when $k = \nu + 1$. Likewise, under re-scaling by $\omega_1$ instead, eq. (\ref{V_plus_eq}) yields, 
\begin{align}
\hat{V}_-(k) = \frac{\hat{\omega}_2}{\omega_1} + \hat{1} = 
\begin{pmatrix}
1 \\ \omega_2/\omega_1
\end{pmatrix} = 
\begin{pmatrix}
1 \\ 1/k
\end{pmatrix}.
\end{align}
\red{The two valid unit bases are} related by the modular transformation, $Q_2 \in Q$ given in eq. (\ref{Q_2_eq}), 
\begin{align}
Q_2\hat{V}_-(k) = Q_2
\begin{pmatrix}
1 \\ 1/k
\end{pmatrix} = 
\begin{pmatrix}
-1/k \\ 1
\end{pmatrix},
\end{align}
\red{provided} 
$k \rightarrow -1/k$ leaves the partition function given in eq. (\ref{Z_cosh_eq}) invariant.\cite{kanyolo2022cationic} Finally, the 
\red{complete} modular invariance (which must also include, $k \rightarrow k + 1$) 
\red{occurs only} for a large number of diffusing cations, $k \rightarrow \infty$. 

\subsubsection{Weight 2 Eisenstein series approximation}

\red{More explicitly, we can make the choice}\cite{kanyolo2022cationic},
\begin{subequations}\label{E2_eq}
\begin{align}
    E_2(k, \nu) = -24\sum_{n = 0}^{\nu + 1}\sigma_1(n)q^n(\tau(k)),\\
    \mathcal{E}(k, \nu) = -\chi(k, \nu) - \theta(k, \nu) = \frac{E_2(k, \nu)}{12} - \frac{1}{4\pi\varepsilon},
\end{align}
\end{subequations}
where $q(\tau(k)) = \exp(2\pi i\tau(k))$, $\tau(k) = k + i\varepsilon$ with $\varepsilon$ an infinitesimal, $E_2(k, \nu \rightarrow \infty) = E_2(k)$ is the 
Eisenstein series \red{of weight 2}, $\sigma_1(0) = -1/24$ and $\sigma_1(n)$ is the sum of divisors of $n \in \mathbb{N}$. 
In the Liouville CFT, 
the energy function, $\mathcal{E}(k, \nu)$ appears in eq. (\ref{E_CFT_eq}) as, $\mathcal{E} = E_{\rm CFT}/E = 2h - c/12 = 2\overline{h} - \overline{c}/12 = \chi + \theta = -1/12 + 2 + \cdots + \infty$, where $q(\tau(k)) = 1$ and $\varepsilon \rightarrow 0$ with $k \in \mathbb{N}$. The infinite ($\infty$) and $-1/12$ terms originate from the vacuum energy term given in eq. (\ref{regularisation_eq}) after regularisation ($-1/12 + 1/4\pi\epsilon = 1 + 2 + 3 \cdots$) whereas $\frac{1}{4\pi}\int_{\mathcal{A}} d^{\,2}r\sqrt{\det(g_{ab})}R_{\rm 2D} = 2 - 2\nu$ is the Euler characteristic of the 2-sphere, with $\nu = 0$. Thus, the marginal condition 
\red{is achieved by} neglecting the 
\red{higher order} contributions given by $k \geq 2$ in the $E_2(k)$ sum
\red{which corresponds to setting $k - 1 = \nu = 0$ as required.} 

\red{Moreover, due to the appearance of the diverging term, $1/\epsilon$ and the Eisenstein series, $E_2$, we find $\mathcal{E}(k)$ transforms as an almost modular form of weight 2 ($\mathcal{E}(-1/k, \nu) = k^2\mathcal{E}(k, \nu)$)}, and \red{only} for a large number of diffusing cations, $\nu \gg 1$, 
\red{and the partition function, equivalent to eq. (\ref{QG_eq2}), is modular invariant}\cite{kanyolo2022cationic},
\begin{multline}
    \mathcal{Z} = \sum_{\nu = 1}^{\infty}2f_{\nu}\cosh(\pi k\mathcal{E}(k, \infty)) \\
    \leftrightarrow \int \mathcal{D}[g_{\mu\nu}]2\cosh(\pi k\mathcal{E}(k, \infty))\\
    \rightarrow \int \mathcal{D}[g_{\mu\nu}]\exp(-\pi k\mathcal{E}(k, \infty)),
\end{multline}
\red{as expected}.
Consequently, eq. (\ref{E2_eq}) guarantees eq. (\ref{partition_CFT_eq}) is invariant under $k \rightarrow -1/k$, \textit{albeit} only for $k \rightarrow \infty$. This \textit{ad hoc} 
\red{construction} of a modular invariant partition function is rigorously justified in the 
\red{succeeding} subsections. 

Thus, we have shown that the honeycomb lattice exhibits modular symmetries generated by $Q \in \rm SL_2(\mathbb{Z})/Z_2 \equiv PSL_2(\mathbb{Z})$, elements of the special linear group, $\rm SL_2(\mathbb{Z})/Z_2$ up to a sign in the cyclic group, $Z_2$. This, together with the fact that re-scaling the length of each bond in the honeycomb lattice by a factor does not alter the honeycomb lattice guarantees scale invariance, 
\red{implies that} the honeycomb lattice and its dual (hexagonal lattice) is an ideal arena to study conformal field theories. 

\subsubsection{Optimal sphere packing/hexagonal lattice}

\red{We are interested in a CFT with modular invariance at all integer values of $k$, especially $k = 1$. This warrants a different consideration for the partition function, which must reduce to eq. (\ref{partition_CFT_eq}) under specific physical conditions}. 
The non-Bravais honeycomb lattice is bipartite, 
\red{comprising} a pair of hexagonal Bravais sub-lattices. We are thus interested in the partition function $\mathcal{Z}$ of 
\red{an individual} hexagonal sub-lattice or generally hexagonal lattices. Since the cations are positively charged, we are dealing with an optimisation problem for 
repelling charges arranged on a 2D surface (Thompson problem\cite{bowick2002crystalline} for cations).
\red{For instance, the problem of finding the optimal arrangement of charged atoms (\textit{e.g.} cations) in $d$ dimensions ($d$D) which minimises their electrostatic energy is a congruent sphere packing problem in mathematics} equivalent to the spinless modular bootstrap for 
\red{CFTs} under the algebra U(1)$^c\times$ U(1)$^c$.\cite{hales2011revision, cohn2017sphere, viazovska2017sphere, zong2008sphere, cohn2009optimality, cohn2014sphere, afkhami2020high, hartman2019sphere}

It is well-known that the 2D hexagonal Bravais lattice saturates the linear programming bound in 2D\cite{hales2011revision, cohn2017sphere, viazovska2017sphere, zong2008sphere, cohn2009optimality, cohn2014sphere, hartman2019sphere}, suggesting a central charge, $c = d/2 = 1$ where,
\begin{align}
    h = \Delta_{\nu}(\vec{x}_i- \vec{x}_j)/2 = -|\vec{x}_i - \vec{x}_j|^2/2\mu = -\nu \in \mathbb{Z},
\end{align}
\red{is the scaling dimension}
proportional to the square of the distance between congruent sphere centres with $\vec{x}_i, \vec{x}_j \in \Lambda_d$ position vectors lying within the $d$ dimensional lattice, $\Lambda_d$ and $\sqrt{\mu} \sim \ell_{\rm P}$ 
\red{of order} lattice constant, $\ell_{\rm P}$. Meanwhile, since the honeycomb lattice is bipartite, 
each sub-lattice is described the CFT given by eq. (\ref{sphere_packing_eq}). Thus, to guarantee that the idealised model, alongside the Liouville CFT previously discussed, appropriately 
\red{describe} the hexagonal packing of cations, we must show that eq. (\ref{sphere_packing_eq}) is equivalent to eq. (\ref{path_int_eq}). 

\red{Since the winding number along the circumference of the cylinder formed by Wick rotation is an integer (eq. (\ref{nu_eq})), we} can consider the Poincar\'{e}-Hopf theorem, which treats such winding numbers as topological charges 
\red{equating them} to the Euler characteristic of some manifold. Thus, using eq. (\ref{AG_eq}), we 
\red{begin by writing} the Poincar\'{e}-Hopf theorem as, 
\begin{subequations}\label{PH_eq}
\begin{align}
    2\nu = r_{\nu}^2/\mu = -\Delta_{\nu}(\vec{x}_i- \vec{x}_j) = 2k - 2,
\end{align}
where $k = \nu + 1$ and, 
\begin{align}
    2 - 2k = \frac{1}{2\pi}\int_{\mathcal{A}(k)}d^{\,2}r\sqrt{\det(g_{ab})}K = \chi(k),
\end{align}
\end{subequations}
is the Gauss-Bonnet theorem. This connection can be made by considering Liouville CFT with the central charge, $c = 1 + 6Q^2 = d/2$ and scaling dimension, $h = -\nu = \alpha(Q - \alpha)$ where $\alpha = Q/2 \pm i\sqrt{\chi(k)/2}$. Thus in 2D, $d = 2$, which requires $Q = 0$ and $\nu = \alpha^2 = -\chi(k)/2 = k - 1$ at the bound, in accordance 
\red{with} eq. (\ref{alpha_eq}). 
\red{In other words, the scaling dimension is the Euler characteristic, $\Delta_{\nu} = 2\alpha(Q - \alpha) = \chi(k)$ which must be bounded for topological reasons by $\chi(k) \in \mathbb{Z} \leq 2$}. However, since $\nu \geq 0$, is the number of vacancies, the highest bound state is given by, $h = -\nu = 1 - k = 0$. Thus the dimensionless energy gap is bounded from above by, $2h = \Delta_{\nu} \leq \Delta_{\nu = 0} = 0$. 
\red{Physically, cationic vacancies, $\nu \neq 0$ cost energy $-\Delta_{\nu}$ to create in the 
electrode, an observation consistent with the Liouville CFT description}. 

\red{Now consider the implications of eq. (\ref{identification_eq}) and eq. (\ref{PH_eq}) on} the path integral given in eq. (\ref{path_int_eq}),
\begin{align}\label{sphere_packing_eq2}
    \mathcal{Z} = \sum_j\sum_{\nu = 0}^{\infty}f_{\nu}\frac{\exp(-i\pi b_j \Delta_{\nu})}{\eta^{2}(b_j)},
\end{align}
where $b_j = (b, \overline{b})$, $b = ik, \overline{b} = -ik$, $\eta(b) = q^{1/24}(b)\prod_{p = 1}^{\infty}(1 - q^p(b))$ is the Dedekind eta function, $q(b) = \exp(2\pi ib)$ is the nome and,  
\begin{multline}\label{partition_Phi_eq}
   \mathcal{Z}_{\Phi}(b) = \eta^{-d}(b)\\
   = \int \mathcal{D}[\Phi]\exp \left(ib_j\int d^{d}r \vec{\nabla}\Phi\cdot\vec{\nabla}\Phi \right),
\end{multline}
is the 
\red{thermal partition function} of the field $\Phi$ treated as a 
\red{finite temperature string theory} in $d = 2$ ($d$ quantum harmonic oscillators with integer frequencies, $p \in \mathbb{N}$).\cite{polchinski1998string2, polchinski1998string1} In other words, the path integral $\mathcal{Z}_{\Phi}$ calculates the thermal partition function of a mass-less spin zero bosonic field with integer frequencies 
\red{and $d$ number of modes},
\begin{multline}\label{strings_eq}
    \mathcal{Z}_{\Phi}(b(k)) = \prod_{l = 1}^{d}\prod_{p = 1}^{\infty}\sum_{n = 0}^{\infty}\exp(-kE_n(p))\\
    = \prod_{l = 1}^{d}\prod_{p = 1}^{\infty}\frac{\exp(-\pi kp)}{1 - \exp(-2\pi kp)}\\
    = \prod_{l = 1}^{d}\frac{\exp(-\pi k \sum_{p = 1}^{\infty} p)}{\prod_{p = 1}^{\infty}(1 - \exp(-2\pi kp))}  = \frac{1}{\eta^{d}(ik)},
\end{multline}
where the energy of each oscillator is given by $E_n(p) = 2\pi p(n + 1/2)$ with $b(k) = ik$ and $k = \beta E$ 
and the vacuum energy given by,
\begin{align}\label{vacuum_eq}
    \zeta(s = -1) = \sum_{p = 1}^{\infty} 1/p^{s = -1} = -1/12,
\end{align}
achieved by zeta function regularisation. 
\red{Consequently, interpreting $f_{\nu}(\lambda)$ as the number of vectors of norm $|\vec{x}_i - \vec{x}_j|^2/\mu = 2\nu$ within lattice $\Lambda_d$, we obtain the lattice theta function\cite{olver2010nist}},
\begin{multline}
    \Theta_{\Lambda_d}(b_j) = \sum_{\nu = 0}^{\infty}f_{\nu}\exp(-i\pi b_j\Delta_{\nu}(\vec{x}_j - \vec{x}_j)))\\
    = \sum_{\vec{x}_j, \vec{x}_j \in \Lambda_d} \exp(-i\pi b_j\Delta_{\nu}(\vec{x}_j - \vec{x}_j)),
\end{multline}
where we recover eq. (\ref{sphere_packing_eq}) for the special case $d = 2 = 2c$. 
\red{Moreover, since the honeycomb lattice and the hexagonal lattice have the same basis vectors, herein they are considered as self-dual. Thus, the above self duality needs to be slightly distinguished from the even unimodular definite condition for lattices which can only be satisfied for some lattices in $\mathbb{R}^{d}$ where $d/8 \in \mathbb{N}$, \textit{e.g.} the $\rm E_8$ root lattice is even unimodular definite.\cite{chenevier2019automorphic}}. 
\red{Thus, linear programming methods\cite{cohn2017sphere, cohn2009optimality, cohn2014sphere, hartman2019sphere} guarantee that the cationic lattices are indeed hexagonal/honeycomb}. 
\red{Finally}, since the Dedekind eta function reduces to $\eta(b) \rightarrow q^{1/24}(b)$ 
\red{when $k \gg 1$ (which corresponds to the zero temperature limit, $1/\beta = T \rightarrow 0$), we recover eq. (\ref{QG_eq2}) as expected.}

Proceeding, the modular symmetries of the honeycomb/hexagonal lattice 
\red{ought to be reflected} 
in the partition function, provided we make the replacements given in eq. (\ref{replacements_eq}). Thus, we need to 
\red{check}
instead 
\red{whether} eq. (\ref{sphere_packing_eq2}) 
\red{is} invariant under $b \rightarrow -1/b$ and/or $b \rightarrow b + 1$. 
\red{In particular, setting $F_{\nu} = f_{\nu}$ for $\nu \geq 0$ and $F_{\nu} = 0$ for $\nu < 0$, the partition function can be transformed into}, 
\begin{align}\label{transform_Z_eq}
    \mathcal{Z} = \sum_j\sum_{\nu = -\infty}^{\infty}F_{\nu}\frac{\exp(-i\pi b_j \Delta_{\nu})}{\eta^{2}(b_j)}.
\end{align}
Thus, making use of the Poisson summation formula\cite{pinsky2008introduction},
\begin{multline}\label{Poisson_eq}
    \int d^{\,d}r'\exp\left (-i2\pi\frac{\vec{r'}\cdot\vec{r}}{\ell_{\rm P}^2} \right )\sum_{\nu' = -\infty}^{\infty}F_{\nu'}\exp \left (i\pi b_j\frac{(\vec{r'}\cdot\vec{r'})_{\nu'}}{\ell_{\rm P}^2}\right )\\
    = vol\sum_{\nu = -\infty}^{\infty}F_{\nu}\exp \left (i\pi b_j\frac{(\vec{r}\cdot\vec{r})_{\nu}}{\ell_{\rm P}^2}\right ),
\end{multline}
where $vol = (\ell_{\rm P})^d$ is the volume of the $d$ dimensional primitive cell (fundamental domain) with $\sqrt{\mu/2} = \ell_{\rm P}$ and $\vec{r}, \vec{r'} \in \mathbb{R}^d$, it is clear that, 
\begin{multline}
    \sum_{\nu = -\infty}^{\infty}F_{\nu}\exp \left (-i\pi b_j^{-1}\frac{(\vec{r}\cdot\vec{r})_{\nu}}{\ell_{\rm P}^2}\right ) \equiv Z(-1/b_j)\\
    = (-ib_j)^{d/2}\sum_{\nu = -\infty}^{\infty}F_{\nu}\exp \left (i\pi b_j\frac{(\vec{r}\cdot\vec{r})_{\nu}}{\ell_{\rm P}^2}\right )\\
    = (-ib_j)^{d/2}Z(b_j). 
\end{multline}
Meanwhile, it is known that the Dedekind eta function under $b_j \rightarrow -1/b_j$ transforms as\cite{siegel1954simple}, 
\begin{align}
    \eta(-1/b_j) = (-ib_j)^{1/2}\eta(b_j). 
\end{align}
\red{Indeed, the ratio}, 
\begin{multline}\label{ratio_eq}
    \eta^{-d}(b_j)Z(b_j) =\\
    \eta^{-d}(b_j)\sum_{\nu = -\infty}^{\infty}F_{\nu}\exp \left (i\pi b_j(\vec{r}\cdot\vec{r})_{\nu}/\ell_{\rm P}^2\right ),
\end{multline}
which appears in the partition function given in eq. (\ref{sphere_packing_eq}) with $d = 2c$ (and hence in eq. (\ref{sphere_packing_eq2}) with $c = d/2 = 1$) is invariant under $b \rightarrow -1/b$, 
\red{as required}. 

However, whilst we have,
\begin{multline}
    Z(b_j + 1) = \sum_{\nu = -\infty}^{\infty}F_{\nu}\exp \left (i\pi (b_j + 1)\frac{(\vec{r}\cdot\vec{r})_{\nu}}{\ell_{\rm P}^2}\right )\\
    = Z(b_j)\exp \left (i\pi\frac{(\vec{r}\cdot\vec{r})_{\nu}}{\ell_{\rm P}^2}\right ) = Z(b_j),
\end{multline}
due to $(\vec{r}\cdot\vec{r})_{\nu}/2\ell_{\rm P}^2 = 2\nu = -\chi(k)$, 
\red{the Dedekind eta function is not invariant under $b_j \rightarrow b_j + 1$, since $\eta(b_j + 1) = \exp(i\pi/12)\eta(b_j)$. Thus, the complete modular invariance of $\eta^{-d}(b_j)Z(b_j)$ under the generators of $\rm PSL_2(\mathbb{Z})$ is unfortunately only guaranteed for dimensions $d = 2c = 24n$, $n \in \mathbb{Z}^+$. For instance, in $d = 24$ dimensions, the optimal lattice is the Leech lattice, which has great utility in string theory and proving the monstrous moonshine.\cite{hartman2019sphere, gannon2006moonshine} 

However, in the case of honeycomb layered oxides, the complete modular invariance cannot be guaranteed for the partition function of the hexagonal lattice ($d = 2c = 2$) given in eq. (\ref{sphere_packing_eq2}). Nonetheless, the bipartite nature of the honeycomb lattice implies we can take the product of two copies of the Virasoro character given in eq. (\ref{ratio_eq}), 
\begin{multline}\label{honeycomb_Z_eq}
    \mathcal{Z}_{\rm hc}(b, \overline{b}) = \frac{\sum_{\nu, \overline{\nu} = 0}^{\infty}f_{\nu}\overline{f}_{\nu'}\exp(\pi i(b\,\nu - \overline{b}\,\overline{\nu}))}{\overline{\eta}^2(\overline{b})\eta^2(b)}, 
\end{multline}
with $\overline{\eta}(\overline{b}) = \overline{q}^{1/24}(\overline{b})\prod_{p = 1}^{\infty}(1 - \overline{q}^p(\overline{b}))$ and $\overline{q}(\overline{b}) = \exp(-2\pi i\overline{b})$, 
which is invariant under the successive transformations, $b \rightarrow b + 1$ and $\overline{b} \rightarrow \overline{b} + 1$, as well as $b \rightarrow -1/b$ and $\overline{b} \rightarrow -1/\overline{b}$.}
The highest weight state 
\red{is proportional to} the genus of the emergent manifold. Due to the assumption that the partition function is connected to a large $\mathcal{N} = \exp(\pi k)$ theory in the limit $k \gg 1$, the manifold can be understood to emerge from the topology of Feynman diagrams of 
\red{a yet unidentified} U($\mathcal{N}$) invariant theory.\cite{kanyolo2021partition} 

\red{Lastly, checking that eq. (\ref{partition_CFT_eq}), where $\chi$ is given by eq. (\ref{E2_eq}), is consistent with the sphere packing partition function given in eq. (\ref{sphere_packing_eq2}) with $k = 1$,}
\red{we shall consider} the transformation of $\mathcal{Z}_{\Phi} = \mathcal{Z}(f_0 = 1)$ when $f_{\nu = 0} = 1$ (using eq. (\ref{def_f_eq})) and $f_{\nu \neq 0} = 0$) under Weyl invariance, $g_{ab}(\Phi(\vec{r} \in \Lambda)) = \exp(2\Phi(\vec{r}))\delta_{ab} = \exp(\gamma b)\delta_{ab}$ at a given value of $\Phi(\vec{r} \in \Lambda) = \gamma b/2$ for $\vec{r}$ in the lattice, $\Lambda$. 
\red{Evidently}, the Weyl transformation is trivial for $b  = i2\pi n/\gamma$ where $n \in \mathbb{Z}$ is an integer and $\gamma$ is a parameter to be determined. 
\red{However}, we shall calculate, 
\begin{multline}\label{derive_E_2_eq}
    \frac{\delta}{\delta b}\ln \left (\mathcal{Z}_{\Phi}\right ) = -i\int d^{\,2}r\left \langle \vec{\nabla}\Phi\cdot\vec{\nabla}\Phi \right \rangle\\
    = \frac{1}{\mathcal{Z}_{\Phi}}\int \mathcal{D}[\Phi]\exp(-iS_{\Phi})(-i)\frac{\delta g_{ab}}{\delta b}\frac{\delta S_{\Phi}}{\delta g_{ab}}\\
    = \frac{1}{\mathcal{Z}_{\Phi}}\int \mathcal{D}[\Phi]\exp(-iS_{\Phi})(-i)\gamma g_{ab}\frac{\delta S_{\Phi}}{\delta g_{ab}}\\
    = 2i\gamma\int d^{\,2}r\sqrt{\det(g_{ab})}\langle g_{ab}T^{ab} \rangle\\
    = -i\gamma\frac{c}{6}\int d^{\,2}r\sqrt{\det(g_{ab})}R_{\rm 2D} = -2\pi i\gamma\frac{c}{3}\chi,
\end{multline}
where, 
\begin{subequations}
\begin{multline}
    S_{\Phi} = \int d^{\,2}r\sqrt{\det(g_{ab})}\,\mathcal{L}_{\Phi} = b\int d^{\,2}r\,\vec{\nabla}\Phi\cdot\vec{\nabla}\Phi\\
    = b\int d^{\,2}r\sqrt{\det(g_{ab})}g^{ab}\,\partial_{a}\Phi\partial_b\Phi,
\end{multline}
and,
\begin{align}
    \sqrt{\det(g_{ab})}\,T^{ab}  = -2\frac{\delta (\sqrt{\det(g_{ab})}\,\mathcal{L}_{\Phi})}{\delta g_{ab}},
\end{align}
\red{and we have applied the Weyl anomaly}\cite{polchinski1998string1},
\begin{multline}
    \langle g_{ab}T^{ab} \rangle = \\
    \frac{1}{Z_{\Phi}}\int \mathcal{D}[\Phi]g_{ab}T^{ab}\exp(-iS_{\Phi}) = -\frac{c}{12}R_{\rm 2D},
\end{multline}
\end{subequations}
with $R_{\rm 2D} = 2K$ the 2D Ricci scalar and $c$ the central charge. Recall that, $\mathcal{Z}_{\Phi}(b) = \eta^{-d}(b) = \Delta^{-d/24}(b)$ where $\Delta (b)$ is the modular discriminant which satisfies\cite{cohen2017modular},
\begin{align}\label{derive_E_2_eq2}
     \frac{\delta}{\delta b}\ln (\mathcal{Z}_{\Phi}) = -\frac{d}{24}\frac{\delta}{\delta b}\ln(\Delta(b)) = -\frac{d}{12}i\pi E_2(b),
\end{align}
where $E_2(b) = -24\sum_{n = 0}^{\infty}\sigma_1(n)q(b)$ is the Eisenstein series of weight 2. Equating eq. (\ref{derive_E_2_eq}) with eq. (\ref{derive_E_2_eq2}), we obtain eq. (\ref{E2_eq}) with a vanishing vacuum energy, $\theta + 1/4\pi\epsilon = 0$, where $b \rightarrow k$, $d = 2c$ and $\gamma = -3$. 
\red{Thus, employing other values of $\gamma$ yields, $\chi = E_2/4\gamma$}.\cite{kanyolo2022cationic} Lastly, the vacuum terms can be re-introduced through the relations,
\begin{align*}
    \sqrt{\det(g_{ab}(b'))}R'_{\rm 2D}
    = \sqrt{\det(g_{ab}(b))}R_{\rm 2D} + \vec{\nabla}b\cdot\vec{\nabla}b,
\end{align*}
and, 
\begin{align*}
    \int \mathcal{D}[b]\exp\left (\frac{\delta \ln (Z_{\Phi}(b'))}{\delta b'} \right ) = \exp\left (\frac{\delta \ln (Z_{\Phi}(b))}{\delta b} \right ), 
\end{align*}
which relate
\red{the coupling constants $b$ and $b'$} with $\chi'(b) = \chi'(b + 1)$ and $\chi(-1/b) = b^2\chi(b)$ which are only \red{either} $T$ invariant 
\red{or} $S$ invariant respectively. 

\subsection{Scale invariance}

\red{\subsubsection{Torus}}

\red{
According to eq. (\ref{EFE_eq}), on the flat-torus where $\rho \propto \rho_{\rm 2D} \propto K = 0$ and the particles are massive $u^{\mu}u_{\nu} = -1$, the theory must be scale invariant since the dilatation current density in Minkowski space-time, $j_{\mu} = x^{\nu}T_{\mu\nu} = \rho\, x^{\nu}u_{\nu}u_{\mu} = 0$ 
vanishes}\cite{francesco2012conformal}, and hence is trivially conserved. Thus, eq. (\ref{EFE_eq}) becomes the vacuum condition, $R_{\mu\nu} = 0$. Non-trivially, on the 2-torus, $T^2$ where $\rho \propto \rho_{\rm 2D} \propto K \neq 0$ and the particles are mass-less, $u^{\mu}u_{\mu} = 0$, the theory remains scale invariant on Minkowski space-time since the dilatation current density, $j_{\mu} = x^{\nu}T_{\mu\nu} = x^{\nu}\rho u_{\nu}u_{\mu} \neq 0$ is conserved 
\red{due to} $\partial^{\mu}j_{\mu} = T_{\mu\nu}g^{\mu\nu} + x^{\nu}\partial^{\mu}T_{\mu\nu} = 0$, where $\partial^{\mu}T_{\mu\nu} = 0$ and $T_{\mu\nu}g^{\mu\nu} = \rho u^{\mu}u_{\mu} = 0$. 
\red{Thus, to preserve scale invariance and hence conformal invariance, the cations either have to be considered mass-less (as is the case on the honeycomb lattice, \textit{e.g.} electrons in graphene\cite{divincenzo1984self, semenoff1984condensed}) or the Gaussian curvature must vanish, $K \propto \rho = 0$.} 
\red{Nonetheless, the latter condition for massive particles is not only trivial, but also overly restrictive, 
and hence, in drawing our conclusions, we shall consider instead the current rather than the current density}.

Thus, the conserved scale invariance operator, $D = ix^{\nu}\partial_{\nu} = x^{\nu}p_{\nu}(x)$ where $p_{\nu}(x) = \int d^{\,3}x'\,T_{\nu}^{\,\,0}(x - x')$ yields instead the current,
\begin{align}
    D^{\mu}(x) = x^{\nu}\int d^{\,3}x'\,u^0(x')T^{\mu}_{\,\,\nu}(x - x'),
\end{align}
whose divergence in Minkowski space-time yields, $\partial^{\mu}j_{\mu} = \int d^{\,3}x'\,u^0(x')T_{\mu}^{\,\,\mu}(x - x') =  -\int d^{\,3}x'\,u^0(x')\rho(x - x') = \chi(k)$. Thus, due to the volume integral, 
this theory with $T^{\nu}_{\,\,\nu} = \rho u^{\nu}u_{\nu} = -\rho = K\sqrt{\det(g_{ab})}/2\pi\beta \neq 0$ and $\int d^{\,2}r\,K\sqrt{\det(g_{ab})} = 2\pi \chi(k) = 2\pi(2 - 2k)$ can be considered scale invariant on the 2-torus with $k = 1$ even for $K \neq 0$. Conversely, all other values, $k \neq 1$ break scale invariance. Indeed, this is the 
\red{condition} we found for the Liouville CFT in eq. (\ref{Invariance_eq}) that guarantees conformal invariance. This realisation is powerful since the theory of cations requires their effective masses in the theory to be given by, $m = 1/\beta$, where we have often taken the zero temperature limit, $\beta \rightarrow \infty$, which is the mass-less case. 
\red{Thus}, for massive cations (finite temperature), scale invariance is maintained \textit{if and only if} $k = g = \nu + 1 = 1$, corresponding to a vacuum with no vacancies. This vacuum is two-fold degenerate, since it corresponds to either the flat-torus ($K = 0$) or the 2-torus ($K \neq 0$), both with genus $g = 1$. For fermionic lattices, this degeneracy can be lifted by a 
\red{phase transition \textit{i.e.} a lattice distortion}, which introduces the $g = 0$ state.\cite{kanyolo2022cationic, masese2023honeycomb} 

\red{Consequently}, $k \neq 1$, which results from finite vacancies/cationic diffusion ($\nu > 0$) or lattice distortion ($\nu = -1$) breaks scale invariance and 
\red{also} conformal invariance. However, 
\red{whilst scale invariance in 2D almost always implies conformal invariance, it is worth mentioning that there are exceptions \textit{e.g.} the 2D field theory of elasticity and membrane theory,} which 
\red{exhibit} scale but not conformal invariance.\cite{mauri2021scale, riva2005scale} Ideally, scale 
invariance in the honeycomb lattice manifests when all bonds between 
\red{cations} are of equal 
\red{length}. 
\red{As a result}, introducing additional bonds that differ in length from the rest is expected to break scale invariance, leading to a monolayer-bilayer phase transition in a cationic lattice of fermions\cite{kanyolo2022cationic, kanyolo2022advances, masese2023honeycomb}, analogous to the Kekul\'{e}/Peierls distortion (2D) in \red{strained graphene}\cite{lee2011band, hou2007electron, ryu2009masses, chamon2000solitons, garcia1992dimerization, peierls1979surprises, peierls1955quantum} 
\red{expected} to generate Dirac masses for the ($1 + 2$)D pseudo-spin cations. 
The mechanisms for 
\red{the monolayer-bilayer phase transition}
in honeycomb layered 
\green{materials} with cationic lattices of fermions 
has been explored in succeeding sections.

\red{Finally, recall that}
\red{the energy,
\begin{multline}\label{activation_E_eq}
    \mathcal{E} = E_{\rm CFT}/E = \int d^{\,3}x' \sqrt{-\det(g_{\mu\nu}(x'))}\,T^{00}(x - x')\\
    = -\int d^{\,3}x'\,J^0(x - x') = -\chi(k),
\end{multline}
given in eq. (\ref{dimensionless_eq}) with $\Delta_{\nu}(x) = -|\vec{x}|^2/\mu = -2\nu = \chi(k) = 2 - 2k$, was appropriately defined to be dimensionless, and in fact corresponds to the Euler characteristic of the 2D manifold, $\mathcal{A}$, which we have showed is equivalent to the scaling dimension of the CFT. Moreover}, 
eq. (\ref{real_eq}) and hence eq. (\ref{CFE_eq}) is invariant under the transformation, $T^{\mu\nu} \rightarrow T^{\mu\nu} + \mathcal{T}^{\mu\nu}$, where $\mathcal{T}^{\mu\nu}$ is a trace-less energy-momentum tensor, $\mathcal{T}_{\mu\nu}g^{\mu\nu} = 0$. Lastly, the finite vacuum energy appearing in eq. (\ref{E_CFT_eq}) can be introduced as the 
\red{cosmological constant}, 
\begin{align}\label{c_lambda_eq}
    \langle \mathcal{T}^{\mu\nu} \rangle  = \frac{c}{12}\mathcal{V}^{-1}g^{\mu\nu},
\end{align}
where $\mathcal{V} = \int d^{\,3}x \sqrt{\det{(g_{\mu\nu})}}g^{00}$ and $c = 1$. 

\subsubsection{Point particles}

\red{We can consider point particles restricted along} trajectories, $x^{\mu}_j(\tau) = (t(\tau), \vec{x}_j(\tau))$. 
\red{Thus, the conserved current that couples to the U($1$) field in eq. (\ref{Real_Imaginary_eq}) corresponds to},
\begin{align}\label{J_eq}
    J^{\mu} = -2\beta^{d - 3}\sum_{j = 0}^g f_j\delta^{\,d}(\vec{x} - \vec{x}_j(t))\frac{dx^{\mu}_j(t)}{dt},
\end{align}
where $d = 2,3$, $\delta^{d}(\vec{x} - \vec{x}(t))$ is the $d$ dimensional Kronecker delta function and $f_{j = 0} = -1$ and $f_{j \geq 1} = 1$ is defined to account for the reference particle at $\vec{r}_{j = 0}$ in the radial distribution function. 
\red{The manifestly Lorentz covariant form of eq. (\ref{J_eq}) is}, 
\begin{align*}
    J^{\mu} = -2\beta^{d - 3}\int d\tau\,\sum_{j = 0}^g f_j\delta^{\,d + 1}(x - x_j(\tau))\frac{dx_j^{\mu}(\tau)}{d\tau},
\end{align*}
where $\delta^{\,d + 1}(x - x_j(\tau)) = \delta(t - t(\tau))\delta^{\,d}(\vec{x} - \vec{x}_j(\tau))$ is the $d + 1$ dimensional Kronecker delta function. 
\red{Meanwhile, the manifestly covariant energy-momentum tensor of the point particles can be defined as},
\begin{align*}
    T^{\mu\nu} = 2\beta^{d - 3}\int d\tau \sum_{j = 0}^g f_j\delta^{\,d + 1}(x - x_j(\tau)) \frac{dx_j^{\mu}(\tau)}{d\tau}\frac{dx_j^{\nu}(\tau)}{d\tau},
\end{align*}
which, after integration over $\tau$, yields,
\begin{align}\label{delta_T_eq}
    T^{\mu\nu} = 2\beta^{d - 3}\sum_{j = 0}^g f_j\delta^{\,d}(\vec{x} - \vec{x}_j(t)) \frac{dx_j^{\mu}(t)}{dt}\frac{dx_j^{\nu}(t)}{d\tau}.
\end{align}
\red{Using} the 
identity,
\begin{align}
    \vec{\nabla}\cdot\left (\delta^{\,d}(\vec{x} - \vec{x}_j(t)) \frac{d\vec{x}_j}{dt} \right) = -\frac{\partial}{\partial t}\delta^{\,d}(\vec{x} - \vec{x}_j(t)),
\end{align}
it is clear that $\partial_{\mu}J^{\mu} = 0$ and $\partial_{\mu}T^{\mu\nu} = 0$ are locally conserved. \red{Consequently},
one can now check that in both cases (eq. (\ref{J_eq}) and eq. (\ref{delta_T_eq})), the density functions are given by,
\begin{align}\label{density_eq}
    \rho_j(x) = 2\beta^{d - 3}\delta^{\,d}(\vec{x} - \vec{x}_j(t))\frac{d\tau}{dt},
\end{align}
where $T^{\mu\nu} = \sum_j\rho_ju_j^{\mu}u_j^{\nu} \equiv \rho u^{\mu}u^{\mu}$, $u^{\mu} = dx^{\mu}/d\tau$,$u_j^{\mu} = dx_j^{\mu}/d\tau$ and we have set $\rho(x) = -g_{\mu\nu}T^{\mu\nu}$ and $u^{\mu}u_{\mu} = -1$. Setting $u_{\mu}^ju^{\mu}_j = -1$, we have,
\begin{align}
    \rho_g = \sum_{j = 0}^gf_j\rho_j.
\end{align}
Moreover, the finite vacuum energy 
\red{appearing in} eq. (\ref{E_CFT_eq}) can be introduced as the Weyl anomaly\cite{polchinski1998string1, polchinski1998string2} by 
\red{the replacement}, $\mathcal{T}^{\mu\nu} \rightarrow \langle \mathcal{T}^{\mu\nu} \rangle$,
\begin{align}\label{Weyl_eq}
    \langle \mathcal{T}^{\mu\nu} \rangle = -\frac{c}{24}\rho_{g = 0}u^{\mu}u^{\nu},
\end{align}
where $c = d/2 = 1$, $\rho_{\rm 2D}\xi^{\mu}u_{\mu} = -\beta\rho_{g = 0}/2$ 
\red{is} the 2D number density related to the 2D Ricci scalar, $R_{\rm 2D}$ via eq. (\ref{K_eq}), eq. (\ref{rho_eq}) and $R_{\rm 2D} = 2K$, and $\xi^{\mu} = (1, \vec{0})$ the time-like Killing vector in eq. (\ref{Killing_eq}). 
\red{Here, $g = 0$ requires only the reference particle to contribute to the vacuum energy}. 

\section{Condensed matter and field theory treatments} 

\red{\subsection{Optimised 3D hexagonal/honeycomb lattices of bosonic cations}}

\begin{figure*}
\begin{center}
\includegraphics[width=\textwidth,clip=true]{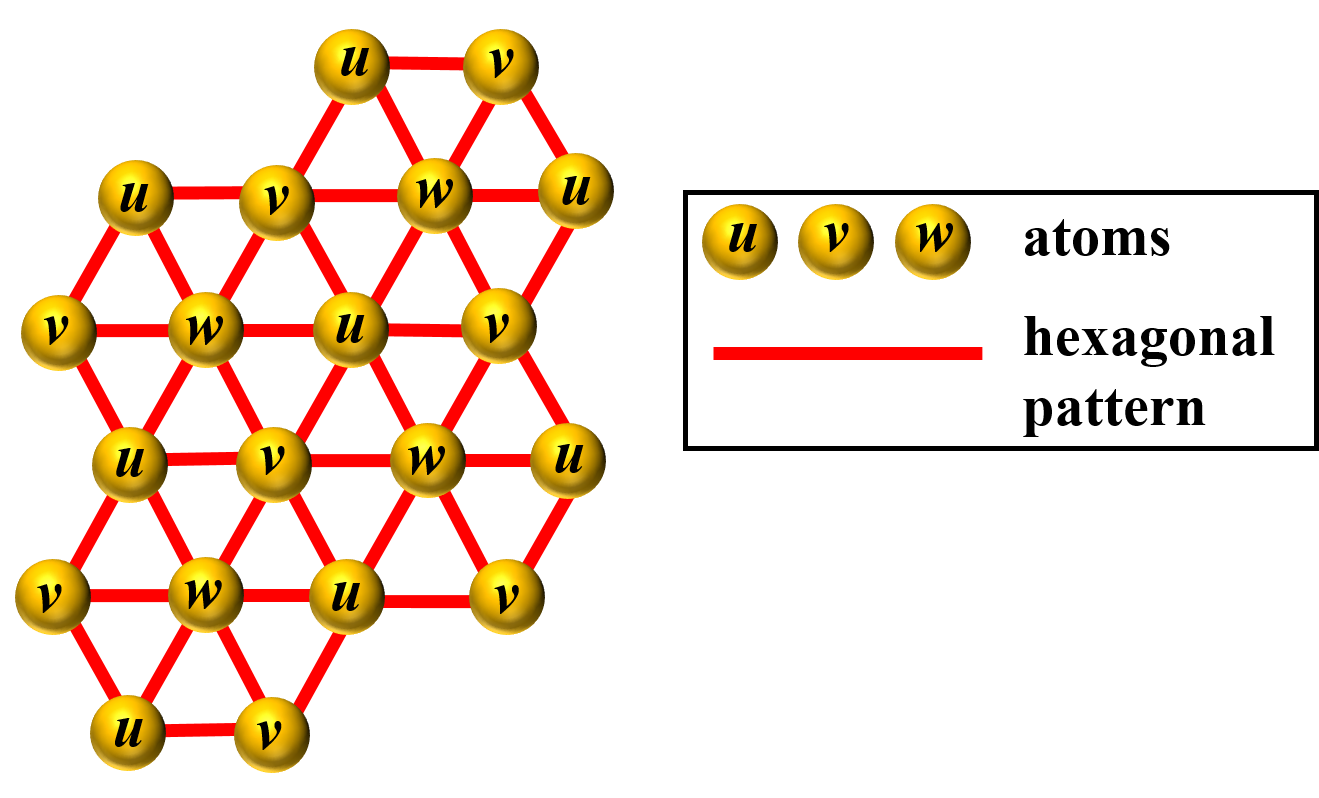}
\caption{The optimised 2D hexagonal 
\red{pattern} with 
\red{three atomic sites labelled as $uvw$.} Thus, there are three possible configurations for the honeycomb lattice, namely, $uv$, $vw$ and $wu$, corresponding to missing 
\red{atoms} in the $u$, $v$ \red{or} $w$ sites respectively.}
\label{Fig_6}
\end{center}
\end{figure*}

\begin{figure*}
\begin{center}
\includegraphics[width=\textwidth,clip=true]{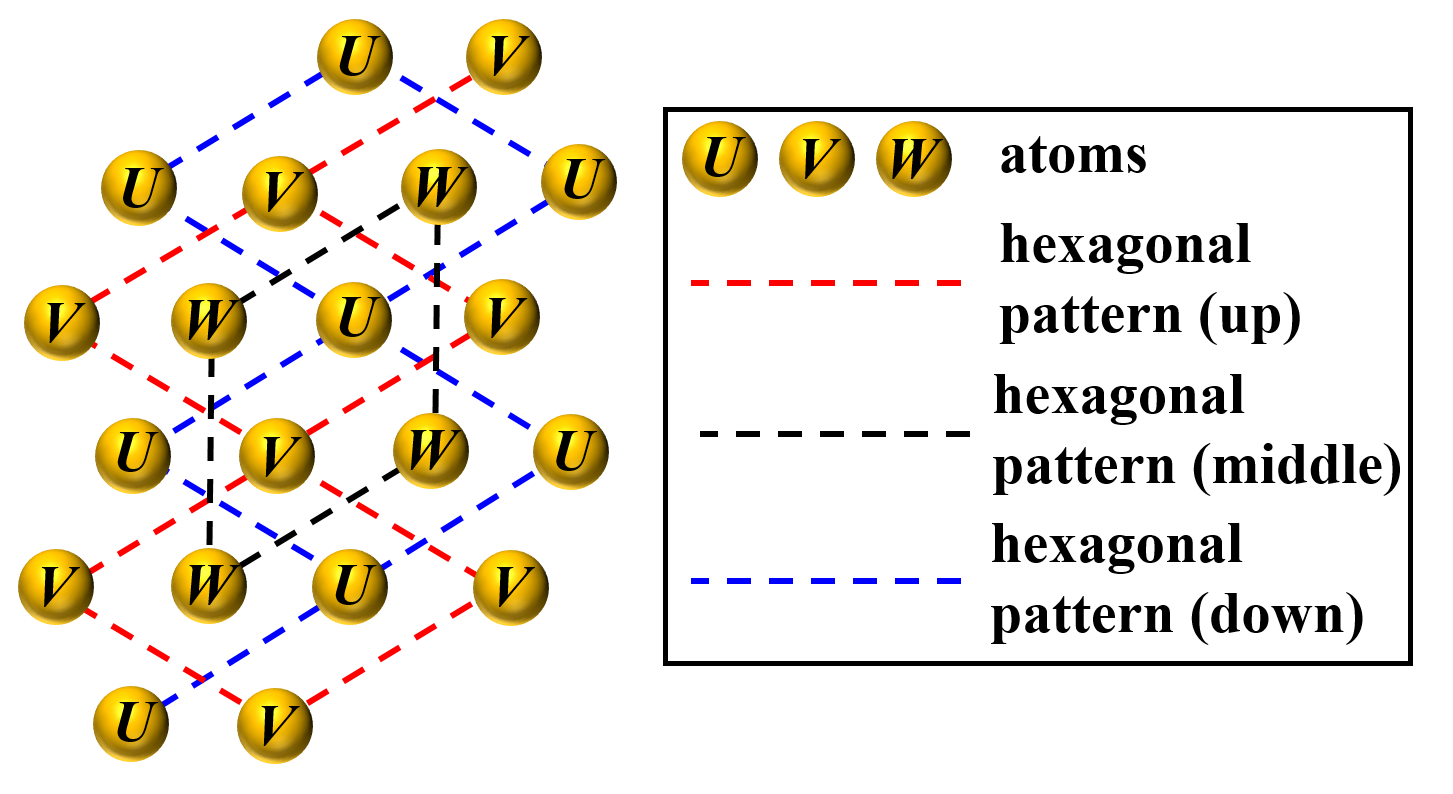}
\caption{The optimised 3D face centred cubic (FCC) packing of atoms labelled as $UVW$. The red, black and blue dashed lines indicate the hexagonal patterns of the up, middle and down lattice of atoms.}
\label{Fig_7}
\end{center}
\end{figure*}

\red{As earlier discussed, the partition function given in eq. (\ref{sphere_packing_eq2}) is optimised when the cations are arranged in a hexagonal pattern. This pattern differs from the honeycomb lattice whereby the centres of the honeycomb pattern are occupied by cations.}  
It is prudent to distinguish the optimised 2D hexagonal pattern from the honeycomb pattern by labelling the three 
\red{atomic} sites in the hexagonal lattice as $uvw$, analogous to the face-centred cubic (FCC) packing labelling of sites (sphere centres) in 3D, as shown in Figure \ref{Fig_6}. 
\red{In this notation, the 2D hexagonal lattice is indicated by $uvw$, whereas} 
there are three possible configurations for the honeycomb lattice analogous to the hexagonal close packing (HCP), namely, $uv$, $vw$ and $wu$, corresponding to missing cations in the $u$, $v$ 
\red{or} $w$ sites respectively. To completely write down the structure of honeycomb layered materials (
\red{that} optimise the sphere packing problem not only in 2D but also in 3D), we shall use the previously introduced small letters ($u$, $v$ and $w$) to indicate the hexagonal 
\red{atomic} sites that lie 
\red{within} 2D and capital letters ($U$, $V$ and $W$) to indicate 
\red{atomic sites in 3D}, as shown in Figure \ref{Fig_7}. Moreover, we shall introduce a lower case index which corresponds to the chemical elements \textit{e.g.} $W_{\rm Ni}$ 
\red{indicates that} nickel, $\rm Ni$ occupies site $W$. 
\red{Whilst we are dealing with atoms with varying ionic radii, we shall assume congruent spheres which might yield inaccurate separation distances between ions but nonetheless gives the correct locations of atomic sites as sphere centres}. 
\red{Consequently}, making such an approximation allows us to apply the Kepler conjecture, which states that congruent spheres can be packed the most efficiently in 3D either by $UVW$ sequences or $UV$, $VW$ or $WU$ sequences, 
\red{corresponding} to FCC packing and HCP packing respectively. 
\red{The Kepler conjecture} was proved by T. Hales via proof by exhaustion involving the checking of many individual cases using complex computer calculations.\cite{hales2011revision}

\begin{table*}
\caption{
Stacking sequences 
adopted by 
\red{select} honeycomb layered oxides.\cite{kanyolo2022advances, kanyolo2021honeycomb} 
}
\label{Table_2}
\begin{center}
\scalebox{1}{
\begin{tabular}{cc} 
\hline
\textbf{honeycomb layered oxide} & \textbf{stacking sequence}\\
\hline\hline

$\rm Li_3Ni_2SbO_6$ & $U_{\rm O}V_{\rm (Ni, Ni, Sb)}W_{\rm O}V_{\rm Li}U_{\rm O}V_{\rm (Ni, Ni, Sb)}U_{\rm O}$\\
$\rm Li_3Co_2SbO_6$ & $U_{\rm O}V_{\rm (Co, Co, Sb)}W_{\rm O}V_{\rm Li}U_{\rm O}V_{\rm (Co, Co, Sb)}U_{\rm O}$\\$\rm Li_3Zn_2SbO_6$ & $U_{\rm O}V_{\rm (Zn, Zn, Sb)}W_{\rm O}V_{\rm Li}U_{\rm O}V_{\rm (Zn, Zn, Sb)}U_{\rm O}$\\
$\rm Li_2Ni_2TeO_6$ & $U_{\rm O}V_{\rm (Ni, Ni, Te)}W_{\rm O}V_{\rm Li}U_{\rm O}V_{\rm (Ni, Ni, Te)}U_{\rm O}$\\
$\rm NaNi_2BiO_6$ & $U_{\rm O}V_{\rm (Ni, Ni, Bi)}W_{\rm O}V_{\rm (Na, Na, -)}U_{\rm O}V_{\rm (Ni, Ni, Bi)}U_{\rm O}$\\
$\rm Na_2Cu_2TeO_6$ & $U_{\rm O}V_{\rm (Cu, Cu, Te)}W_{\rm O}V_{\rm Na}U_{\rm O}V_{\rm (Cu, Cu, Te)}U_{\rm O}$\\
$\rm Na_3Co_2SbO_6$ & $U_{\rm O}V_{\rm (Co, Co, Sb)}W_{\rm O}V_{\rm Na}U_{\rm O}V_{\rm (Co, Co, Sb)}U_{\rm O}$\\
&\\
$\rm Na_2Zn_2TeO_6$ & $U_{\rm O}V_{\rm (Zn, Zn, Te)}W_{\rm O}V_{\rm (Na, Na, -)}W_{\rm O}V_{\rm (Zn, Zn, Te)}U_{\rm O}V_{\rm Na}$\\
$\rm Na_2Co_2TeO_6$ & $U_{\rm O}V_{\rm (Co, Co, Te)}W_{\rm O}V_{\rm Na}W_{\rm O}V_{\rm (Co, Co, Te)}U_{\rm O}V_{\rm Na}$\\
$\rm Na_2Ni_2TeO_6$ & $U_{\rm O}V_{\rm (Ni, Ni, Te)}W_{\rm O}V_{\rm (Na, Na, -)}W_{\rm O}V_{\rm (Ni, Ni, Te)}U_{\rm O}V_{\rm (Na, Na, -)}$\\
$\rm K_2Ni_2TeO_6$ & $U_{\rm O}V_{\rm (Ni, Ni, Te)}W_{\rm O}V_{\rm (K, K, -)}W_{\rm O}V_{\rm (Ni, Ni, Te)}U_{\rm O} V_{\rm (K, K, -)}$\\
&\\
$\rm Ag_3Co_2SbO_6$ & $U_{\rm O}W_{\rm (Co, Co, Sb)}V_{\rm O}V_{\rm Ag} V_{\rm O}U_{\rm (Sb, Co, Co)}W_{\rm O}W_{\rm Ag}W_{\rm O}V_{\rm (Co, Co, Sb)}U_{\rm O}U_{\rm Ag}$\\
$\rm Ag_3Ni_2BiO_6$ & $U_{\rm O}V_{\rm (Ni, Ni, Bi)}W_{\rm O}W_{\rm Ag}W_{\rm O}V_{\rm (Ni, Ni, Bi)}U_{\rm O}$\\
\hline
\end{tabular}}
\end{center}
\end{table*}

\begin{figure}
\begin{center}
\includegraphics[width=0.8\columnwidth,clip=true]{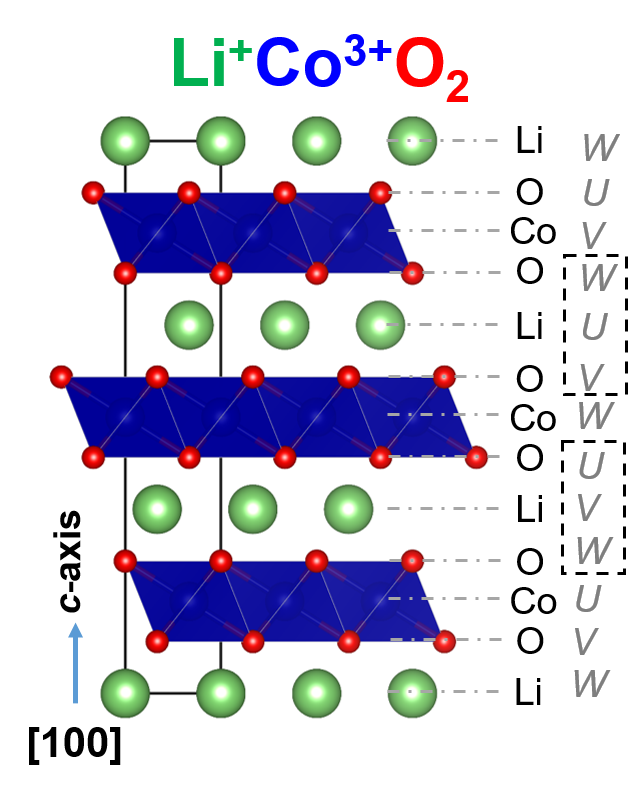}
\caption{Crystal structure of $\rm LiCoO_2$ (rhombohedral polytype) visualised along [100] projection showing the arrangement of constituent atoms. Li atoms are shown in green, 
\red{polyhedra enclosing Co} in blue, and O in red. The unit cell is shown 
\red{as the black solid rectangle}, whilst the sequences indicating octahedral coordination of Li to O atoms, $W_{\rm O}U_{\rm Li}V_{\rm O}$ and $U_{\rm O}V_{\rm Li}W_{\rm O}$ are enclosed by the black dashed rectangle.}
\label{Fig_LiCoO2}
\end{center}
\end{figure}

\red{For illustration purposes, we first consider the case for $\rm Li^{1+}Co^{3+}O_2^{2-}$, which alongside $\rm Li_2CO_3$ and $\rm CoCO_3$ or $\rm Co_3O_4$ can be used as a starting reagent for the preparation of these materials}.\cite{mizushima1980} Considering the rhombohedral polytype in our FCC notation, the $\rm [CoO_2]^{1-}$ octahedral slabs have either the notation $U_{\rm O}V_{\rm Co}W_{\rm O}$, $V_{\rm O}W_{\rm Co}U_{\rm O}$ or $W_{\rm O}U_{\rm Co}V_{\rm O}$ which signifies the locations of each atom. 
\red{The 3D lattice is arranged as},
\begin{align}\label{LiCoO_2_FCC_eq}
    U_{\rm O}V_{\rm Co}W_{\rm O}U_{\rm Li}V_{\rm O}W_{\rm Co}U_{\rm O}V_{\rm Li}W_{\rm O}U_{\rm Co}V_{\rm O}W_{\rm Li}U_{\rm O}, 
\end{align}
as 
\red{illustrated} in Figure \ref{Fig_LiCoO2}, where 
\red{the pattern is repeated to form the material}. The utility of this FCC notation is that the coordination of O with the cations as well as information about the location of ions in the 2D lattice layers is already 
\red{implicit}. 
\red{Herein, we shall be concerned with typical honeycomb layered oxides which exhibit linear, octahedral and prismatic coordination of cations to oxygen atoms}, as displayed in Table \ref{Table_2}. The 2D lattices are given by the notation $U = uvw, V = uvw, W = uvw$ which corresponds to the 2D hexagonal lattice pattern. 
\red{In other words, the capital letters are shorthand notations for, \textit{e.g.} $V_{\rm Co} = u_{\rm Co}v_{\rm Co}w_{\rm Co}$, which are the three independent sites of the 2D hexagonal cobalt lattice}. Moreover, whenever one or two of the hexagonal lattice sites are occupied by a different atom than the other(s), for instance in $A_2\rm Ni_2TeO_6$ ($A =$ Na, K) whereby the slabs are formed by either Ni and Te octahedrally coordinated to O, we shall employ the notation $U_{\rm NiNiTe} = u_{\rm Ni}v_{\rm Ni}w_{\rm Te}$, $V = u_{\rm Ni}v_{\rm Ni}w_{\rm Te}$ or $W = u_{\rm Ni}v_{\rm Ni}w_{\rm Te}$, which indicates 
\red{which indicates the 2D lattice of Ni is honeycomb with Te at the centres}. Lastly, whenever an atom is missing in any one of the 2D hexagonal sites, we shall use a hyphen \textit{e.g.} $U_{(A, A, -)} = u_Av_A$, $U_{(-, A, A)} = v_Aw_A$ or $U_{(A, -, A)} = u_Aw_A$ indicating the 2D honeycomb lattice of $A$ atoms. 
\red{Consequently}, in the case of the general chemical formula, $A_aM_mD_d\rm O_6$ where $A$ is a bosonic cation, and $M$ and $D$ are transition metal atoms, the cation coordination to oxygen in,
\begin{subequations}
\begin{align}
    \underbrace{U_{\rm O}V_{(M, M, D)}W_{\rm O}}_{\rm top\,\,slab}\underbrace{W_A}_{\rm cation}\underbrace{V_{\rm O}W_{(M, M, D)}U_{\rm O}}_{\rm bottom\,\,slab},
\end{align}
is tetrahedral. This is because of the coordination sequence, $W_{\rm O}W_AV_{\rm O}$ 
\red{requiring} an $A$ atom 
be sand-witched directly below one O atom and directly above the centre of triangle formed by the three O atoms below it.
\red{Likewise, the cation coordination to oxygen in},
\begin{align}
    U_{\rm O}V_{(M, M, D)}W_{\rm O}W_AW_{\rm O}V_{(M, M, D)}U_{\rm O},\\
    U_{\rm O}V_{(M, M, D)}W_{\rm O}V_AU_{\rm O}V_{(M, M, D)}W_{\rm O},\\
    U_{\rm O}V_{(M, M, D)}W_{\rm O}V_AW_{\rm O}V_{(M, M, D)}U_{\rm O},
\end{align}
\end{subequations}
is linear/dumbbell ($W_{\rm O}W_AW_{\rm O}$), octahedral ($W_{\rm O}V_AU_{\rm O}$) and prismatic ($W_{\rm O}V_AW_{\rm O}$) respectively, whereas $M$ and $D$ atoms are octahedrally coordinated to oxygen atoms where $V_{(M, M, D)} = u_Mv_Mw_D$ indicates that $M$ are arranged in a 2D honeycomb lattice with the sites at the centres of the honeycomb pattern occupied by $D$ atoms. Likewise, all the other subscripts \textit{e.g.} $V_A = V_{(A, A, A)} = u_Av_Aw_A$ are written in this short hand. 
\red{Thus, we shall only indicate the subscript components where the 2D lattice is not hexagonal \textit{e.g.} in $\rm K_2Ni_2TeO_6$ where}, 
\begin{align}
    U_{\rm O}V_{\rm (Ni, Ni, Te)}W_{\rm O}V_{\rm (K, K, -)}W_{\rm O}V_{\rm (Ni, Ni, Te)}U_{\rm O},
\end{align}
with $V_{\rm (Ni, Ni, Te)} = u_{\rm Ni}v_{\rm Ni}w_{\rm Te}$ the honeycomb lattice of Ni atoms whose honeycomb centres are occupied by Te atoms and $V_{\rm (K, K, -)} = u_{\rm K}v_{\rm K}$ the 2D honeycomb lattice of K atoms. Moreover, as per our FCC notation, 
\red{Ni and Te} are octahedrally coordinated to oxygen whilst K are prismatically coordinated. \red{A summary of selected honeycomb layered oxides and their FCC notation has been availed in Table \ref{Table_2}}.

Since the 
\red{relevant diffusion theory} for the $A$ cations in $A_aM_mD_d\rm O_6$ is 
\red{characterised} by the idealised model, our sphere packing treatment suggests the partition function is given by eq. (\ref{sphere_packing_eq}), and by extension (\ref{sphere_packing_eq2}), 
\red{even for cases such as $\rm K_2Ni_2TeO_6$ where the lattice of cations is honeycomb}, implying we have to consider additional perturbative interactions not captured by the partition function. For instance, for $A_aM_mD_d\rm O_6$ systems with $M_mD_d\rm O_6$ octahedra forming the slabs and a prismatic coordination of $A$ atoms to oxygen atoms, defining the valency of $M$ and $D$ respectively as $\mathcal{V}_M$ and $\mathcal{V}_D$, an 
\red{empirical observation obtains the inequality},
\begin{subequations}\label{inequality_eq}
\begin{align}
    m\times \mathcal{V}_M + d\times \mathcal{V}_D > 9,
\end{align}
such as in $A^{1+}_2\rm Ni^{2+}_2Te^{6+}O^{2-}_6$ ($A$ = Na, K), appears to render cationic sites directly below and above the $D$ atoms unfavourable for occupation due to high electrostatic repulsion
\red{, thus producing a honeycomb pattern instead of the expected hexagonal lattice of cations}. 
 
\red{Meanwhile}, the hexagonal lattice is the optimised choice when,
\begin{align}
    m\times \mathcal{V}_M + d\times \mathcal{V}_D = 9,
\end{align}
\end{subequations}
such as $\rm LiCoO_2 = Li_3^+Co_3^{3+}O_6^{-2}$ given in eq. (\ref{LiCoO_2_FCC_eq}) and $\rm A^{1+}_3Ni^{2+}_2Bi^{5+}O^{2-}_6$ ($A =$ Na, K) given \red{in our FCC notation} by, 
\begin{align}
     U_{\rm O}V_{\rm (Ni, Ni, Bi)}W_{\rm O}V_AW_{\rm O}V_{\rm (Ni, Ni, Bi)}U_{\rm O}. 
\end{align}
For instance, the validity of eq. (\ref{inequality_eq}) can be checked 
whereby, $\rm Na_3^{1+}Ni_2^{2+}Bi^{5+}O_6^{2-}$ \red{(intercalated)} has Na atoms in a hexagonal lattice, whilst $\rm Na^{1+}Ni_2^{3+}Bi^{5+}O_6^{2-}$ \red{(de-intercalated)} has Na atoms arranged in a honeycomb lattice.\cite{seibel2013structure, bhange2017, seibel2014} 
\red{Consequently}, 
higher order interactions \red{such as (de-)intercalation processes} of the lattice of cations with the atoms in the slab 
are expected to 
\red{somewhat} affect the observed lattice patterns, perturbing the stable configurations from the discussed optimised cases that saturate the sphere packing linear programming bound.\cite{hartman2019sphere, cohn2014sphere, cohn2017sphere} 

Since these perturbations in $A_aM_mD_d\rm O_6$ can not only include changes in valence states of $A$ and $M$ 
\red{ions}, but also introduce lattice 
\red{shear transformations} and distortions, 
\red{we shall refer to the mechanism responsible for these higher order interactions} as \textit{Jenga} mechanism\cite{kanyolo2021honeycomb, masese2021math}, 
\red{in analogy with} the rearrangement of slabs in the popular game of the same name.\cite{walsh2005timeless} 
\red{Excluding the mixed alkali and other hybrids, only the hexagonal and honeycomb monolayers of cations have been observed to date
especially for the linear and prismatic coordinations to oxygen atoms}, suggesting some underlying universality of the valence bond and 
\red{conformal field theories governing the formation and stability of the cationic lattices}.\cite{odor2004universality} Moreover, unlike the carbon atoms in a honeycomb lattice of graphene\cite{mecklenburg2011spin, georgi2017tuning}, the cations in honeycomb layered oxides can be mobilised when a relatively low activation energy of $< 1$ eV (Li, Na, K) per cation is available\cite{sau2022insights, matsubara2020magnetism}, suggesting a more elaborate charge transport theory 
\red{for} positive ions, compared to the electron transport in graphene \red{which is} restricted to localised carbon atoms, \textit{albeit} both lattices expected to share particular properties such as pseudo-spin \red{inherited from} 
the honeycomb lattice.\cite{kanyolo2022cationic, mecklenburg2011spin, georgi2017tuning}

\red{Such universality has motivated the reformulation of the 2D 
molecular dynamics of cations in terms of an idealised model, whereby the (de-)intercalation process of a honeycomb layered oxide cathode is captured by Liouville CFT with $c = 1$, 
corresponding to the aforementioned spinless
modular bootstrap for CFT in the $d = 2c$ sphere packing problem.}\cite{kanyolo2020idealised, kanyolo2022cationic, nakayama2004liouville, polchinski1998string2, zamolodchikov1996conformal, afkhami2020high, hartman2019sphere} In particular, due to charge conservation, each extracted cation 
\red{creates} a neutral vacancy, 
\red{\textit{albeit}} with a pseudo-magnetic moment, at each cationic site 
during the de-intercalation process, whereby the charged cation acquires an Aharanov-Casher phase as it diffuses around the vacancies, along the honeycomb pathways shown in Figure \ref{Fig_2}(b).\cite{aharonov1984topological} Consequently, these vacancies correspond to $\nu$ number of topological defects, where $\nu$ is the first Chern number. As a result, these defects can be treated as topological charges satisfying the Poincar\'{e}-Hopf theorem, where the 
\red{number density of the vacancies (proportional to the charge density of the cations)} 
\red{corresponds} to the Gaussian curvature of an emergent 
2D closed manifold of genus $g = \nu$ with a \red{conformal} metric, $dt^2 = \exp(2\Phi)(dx^2 + dy^2)$ where $\Phi$ is a potential satisfying Liouville's equation.\cite{nakayama2004liouville, polchinski1998string2, zamolodchikov1996conformal} Thus, the quantum state with no vacancies ($g = 1$) 
\red{corresponds to} the 2-torus, invariant under the operation of the generators, $S$ and $T$ of the modular group, $\rm PSL_2(\mathbb{Z})$ as expected.\cite{kanyolo2022cationic, cohen2017modular} 

\begin{figure*}
\begin{center}
\includegraphics[width=0.9\textwidth,clip=true]{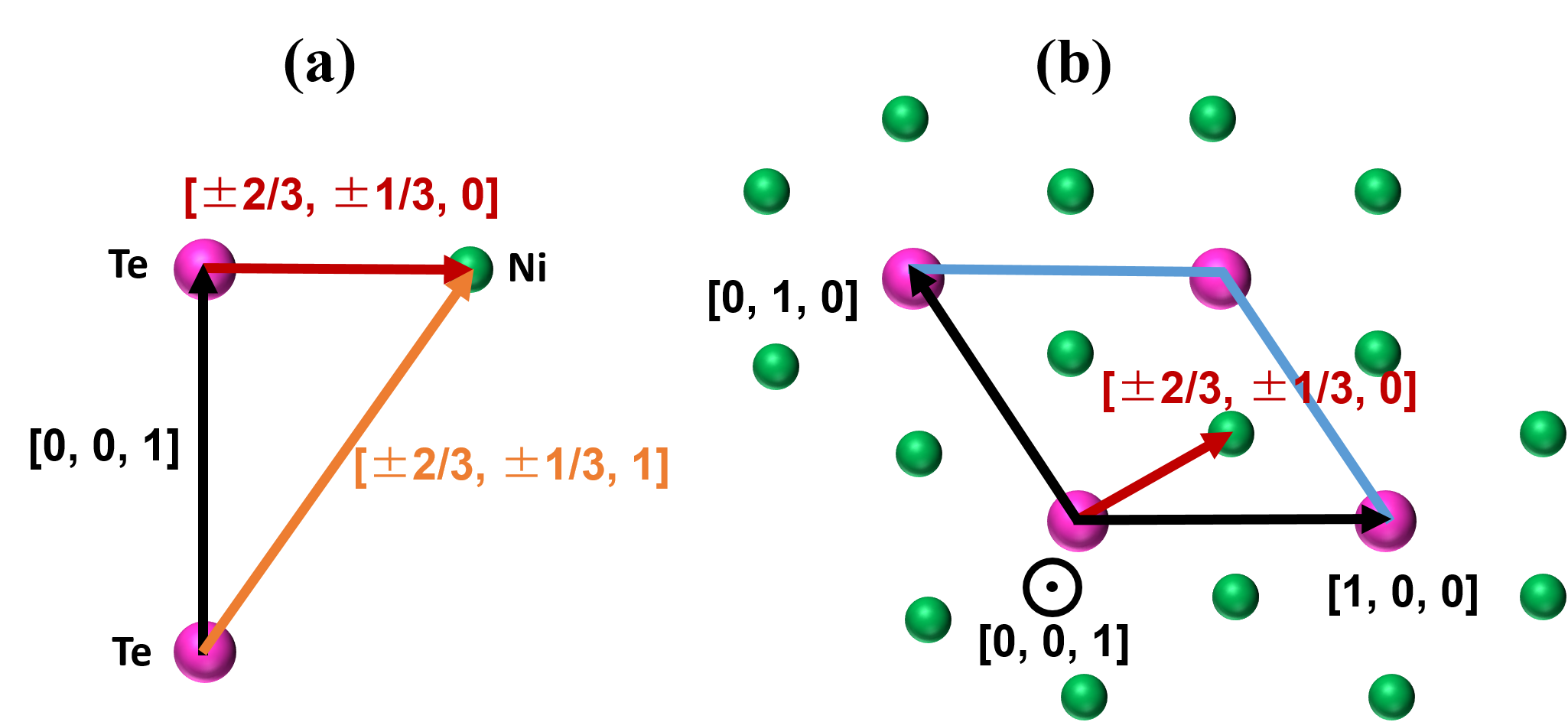}
\caption{The basis vectors defined on the unit cell relative to the shear transformation \red{in eq. (\ref{burgers_eq})}. (a) The {\it c}-zone axis unit basis vector $[0, 0, 1]$, its transformed vector $[\pm 2/3, \pm 1/3, 1]$ under shear transformation, $S$ and the corresponding Burgers vector $[\pm 2/3, \pm 1/3, 0]$ depicting the Te ion occupying the relative position of Ni. (b) The shear transformation in (a) as viewed from the $[001]$ axis ({\it ab} plane) showing the unit basis vectors $[0, 0, 1]$, $[0, 1, 0]$ and $[1, 0, 0]$, and the Burgers vector, $[\pm 2/3, \pm 1/3, 0]$.\cite{masese2021mixed}}
\label{Fig_9}
\end{center}
\end{figure*}

Moreover, the 
\red{classical Liouville CFT} can be recast in terms of 
\red{Einstein's theory of general relativity in ($1+3$)D} by imposing a space-like Killing vector along one of the 
\red{spacial} directions perpendicular to the honeycomb lattice as well as a time-like Killing vector\cite{kanyolo2020idealised}, with the resultant field equations having found applications in quantum black hole information theory.\cite{kanyolo2022local} \red{Thus,} the gravity field equations can be derived from 
\red{the typical Einstein-Hilbert action} with a torsion-free connection\cite{thorne2000gravitation}, $\Gamma_{\,\,\mu\nu}^{\rho} = \Gamma_{\,\,\nu\mu}^{\rho}$, where the topological defects 
\red{are} non-vanishing 
\red{under} a torsion-free manifold 
\red{\textit{albeit} with} the Gauss-Bonnet term 
also present 
\red{in the action}.\cite{lovelock1971einstein} 
Thus, cationic diffusion in honeycomb layered oxides effectively serve as vital testing grounds for theories of $d = 2, 3$, $(d + 1)D$ 
\red{emergent} gravity.\cite{gross1991two, holz1988geometry}

\red{Challenges} describing the intricate (de-)intercalation processes within the context of the idealised model 
\red{exist, since} \textit{Jenga} mechanism 
\red{not only includes topological effects but also stress and strain effects}. 
\red{Nonetheless,} the Einstein Field Equations given in eq. (\ref{EFE_eq}) 
\red{offer a remedy since we can include an appropriate} trace-less stress-energy-momentum tensor 
in eq. (\ref{EFE_eq}), which should yield a metric that effectively characterises these processes, whilst topological effects will 
\red{be captured as before} by $T^{\mu\nu} = |\Psi|^2u^{\mu}u^{\nu}$ term. However, since this approach is far beyond the 
\red{intended difficulty level} of this work, we 
\red{shall give an exemplar of a shear transformation that can affect the observed pattern of the cationic lattice in honeycomb layered oxides.}

\red{For instance}, in the case of the mixed alkali ${\rm K}_{2 - a}{\rm Na}_a\rm Ni_2^{2+}Te^{6+}O^{2-}_6$ or ${\rm K}_{2 - a}{\rm Na}_a\rm Ni_2^{3+}Te^{4+}O^{2-}_6$ with $m\times \mathcal{V}_{\rm Ni} + d\times \mathcal{V}_{\rm Te} = 10 > 9$ ($0 \leq a < 2$), 
\red{where} the crystal structure 
\red{consists of} 
\red{alternating monolayers} of Na and K lattices, the Na lattice was found to be hexagonal, which differs from the K lattice 
\red{that} retains its 
\red{expected} honeycomb pattern.\cite{masese2021mixed, berthelot2021stacking} 
\red{The differing lattice of Na from K} is attributed to the appearance of an edge dislocation which exchanges the relative position of Ni and Te along a Burgers vector, $S\hat{z} - \hat{z} = [\pm 2/3, \pm 1/2, 0]^{\rm T}$ as shown in Figure \ref{Fig_9}, implemented by the shear matrix,
\begin{align}\label{burgers_eq}
S = \begin{pmatrix}
1 & 0 & \pm 2/3\\ 
0 & 1 & \pm 1/3\\ 
0 & 0 & 1
\end{pmatrix},
\end{align}
acting on the unit vector pointing in the $z$ direction given by the transpose, $\hat{z}^{\rm T} = [001]$,
\red{thus lowering} the Te-Te' Coulomb repulsion (by increasing the Te-Te' separation distance) across the interlayers containing Na cations. In our FCC notation, the mixed alkali ${\rm K}_{2 - a}{\rm Na}_a \rm Ni_2TeO_6$ is written as, 
\begin{multline}
     U_{\rm O}V_{\rm (Ni, Ni, Te)}W_{\rm O}V_{\rm (K, K, -)}W_{\rm O}V_{\rm (Ni, Ni, Te)}U_{\rm O}V_{\rm Na}\times\\
     U_{\rm O}V_{\rm (Ni, Te, Ni)}W_{\rm O}V_{\rm (K, -, K)}W_{\rm O}V_{\rm (Ni, Te, Ni)}U_{\rm O}V_{\rm Na}U_{\rm O}. 
\end{multline}
Since Na has a smaller ionic radius compared to K, 
\red{in order to form stable structures, the crystal achieves more stability by distortion than by the exclusion of Na by Te-Te'}. 
\red{However}, under cell cycling, the Na-rich mixed alkali state, $\rm Na_2Ni_2TeO_6$ with $a = 2$ maintains the hexagonal lattice even for cases where the Burgers vectors was 
\red{absent}.\cite{masese2021mixed} In short, 
\red{one obtains both},
\begin{subequations}
\begin{multline}\label{Burgers_Na_eq}
    U_{\rm O}V_{\rm (Ni, Ni, Te)}W_{\rm O}V_{\rm Na}W_{\rm O}V_{\rm (Ni, Ni, Te)}U_{\rm O}V_{\rm Na}\times\\
     U_{\rm O}V_{\rm (Ni, Te, Ni)}W_{\rm O}V_{\rm Na}W_{\rm O}V_{\rm (Ni, Te, Ni)}U_{\rm O}V_{\rm Na}U_{\rm O}, 
\end{multline}
and, 
\begin{align}\label{No_Burgers_Na_eq}
    U_{\rm O}V_{\rm (Ni, Ni, Te)}W_{\rm O}V_{\rm Na}W_{\rm O}V_{\rm (Ni, Ni, Te)}U_{\rm O}.
\end{align}
\end{subequations}
\red{Nonetheless}, it is expected that eq. (\ref{Burgers_Na_eq}) with the Burgers vector, $[\pm 2/3, \pm 1/3, 0]$ is the more stable structure, whereby eq. (\ref{No_Burgers_Na_eq}) transforms into eq. (\ref{Burgers_Na_eq}) (\textit{i.e.} $V_{\rm Na} \rightarrow V_{\rm (Na, Na, -)}$) 
\red{during synthesis}.

\subsection{Coinage metal atom lattices (fermionic lattices)}

Lattices of coinage metal atoms in layered materials tend to vastly differ from the typical lattice patterns 
\red{observed in} bosonic lattices. 
\red{Particularly}, coinage metal atoms 
\red{not only} 
\red{exhibit fermionic behaviour}, but also 
some peculiar properties such as pseudo-spin, anionic behaviour, metallophilicity and sub-valency.\cite{kanyolo2022cationic, mecklenburg2011spin, georgi2017tuning, minamikawa2022electron, ho1990photoelectron, dixon1996photoelectron, schneider2005unusual, jansen2008chemistry, derzsi2021ag, kovalevskiy2020uncommon, ahlert2003ag13oso6, jansen1992ag5geo4, jansen1990ag5pb2o6, argay1966redetermination, beesk1981x, bystrom1950crystal, schreyer2002synthesis, wedig2006studies, yoshida2006spin, taniguchi2020butterfly, eguchi2010resonant, johannes2007formation} 
Moreover, a defining characteristic of 
\red{such fermionic lattices} is the unexpected stability of bilayers, whose defining features can be summarised as follows\cite{masese2023honeycomb}:\red{
\begin{enumerate}[(i)]
    \item Stable bonds between like charges of coinage metal atoms due to metallophilic interactions; 
    \item Subvalent states (specifically reported for Ag atoms) in most reported bilayered 
    \green{materials};
    \item Bilayers comprising 
    a bifurcated bipartite honeycomb lattice.
\end{enumerate}
}
Thus, to effectively discuss 
\red{lattices of coinage metal atoms}, we can focus on bilayered 
\green{materials}, 
\red{specifically the relevant factors responsible for the monolayer-bilayer phase transition.} 
\red{Moreover, to guarantee theoretical consistency, it is also imperative to reproduce the topological features already discussed within the context of the idealised model}. This is embarked upon in the succeeding subsections. 

\subsubsection{Metallophilicity}

\red{Metallophilicity refers to a non-covalent interaction between heavy metal atoms which forms bonds with similar strength to hydrogen bonds}.\cite{hunks2002supramolecular, wan2021strong, assadollahzadeh2008comparison, runeberg1999aurophilic, schmidbaur2000aurophilicity} 
Whilst there is no consensus to the origin of such interactions, recent progress particularly from the crystallographic analysis coupled with density functional theory (DFT) and coupled-cluster singles and doubles with perturbative triples (CCSD-T) computations pertaining organometallic complexes suggests $(n + 1)s-nd$ hybridisation results in a Pauli exclusion principle repulsion which is suppressed by $(n + 1)p-nd$ hybridisation.\cite{wan2021strong} Metallophilic interactions can be intra-molecular as well as inter-molecular. However, inter-molecular metallophilicity tends to be weaker, and hence can readily be disrupted by 
\red{solvation}, 
\red{for instance as observed in the disruption of} luminescence in gold(I) nanoparticles, a property attributed to Au-Au' aurophilic interactions.\cite{schmidbaur2000aurophilicity} 
\red{Moreover, the strength of metallophilic interactions \textit{e.g.} in group 11 elements (Cu, Ag, Au) tends to increase with increasing relativistic effects from top to bottom of the group}.\cite{assadollahzadeh2008comparison} 
In subsequent sections, we shall refer to the metallophilicity of the coinage metal atoms as numismophilicity, based on the latin word `numisma' for coin.\cite{vicente1993synthesis} 
We shall focus on Ag-based honeycomb layered materials and the effects such as subvalent states and bilayers\cite{schreyer2002synthesis, yoshida2006spin, beesk1981x, johannes2007formation, taniguchi2020butterfly}, attributed to 
\red{numismophilic} (argentophilic) interactions. 

\subsubsection{Pseudo-spin, isospin and sub-valency}

\red{Despite having equal positive charges, Ag atoms in these compounds form idiosyncratic structural 
\green{materials} with cluster-like agglomerates of conspicuously short $\rm Ag^{+1} - Ag^{+1}$ interatomic distances akin to those of elemental Ag metal, suggestive} of unconventional weak attractive interactions between $d$-orbitals of monovalent Ag atoms ($d^{\rm 10}$-$d^{\rm 10}$ orbital interactions), what is referred to in literature as argentophilic interactions.\cite{jansen1980silberteilstrukturen} This postulation for the origin of weak attractive argentophilic interactions between Ag cations stems from diffuse reflectance spectroscopy measurements performed in a series of Ag-rich ternary oxides, which indicate a special electronic state of $\rm Ag^{+1}$ in the ultraviolet-visible regime.\cite{kohler1985electrical} The unique structural features are accompanied by the formation of an empty orbital band of mainly Ag-$5s$ orbital character near the Fermi level, capable of accomodating additional electrons, which translates to a range of anomalous subvalent states in Ag cations.\cite{schreyer2002synthesis} 
For instance, the sub-valency of Ag ($\rm Ag^{1/2+}$) in $\rm Ag_2NiO_2$ was demonstrated using X-ray absorption spectroscopy, resonant photoemission spectroscopy, magnetic susceptibility measurements and quantum chemical calculations.\cite{schreyer2002synthesis, wedig2006studies, yoshida2006spin, eguchi2010resonant, johannes2007formation} 
\red{The conductivity of such layered 
\green{materials} ranges from metallic behaviour in} 
\red {$\rm Ag_5Pb_2O_6$, $\rm Ag_3O$ and $\rm Ag_2F$} to semiconducting behaviour in 
\red {$\rm Ag_5GeO_4$ and $\rm Ag_5SiO_4$}.\cite{schreyer2002synthesis}

\begin{figure*}
\begin{center}
\includegraphics[width=0.8\textwidth,clip=true]{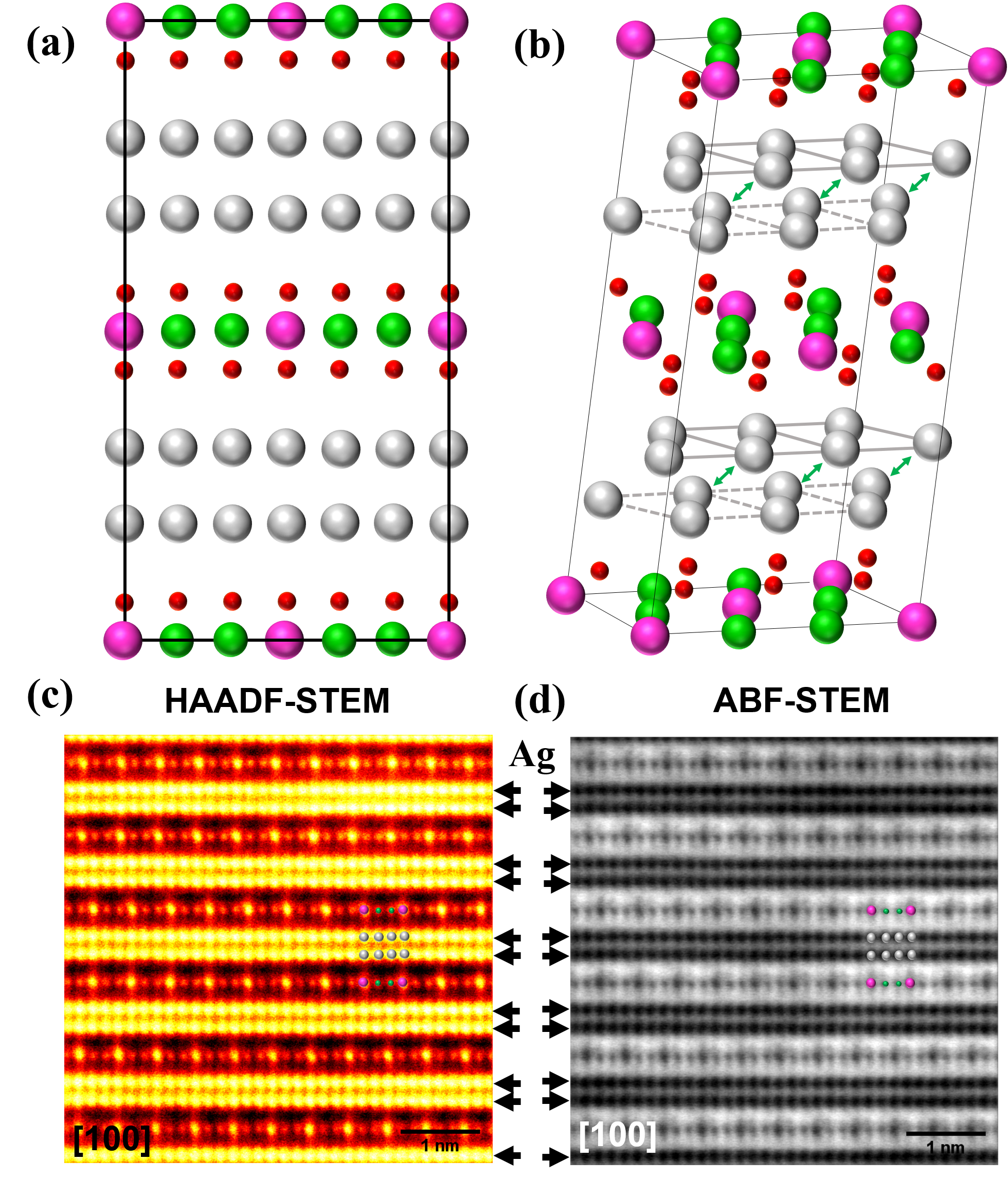}
\caption{
\red{Visualisation of Ag atom bilayers within the Ag-rich crystalline domain (\ce{Ag6Co2TeO6}) of \ce{Ag2Co2TeO6}.}
(a) A unit cell of \ce{Ag6Co2TeO6} showing the arrangement of atoms as viewed in the [100] zone axis. Ag atoms are shown in grey whilst Co and Te atoms are shown in green and pink, respectively. Oxygen atoms are shown in red. (b) A perspective view of the unit cell of \ce{Ag6Co2TeO6} showing the arrangement of the atoms. The Ag layers form two triangular lattices (drawn as dashed grey lines (down) or solid grey lines (up)) comprising a single bilayer. The green double arrows show a few pairs of cations which belong in the same primitive cell, as indicated by the green solid line in Figure \ref{Fig_13}(c). (c) HAADF-STEM image of \ce{Ag6Co2TeO6} taken along [100] zone axis and, (d) the corresponding ABF-STEM image. The contrast ($I$) of the high-angle annular dark-field (HAADF) STEM image is proportional to the atomic number ($Z$) of elements (where $I \propto Z^{1.7} \approx Z^2$). Bilayer planes of Ag atoms ($Z = 47$), marked by the brighter and larger golden spots, are located between the layers of Co atoms ($Z = 27$) represented by the darker amber spots, and Te atoms ($Z = 52$) denoted by the smaller golden spots. As for annular bright-field (ABF) STEM imaging, $I \propto Z^{1/3}$, which means elements such as O ($Z = 8$) can also be discerned. The Te-Co-Co-Te sequential arrangement of the Te atoms and Co atoms, typical in honeycomb layered 
\green{materials}, is succinctly visualised. Ag atoms bilayers (highlighted by arrows) are sandwiched between the honeycomb slabs. An atomistic model has been embedded on the STEM image as a guide. Ag atoms have been shown in blue in the atomistic model for clarity.}
\label{Fig_10}
\end{center}
\end{figure*}

Proceeding, \red{due to the availability of experimental results,} we shall focus on Ag-based layered materials with cations arranged in hexagonal or honeycomb lattices. 
\red{W}e specifically consider the conditions that lead to 
\red{bilayers of} cations in the honeycomb sub-lattices of honeycomb layered tellurates such as ${\rm Ag_2}M_2\rm TeO_6$ (where $M$ is a transition metal 
\red {(such as Co, Ni, Cu and Zn)} or alkaline-earth metal (such as Mg), or a combination of multiple transition metals), layered binary and ternary oxides such as 
\red {${\rm Ag_2}M\rm O_2$ ($M$ = Rh, Mn, Fe, Cu, Ni, Cr, Co) and $\rm Ag_3O$ (or equivalently as $\rm Ag_6O_2$)}, $\rm Ag_3Ni_2O_4$, and layered halides such as $\rm Ag_2F$ \cite{allen2011electronic,schreyer2002synthesis,matsuda2012partially, ji2010orbital,yoshida2020static, yoshida2011novel, yoshida2008unique, yoshida2006spin, masese2023honeycomb,argay1966redetermination, beesk1981x, taniguchi2020butterfly}, which requires features such as pairing of pseudo-spins and conformal symmetry breaking, resulting in a cation monolayer-bilayer phase transition. In this description, the relevant CFT exists at the critical point of the phase transition.\cite{domb2000phase} \red{Occasionally, bilayers of cations can manifest sporadically within otherwise monolayered honeycomb layered frameworks.\cite{masese2023honeycomb}} As an exemplar, 
\red{considering the case of} honeycomb layered $\rm Ag_2Co_2TeO_6$, 
\red{manifesting} a Ag-rich composition of $\rm Ag_6Co_2TeO_6$, whose crystallographic structure and scanning transmission electron microscope (STEM) images are displayed in Figure \ref{Fig_10}.\red {The Ag-rich compositional domain of $\rm Ag_6Co_2TeO_6$ was ascertained explicitly using TEM augmenting energy dispersive X-ray spectroscopy (TEM-EDX).} Figure \ref{Fig_10} (a) depicts a unit cell of \red {$\rm Ag_6Co_2TeO_6$} showing the alignment of the atoms as viewed in the [100] crystallographic axis. Ag atoms are drawn in grey, Te atoms in pink, Ni atoms in green and O atoms in red. In addition, a perspective view of the unit cell is shown in Figure \ref{Fig_10} (b). The Ag layers form two triangular lattices (drawn as dashed grey lines (down) or solid grey lines (up)) comprising a single bilayer. Figure \ref{Fig_10} (c) shows a high-angle annular dark-field (HAADF) STEM image of 
\red {$\rm Ag_6Co_2TeO_6$} crystallite revealing bilayer planes of Ag atoms (marked by the brighter and larger golden spots) located between the layers of Co atoms (represented by the darker amber spots) and Te atoms (denoted by the smaller golden spots). The corresponding annular bright-field (ABF) STEM imaging (shown in Figure \ref{Fig_10} (d)), reveals also the atomic position of O atoms, affirming the atomistic model shown in Figure \ref{Fig_10} (a). The rather confounding result that the Ag cations are arranged in bilayers is understood herein to be reflective of the underlying 
\red{SU($2$)$\times$U($1$) symmetry between the pseudo-spins of the honeycomb lattice and electric charge of the cations forming effective charges and subvalent states}.\cite{masese2023honeycomb, kanyolo2022advances} 

\red{We shall begin by considering carbon hybridisation, making comparisons with graphene-based systems exhibiting the bipartite honeycomb lattice. In the case of graphene-based systems}\cite{allen2010honeycomb, mecklenburg2011spin, georgi2017tuning, kvashnin2014phase}, 
$2sp^2$ hybridisation in carbon with valency $4+$ leads to three $\sigma$-bonds and a leftover $p_z$ orbital electron 
\red{that forms} a $\pi$-bond with an adjacent carbon atom, leading to a trigonal planar geometry. This 
\red{leftover} $p_z^1$ orbital electron is responsible for the rather differing properties of graphene and graphite compared to diamond, whose hybridisation instead is $2sp^3$.\cite{luo2022coherent} Of particular interest is the excellent conduction of carbon atoms in graphene, facilitated by the 
\red{itenerant}
$p_z^1$ orbital electron moving at 
\red{the Fermi velocity} in graphene layers with two helicity states 
\red{that can be associated with} the pseudo-spin degrees of freedom at the Dirac point.\cite{mecklenburg2011spin} Essentially, the bipartite nature of the honeycomb lattice in graphene requires the wavefunction of the conduction electron at the Dirac points to be described by a 2D mass-less Dirac spinor, with each component representing the helicity states known as pseudo-spins.\cite{kanyolo2022cationic, mecklenburg2011spin, georgi2017tuning} An analogous situation 
\red{is thought to occur} for honeycomb lattices of Cu, Ag and Au ($n =$ 3, 4 and 5 respectively) cations with closed $nd^{10}$ and half-filled $(n + 1)s^1$ orbitals.\cite{masese2023honeycomb} Like carbon atoms in graphene, these coinage metal atoms are known to be excellent conductors, whereby the $(n+1)s^1$ valence electrons are delocalised. Due to electrostatic screening of the 
\red{electric charge of the nucleus} and other factors\cite{schwarz2010full}, the $(n + 1)s^1$ orbital energy level is located 
\red{at close proximity to} the degenerate $nd^{10}$ orbitals, which encourages $sd_{z^2}$ hybridisation\cite{aristov1997indirect, ruderman1954indirect, kasuya1956prog, yosida1957magnetic} 
\red{resulting in} two states, $(n + 1)s^2$, $nd_{z^2}^1$ in addition to $(n + 1)s^1$, $nd_{z^2}^2$ state. It is the $nd_{z^2}^1$ orbital that is analogous to the $p_z^1$ of carbon in graphene, responsible for the pseudo-spin degree of freedom in coinage metal atoms.

\red{Electronically, there are three coinage metal atom states, depending on the occupancy of the $nd$ and $(n + 1)s$ orbitals. Due to the odd number of electrons, the neutral atom is a fermion with its spin state inherited from the spin of the valence electron. Due to $sd_{z^2}$ hybridisation, \red{a single spin up or down electron} can either be in the $nd_{z^2}^1$ orbital or the $(n + 1)s^1$ orbital with all the remaining lower energy orbitals fully occupied. Nonetheless, the valency corresponds to the number of electrons in the $ns$ orbital ($ns^2$ or $ns^1$). This results in two valence states, $A^{2+}$ and $A^{1+}$ ($A = \rm Cu, Ag, Au$). Moreover, in order to become closed shell in chemical reactions, the coinage metal atom can either be an electron donor with valency $2+, 1+$ as discussed, or a receptor with valency $1-$, whereby the receptor $A^{1-}$ achieves closed shell $nd_{z^2}^2$ and $(n + 1)s^2$ orbitals forming stable bonds. Indeed, this anion state has been observed in coinage metal cluster ions as $\rm Ag_N^{1-} (N \in \mathbb{Z}^+)$\cite{minamikawa2022electron, ho1990photoelectron, dixon1996photoelectron, schneider2005unusual}, whereas the isolated anion state ($N = 1$) is readily observed in compounds such as $\rm CsAu\cdot NH_3$ due to enhanced relativistic effects of Au.\cite{jansen2008chemistry} Thus, the $A^{1+}$ and $A^{1-}$ valence states are related by isospin rotation (SU($2$)) with the isospin given by $I = \mathcal{V}_A/2$ where $\mathcal{V}_A = 1+, 1-$ are the valence states, and $Y = 0$ is the electric charge of the neutral atom. Meanwhile, the $A^{2+}$ state is an isospin singlet with electric charge, $Y = \mathcal{V}_A = 2+$. Nonetheless, these three cation states $A^{2+}, A^{1-}$ and $A^{1+}$ must have an effective charge, $Q = +2, -1$ and $Q = +1$ respectively, obtained by the Gell-Mann–Nishijima formula\cite{zee2010quantum},
\begin{align}\label{Gell-Mann–Nishijima_eq2}
    Q = 2I + Y,
\end{align}
where $Y$ is the U($1$) electric charge (playing the role of hypercharge), $I$ is the $z$-component of SU($2$) pseudo-spin (isospin) and $Q$ is the effective charge}. These states must be treated as independent ions related to each other by $\rm SU(2)\times U(1)$, forming the basis for fractional valent (subvalent) states.\cite{masese2023honeycomb} Moreover, due to $sd$ hybridisation, all these three states are degenerate on the honeycomb lattice. The degeneracy between $\rm Ag^{2+}$ and $\rm Ag^{1-}$ corresponds to right-handed and left-handed chirality states of $\rm Ag$ fermions on the honeycomb lattice, 
\red{distinguishable by their opposite pseudo-spin degrees of freedom}.\cite{masese2023honeycomb}

Based on our discussions above, the simple case of the bilayered $\rm Ag_2NiO_2$, which requires the existence of the subvalent state $\rm Ag^{1/2+}$ to be electronically neutral is replaced by $\rm Ag_2NiO_2 = Ag^{2+}Ag^{1-}Ni^{3+}O_2^{2-}$ instead, which already implies bifurcation of the honeycomb lattice. Hybrids with a stable honeycomb monolayer and hexagonal bilayer arranged along the [001] plane in an alternating fashion can also be explained, $\rm Ag_3Ni_2O_4 = \frac{1}{2}(Ag_2^{1+}Ni_2^{3+}O_4^{2-})(Ag_2^{2+}Ag_2^{1-}Ni_2^{3+}O_4^{2-})$ with a subvalent state, $\rm Ag^{2/3+}$. Likewise, the Ag lattices of $\rm Ag_2F = Ag^{2+}Ag^{1-}F^{1-}$, $\rm Ag_4Co_2TeO_6 = Ag_2^{2+}Ag_2^{1-}Co_2^{2+}Te^{6+}O_6^{-2}\,\,or\,\,Ag_2^{2+}Ag_2^{1-}Co_2^{3+}Te^{4+}O_6^{-2}$, $\rm Ag_6Co_2TeO_6$ = $\rm Ag_3^{2+}Ag_3^{1-}Co_2^{2+}Te^{5+}O_6^{-2}$ must be hexagonal bilayers. Here, ${\rm Te^{5+}}$ has a propensity to disproportionate into mixed valency states of ${\rm Te^{6+}}$ and ${\rm Te^{4+}}$, as reported in $\rm Ag_2Te_2O_6$ ($\rm Ag_2Te^{6+}Te^{4+}O_6$) and $\rm Te_2O_5$ ($\rm Te^{6+}Te^{4+}O_5$), amongst others.\cite {siritanon2009, lindqvist1973, loopstra1986, minimol2005, barrier2006, klein2005neue}

\red{Thus, summarising the possible fractional subvalent states of Ag in these materials is a matter of considering the various ratios of coinage metal atoms in the possible lattices. In this case, the lattice with $1:1$ left-right chiral ($A^{2+}, A^{1-}$) is bilayered (bifurcated honeycomb) with sub-valency $1/2+$, whilst the lattice with $1:1$ left chiral ($A^{1+}, A^{1+}$) is hexagonal with valency $1+$.} 
\red{Moreover}, $sd_{z^2}$ hybridisation tends to occur efficiently whenever the $d_z^2$ orbital is isolated from the rest of the $d^{10}$ orbitals by crystal field splitting. Thus, the bifurcation mechanism is favoured in layered crystal structures whose Ag atoms 
\red{are prismatically or linearly coordinated} to O atoms \red{before bifurcation}, since these systems would have an isolated $nd_{z^2}$ orbital according to crystal field splitting theory.\cite{burns1993mineralogical, ballhausen1963introduction, jager1970crystal, de19902} 
\red{T}he bifurcation of the honeycomb lattice is 
\red{analogous to} Peierls distortion which, \textit{e.g.} in the dimerisation of 
polyacetylene\cite{garcia1992dimerization, peierls1979surprises, peierls1955quantum}, results in a metal-insulator phase transition.\cite{stewart2012evidence} 

Finally, more complicated structures may have different ratios and combinations leading to sub-valence states, $+1/3$ ($\rm 2Ag^{1+}, Ag^{1-}$) or $+4/5$ ($\rm Ag^{2+}, Ag^{1-}, 3Ag^{1+}$), provided $sd_{z^2}$ hybridisation is guaranteed.\cite{pettifor1978theory, lacroix1981density, manh1987electronic, gallagher1983positive, horn1979adsorbate}
In principle, subvalent Ag cations have also been reported in Ag-rich oxide compositions such as 
\red {$\rm Ag_{16}B_4O_{10}$, $\rm Ag_3O$, $\rm Ag_{13}OsO_6$, $\rm Ag_5Pb_2O_6$, $\rm Ag_5GeO_4$, $\rm Ag_5SiO_4$,} the halides such as $\rm Ag_2F$, and the theoretically predicted $\rm Ag_6Cl_4$.\cite{derzsi2021ag, kovalevskiy2020uncommon, ahlert2003ag13oso6, jansen1992ag5geo4, jansen1990ag5pb2o6, argay1966redetermination, beesk1981x, bystrom1950crystal} Based on all the aforementioned theoretical advances in understanding the nature of bilayers in silver-based layered materials with honeycomb lattices, a \red{fairly} complete 
\red{treatise} that tackles their characterisation 
\red{has been presented in the succeeding sections.} 
 
\subsubsection{1D Ising model and bilayers}

\begin{figure*}
\begin{center}
\includegraphics[width=\textwidth,clip=true]{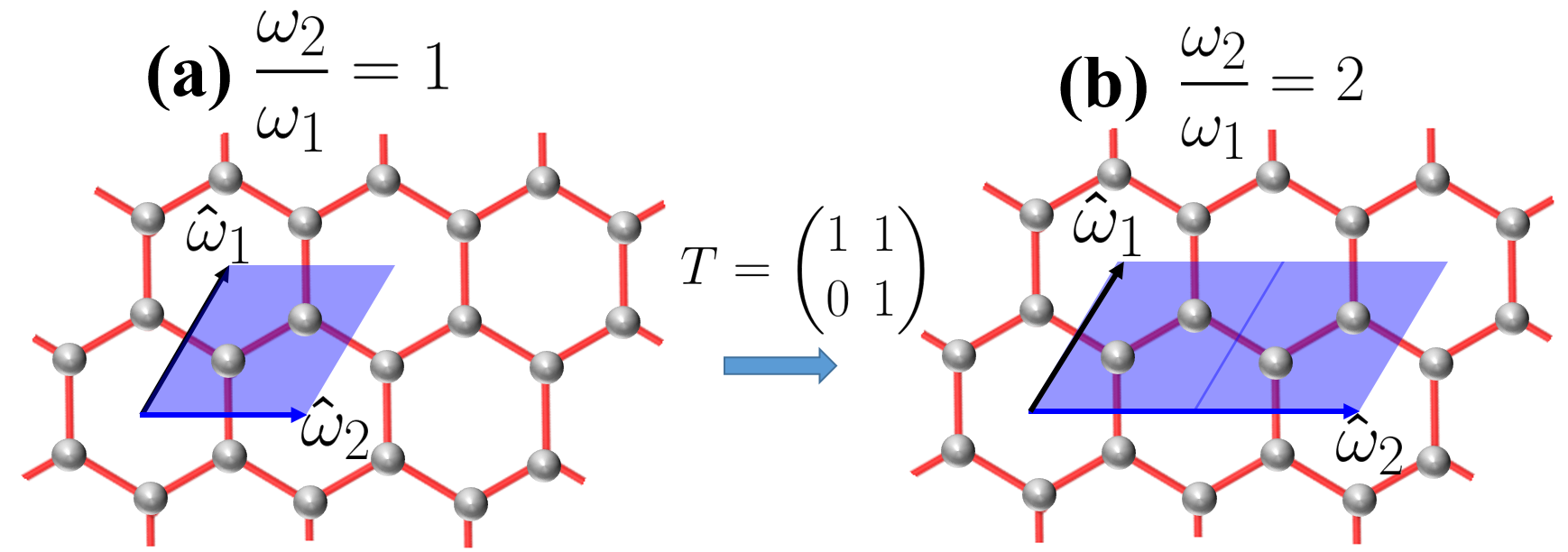}
\caption{The honeycomb lattice of cations depicting the action of the $Q_1 = T$ generator of $\rm SL_2(\mathbb{Z})$ on the primitive vectors. (a) The primitive vectors $\hat{\omega}_1$ and $\hat{\omega}_2$ of the honeycomb lattice, where $\omega_2/\omega_1 = 1$ is the number of pairs of cations/cationic sites enclosed within the primitive cell. (b) $Q_1 = T$ transformation corresponding to the re-scaling of the primitive vector, $\hat{\omega}_2$ and hence an expansion of the unit cell, $\omega_2/\omega_1 = 2$.
\red{The honeycomb lattice of cations depicting the action of the $Q_1 = T$ generator of $\rm SL_2(\mathbb{Z})$ on the primitive vectors. (a) The primitive vectors $\hat{\omega}_1$ and $\hat{\omega}_2$ of the honeycomb lattice, where $\omega_2/\omega_1 = 1$ is the number of pairs of cations/cationic sites enclosed within the primitive cell. (b) $Q_1 = T$ transformation corresponding to the re-scaling of the primitive vector, $\hat{\omega}_2$ and hence an expansion of the unit cell, $\omega_2/\omega_1 = 2$}.}
\label{Fig_11}
\end{center}
\end{figure*}

\begin{figure*}
\begin{center}
\includegraphics[width=\textwidth,clip=true]{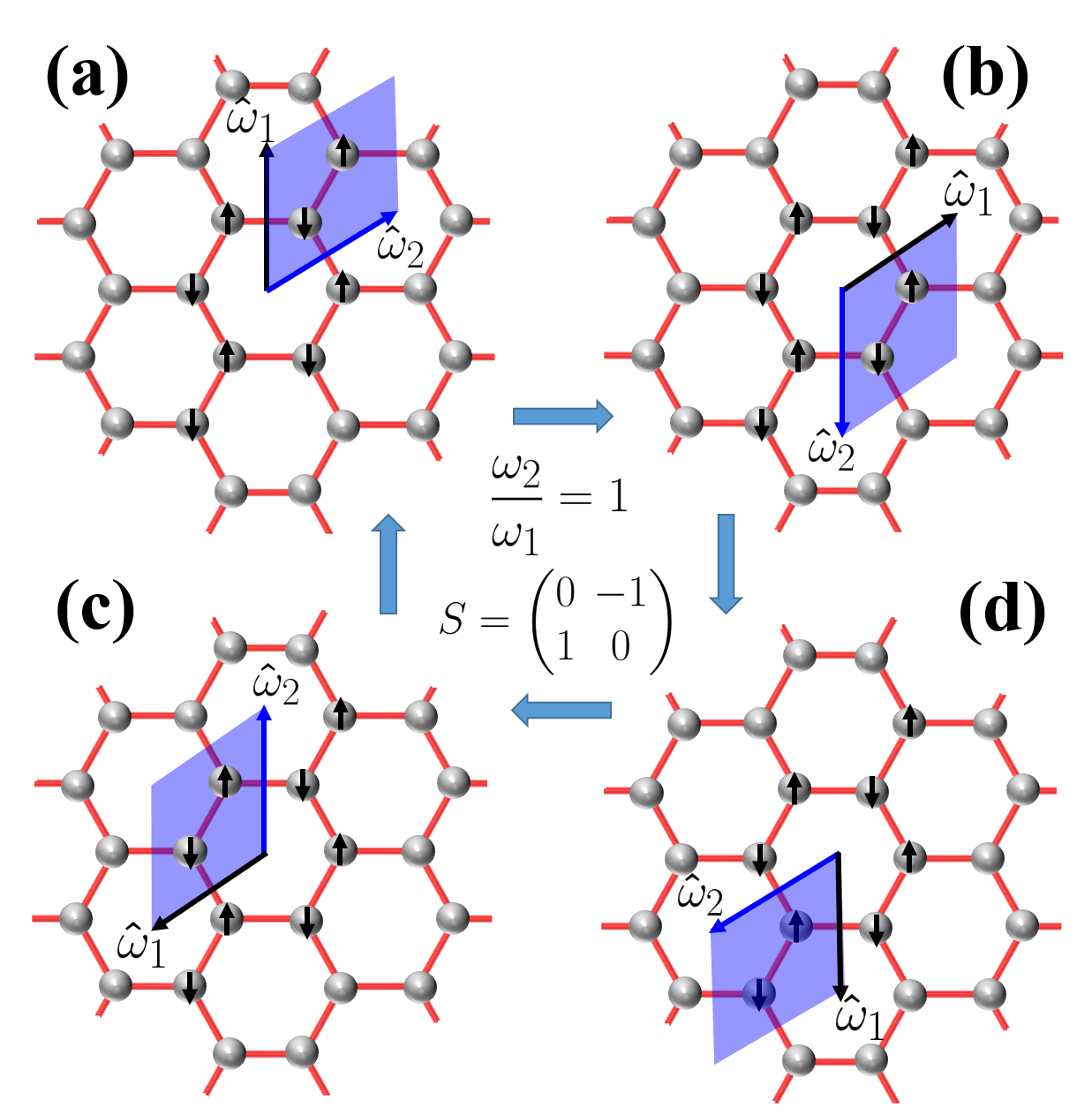}
\caption{The honeycomb lattice of cations and discrete rotations ($\omega_2/\omega_1 = 1$) generated by $Q_2 = S \in \rm SL_2(\mathbb{Z})$ acting on the primitive cell. Some pseudo-spin states of the honeycomb lattice have been included for clarity. The primitive cell is rotated as shown in (a), (b), (c) and (d) by the application of $Q_2 = S$ transformation, such that $Q_2^2 = S^2 = -\sigma_0$ corresponds to inversion of (a), (b) to (d), (c) respectively, where $\sigma_0$ is the $2\times 2$ identity matrix, requiring that $Q_2^4 = \sigma_0$. The inversion corresponds to the honeycomb lattice originally viewed from above to appear viewed from below, equivalent to the exchange of the two pseudo-spin states relative to the primitive basis vectors.}
\label{Fig_12}
\end{center}
\end{figure*}

\red{Proceeding, we shall introduce the basis vectors of the unit cell}, 
\begin{align}\label{basis_vector_eq}
   \hat{\omega}_1 = \omega_1\begin{pmatrix}
1\\0
\end{pmatrix} = \omega_1|\uparrow \,\rangle, \,\,
\hat{\omega_2} = \omega_2\begin{pmatrix}
0\\1
\end{pmatrix} = \omega_2|\downarrow \,\rangle,
\end{align}
which requires that $\omega_1/\omega_2 = 1$. Here, we have introduced a pseudo-spin degree of freedom, corresponding to the unit basis vector. 
\red{Moreover, as already discussed}, there exists discrete/modular symmetry generators $Q_1 = T, Q_2 = S \in Q$ where $Q$ is an element of $\rm PSL_2(\mathbb{Z})$, and, 
\begin{align}
T = \begin{pmatrix}
1 & 1\\ 
0 & 1
\end{pmatrix},\,\,
S = \begin{pmatrix}
0 & -1\\ 
1 & 0
\end{pmatrix},
\end{align}
act on the basis, $\hat{\omega}_1 + \hat{\omega}_2$, which map unit/primitive cells by either discrete re-scaling ($T$) or rotations ($S$) as shown in Figure \ref{Fig_11} and Figure \ref{Fig_12} respectively. 
\red{This emphasises the conformal nature of the ground state of the theory without cationic vacancies}. In particular, discrete expansions are given by, 
\begin{subequations}
\begin{align}\label{Tk_eq}
T^k = \begin{pmatrix}
1 & k\\ 
0 & 1
\end{pmatrix},
\end{align}
where $\omega_1/\omega_2 = k = N/2 \in \mathbb{N}$ is an integer and $N$ is the number of cationic sites within the region bounded by the basis vectors defining the unit cell.

\red{Meanwhile, consider $S$ invariance implemented by $Q_2 \in Q$}. Note that, we have $Q_2^2 = -\sigma_0$, where,
\begin{align}
\sigma_0 = \begin{pmatrix}
1 & 0\\ 
0 & 1
\end{pmatrix},
\end{align}
\end{subequations}
is the identity, which implies the unit vector acquires a minus sign. Since $Q_2$ 
\red{exchanges} the basis vectors, it corresponds to a discrete rotation when acting on the primitive cell. There are 4 such discrete rotations such that $Q_2^{4n} \in Q$ correspond to complete $2\pi n$ rotations of the primitive cell, where $n \in \mathbb{N}$ as shown in Figure \ref{Fig_12}. 
\red{In addition}, $Q_2^{4n + 2} \in Q$ exchanges one cationic site in the primitive cell with the other. 
\red{This gives rise to a pseudo-degree of freedom we shall refer to as pseudo-spin}. Indeed, it is well-known that the 2D honeycomb structure of graphene requires an additional degree of freedom to describe the orbital wave functions sitting in two different triangular sub-lattices, 
known as a pseudo-spin.\cite{mecklenburg2011spin}

\red{Thus, under the exchange of two cations belonging to the same primitive cell, their wave function picks up a minus sign. This implies that, the wave function of the primitive cell is anti-symmetric}, 
\begin{align}
    \psi(\vec{r}_1, \vec{r}_2) = \psi_1(\vec{r}_1)\psi_2(\vec{r}_2) - \psi_2(\vec{r}_1)\psi_1(\vec{r}_2),
\end{align}
with respect to exchange of the positions $r_1$ and $r_2$ of the cations implemented by $S$ transformation, $Q_2^2$,
where $\psi_1$ and $\psi_2$ are the wave functions of the cations in the primitive cell. 
Within our formalism, the sphere packing partition function in eq. (\ref{sphere_packing_eq2}) requires modification 
\red{to},
\begin{multline}\label{pseudo_partition_eq}
    \mathcal{Z} = \sum_j\sum_{\nu = 0}^{\infty}f_{\nu}\frac{\exp(-i\pi b_j\Delta_{\nu})}{\eta^2(b)}\\
    = \sum_{\nu = 0}^{\infty}2f_{\nu}\frac{\cos(\pi b\Delta_{\nu})}{\eta^2(b)},
\end{multline}
by the transformation, $\eta(b_j) \rightarrow \eta(b)$, where $b = -\overline{b} = ik$ as before. 
\red{This modification is performed in order for $\mathcal{Z}$ to be invariant under $Q_2^{2n}$ instead of $Q_2^n$ ($n \in \mathbb{N}$), which can be checked using the Poisson summation formula}. 
\red{We conclude that}, the two partition functions differ in that eq. (\ref{sphere_packing_eq2}) describes 
\red{(bosonic) hexagonal lattice as the most efficient packing} 
whereas eq. (\ref{pseudo_partition_eq}) describes 
\red{(fermionic) honeycomb lattices} as the most efficient packing. 

\red{Explicitly, consider the 1D Ising Hamiltonian}\cite{baxter1982inversion}, 
\begin{align}\label{Ising_eq}
    \mathcal{H}_{\nu} = -\frac{1}{2}\sum_{j'j} J_{jj'} S_j\cdot S_{j'} + \pi E\Delta_{\nu}\sum_j S_j
\end{align}
where $j, j' = 1, 2$, $2S_j = \sigma_z^j$ is the $z$ component of the Pauli vector acting on the $j$-th pseudo-spin state, the Heisenberg matrix term is given by, 
\begin{align}
    J_{jj'} = \begin{pmatrix}
 0 & J(k) \\
 J(k) & 0 \\
 \end{pmatrix},\, J(k) = \pm \beta^{-1}\ln(\mathcal{Z}_{\Phi}(k))
\end{align}
$-\Delta_{\nu} = 2\nu$ is the pseudo-magnetic field\cite{georgi2017tuning} in the $z$-direction interacting with the pseudo-spins, whilst $J(k)$ is equivalent to the 
free energy (up to the $\pm$ sign) of $\Phi$ degrees of freedom representing the exchange interaction between the pseudo-spins.
\red{This exchange interaction can be considered as a pairing interaction due to mass-less phonons in the 2D crystal treated as a real scalar field, $\Phi$ with discrete frequencies.} This 1D Ising model is exactly 
\red{solvable}, where a standard calculation for the partition function yields\cite{grosso2013solid}, 
\begin{multline}
    \mathcal{Z}_{\nu} = {\rm Tr}\left (\exp(-\beta \mathcal{H}_{\nu})\right ) = {\rm Tr}(P)\\
    = \lambda_{\nu}^+ + \lambda_{\nu}^- = 2\exp(\beta J(k))\cosh(\pi k\Delta_{\nu})
\end{multline}
where $\rm Tr$ is the trace over the two pseudo-spins, $S_j = \sigma_j^z$, $\beta E = k$ and $\lambda_{\nu}^{\pm}$ are the eigenvalues of the transfer matrix,
\begin{align}
P = \begin{pmatrix}
\exp(-\beta E_{\uparrow}) & \exp(-\beta E_{\uparrow\downarrow})\\
 \exp(-\beta E_{\downarrow\uparrow}) & \exp(-\beta E_{\downarrow})\\
\end{pmatrix},
\end{align}
given by, 
\begin{multline}
    \lambda_{\nu}^{\pm} = \exp(\beta J)\cosh(\pi k\Delta_{\nu})\\
    \pm \sqrt{\exp(2\beta J)\sinh^2(\pi k\Delta_{\nu}) + \exp(-2\beta J)},
\end{multline}
with,
\begin{align}
   E_{\uparrow\downarrow} = E_{\downarrow\uparrow} = J,
\end{align}
and,
\begin{align*}
    E_{\uparrow}(\nu) = \pi E\Delta_{\nu} - J,\,\, E_{\downarrow}(\nu) = -(\pi E\Delta_{\nu} + J).
\end{align*}
\red{Thus, taking the weighted sum of $\mathcal{Z}_{\nu}$ with probabilities $f_{\nu}$ for different topologies yields eq. (\ref{pseudo_partition_eq}) with $b = ik$, provided we restrict ourselves to the ferromagnetic case with,
\begin{align}\label{J_great_zero_eq}
    J = \beta^{-1}\ln (\mathcal{Z}_{\Phi}(k)) > 0.
\end{align}
} 
\red{The components in the exponent of the transfer matrix correspond to pseudo-spin energy states, with $E_{\uparrow} - E_{\downarrow} = 4\pi E\Delta_{\nu}$ a gapped phase where the honeycomb lattice is expected to bifurcate into bilayers with energies $E_{\downarrow}$ and $E_{\uparrow}$ at finite pseudo-magnetic field, $-\Delta_{\nu} \neq 0$. Meanwhile, $E_{\uparrow\downarrow} = E_{\downarrow\uparrow} = J$ corresponds to the ferromagnetic ($J > 0$) and anti-ferromagnetic ($J < 0$) alignment of the pseudo-spins as usual}.

\red{Magnetisation, $\tilde{m}(\nu)$ and entropy, $S(\nu)$ can be calculated from the expression}, 
\begin{align}
    dF_{\nu} = (S(\nu)/\beta^2)d\beta + \tilde{m}(\nu)\pi Ed\Delta_{\nu},
\end{align}
where $F_{\nu} = -\beta^{-1}\ln(\mathcal{Z}_{\nu})$ is the free energy for the system with $\nu$ vacancies. Specifically focusing on the magnetisation, we find, 
\begin{align}
    \tilde{m} = \left \langle \sum_j S_j \right \rangle = -\frac{1}{\pi k}\frac{\partial Z_{\nu}}{Z_{\nu}\partial \Delta_{\nu}} = -\tanh(\pi k\Delta_{\nu}), 
\end{align}
where we have used $\beta E = k$. 
\red{Thus, since $\tanh(x)$ approaches $\pm 1$ for $x \rightarrow \pm \infty$, when $\Delta_{\nu} \neq 0$ at finite temperature, $k \neq 0$, we have ferromagnetic behaviour, with parallel pseudo-spins}. Anti-parallel ferromagnetic or anti-ferromagnetic behaviour is only possible at finite temperature when $\Delta_{\nu} = \chi(\nu) = 0$, corresponding to the 2-torus. In other words, to avoid pseudo-spin frustration when $-\Delta_{\nu} = 2\nu = 0$, the honeycomb lattice pseudo-spins must be anti-parallel, described by the singlet bound state ($S = 0, \tilde{m} = 0$),
\begin{subequations}\label{singlet_eq}
\begin{align}
    \frac{1}{\sqrt{2}}(|\uparrow\downarrow\rangle - |\uparrow\downarrow\rangle),\\
    \left\langle S_1S_2\right\rangle = -3/4, \,\, \tilde{m} = \left \langle \sum_j S_j \right \rangle = 0,
\end{align}
\end{subequations}
where $J = -\beta^{-1}\ln Z_{\Phi} < 1$, $S_1 = 1/2$, $S_2 = -1/2$, $S(S + 1) = S_1^2 + S_2^2 + 2\langle S_1S_2 \rangle$ and $S_j^2 = |S_j|(|S_j| + 1)$.\cite{blundell2003magnetism} 
\red{However, this is not the condition that reproduces pseudo-spin behaviour depicted by the sphere packing partition function in eq. (\ref{pseudo_partition_eq})}, since it violates eq. (\ref{J_great_zero_eq}). 

\begin{figure*}
\begin{center}
\includegraphics[width=\textwidth,clip=true]{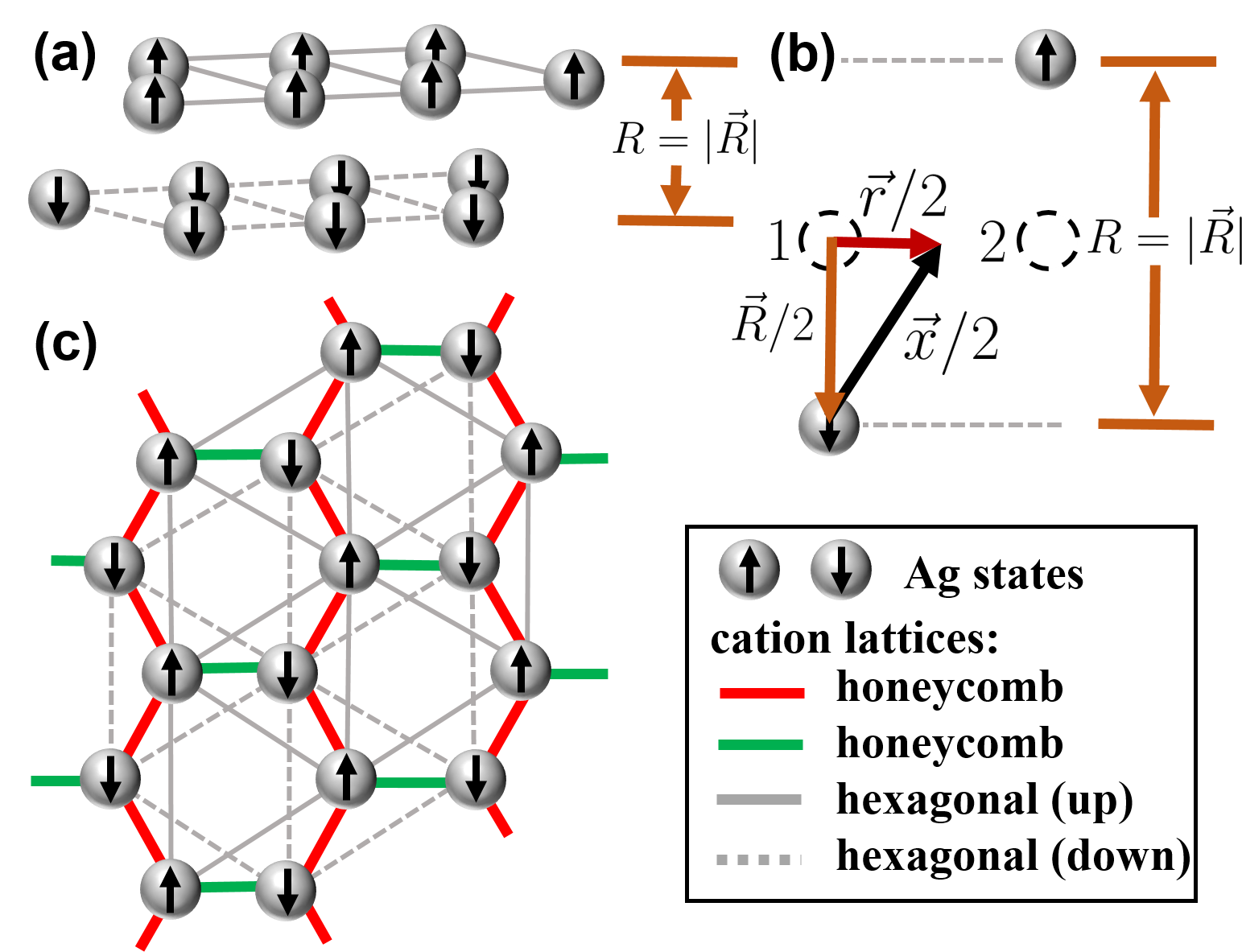}
\caption{Bifurcation of Ag cations in honeycomb layered 
\green{materials}. (a) The bifurcated honeycomb lattice showing pseudo-spin up and down orientations and the displacement, $R = |\vec{R}|$. (b) An illustration of the displacement vector, $\vec{R} = \vec{x} - \vec{r}$. 
\red{The dashed Ag atoms indicates the \red{pre-bifurcation} positions $\vec{r}_1$ and $\vec{r}_2$ ($\vec{r} = \vec{r}_1 - \vec{r}_2$) of the pseudo-spin down and up respectively}. (c) The honeycomb lattice of cations (grey spheres connected by red and green lines), as viewed in the [001] direction. The cation honeycomb lattice comprises a pair of up and down triangular sub-lattices (grey solid and dashed lines respectively), assigning opposite pseudo-spins (indicated by black up or down arrows) to each cation in the unit cell, where the green line indicates part of the honeycomb lattice within a primitive cell.}
\label{Fig_13}
\end{center}
\end{figure*}

Nonetheless, the anti-ferromagnetic case encourages opposite pseudo-spin pairing, 
\red{and instead is interpreted as the pseudo-spin description for} the sphere packing partition function given in eq. (\ref{sphere_packing_eq2}) with bosonic behaviour, where 
\red{the Cooper pairing} Hamiltonian is given by,
\begin{align}\label{Cooper_eq}
   \mathcal{H}_{\rm Cooper} = -\frac{1}{m}\nabla^2 - \mathcal{H}_{\nu}\delta(\vec{r}),
\end{align}
where the first term is the kinetic energy of two pseudo-spin cations where $\vec{\nabla} = \partial/\partial \vec{r}$ and $\vec{r} = \vec{r}_1 - \vec{r}_2$ and the last term is the phonon-mediated delta interaction, arising from the anti-ferromagnetic constraints given in eq. (\ref{singlet_eq}) applied to the Ising Hamiltonian given in eq. (\ref{Ising_eq}). In momentum space, $\vec{k}$ the wave function 
corresponds to,
\begin{align}
    \Psi_{S = 0}^{\tilde{m} = 0}(\vec{k}) = \frac{-3J/4}{\mathcal{E}_{\rm Cooper} - k^2/m}\int_{0}^{k_{\rm F}}\frac{d^{\,2}k'}{(2\pi)^2}\Psi_{S = 0}(\vec{k'}), 
\end{align}
where $k_{\rm F}$ is the Fermi wave vector. Integrating over $\vec{k}$ and crossing out the wave function term from both sides yields, 
\begin{multline}
    -4/3J = \frac{1}{(2\pi)^2}\int_0^{k_{\rm F}} \frac{d^{\,2}k}{\mathcal{E}_{\rm Cooper} - k^2/m}\\
    = \frac{1}{\Omega}\int_{0}^{\mathcal{E}_{\rm F}} \frac{d\mathcal{E}_k}{\mathcal{E}_{\rm Cooper} - 2\mathcal{E}_k}\\
    = -\frac{1}{2\Omega}\ln \left (\frac{\Delta_{\rm Cooper}}{\mathcal{E}_{\rm Cooper}}\right ),
\end{multline}
where $\mathcal{E}_k = k^2/2m$ is the kinetic energy of a single cation, $\Omega = 2\pi/m$ is the 2D constant density of states, $\Delta_{\rm Cooper} = \mathcal{E}_{\rm Cooper} - 2\mathcal{E}_F$ is the 
\red{energy} gap and $\mathcal{E}_{\rm Cooper} \simeq -2\omega_{\rm D}$ with $\omega_{\rm D}$ the Debye frequency. Rearranging, we find the 
\red{energy} gap, 
\begin{align}
    \Delta_{\rm Cooper}^{S = 0, \tilde{m} = 0}(\nu = 0) = -2\omega_{\rm D}\exp\left (8\Omega/3J \right), 
\end{align}
as expected, where $J < 0$. Likewise, the partition in eq. (\ref{pseudo_partition_eq}) corresponds to a bound state (triplet state, $S = 1, \tilde{m} = 0$) with $J > 0$. At the critical point where the system is scale invariant, $\Delta_{\nu} = 0$, 
and the triplet wave function is given by, 
\begin{align}
    \Psi_{S = 1}^{\tilde{m} = 0}(\vec{k}) = \frac{J/4}{\mathcal{E}_{\rm Cooper} - k^2/m}\int_{0}^{k_{\rm F}}\frac{d^{\,2}k'}{(2\pi)^2}\Psi_{S = 1}(\vec{k'}), 
\end{align}
where, 
\begin{subequations}\label{triplet_eq}
\begin{align}
    \frac{1}{\sqrt{2}}(|\uparrow\downarrow\rangle + |\uparrow\downarrow\rangle),\\
    \left\langle S_1S_2\right\rangle = 1/4, \,\, \tilde{m} = \left \langle \sum_j S_j \right \rangle = 0,
\end{align}
\end{subequations}
where $J = \beta^{-1}\ln Z_{\Phi} > 1$, $S_1 = 1/2$, $S_2 = -1/2$, $S(S + 1) = S_1^2 + S_2^2 + 2\langle S_1S_2 \rangle$ and $S_j^2 = |S_j|(|S_j| + 1)$.\cite{blundell2003magnetism} This yields 
\red{an energy} gap given by, 
\begin{align}
    \Delta_{\rm Cooper}^{S = 1, \tilde{m} = 0}(\nu = 0) = -2\omega_{\rm D}\exp(-8\Omega/J),
\end{align}
where $J > 0$. 
\red{Lastly}, the triplet state, $S = 1, \tilde{m} = +1$ is the more interesting case since it leads to a monolayer-bilayer phase transition.\cite{masese2023honeycomb, kanyolo2022advances, kanyolo2022cationic} The pseudo-spins are anti-parallel and hence the magnetisation is finite for a finite pseudo-magnetic field, $-\Delta_{\nu} > 0$ and finite temperature, $k > 0$ where, 
\begin{subequations}\label{triplet_eq2}
\begin{align}
    |\uparrow\uparrow \rangle,\\
    \left\langle S_1S_2\right\rangle = 1/4, \,\, \tilde{m} = \left \langle \sum_j S_j \right \rangle = +1,
\end{align}
\end{subequations}
$J = \beta^{-1}\ln Z_{\Phi} > 1$, $S_1 = 1/2$, $S_2 = 1/2$, $S(S + 1) = S_1^2 + S_2^2 + 2\langle S_1S_2 \rangle$ and $S_j^2 = |S_j|(|S_j| + 1)$.\cite{blundell2003magnetism} 
\red{Thus, the finite pseudo-magnetic field $-\Delta_{\nu} > 0$ lifts the degeneracy with the bound state ($\tilde{m} = 0$, $S = 1$) leading to the bifurcation of the honeycomb lattice into two hexagonal sub-lattices with a cation and a vacancy at each unit cell}. The two sub-lattices are shown in Figure \ref{Fig_13} as grey solid and dashed lines, where each cationic site has a pseudo-spin up or pseudo-spin down degree of freedom whilst the honeycomb lattice of cations is drawn as the red and green solid lines corresponding to the colour coding (same as in Figure \ref{Fig_16}). 

In particular, in each sub-lattice, the bifurcation 
can be interpreted as a unique honeycomb lattice with $\nu$ vacancies and $\nu$ number of cations as shown in Figure \ref{Fig_13}(c), thus forming a 
hexagonal close packing (HCP) with two hexagonal layers of cations. The 
\red{energy} gap is given by, 
\begin{align}\label{energy_gap_eq}
    \Delta_{\rm Cooper}^{S = 1, \tilde{m} = 1}(\nu) = -2\omega_{\rm D}\exp(2\Omega/\mathcal{E}_{\nu}),
\end{align}
where $\mathcal{E}_{\nu} = -|J|/4 + \pi E\Delta_{\nu} \leq -|J|/4$. This means that a finite pseudo-magnetic field corresponding to $\nu \neq 0$ leads to a larger energy gap and is preferred over the 2-torus state, $\Delta_{\nu} = 0$. In other words, the $S = 1, \tilde{m} = 0$ state is unstable. However, in comparison to the singlet state ($S = 0, \tilde{m} = 0$), only the condition, $-\pi E\Delta_{\nu} > |J|/2$ that guarantees $\mathcal{E}_{\nu} < -3|J|/4$ leads to an instability whereby magnetisation $\tilde{m} = 1$ spontaneously arises to stabilise the honeycomb lattice preferring the triplet state ($S = 1$, $\tilde{m} = 1$). 
\red{Thus}, based on 
\red{the conditions for scale invariance given} in previous section, only the flat torus (with a vanishing Gaussian curvature) and the 2-torus with a vanishing Euler characteristic, $\chi = \Delta_{\nu} = 0$ preserve scale invariance and hence conformal invariance. 

\red{Spontaneous} symmetry breaking of scale invariance leads to a finite magnetisation and hence the lifting of 
\red{triplet state degeneracy}, thus lowering the energy of the $\tilde{m} = 1$ state. In other words, there is a `Zeeman splitting' of the $S = 1, \tilde{m} = 0$ and $S = 1, \tilde{m} = 1$ states given by $-\pi E\Delta_{\nu}$. This occurs when the triplet $S = 1, \tilde{m} = 1$ state falls below the singlet $S = 0, \tilde{m} = 0$ state, a condition given by $-\pi E\Delta_{\nu} > |J|/2$. However, since the partition function in eq. (\ref{pseudo_partition_eq}) is only valid for the ferromagnetic case ($J = \beta^{-1} \ln \mathcal{Z}_{\Phi} > 0$) with $S = 1$ as its ground state, this latter condition is not necessary. 

\subsubsection{Distinguishing vacancy creation and bifurcation}

\red{The Ising model introduced above appears to suggest that creating cationic vacancies in the honeycomb lattice ($-\Delta_{\nu} = 2\nu > 0$) is directly responsible for the bifurcation of the honeycomb lattice into its bipartite hexagonal sub-lattices}. Since vacancy creation 
\red{occurs discretely} costing activation energy proportional to the number of vacancies, $\nu$ whilst bifurcation corresponds to 
\red{a phase transition which spontaneously creates a vacancy and a cation at each unit cell}, the finite pseudo-magnetic field responsible for the two processes need to differ 
\red{quantitatively}. In fact, since a bifurcated lattice is the more stable structure, we should expect the activation energy (eq. (\ref{activation_E_eq})) $\mathcal{E} = 2\nu = -\Delta_{\nu} = -\chi(g) = 2g - 2 < 0$ to be negative, whilst for vacancy creation is positive, $\mathcal{E} > 0$. Since $\chi(g)$ is the Euler characteristic, this implies that for bifurcation, we have, $g = \nu + 1 = 0$ ($\chi(g = 0) = 2$, the Euler characteristic of the 2-sphere), 
\red{exploiting} the last remaining degree of freedom with $\nu = -1$, in order to spontaneously create a pseudo-magnetic field. 

Specifically, we are interested in replacing $\Delta_{\nu}$ in eq. (\ref{Ising_eq}) with another suitable scaling dimension, $\Delta_{\nu} \rightarrow \Delta (\nu)$ such that $\Delta(\nu \geq 0) = 0$, and $\Delta(\nu = -1) \neq 0$, thus 
\red{consistently} distinguishing the two scenarios. To identify the nature of $\Delta(\nu)$, it is important to obtain cues from the bifurcation phenomenon, particularly the fact that cationic interactions due to $\mathcal{Z}_{\Phi}$ will 
\red{be 3D, and no longer 2D}. At the critical point of the phase transition, this new scaling dimension 
\red{must also} depend on the 
\red{dimensionality}.
\red{In fact}, the scaling dimension we are interested in is for 
\red{a mass-less scalar conformal field theory}\cite{francesco2012conformal} in $d$D ($\Phi(\Omega \vec{r}) = \Omega^{-\Delta(d)} \Phi(\vec{r})$, $\Omega$ is a 
\red{scaling factor}), given by $\Delta(d) = (d - 2)/2$, which vanishes for $d = 2$, but is finite for $d = 3$ \textit{i.e.} the Liouville action,
\begin{align}\label{Newt_gravity_S_eq}
    S_b = \beta M^{2\Delta^*(d)}\int d^{d}r\left(\vec{\nabla}\Phi\cdot\vec{\nabla}\Phi + K\exp(2\Phi)\right),
\end{align}
where $\Delta^*(d) = (1 - 2\Delta(d))/2$ and $M = b/\beta$ (confer, eq. (\ref{Liouville_action_eq2})), is scale invariant in 2D for $K \propto \Delta_{\nu} \rightarrow \Delta(d)$ , whereas scale invariance is broken in 3D by the second term since $\Delta(d = 3) = 1/2$ ($K \neq 0$). In other words, scale invariance is broken in 3D only when the equations of motion take the form, 
\begin{align}\label{Newt_limit_eq}
    \nabla^2\Phi(\vec{r}) = 4\pi GM\rho(\vec{r})u^{0}(\vec{r})u^{0}(\vec{r}), 
\end{align}
where $K \equiv -R/2 = -4\pi G\rho(\vec{r})$ and $u^{0}(\vec{r}) \equiv \exp(\Phi(\vec{r}))$, introducing a theory reminiscent of Newtonian gravity. This is consistent with the idealised model where space-like Killing vector, $n^{\mu} = (\vec{0}, 1)$ no longer exists, whereas the time-like Killing vector $\xi^{\mu} = (1, \vec{0})$ exists and is related to the four-velocity by $u^{\mu} = \exp(\Phi)\xi^{\mu}$.\cite{kanyolo2020idealised} Consequently, bifurcation defined this way is equivalent to introducing a Newtonian theory of gravity. Thus, whilst 
\red{the replacement of $\Delta_{\nu}$ with $\Delta(d)$}
might 
\red{appear} rather \textit{ad hoc}, it is not only physically well-motivated as discussed, but it also introduces 
\red{a rather intriguing question:}
\textit{what is the 
\red{connection} between $\Delta_{\nu}$ and $\Delta(\nu) = \Delta(d)$}? 

We shall answer this question by 
\red{introducing the \textit{ansatz}}, 
\begin{align}\label{ansatz_eq}
    \langle \langle \Delta_{\nu} \rangle_{\Theta_{\Lambda}}\rangle_{\Gamma(-2g)} = \Delta(d),
\end{align}
where the ordered quantum average is 
\red{carried out using appropriate} partition functions, $\Theta_{\Lambda}$ and $\Gamma(-2g)$ respectively, to be defined below. 
\red{We shall consider the lattice theta function}\cite{olver2010nist},
\begin{subequations}
\begin{multline}\label{theta_eq}
    \Theta_{\Lambda}(b) = \sum_{0 \neq \vec{x} \in \Lambda_d}\exp(i\pi b|\vec{x}|^2)\\
    = \sum_{\nu = 1}^{\infty}f_{\nu}\exp(\pi k \Delta_{\nu})\\
    = \frac{1}{\exp(2\pi k) - 1},
\end{multline}
with $f_{\nu} = 1$, $|\vec{x}|^2/2 = \nu \in \mathbb{N} > 0$, $b = ik$ and $\Delta_{\nu} = -2\nu$ (as already defined \textit{e.g.} in eq. (\ref{sphere_packing_eq})), 
and the Gamma function\cite{olver2010nist}, 
\begin{align}
    \Gamma(s) = \int_{0}^{\infty}\frac{dt}{t}2^{-s}t^s\exp(-2t),
\end{align}
\end{subequations}
with $t = \pi k$. The first average over $\Theta_{\Lambda}(t)$ is given by, 
\begin{multline}
    \langle \Delta_{\nu} \rangle_{\Theta_{\Lambda}} = \frac{\sum_{\nu = 1}^{\infty}\Delta_{\nu}\exp(t\Delta_{\nu})}{\Theta_{\Lambda}(t)} = \frac{\Theta_{\Lambda}'(t)}{\Theta_{\Lambda}(t)}\\
    = \frac{\partial}{\partial t}\ln\left(\Theta_{\Lambda}(t)\right)
    = \frac{-1}{1 - \exp(-2t)}.
\end{multline}
Proceeding, we calculate the second quantum average, 
\begin{multline}\label{delta_delta_eq}
    \Delta(d) = \langle \langle \Delta_{\nu} \rangle_{\Theta_{\Lambda}}\rangle_{\Gamma(-2g)} = \left \langle \frac{-1}{1 - \exp(-2t)} \right \rangle_{\Gamma(-2g)}\\
    = \frac{2^{2g}}{\Gamma(-2g)}\int_0^{\infty}\frac{dt}{t}t^{-2g}\exp(-2t)\left ( \frac{-1}{1 - \exp(-2t)}\right )\\
    = \frac{-2^{2g}}{\Gamma(-2g)}\int_0^{\infty}\frac{dt}{t}\frac{t^{-2g}}{\exp(t) - 1} = -2^{2g}\zeta(-2g),
\end{multline}
where, 
\begin{subequations}
\begin{align}
    \zeta(s) = \sum_{n = 1}^{\infty}\frac{1}{n^s},
\end{align}
\red{is the Riemann zeta function,} 
convergent for ${\rm real} (s) > 1$.\cite{karatsuba2011riemann} 
\red{Fortunately, $s$ can be extended} \red{by analytic continuation}, 
\begin{align}\label{operation_zeta_eq}
     \zeta(s) = 2^{s}\pi^{s - 1}\sin(\pi s/2)\Gamma(1 - s)\zeta(1 - s),
\end{align}
\end{subequations}
to all relevant values of $\Delta_{\nu}$, especially the desired values $\Delta_{\nu} = -2\nu = \chi(g) = 2 - 2g$. 
Thus, by design, the trivial zeros of the Riemann zeta function\cite{broughan2017equivalents, karatsuba2011riemann}, $\Delta(d = 2) = -2^{2g}\zeta(-2g) = 0$ correspond to $g > 0$, which is the 2D theory without bifurcation (of the honeycomb lattice). 
\red{Conversely}, bifurcation 
\red{occurs} spontaneous for $g = 0$, which leads to $\Delta(d = 3) = -\zeta(0) = 1/2$ 
\red{and} the lowering of the energy as 
\red{earlier discussed}. 

In other words, 
\textit{given a 2D bipartite honeycomb lattice of weakly interacting fermionic atoms with degenerate pseudo-spin states, the system can always lower its energy by geometrically (topologically) lifting this degeneracy by 
\red{bifurcation}}. This theorem is analogous to the Peierls and Jahn–Teller theorems in 1D and 3D systems.\cite{garcia1992dimerization, jahn1937stability} Thus, due to pseudo-spin frustration, spontaneous magnetisation state ($S = 1, \tilde{m} = 1$) is only possible through bifurcation, which introduces a finite displacement $|\vec{x} - \vec{r}| = |\vec{R}| = R$ along the $z$-coordinate, where $\vec{r} = \vec{r}_1 - \vec{r}_2$ lies within the 2D lattice, $\vec{r}_j \in \mathbb{R}^2$ and $\vec{x} \in \mathbb{R}^3$ lies within the 3D lattice, as depicted in Figure \ref{Fig_13}(b). 
It is not necessarily 
\red{the case} that the energy gap given in eq. (\ref{energy_gap_eq}) leads to superconductivity. Experimentally, crystallographic distortions such as 
the dimerisation of 
polyacetylene and Kekul\'{e} distortion in graphene-based systems\cite{garcia1992dimerization, peierls1979surprises, peierls1955quantum, lee2011band}, 
\red{result in} a metal-insulator phase transition, where eq. (\ref{energy_gap_eq}) 
\red{can be interpreted as the} energy gap between the valence band and conduction band, which generally accounts for the semi-conducting/\red{insulating} 
behaviour in 1D 
\red{systems}.\cite{gruner1988dynamics} 

It is worth restating that the relation between $\Delta_{\nu}$ and $\Delta(d)$ is some operation (mathematically equivalent to a Mellin transform\cite{weisstein2004mellin}, normalised by the Gamma function) given in eq. (\ref{delta_delta_eq}), where $\zeta(s)$ is the Riemann zeta function given in eq. (\ref{delta_delta_eq}), which makes use of the locations of the trivial zeros. This relation appears to relate the scaling dimension, $\Delta$ of a high energy theory to another at lower energy, in the spirit of renormalisation group flows.\cite{zamolodchikov1986irreversibility, jack1990analogs, osborn1989derivation} Indeed, renormalisation group flows in the context of C-theorem relate the central charge, $c$ of a CFT at high energy (ultra-violet, $\rm UV$), $c_{\rm UV}$ to another at low energy (infra-red, $\rm IR$) by $c_{\rm UV} > c_{\rm IR}$, suggesting the Riemann zeta function above serves a similar role here. In our case, the scaling dimension scales with the genus, and the fixed points correspond to the zeros, $\zeta(s = -2g) = 0$ of the Riemann zeta function.\cite{karatsuba2011riemann} Finally, this suggests that another CFT lives at the non-trivial/essential zeros, conjectured in the famous hypothesis by Riemann to occur at $s = \Delta(d = 3) \pm i\gamma_n$, with real values of $\gamma_n$.\cite{conrey2003riemann}


\red{\subsubsection{Isospin, pseudo-spin and electromagnetic interactions 
}}

\red{The Aufbau principle typically employed in standard chemistry to determine the electron occupation of orbital energy levels and their valencies in atoms can be violated for transition metals due to electrostatic shielding of electric charge of the nucleus and other factors.\cite{schwarz2010full} In the case of elements in group 11, the close proximity of the $nd^{10}$ and $(n + 1)s$ orbitals (energy gap of order $< 3.5$ eV)\cite{blades2017evolution} encourages $sd$ hybridisation, leading to degenerate states. Considering the Ag atom, these degenerate states are $4d^{10}5s^1$ ($\rm Ag^{1+}$) and $4d^95s^2$ ($\rm Ag^{2+}$, $\rm Ag^{1-}$), where $1+$, $2+$ and $1-$ denotes the valence states, corresponding to the number of electrons lost to achieve chemical stability. Due to the odd number of electrons, the neutral atom is a fermion (as expected) with its spin state inherited from the spin of the valence electron. Thus, due to $sd_{z^2}$ hybridisation, \red{a single spin up or down electron} can either be in the $nd_{z^2}^1$ or $(n + 1)s^1$ orbital with all the remaining lower energy orbitals fully occupied. Nonetheless, the valency corresponds to the number of electrons in the $ns$ orbital ($ns^2$ or $ns^1$). This results in two valence states, $A^{2+}$ and $A^{1+}$ ($A = \rm Cu, Ag, Au$). Moreover, in order to become closed shell in chemical reactions, the coinage metal atom can either be an electron donor with valency $2+, 1+$ or a receptor/anion with valency $1-$, whereby the receptor $A^{1-}$ achieves closed shell $nd_{z^2}^2$ and $(n + 1)s^2$ orbitals forming stable bonds. 

Indeed, this anion state has been observed in coinage metal cluster ions as $\rm Ag_N^{1-} (N \in \mathbb{Z}^+)$\cite{minamikawa2022electron, ho1990photoelectron, dixon1996photoelectron, schneider2005unusual}, whereas the isolated anion state ($N = 1$) is readily observed in compounds such as $\rm CsAu\cdot NH_3$ due to enhanced relativistic effects of Au.\cite{jansen2008chemistry} Thus, the $A^{1+}$ and $A^{1-}$ valence states are related by isospin rotation (SU($2$)) with the isospin given by $I = \mathcal{V}_A/2$ where $\mathcal{V}_A = 1+, 1-$ are the valence states, and $Y = 0$ is the electric charge of the neutral atom. Meanwhile, the $A^{2+}$ state is an isospin singlet ($I = 0$) with electric charge, $Y = \mathcal{V}_A = 2+$. Nonetheless, these three cation states $A^{2+}, A^{1-}$ and $A^{1+}$ must have an effective charge, $Q = +2, -1$ and $Q = +1$ respectively, obtained by the Gell-Mann–Nishijima formula (eq. (\ref{Gell-Mann–Nishijima_eq2})) and \red{are} treated as independent ions related to each other by $\rm SU(2)\times U(1)$, forming the basis for fractional valent (subvalent) states. Due to $sd$ hybridisation, all these three states are degenerate on the honeycomb lattice. The degeneracy between $\rm Ag^{2+}$ and $\rm Ag^{1-}$ corresponds to right-handed and left-handed chirality of $\rm Ag$ fermions on the honeycomb lattice, treated as the pseudo-spin.\cite{masese2023honeycomb}} 
\red{Focusing on Ag, this degeneracy is lifted by SU($2$)$\times$U($1$) symmetry breaking, whereby the pseudo-spin states of Ag gain mass. The transition between $\rm Ag^{1+}$ and $\rm Ag^{1-}$ corresponds to the excitation of the valence electron by a photon between the $4d_{z^2}$ and $5s$ orbitals. We shall treat the entangled state of the electron and the photon as a charged $W^{\mu}_{\pm} = \frac{1}{\sqrt{2}}(W_1^{\mu} \pm iW_2^{\mu})$ gauge boson, responsible for this transition, where $W^{\mu} = W^{\mu}_a\tau_a$ is a gauge field transforming under SU($2$) ($\tau_a = \sigma_a/2$, $\sigma_a = (\sigma_1, \sigma_2, \sigma_3)$ are the Pauli matrices). This transition has to be of the order $< 3.5$ eV, corresponding to the mass of $W_{\pm}^{\mu}$. Moreover, there will be screened and un-screened 
electromagnetic potentials, $Z_{\mu}$ and $p_{\mu}$ respectively responsible for self-interactions between the Ag atoms.}

Now consider the relativistic action for spin-1/2 silver cations coupled to SU($2$)$\times$U($1$) gauge fields in $3 + 1$-dimensions\cite{zee2010quantum}, 
\begin{multline}\label{Argentophilic_action_eq}
    S = \int dt\int d^{\,d}x\,i\overline{\psi}_{\rm R}\gamma^{\mu}\left (\partial_{\mu} + i\frac{Y}{2}q_{\rm e}A_{\mu}\right )\psi_{\rm R}\\
    + \int dt\int d^{\,d}x\,i\overline{\psi}_{\rm L}\gamma^{\mu}\left (\partial_{\mu} + iq_{\rm w}\vec{I}\cdot\vec{W}_{\mu} + i\frac{Y}{2}q_{\rm e}A_{\mu}\right )\psi_{\rm L}\\
    + \int dt\int d^{\,d}x\,y_{\rm c}\left (\overline{\psi}_{\rm R}\phi^{\dagger}\psi_{\rm L} + \overline{\psi}_{\rm L}\phi\psi_{\rm R} \right )\\
    + S_{\rm SU(2), U(1),\phi},
\end{multline}
where $d = 2, 3$ dimensions, $\psi_{\rm L}^{\rm T} = \rm (Ag^{1+}, Ag^{1-})$ is the doublet, $\psi_{\rm R} = \rm Ag^{2+}$ is the singlet, 
$\vec{I} = \frac{1}{2}\vec{\sigma}$ is the weak isospin vector with $\vec{\sigma}$ the Pauli vector, $\gamma_{\mu}$ are gamma matrices, $\phi^{\rm T} = (0, |\Psi(\vec{x})|/m^{\Delta(d)})$ is a mass-inducing scalar field with $y_{\rm c}$ \red{a Yukawa coupling} and $\Delta(d) = (d - 2)/2$ is the scaling dimension, 
and $\vec{R} = \vec{x} - \vec{r}$ is defined as a displacement vector in 3D (Figure \ref{Fig_13}) whose magnitude will appear in eq. (\ref{RKKY_eq}) with the approximation $|\vec{x}| \gg |\vec{r}|$ (the separation distance, $|\vec{r}|$ of cations in the honeycomb lattice is greatly smaller than their separation distance $|\vec{x}|$ after bifurcation) such that $\vec{R} \simeq \vec{x}$ in $d = 3$. 
\red{Note that}, $|\Psi(\vec{R})|^2 \simeq \rho(\vec{R})$ plays the role of charge density, normalised as, 
\begin{align}\label{psi_norm_eq}
    \left \langle\left \langle\int d^{d}x\,|\Psi(\vec{R})|^2 \right \rangle\right \rangle = -\langle \langle \Delta_{\nu} \rangle\rangle = -2\Delta(d),
\end{align}
which is 
equivalent to eq. (\ref{ansatz_eq}). 
Thus, $|\Psi| = 0$ in 2D but finite ($|\Psi| \neq 0$) in 3D. Moreover, $S_{\rm SU(2), U(1), \phi}$ is the action for the scalar field, $\phi$, the SU($2$)$\times$U($1$) gauge fields $\vec{W}_{\mu} = (W_{\mu}^1, W_{\mu}^2, W_{\mu}^3)$ and $A_{\mu}$ respectively, 
\begin{multline}\label{S_electro_weak_eq}
    S_{\rm SU(2), U(1),\phi} =  - \frac{\lambda}{2}\int dt\int d^{\,d}x\,\left (\phi^{\dagger}\phi - m^2/y_{\rm c}^2 \right )^2\\
    + \frac{1}{2}\int dt\int d^{\,d}x\,\left |\left (\partial_{\mu} + iq_{\rm w}\vec{I}\cdot\vec{W}_{\mu} + i\frac{Y}{2}q_{\rm e}A_{\mu}\right )\phi\right |^2\\
    -\frac{1}{4}\int dt \int d^{\,3}x \left (\vec{W}_{\mu\nu}\cdot\vec{W}_{\mu\nu} + F_{\mu\nu}F^{\mu\nu}\right ),
\end{multline}
with $\vec{W}_{\mu\nu} = \partial_{\mu}\vec{W}_{\nu} - \partial_{\nu}W_{\mu} - q_{\rm w}\vec{W}_{\mu}\times\vec{W}_{\nu}$, $F_{\mu\nu} = \partial_{\mu}A_{\nu} - \partial_{\nu}A_{\mu}$ and $V(\phi) = \left (\phi^{\dagger}\phi - m^2/y_{\rm c}^2 \right )^2$ the Mexican hat potential responsible for the finite value, $\phi \neq 0$ where $\lambda$ and $m$ are constants. Thus, this action takes the form analogous to electroweak interactions of the electron and left-handed neutrino.\cite{weinberg1967model} This means that calculations run parallel to electroweak symmetry breaking and need not be 
\red{comprehensively treated} herein.\cite{zee2010quantum}

Nonetheless, for our purposes, it is important to consider the mixing of the gauge fields\cite{zee2010quantum}, 
\begin{align}\label{Weinberg_eq}
    \begin{pmatrix}
A_{\mu}\\ W^3_{\mu}
\end{pmatrix}
= 
\frac{1}{q_{\rm e}q_{\rm w}}
\begin{pmatrix}
q_{\rm w} & -q_{\rm e}\\ 
q_{\rm e} & q_{\rm w}
\end{pmatrix}
\begin{pmatrix}
p_{\mu}\\ q_{\rm eff} Z_{\mu}
\end{pmatrix},
\end{align}
where $q_{\rm eff} = q_{\rm w}q_{\rm eff}/\sqrt{q_{\rm w}^2 + q_{\rm e}^2}$ is the effective charge, which leads to a mass-less $p_{\mu}$ and massive $W_{\mu}^{\pm} = (W_{\mu}^1 \mp iW_{\mu}^2)/\sqrt{2}$ and $Z_{\mu}$ gauge fields (Here, the mass-less field has been re-scaled as $p_{\mu} \rightarrow p_{\mu}/q_{\rm eff}$ for later convenience).\cite{zee2010quantum, masese2023honeycomb} 

\red{Moreover}, we are interested in 
\red{the coupling of the fields to} $p_{\mu}$ and $\phi$ in 3D. Plugging in eq. (\ref{Weinberg_eq}) into 
eq. (\ref{Argentophilic_action_eq}), we find the terms, 
\begin{multline}
    S = \int dt\int d^{\,3}x\,\left (i\overline{\psi}\gamma^{\mu}\left (\partial_{\mu} + ip_{\mu}\right ){\psi} - m(\vec{x})\overline{\psi}\psi \right )\\
    - \frac{1}{4}\int dt\int d^{\,3}x\,\, (\partial_{\mu}p_{\nu} - \partial_{\nu}p_{\mu})^2\\
    + \int dt\int d^{\,3}x\,\,\overline{{\rm Ag}^{1+}}\gamma^{\mu}i\left (\partial_{\mu} - ip_{\mu}\right){\rm Ag}^{1+} + \cdots,
\end{multline}
where we have introduced a massive Dirac spinor for the charged cations, $\psi^{\rm T} = \rm (Ag^{2+}, Ag^{1-})$ with $m(\vec{x}) = -y_{\rm c}|\Psi(\vec{x})|/\sqrt{m}$, 
\red{whilst the left-handed spinor ${\rm Ag}^{1+}$ 
remains mass-less} even in 3D. Note that $|\Psi(\vec{R})|$ 
can be written as a critical exponent\cite{masese2023honeycomb, domb2000phase},
\begin{align}\label{critical_exp_eq}
   |\Psi(\vec{x})| = 2\Delta(d)m^{\Delta(d)}\frac{T(\vec{x}) - m}{y_{\rm c}}\geq 0,
\end{align}
where the fermion mass, $m$ is the transition temperature, $T(\vec{x})$ is a temperature gradient and $m(\vec{x}) = m - T(\vec{x})$ is a perturbation from the equilibrium solution of the Mexican hat potential, $\phi^{\dagger}\phi = m^2(\vec{x})/y_{\rm c}^2 = m^2/y_{\rm c}^2$. 

\red{Thus, the bonding potential between $\rm Ag^{2+}$ and $\rm Ag^{1-}$ corresponds to},
\begin{align}\label{potential_eq}
    \mathcal{L}_{\rm mass}(\overline{\psi}, \psi)  = -\int d^{\,3}x\,m(\vec{x})\overline{\psi}\psi,
\end{align}
\red{which corresponds to the argentophilic interaction. The critical exponent in eq. (\ref{critical_exp_eq}) guarantees that the phase transition occurs at $T \geq m$ for $d = 3$ as required. Conversely, for $T(\vec{x}) < m$, $|\Psi(\vec{x})|$ will be negative unless $\Delta(d) = 0$, which corresponds to $d = 2$ and $|\Psi(\vec{x})| = 0$, as expected.}
\red{Moreover, the mass $m_W$ and $m_Z$ of the $W_{\pm}^{\mu}$ bosons (responsible for transitions between $\rm Ag^{1+}$ and $\rm Ag^{1-}$) and the $Z^{\mu}$ boson respectively are constrained by}, 
\begin{align}
    3.5 \, {\rm eV} \geq m_W = \frac{q_{\rm w}}{\sqrt{q_{\rm e}^2 + q_{\rm w}^2}}m_Z,
\end{align}
whereas $p_{\mu}$ is mass-less.\cite{weinberg1967model, zee2010quantum}

Introducing the gravitational field by minimal coupling 
yields, 
\begin{multline}
    S = \int d^{\,4}x\sqrt{-\det(g_{\mu\nu})}\,\,i\overline{\psi}\gamma^{\mu}\left (D_{\mu}+ ip_{\mu} \right ){\psi}\\
    - \frac{1}{4}\int d^{\,4}x\sqrt{-\det(g_{\mu\nu})}\,\, p_{\mu\nu}p^{\mu\nu}\\
    - \int d^{\,4}x\sqrt{-\det(g_{\mu\nu})}\,\, m(\vec{R})\overline{\psi}\psi\\
    +  \int d^{\,4}x\sqrt{-\det(g_{\mu\nu})}\,\,\overline{{\rm Ag}^{1+}}\gamma^{\mu}iD_{\mu}{\rm Ag}^{1+}\\
    + \frac{\kappa}{2\mu}\int d^{\,4}x\sqrt{-\det(g_{\mu\nu})}\,\, (R - 2\Lambda), 
\end{multline}
where $p_{\mu\nu} = \partial_{\mu}p_{\nu} - \partial_{\nu}p_{\mu}$, $\gamma_{\mu}$ are the gamma matrices in curved space-time satisfying $\gamma_{\mu}\gamma_{\nu} + \gamma_{\nu}\gamma_{\mu} = 2g_{\mu\nu}$, $\overline{\psi} = \psi^{*\rm T}(\gamma^0)^{-1}$, $({\rm Ag^{1+}})^{*\rm T}$ are the Dirac adjoint spinors, $\gamma^{\mu} = e^{\mu}_{\,\,\overline{a}}\gamma^{\overline{a}}$ with $\gamma^{\overline{a}}$ the Dirac matrices in Minkowski space-time satisfying $\gamma_{\overline{a}}\gamma_{\overline{b}} + \gamma_{\overline{b}}\gamma_{\overline{a}} = 2\eta_{\overline{a}\overline{b}}$, and $e^{\overline{a}}_{\,\,\mu}, e^{\mu}_{\,\,\overline{a}}$ are tetrad fields satisfying $e^{\overline{a}}_{\mu}e_{\overline{a}\nu} = g_{\mu\nu}$ and $e^{\mu}_{\,\,\overline{a}}e_{\mu\overline{b}} = \eta_{\overline{a}\overline{b}}$ with $\eta_{\overline{a}\overline{b}}$ the Minkowski metric tensor, 
$D_{\mu} = \partial_{\mu} - \frac{1}{4} \omega_{\mu}^{\overline{a}\overline{b}}\gamma_{\overline{a}}\gamma_{\overline{b}}$, $\omega_{\mu}^{\overline{a}\overline{b}} = e^{\overline{a}}_{\,\,\alpha}\partial_{\mu}e^{\overline{b}\alpha} + \Gamma^{\beta}_{\,\,\mu\alpha}e^{\overline{a}}_{\,\,\beta}e^{\overline{b}\alpha}$ is the spin connection, $\Gamma^{\alpha}_{\,\,\mu\nu} = \frac{1}{2}g^{\alpha\beta}(\partial g_{\mu\beta}\partial x^{\nu} + \partial g_{\beta\nu}/\partial x^{\mu} - \partial g_{\mu\nu}/\partial x^{\beta})$ are the Christoffel symbols, 
$x^{\mu} = (t, \vec{R})$, $\mu = 8\pi G$ is the mobility, $\kappa = 1/\nu$ and $\Lambda$ is a cosmological constant obtained from the rest of the gauge fields, $W_{\mu}^{\pm}$ and $Z_{\mu}$ by assuming they are in their quantum mechanical vacuum state. The vacuum energy is achieved by assuming extremely large 
\red{mass gaps} for the $W_{\pm}^{\mu}$ and $Z^{\mu}$ bosons, 
\red{$m_W \gg m$,} rendering their 
\red{excited} states unobserved.

Varying with respect to $g_{\mu\nu}$ yields the Einstein Field equations given in eq. (\ref{EFE_eq}) with $\beta = 8\pi GM$, $M = m/\kappa = m\nu$ and the energy-momentum given by, 
\begin{multline}
    T^{\mu\nu} = \frac{1}{4mi}\left (\overline{\psi}\gamma^{\mu}\left (D^{\nu} + ip^{\nu}\right )\psi + \overline{\psi}\gamma^{\nu}\left (D^{\mu} + ip^{\mu}\right )\psi \right )\\
    + \frac{1}{4mi}\left (\left (\left (\tilde{D}^{\nu} + ip^{\nu}\right )\overline{\psi}\right )\gamma^{\mu}\psi + \left (\left (\tilde{D}^{\nu} + ip^{\nu}\right )\overline{\psi}\right )\gamma^{\mu}\psi \right )\\
    + \frac{1}{4mi}\left (\overline{\rm Ag^{1+}}\gamma^{\mu}\left (D^{\nu} - ip^{\nu}\right ){\rm Ag^{1+}} + \overline{\rm Ag^{1+}}\gamma^{\nu}\left (D^{\mu} - ip^{\mu}\right ){\rm Ag^{1+}}\right )\\
    \frac{1}{4mi}\left (\left (\left (\tilde{D}^{\mu} - ip^{\mu}\right )\overline{\rm Ag^{1+}}\right)\gamma^{\nu}{\rm Ag^{1+}} + \left (\left (\tilde{D}^{\nu} - ip^{\nu}\right )\overline{\rm Ag^{1+}} \right)\gamma^{\mu}{\rm Ag^{1+}}\right )\\
    + \frac{1}{m}\left (p^{\mu\alpha}p_{\,\,\alpha}^{\nu} - \frac{1}{4}p^{\alpha\beta}p_{\alpha\beta}g^{\mu\nu}\right ) - \frac{\Lambda}{\beta}g^{\mu\nu},
\end{multline}
where $\tilde{D}_{\mu}\overline{\psi} = -\partial_{\mu}\psi - \frac{1}{4}\overline{\psi}\omega_{\mu}^{\overline{a}\overline{b}}\gamma_{\overline{a}}\gamma_{\overline{b}}$. Proceeding, varying the action with respect to $\psi$ and $\rm Ag^{1-}$ yields the Dirac equations, 
\begin{subequations}\label{dirac_psi_eq}
\begin{align}
    \left (D_{\mu} + ip_{\mu}\right)\gamma^{\mu}\psi = -im\psi,\\
    \left (\tilde{D}_{\mu} + ip_{\mu}\right)\overline{\psi}\gamma^{\mu} = -im\overline{\psi}
\end{align}
\end{subequations}
and, 
\begin{subequations}\label{dirac_Ag_eq}
\begin{align}
    i\left (D_{\mu} - ip_{\mu}\right )\gamma^{\mu}{\rm Ag^{1+}} = 0,\\
    i\left (\tilde{D}_{\mu} - ip_{\mu}\right )\overline{\rm Ag^{1+}}\gamma^{\mu} = 0,
\end{align}
\end{subequations}
respectively. Lastly, varying with respect to $p_{\mu}$ yields, 
\begin{align}\label{maxwell_p_eq}
    \nabla_{\mu}p^{\mu\nu} = - J^{\nu},
\end{align}
where $J^{\mu} = \overline{\psi}\gamma^{\mu}\psi$. 

To make connections with the idealised model (eq. (\ref{CFE_eq})) and Liouville CFT discussed earlier, we shall make some approximations and assumptions, and take the averages, $\langle T^{\mu\nu} \rangle$ and $\left \langle J^{\mu} \right \rangle$ over the Dirac fields to yield,
\begin{subequations}
\begin{align}
    \langle T^{\mu\nu} \rangle = \rho u^{\mu}u^{\nu} + \rho_{\rm Ag^{1-}}v^{\mu}v^{\nu} + T_{U(1)}^{\mu\nu} - \frac{\Lambda}{\beta}g^{\mu\nu},\\
    T_{U(1)}^{\mu\nu} = \frac{1}{m}\left (p^{\mu\alpha}p_{\,\,\alpha}^{\nu} - \frac{1}{4}p^{\alpha\beta}p_{\alpha\beta}g^{\mu\nu}\right ),
\end{align}
\end{subequations}
and, 
\begin{align}
    \left \langle J^{\mu} \right \rangle = \rho u^{\mu},
\end{align}
where $\Lambda = -\beta c/12\mathcal{V}$ corresponds to eq. (\ref{c_lambda_eq}), $\rho = \langle \overline{\psi}\psi \rangle$, $\rho_{\rm Ag^{1+}} = \langle \overline{\rm Ag^{1+}}{\rm Ag^{1+}} \rangle$, $u^{\mu}u_{\mu} = -1$ and $v^{\mu}v_{\mu} = 0$ with $u^{\mu} = \langle \overline{\psi}\gamma^{\mu}\psi \rangle/\langle \overline{\psi}\psi \rangle$ and $v^{\nu} = \langle \overline{\rm Ag^{1+}}\gamma^{\nu}{\rm Ag^{1+}} \rangle/\langle \overline{\rm Ag^{1+}}{\rm Ag^{1+}} \rangle$ the four-velocities of the massive and mass-less Dirac fields respectively. Using the field equations given in eq. (\ref{dirac_psi_eq}), eq. (\ref{dirac_Ag_eq}) and eq. (\ref{maxwell_p_eq}), the trace of the energy-momentum tensor yields, $T^{\mu\nu}g_{\mu\nu} = -\overline{\psi}\psi - 4\Lambda/\beta$, which corresponds to $g_{\mu\nu}\langle T^{\mu\nu} \rangle = -\rho - 4\Lambda/\beta$ as 
\red{required}. The idealised model then corresponds to the equation of motion in eq. (\ref{CFE_eq}) where $K_{\mu\nu} = R_{\mu\nu} + ip_{\mu\nu}$. 

Finally, provided $c = 1$, it is clear that the mass term in $\psi$ breaks conformal symmetry since the dilatation current, $j^{\nu} = x^{\mu}(\rho u_{\mu}u^{\nu} + T_{U(1)\mu}^{\,\,\nu})$ is no longer conserved in Minkowski space-time, $\partial_{\mu}j^{\mu} = \rho u^{\mu}u_{\mu} = -\rho$, where we have applied $\partial_{\mu}(\rho u^{\mu}u^{\nu} + T_{U(1)}^{\mu\nu}) = 0$ and used $\partial_{\mu}x^{\nu} = \delta^{\nu}_{\mu}$ and $u^{\mu}u_{\mu} = -1$.  


\red{\subsubsection{Numismophilicity as cationic RKKY interactions due to the effective mass term}}

Since the mass term breaks conformal symmetry, we shall calculate its form and show that it corresponds to the RKKY interaction for the Ag cations in the non-relativistic limit, where the metric tensor is assumed Minkowski. In particular, we seek to calculate, 
\begin{multline}\label{S_mass_eq}
    S_{\rm mass} = -\int dt \int d^{\,3}R\,\,m(\vec{R})\left \langle \overline{\psi}(\vec{R}, t_{\rm })\psi(\vec{R}, t)\right \rangle\\
   = -\int dt \int d^{\,3}R\,\,m\left (\rho(\vec{R}, t) - \frac{T(\vec{R})}{m}\rho(\vec{R}, t)\right )\\
   = -\int dt \int d^{\,3}R\,\,m\left (\rho(\vec{R}, t) + \delta\rho(\vec{R}, t)\right ),
\end{multline}
where $\vec{x} \simeq \vec{R}$, $m(\vec{R}) = m - T(\vec{R})$ is the effective mass of the cations at each honeycomb unit cell acquired from the interaction with the scalar field, $\phi$ in the previous section, the density and density fluctuation are given by,
\begin{align}
    \rho(\vec{R}, t) = \left \langle \overline{\psi}(\vec{R}, t_{\rm })\psi(\vec{R}, t) \right \rangle,\\
    \delta\rho(\vec{R}, t) = -\frac{T(\vec{R})}{m}\rho(\vec{R}, t),
\end{align}
respectively.

\red{The scalar field, $T(\vec{R})$ is interpreted as the temperature dependent on $\vec{R}$, in order for $m(R) = 1/\beta(R)$ and $m = 1/\beta$ to be the equilibrium temperature. Thus, its quanta is given by thermal phonons (analogue to the Higgs boson\cite{zee2010quantum}) that can propagate from hot to cold areas of the material. In fact, it is likely that the simplest equation of motion for $T(\vec{R})$ will not exactly satisfy the perturbed relativistic scalar equations of motion from eq. (\ref{S_electro_weak_eq}), but rather, the first order non-relativistic terms take the form of the heat conduction equation}\cite{hahn2012heat}, 
\begin{align}\label{heat_eq}
    \frac{\partial T(\vec{R}, t)}{\partial t} = \alpha\nabla_{\vec{R}}^2T(\vec{R}, t) + \mathcal{P}(\vec{R}), 
\end{align}
where $\alpha$ is the thermal diffusivity and $P(\vec{R})$ is the internal heat generation term  at point $\vec{R}$ responsible for the heating up of the cathode during (de-)intercalation processes.\cite{maleki2004role} Considering the 
\red{steady-state} solution ($\partial T(\vec{R}, t)/\partial t = 0$), 
\begin{align}\label{static_P_eq}
    T(\vec{R}) = \frac{1}{4\pi\alpha}\int d^{\,3}R'\,\frac{\mathcal{P}(\vec{R'})}{|\vec{R} - \vec{R'}|}.
\end{align}
The component form of the wave function is given by, 
\begin{align}
    \psi = \begin{pmatrix}
\psi_{s_1}(\vec{R}, t)\\ \psi_{s_2}'(\vec{R}, t)
\end{pmatrix}
= \begin{pmatrix}
{\rm Ag}_{s_1}^{2+}\\ {\rm Ag}_{s_2}^{1-}
\end{pmatrix},
\end{align}
where $s_1, s_2$ are the pseudo-spin degrees of freedom of the cations, $\psi = {\rm Ag}^{2+}$ and $\psi' = {\rm Ag}^{1-}$. In momentum, $\vec{k}$ space, the quantum average of the temperature dependent term in eq. (\ref{S_mass_eq}) is given by, 
\begin{align}\label{cal_RKKY_eq1}
    \int dt \int \frac{d^{\,3}k}{(2\pi)^3}\frac{d^{\,3}k'}{(2\pi)^3}T(k + k')\left \langle\overline{\psi}(\vec{k}, t)\psi(\vec{k}', t)\right\rangle, 
\end{align}
where, 
\begin{align}
    T(k + k') = \int d^{\,3}R\,\,T(\vec{R})\exp(i(\vec{k} + \vec{k'})\cdot \vec{R}),
\end{align}
and, 
\begin{align}
    \overline{\psi}(\vec{k},t) = \psi^{*\rm T}(\vec{k},t)(\gamma^0)^{-1}\\
    \psi(\vec{k}, t) = \begin{pmatrix}
\psi_{s_1}(\vec{k}, t)\\ \psi'_{s_2}(\vec{k}, t)
\end{pmatrix},
\\
(\gamma^0)^{-1} = 
\begin{pmatrix}
0 & 1\\ 1 & 0
\end{pmatrix}
= \gamma^0,
\end{align}
with,
\begin{align}
    \psi_s(\vec{k}, t) = \int \frac{d\omega}{2\pi T}\frac{ i\psi_s(\vec{k})}{\omega + iE_{|\vec{k}|}}\exp(-\omega t),\\
    \psi'_s(\vec{k}, t) = \int \frac{d\omega}{2\pi T}\frac{ i\psi'_s(\vec{k})}{\omega + iE_{|\vec{k}|}}\exp(-\omega t),\\
    \psi_s(\vec{R}, t) = \int \frac{d^{\,3}k}{(2\pi)^3}\psi_s(\vec{k}, t)\exp(i\vec{k}\cdot\vec{R}),
\end{align}
$\vec{q} = \vec{k} + \vec{k'}$, $s = s_1, s_2$ and $E_{|\vec{k}|}, E_{|\vec{q} - \vec{k}|}$ are the energy eigenvalues. 
\red{Here, the wave functions of the fermions have been written in terms of the thermal Green's function, under the transformation}, 
\begin{align}
    \sum_n \rightarrow \int \frac{d\omega}{2\pi T},\\
    \omega_n = \pi T(2n + 1) \rightarrow \omega, 
\end{align}
where $\omega_n$ is the fermion Matsubara frequency ($k_{\rm B} = 1$).\cite{abrikosov2012methods} 

Note that, for interactions with opposite momenta and spins, we shall have $(\vec{q}, s_1, s_2) = (0, \pm 1/2, \mp 1/2)$. 
\red{However, we wish to encompass all scattering interactions} with total momentum, $\vec{q}$ and spin, $s$. Moreover, setting,
\begin{align}
    \psi'_s(\vec{k}) = \psi_s(\vec{k'}) = \psi_s(\vec{q} - \vec{k}).  
\end{align}
we can finally write the spinor as, 
\begin{align}
    \psi(\vec{k}, t) = \int \frac{d\omega}{2\pi T}\frac{i\psi(\vec{k})}{\omega + iE_{|\vec{k}|}}\exp(-\omega t),
\end{align}
where, 
\begin{align}
    \psi(\vec{k}) = 
    \begin{pmatrix}
    \psi_{s_1}(\vec{k})\\
    \psi_{s_2}(\vec{q} - \vec{k})
    \end{pmatrix},
\end{align}
depends only on the momenta. Thus, plugging in $\overline{\psi}(\vec{k}, t)$ and $\psi(\vec{k'}, t)$ into eq. (\ref{cal_RKKY_eq1}), and performing the transformation, $i\omega = z, -i\omega = \overline{z}$, we find, 
\begin{multline}\label{cont_calculation_eq}
    -\int d^{\,3}R\,\,\frac{T(R)}{T^2}\int \frac{d^{\,3}k}{(2\pi)^3}\frac{d^{\,3}k'}{(2\pi)^3}\exp(i(\vec{k} + \vec{k})\cdot\vec{R})\times\\
    \int \frac{dzd\overline{z}}{(2\pi)^2}\int dt\frac{\left \langle \psi^{*\rm T}(\vec{k})\gamma^0\psi(\vec{k'}) \right \rangle\exp(i(z - \overline{z})t)}{(\overline{z} - E_{|\vec{k}|})(z - E_{|\vec{k'}|})}. 
\end{multline}

Moreover, using $\gamma^0$, we write, 
\begin{multline}
    \left \langle \psi^{*\rm T}(\vec{k})\gamma^0\psi(\vec{k'}) \right \rangle\\
    = \left \langle \psi_{s_1}^*(\vec{k})\psi_{s_2}(\vec{k}) \right \rangle + \left \langle \psi_{s_2}^*(\vec{k'})\psi_{s_1}(\vec{k'}) \right \rangle.
\end{multline}
These correlation functions will explicitly depend on the equation of motion for the fields. 
\red{Relating the correlation function to} the density function, $\rho(R) = \langle \psi^*(R)\psi(R) \rangle$ by,
\begin{multline}\label{correlation_relate_eq}
    \left \langle \psi_{s_1}^*(\vec{k})\psi_{s_2}(\vec{k}) \right \rangle\\
    \equiv 4\langle S_1S_2 \rangle (2\pi)^{3/2}\int d^{\,3}R\,\exp(i\vec{k}\cdot\vec{R})\beta\rho(\vec{R}),
\end{multline}
we shall consider the case where the density satisfies the Poisson equation,  
\begin{align}\label{Newton_eq}
    \nabla^2\Phi(\vec{R}) = \beta\rho(\vec{R}),
\end{align}
which corresponds to the Newtonian limit of eq. (\ref{EFE_eq}) with $T^{\mu\nu} = \rho u^{\mu}u^{\nu}$, where $2S_1 = \sigma_z^1$ and $2S_2 = \sigma_z^2$ is the $z$ component of the Pauli vector acting on the pseudo-spin states $1$ and $2$.

To make further progress, we are interested in a density function of the form, 
\begin{align}\label{norm_rho_eq}
    \rho(\vec{R}) = \rho_0g(\vec{R}),\\
    \int d^{\,3}R\,\,\rho(\vec{R}) = -\Delta_{\nu} = 2\nu. 
\end{align}
with $\Delta_{\nu} = -2\beta M  = \chi(g)$, which can be interpreted as the number density where,
\begin{align}\label{radial_eq}
    g(\vec{R}) = \exp(\Delta_{\nu}\Phi(\vec{R})),
\end{align}
is the radial distribution function interpreted as a Boltzmann factor, $2M\Phi(\vec{R})$ is the potential energy governing the dynamics of the cations (\textit{i.e.} the average work needed to bring the two cations from infinite separation to a distance $\vec{R}$ apart whilst assuming a cavity distribution function of order unity\cite{hansen2013theory}), $\rho_0$ is the bulk density and $\chi(g) = 2 - 2g$ is the Euler characteristic and $g$ is number of cations. For instance, for a 2D system, the $z$-like Killing vector, $n^{\nu} = (\vec{0}, 1)$ is imposed, $n^{\mu}\partial_{\mu}\Phi(\vec{R}) = 0$, which reduces eq. (\ref{Newton_eq}) to Liouville's equation given in eq. (\ref{Liouville_eq}) with Gaussian curvature, $K(\vec{R}) = \beta\rho_0\exp(-2g\Phi(\vec{R}))$. \red{Thus, eq. (\ref{Newton_eq}) in 3D corresponds to the Emden-Chandrasekhar equation whose solution is given by\cite{chandrasekhar1949isothermal, kanyolo2021reproducing},} 
\begin{subequations}\label{solutions_eq}
\begin{align}
    \Phi(R) = \frac{1}{\nu}\ln(2k_{\rm F}R),\\
    \rho_0 = 4\nu k_{\rm F}^2/\beta,\\
    \rho(R) = \frac{1}{\beta\nu R^2},
\end{align}
\end{subequations}
where we have introduced the Fermi wave vector, $k_{\rm F}$ to be later justified. Using the number density, $\rho(R)$ in eq. (\ref{solutions_eq}), the normalisation of the 3D number density in eq. (\ref{norm_rho_eq}) corresponds to, 
\begin{subequations}\label{cut_off_delta_eq}
\begin{align}
    \Delta_{\nu} = -\int_0^{R_{\rm c}} dR\,4\pi R^2\rho(R) = -\frac{4\pi}{\beta\nu}\int_0^{R_{\rm c}} dR, 
\end{align}
\red{which requires} the cut-off scale, 
\begin{align}
    R_{\rm c} = \beta\nu^2/2\pi,
\end{align}
\end{subequations}
for consistency. Physically, this cut-off scale requires the radial distribution function of the cations to not extend beyond a width $R_{\rm c}$. 
\red{Since R is the displacement given in Figure \ref{Fig_13}}, $R_{\rm c}$ corresponds to the maximum displacement, $R_{\rm c} = \pi\nu/k_{\rm F}$ which defines the Fermi wave vector as $k_{\rm F} = 2\pi^2/\nu\beta$ ($2k_{\rm F}R_{\rm c} = 2\pi\nu$ where $\nu \in \mathbb{N}$ are the Brillouin zones). 

Proceeding, we can now obtain the correlation functions using eq. (\ref{correlation_relate_eq}) and $\rho(R)$ in eq. (\ref{solutions_eq}). The 
\red{Fourier transform} yields, 
\begin{align}
    \left \langle \psi_{s_1}^*(\vec{k})\psi_{s_2}(\vec{k}) \right \rangle = 2J^2(|\vec{k}|)\left \langle S_1S_2 \right\rangle,
\end{align}
where, 
\begin{align}
    J^2(|\vec{k}|) = -\frac{64\pi^3}{\Delta_{\nu}\sqrt{\vec{k}\cdot\vec{k}}}.
\end{align}
Thus, assuming the temperature, $T(R)$ varies slowly over $\vec{R}$ with $J^2(|\vec{k}|) = J^2(|\vec{q} - \vec{k}|) \simeq J^2(k_{\rm F})$ evaluated at the Fermi surface, we can make the approximation $T(\vec{R}) \simeq T$ corresponding to $\mathcal{P}(\vec{R}) = \alpha T4\pi |\vec{R} - \vec{R'}|\delta^3(\vec{R} - \vec{R'})$ in eq. (\ref{static_P_eq}) and integrate eq. (\ref{cont_calculation_eq}) by $t$ and $\overline{z}$ to find, 
\begin{align}
    -iT^{-1}4J^2(k_{\rm F})\left \langle S_1S_2 \right\rangle\int d^{\,3}R\int \frac{d\omega}{2\pi}\,G^2(i\omega, \vec{R}),
\end{align}
where, 
\begin{align}
    G(i\omega, \vec{R}) = \int \frac{d^{\,3}k}{(2\pi)^3}\frac{\exp(i\vec{k}\cdot\vec{R})}{i\omega - E_{|\vec{k}|}},
\end{align}
we have used $\langle S_1S_2 \rangle = \langle S_2S_1\rangle$ and $G(i\omega, \vec{R})$ is the thermal Green's function\cite{aristov1997indirect}. Integrating over $\omega$ and $|\vec{k}| \leq k_{\rm F}$, $|\vec{q} - \vec{k}| \geq k_{\rm F}$, and setting $E_{|\vec{k}|} = |\vec{k}|^2/2m + m$ as the non-relativistic energy of the cations yields\cite{aristov1997indirect},
\begin{align}\label{S_RKKY_eq}
    S_{\rm RKKY} = i\frac{4J^2(k_{\rm F})}{T}\langle S_1S_2 \rangle \int d^{\,3}R\,\frac{mk_{\rm F}^4}{(2\pi)^3}f(R),
\end{align}
where we have defined $S_{\rm RKKY} = -m\int \delta \rho(R)$ from eq. (\ref{S_mass_eq}) and,
\begin{align}
    f(R) = \frac{(2k_{\rm F}R\cos(2k_{\rm F}R) - \sin(2k_{\rm F}R))}{(2k_{\rm F}R)^4},
\end{align}
is the susceptibility, which 
\red{is proportional} to the Fourier transform of the well-known Lindhard function.\cite{yosida1957magnetic, aristov1997indirect, beal1987ruderman, fischer1975magnetic} 

Note that, eq. (\ref{S_RKKY_eq}) can be written as $S_{\rm RKKY} = \int_0^{i\beta} dt\,\mathcal{H}_{\rm RKKY}$, where $\beta = 1/T$ and $\mathcal{H}_{\rm RKKY}$ is the RKKY interaction Hamiltonian.\cite{yosida1957magnetic, aristov1997indirect, beal1987ruderman, fischer1975magnetic} Thus, considering the static case in 3D, 
\begin{align}
    S_{\rm RKKY}(R_{\rm c}) = i\frac{4J^2(k_{\rm F})}{T}\langle S_1S_2 \rangle \frac{mk_{\rm F}^4}{(2\pi)^3}f(R)\frac{4\pi R_{\rm c}^3}{3},
\end{align}
where $R_{\rm c} = -\pi\Delta_{\nu}/2k_{\rm F}$, the RKKY indirect exchange interaction of the pseudo-spins given by the fluctuation term, $\delta\rho$ yields\cite{aristov1997indirect}, 
\begin{align}
   \mathcal{H}_{\rm RKKY} = \sigma f(R)\langle S_1S_2 \rangle = -J_{\rm RKKY}\langle S_1S_2 \rangle,
\end{align}
where $J_{\rm RKKY}$ is the Ruderman–Kittel–Kasuya–Yosida (RKKY) interaction, 
\begin{align}\label{RKKY_eq}
    J_{\rm RKKY} = -m\sigma\frac{2k_{\rm F}R\cos(2k_{\rm F}R) - \sin(2k_{\rm F}R)}{(2k_{\rm F}R)^4},
\end{align}
and $\sigma = 16\pi^4\Delta_{\nu}^2/3$. Moreover, the density term, $\rho = 2/\beta\Delta_{\nu}R^2$ evaluated in 3D yields eq. (\ref{norm_rho_eq}). 
\red{However, since we wish to interpret the density term as the mass term on the honeycomb lattice in 2D, we shall evaluate the density term in 2D, differing from the fluctuation term evaluated in 3D}. Thus, the static approximation for the density term in 2D becomes, 
\begin{align}
    \rho(R)\pi R_{\rm c}^2 = -m\Delta_{\nu}2\pi^3/(2k_{\rm F}R)^2,
\end{align}
where $\beta = 1/m$ is the cut-off along the $z$ coordinate and $R_{\rm c} = -\pi\Delta_{\nu}/2k_{\rm F}$ the cut-off along the $x$-$y$ plane. 

Thus, like stretched, strained or folded graphene, the 2D density of the cations can interact with pseudo-spins as a pseudo-magnetic field.\cite{kanyolo2022cationic, mecklenburg2011spin, georgi2017tuning} Thus, the total energy corresponds to $\mathcal{F} = -(m - T(R))\left \langle\overline{\psi}\psi \right \rangle$, where $\mathcal{F}$ is a free energy of the form (1D Ising Hamiltonian), 
\begin{multline}\label{free_energy_pseudo_eq}
    \mathcal{F} = -\frac{\mathcal{B}_z}{2m}\left \langle \sum_jS_j \right\rangle - J_{\rm RKKY}\langle S_1S_2 \rangle\\
    = -m\Delta_{\nu}2\pi^3\varphi^2 + \frac{m\sigma}{4}\varphi^3,  
\end{multline}
where $\varphi = 1/2k_{\rm F}R$ with the first term the pseudo-magnetic field in the $z$ direction,
\begin{align}\label{pseudo_mag_eq}
    \mathcal{B}_z \equiv \vec{\mathcal{B}}\cdot\vec{n} = -4\pi^3m^2\Delta_{\nu}/(2k_{\rm F}R)^2 \geq 0,
\end{align}
$\vec{n} = (0, 0, 1)$ is the vector normal to the honeycomb lattice and the second term (the RKKY interaction) given in eq. (\ref{RKKY_eq}), with $\sigma = 16\pi^4\Delta_{\nu}^2/3$ a constant, $\Delta_{\nu}$ interpreted as the 
`Zeeman splitting' of the $S = 1, \tilde{m} = 0$ and 
$S = 1, \tilde{m} = \pm 1$ states (eq. (\ref{triplet_eq}) and eq. (\ref{triplet_eq2})), the spin averages are evaluated for the triplet state (eq. (\ref{triplet_eq2})) and $2k_{\rm F}R = 2\pi\nu$ evaluated at the Brillouin zones, $\nu$. Minimising the free energy with respect to $\varphi = 1/2k_{\rm F}R$ yields, 
\begin{align}
    m\varphi\left(-4\pi^3\Delta_{\nu} + \frac{3}{4}\sigma\varphi^2 \right ) = 0, 
\end{align}
which has a solution for $\Delta_{\nu} = 2\nu \geq 0$. Note that, for the other states with vanishing magnetisation, $\tilde{m} = 0$ (eq. (\ref{singlet_eq}) and eq. (\ref{triplet_eq})), the free energy,  $\mathcal{F} \propto \Delta_{\nu}^2\varphi^3$ in eq. (\ref{free_energy_pseudo_eq}) is minimised for $\Delta_{\nu}^2/(2k_{\rm F}R)^2 = 0$ where $2k_{\rm F}R = 2\pi\nu$. 

However, since $\Delta_{\nu} = 2\pi \nu$, $\Delta_{\nu}^2/(2k_{\rm F}R)^2 = 1$, and this is never satisfied. Thus, we must have a finite magnetisation, $\tilde{m} = 1$ for this free energy. Moreover, as discussed earlier, 
\red{the displacement vanishes} ($R = 0$) for the 2-torus, $\nu = 0$ whereas $\nu \neq 0$ leads to a bifurcation, $R \neq 0$. Experimentally, such a bifurcation has been found in layered materials with the coinage metal atom, Ag which satisfy rather general properties such as the existence of Ag-Ag' 
\red{numismophilicity} 
\red{assumed to give} rise to $J_{\rm RKKY} \neq 0$.\cite{schreyer2002synthesis, yoshida2006spin, beesk1981x, johannes2007formation} This result, and 
\red{the nature of conventional 
\red{numismophilicity} as electronic RKKY interactions} in coinage metal atoms will be 
\red{explored} in subsequent sections. 
\red{Finally, since the free energy originates from the mass term associated with the spontaneous symmetry breaking of SU($2$)$\times$ U($1$) which also breaks scale invariance, using eq. (\ref{ansatz_eq}), we can replace $\Delta_{\nu}$ with $2\Delta(d) = d - 2$, thus differentiating cationic vacancy creation from bifurcation}. 

\subsubsection{Newtonian gravity dual}

We shall consider the energy density from $S_{\rm mass} = \int_0^{i\beta} dt \int_{\mathcal{V}} d^{\,3}R\,\,\mathcal{E}(R)$ in eq. (\ref{S_mass_eq}) instead of the energy, where $\mathcal{E}$ is the energy density. We are interested in 
\red{the energy} $a(R) = \mathcal{V}\mathcal{E}(R)$, when the 3D volume is defined as,
\begin{align}\label{mathcal_V_eq}
    \mathcal{V} = \Delta_{\nu}\beta^3/2. 
\end{align}
\red{The energy corresponds to},
\begin{multline}\label{gravity_dual_acceleration_eq}
    a(R) = -\frac{\Delta_{\nu}\beta^2}{2}\rho(R) + \Delta_{\nu}\beta^2J^2(k_{\rm F})2\left\langle S_1S_2\right\rangle \frac{k_{\rm F}^4}{(2\pi)^3}f(R)\\
    a_-(R) + a_+(R) = -\frac{\mu M}{R^2} + \frac{L^2}{m^2R^3}, 
\end{multline}
where $\beta = 1/m = 8\pi GM$, $\mu = \beta/M$, $M = \Delta_{\nu}m/2$ and we have set, $2k_{\rm F}R = \pi\Delta_{\nu}$ and $L^2 = 2\left\langle S_1S_2\right\rangle = \left\langle S^2 - S_1^2 - S_2^2 \right\rangle$ with $S = S_1 + S_2$. With our definitions, this energy takes the form of the Newtonian gravitational acceleration $a_-(R) = -\partial V_-(R)/m\partial R$ with a reduced Newton's constant, $\mu = 8\pi G$ and a repulsive centrifugal acceleration, $a_+(R) = -\partial V_+(R)/m\partial R$ where $L$ plays the role of the angular momentum of the particle of mass, $m$. 

Thus, the gravitational term arises from the CFT description satisfying the Emden-Chandrasekhar equation in eq. (\ref{Newton_eq}) whereas the angular momentum term arises from the RKKY interaction. 
\red{This represents a duality between the Newtonian gravitational description and the 1D Ising model formalism captured by eq. (\ref{Newt_gravity_S_eq}). Surprisingly, pseudo-spin correlations are interpreted as the angular momentum-squared of the cations.} 
\red{Qualitatively}, the problem of finding stable separation distances $R$ between bifurcated honeycomb layers is mapped to the problem of finding stable orbits in Newtonian gravity, with the exception that $\mathcal{V}$ is defined differently from eq. (\ref{mathcal_V_eq}) in the actual calculation ($\mathcal{V}_{\rm 2D} = \beta\pi(\pi\Delta_{\nu}/2k_{\rm F})^2$, $\mathcal{V}_{\rm 3D} = 4\pi(\pi\Delta_{\nu}/2k_{\rm F})^3/3$ in eq. (\ref{S_mass_eq})). In other words, for finite $R$ with eq. (\ref{mathcal_V_eq}), we are interested in the vanishing point, $\partial V(R)/m\partial R = a(R_{\rm c}) = 0$, which corresponds to $R_{\rm c} = 2L^2/\mu Mm^2 = 2\langle S_1S_2 \rangle/m \geq 0$. Thus, qualitatively, there are no stable orbits for the $S = 0$ state since the centrifugal force becomes attractive. Moreover, since the gravitational term arises from a finite magnetisation, $\tilde{m} = 1$, this duality appears valid only for the state $\tilde{m} = 1, S = 1$. 

\subsubsection{Electronic RKKY interactions}

\red{It is clear that the two silver cations in the honeycomb unit cell can interact via the RKKY and pseudo-magnetic field terms given by the free energy (1D Ising model) corresponding to eq. (\ref{free_energy_pseudo_eq}). Whilst the interactions are between two Ag cations, $\psi^{\rm T} = ({\rm Ag}^{2+}_{s_1}, {\rm Ag}^{1-}_{s_2})$, electronically, it is the spin of the electron in the half-filled $\rm Ag^{2+}$ $4d_{z^2}$ orbital that interacts with the spin of the half-filled $\rm Ag^{1+}$ $5s$ orbital via the RKKY interaction mediated by conduction electrons that can move freely from the valency band (all Ag electrons in the $5s$ orbitals) to the conduction band (all Ag electrons in the $4d_{z^2}$ orbitals). In other words, this is the so-called $sd$ interaction.\cite{aristov1997indirect}
Indeed, even when the cations in the lattice are localised, the electrons are mobile and hence responsible for additional interactions that form bonds that stabilise the bilayers.}

To elucidate this, we shall first consider the electronic RKKY interactions for 3D materials with the coinage metal atoms without incorporating the physics of phase transitions,
\begin{align}
    J_{\rm RKKY}^A = -4J^2(k_{\rm F}^{\rm e})\int d^{\,3}R\,\frac{m_{\rm e}^{\rm rel.}k_{\rm F}^{\rm e\,4}}{(2\pi)^3}f(R),
\end{align}
where, 
\begin{align}
    \left \langle e_{s_1}^*(\vec{k})e_{s_2}(\vec{k}) \right \rangle = 2J^2(|\vec{k}|)\left \langle S_1S_2 \right\rangle,
\end{align}
$e_s$ are the conduction electron wave functions, $s = s_1, s_2$ are their spins, $A =$ Cu, Ag or Au coinage metal atoms, $k_{\rm F}^{\rm e}$ is the electron Fermi vector and,
\begin{align}
    m_{\rm e}^{\rm rel.}(Z, n) = \frac{m_{\rm e}^*}{\sqrt{1 - (Z\alpha/n)^2}},
\end{align}
is the relativistic mass of the conduction electrons with $m_{\rm e}^*$ the effective non-relativistic electron mass in the material, $\alpha \simeq 1/137$ the fine structure constant, $(Z/n) = (29/3 \simeq 9.67), (47/4 = 11.75), (79/5 = 15.8)$ the fraction (atomic number/principal quantum number) of the number of protons of Cu, Ag, Au coinage metal atoms respectively and the principal quantum number $n$ of their $ns$ orbital. In the presence of 
a magnetic field, $B_z = \vec{B}\cdot\vec{n}$, the total interaction Hamiltonian is given by the 1D Ising model,
\begin{align}\label{electron_Ising_eq}
    \mathcal{H}_{\rm Ising} = -J_{\rm RKKY}^AS_1\cdot S_2 - \frac{B_z}{2m_{\rm e}^*}(S_1 + S_2).
\end{align}
Due to the relativistic mass appearing in the expression, we obtain $|J_{\rm RKKY}^{\rm Cu}| < |J_{\rm RKKY}^{\rm Ag}| < |J_{\rm RKKY}^{\rm Au}|$ 
\red{where} the Ag 
\red{numismophilic} (argentophilic) interaction is greater than Cu 
\red{numismophilic} (cuprophilic) interaction but smaller than Au 
\red{numismophilic} (aurophilic) interaction 
\red{in the absence of a magnetic field,} $\vec{B} = 0$. 

\red{Nonetheless}, a finite pseudo-magnetic field can be incorporated using phonon interactions even for $\vec{B} = 0$. In particular, one can consider an attractive phonon interaction term of the form, 
\begin{align}\label{phonon_electron_eq}
    E_{\rm elec.-phon.} = \int d^{\,3}R\,d^{\,3}R'\rho(\vec{R})G(\vec{R} - \vec{R'})\rho(\vec{R'}), 
\end{align}
where $\rho(R) \geq 0$ is the 3D electron density,  
\begin{align}\label{phonon_Green_eq}
    G(\vec{R} - \vec{R'}) = -\frac{g_{\rm phon.}^2}{2}\int d^{\,3}q\,\frac{\exp(i\vec{q}\cdot(\vec{R} - \vec{R'}))}{|\vec{q}\,\,|^2},
\end{align}
is the mass-less phonon mediated interaction (Green's function), $g_{\rm phon.}$ is the electron-phonon coupling constant and $\vec{q}$ is the phonon momentum. Whilst the phonons are mass-less, the phonon-mediated electron-electron interaction can be restricted to the vicinity of the Fermi sphere such that $|\vec{k} + \vec{k'}| = |\vec{q}\,| = 2k_{\rm F}^{\rm e}$, where $\vec{k}$ and $\vec{k'}$ are the wave vectors of the interacting electrons and $\vec{k}_{\rm F}^{\rm e}$ is the Fermi wave vector. This is akin to introducing a pairing mechanism for the electrons via the phonon interaction, where opposite spins form Cooper-pairs on the Fermi sphere.\cite{tinkham2004introduction} 

Moreover, introducing the cut off length scale of phonon interactions, $R_{\rm c}$ (consistent with eq. (\ref{cut_off_delta_eq})) to the Green's function ($G(\vec{R} - \vec{R'}) \rightarrow G(\vec{R}_{\rm c} - \vec{R'})$) and performing the first integral over $R'$ in eq. (\ref{phonon_electron_eq}) yields, 
\begin{align}
    E_{\rm elec.-phon.} = -\frac{g_{\rm phon.}^2\rho(\vec{R}_{\rm c})}{(2k_{\rm F}^{\rm e})^2}\int d^{\,3}R\,\rho(\vec{R}). 
\end{align}
Thus second integral over $\vec{R}$, corresponding to eq. (\ref{cut_off_delta_eq}) yields the conformal dimension resulting in,
\begin{align}
    E_{\rm elec.-phon.} = \frac{g_{\rm phon.}^2}{(2k_{\rm F}^{\rm e})^2}\rho(\vec{R}_{\rm c})\Delta_{\nu} = -\mathcal{B}_z/2m,
\end{align}
with $\Delta_{\nu} = -2\nu \leq 0$ and $\beta = 1/m$. Consequently, the electron-phonon interaction is proportional to the electron density and the conformal dimension, and takes the form of the pseudo-magnetic field, $\mathcal{B}_z$ given in eq. (\ref{pseudo_mag_eq}) where $\rho(\vec{R}) = 1/\beta\nu R^2$ (eq. (\ref{solutions_eq})) and the coupling yields $g_{\rm phon.} = \nu = \sqrt{2\pi R_{\rm c}/\beta}$, where we have used eq. (\ref{cut_off_delta_eq}). Nonetheless, it is imperative to show that $\mathcal{B}_{\rm z}$ indeed behaves as a pseudo-magnetic field, namely that it arises from a vector potential, $\vec{\mathcal{B}} = \vec{\nabla}\times\vec{\mathcal{A}}$. It turns out this is the case only if the 3D system exhibits features of a 2D system, namely the current density couples to a Chern-Simons theory.\cite{dunne1999aspects} Indeed, this is the case for a bifurcated honeycomb lattice (3D) provided there are no charge-exchange interactions between the sub-lattices (2D). 

In particular, local energy-momentum conservation in 2D, $\partial_{\mu}p^{\mu} = 0$ must take the Chern-Simons form\cite{dunne1999aspects}, 
\begin{multline}
    p^{\mu} = \int d^{\,3}R\,d^{\,3}R'j^{\mu}(\vec{R})G(\vec{R} - \vec{R'})j^0(\vec{R'})\\
    = -\frac{\beta|\tilde{m}|}{2}\varepsilon^{\mu\nu\sigma}\partial_{\nu}\mathcal{A}_{\sigma}, 
\end{multline}
where $j^{\mu} = (\rho, \vec{j})$ is the current density, $\mathcal{A}_{\sigma} = (\mathcal{A}^0, \vec{\mathcal{A}})$ is a U(1) pseudo-vector potential, $\varepsilon^{\mu\nu\sigma} = \varepsilon^{\mu\nu\sigma\rho}n_{\nu}$, $n_{\nu} = (0, \vec{n})$ is the 4-vector normal to the 2D layers with $\vec{n} = (0, 0, 1)$, $\tilde{m} = \left \langle S_1 + S_2 \right \rangle$ is the magnetisation and $\varepsilon^{\mu\nu\sigma\rho}$ is the 4D Levi-Civita symbol. The form of the current density, $j^{\mu}$ is determined by the influence of the cation lattice and its Brillouin zones. Moreover, like in graphene, the electron dynamics at the Dirac \red{points} 
on the honeycomb vertices 
\red{is mass-less}. This means that the mass term $m = 1/\beta$ has to arise entirely from symmetry breaking. Indeed, this is the case for the cation dymanics whereby $m = 1/\beta$ is the cationic mass, which corresponds to the 2D-3D transition temperature for the cationic lattice. Since the electron dynamics on the honeycomb lattice are governed by the same phase transition, the electrons also acquire the same mass, $m = 1/\beta$. This means that the mass terms for the electrons and the cations can collectively be written as, 
\begin{subequations}
\begin{multline}
    \mathcal{H}_{\rm Ising}(e, \psi) = 2\Delta(d)\int d^{\,d}R\,\,\beta^{-2\Delta(d)}(\vec{R})\left \langle \overline{e}{e} \right\rangle\\
    + 2\Delta(d)\int d^{\,d}R\,\,\beta^{-2\Delta(d)}(\vec{R})\left\langle \overline{\psi}\psi \right\rangle, 
\end{multline}
\end{subequations}
where $\Delta(d) = (d - 2)/2$ is the scaling dimension, $\beta(\vec{R}) = 1/(T(\vec{R}) - m) \geq 0$,
\begin{subequations}
\begin{align}
    \overline{e}(\vec{k},t) = e^{*\rm T}(\vec{k},t)(\gamma^0)^{-1},\\
    e(\vec{k}, t) = \begin{pmatrix}
e_{s_1}(\vec{k}, t)\\ e_{s_2}(\vec{q} - \vec{k}, t)
\end{pmatrix},
\end{align}
\end{subequations}
and $e_{s}(\vec{k}, t)$ is the electron wave function of momentum $\vec{k}$ and spin $s$, which can be calculated as before to yield the pseudo-magnetic term, where one must distinguish between the electron and the cation Fermi vectors, $\vec{k}_{\rm F}^{\rm e}$ and $k_{\rm F}$ respectively. Finally, $\beta(\vec{R})$ is responsible for a monolayer-bilayer phase transition and hence a finite mass term for $T(\vec{R}) \geq m$ and $d = 3$, as required, leading to no distinctions between the electronic and cationic pseudo-magnetic fields. Since the pseudo-magnetic field is the analogue of Newtonian force of gravity, this statement can be viewed as the analogue of the equivalence principle, whereby all objects (cations and electrons) fall at the same rate in a gravitational field.\cite{thorne2000gravitation} 

\begin{figure*}
\begin{center}
\includegraphics[width=\textwidth,clip=true]{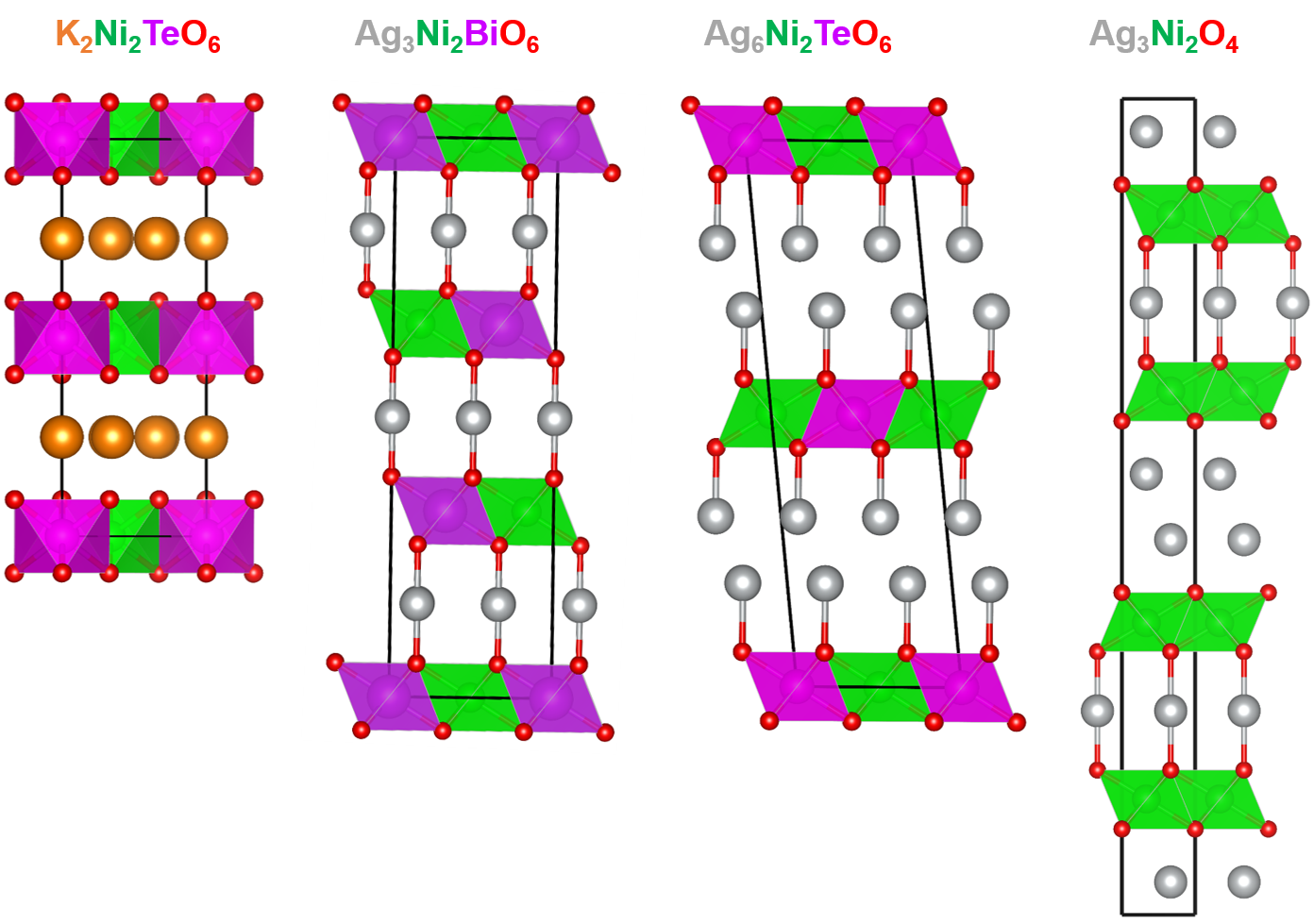}
\caption{Crystal structures of $\rm K_2Ni_2TeO_6$, $\rm Ag_3Ni_2BiO_6$, $\rm Ag_6Ni_2TeO_6$ and $\rm Ag_3Ni_2O_4$ visualised along [100] projection showing the arrangement of their constituent atoms. K atoms are shown in brown, Ag atoms are shown in grey, O atoms are shown in red, Ni octahedra shown in green and Te/Ni octahedra shown in purple. Thus, their cationic lattices are monolayered, monolayered, bilayered and hybrid respectively. Other differing structures and properties have been tabulated in Table \ref{Table_3}.} 
\label{Fig_14}
\end{center}
\end{figure*}

\begin{table*}[!t]
    \caption{Tabulated properties of cations and their lattices in the honeycomb layered oxides, $\rm K_2Ni_2TeO_6$, $\rm Ag_3Ni_2BiO_6$, $\rm Ag_6Ni_2TeO_6$ and $\rm Ag_3Ni_2O_4$ with crystal structures given in Figure \ref{Fig_14}.}\label{Table_3}
    \centering
    \resizebox{\textwidth}{!}{
    \begin{tabular}{|c|c|c|c|c|}
    \hline
    &&&&\\
    \textbf{property $\diagdown$ material} & $\rm K_2^{1+}Ni_2^{2+}Te^{6+}O_6^{2-}$ & ${\rm Ag_3^{1+}Ni_2^{2+}Bi^{5+}O_6^{2-}}$ & ${\rm Ag_6^{2+}Ag_6^{1-}Ni_4^{2+}Te^{4+}Te^{6+}O_{12}^{2-}}$ & $\rm Ag^{2+}Ag^{1-}Ag^{1+}Ni_2^{3+}O_4^{2-}$\\
     & (monolayer) & (monolayer) & (bilayer) & (hybrid) \\
    \hline \hline
    valency & 1+ & 1+ & 1/2+ & 2/3+\\
    spin & integer & half-integer & half-integer & half-integer\\
    isospin & undefined & $+1/2$ & $-1/2$, $0$ & $0$, $\pm 1/2$\\
    effective charge, $Q$ & $+1$ & $+1$ & $-1$, $+2$ & $-1$, $+1$, $+2$\\
    chirality & undefined & left & left, right & left, right\\
    pseudo-spin & undefined & $0$ & $\pm 1/2$ & $0$, $\pm 1/2$\\
    scale invariant & yes & yes & no & no (bilayers)\\
    metallophilic bond & no & no & yes & yes (bilayers)\\
    \hline
    \end{tabular}}
\end{table*}

\subsubsection{Bilayered hybrids}


\red{The triumph herein is that the formalism satisfies conditions for observation of bilayers, (i), (ii) and (iii) stated explicitly in the beginning. Specifically, (i) is satisfied by the existence of a mass gap for $\rm Ag$ atoms in a bifurcated lattice \textit{i.e.} a chiral symmetry breaking term between $\rm Ag^{2+}$ and $\rm Ag^{1-}$ degenerate states; (ii) is satisfied for instance by writing\cite{kanyolo2022advances}, $\rm Ag_2^{1/2+}Ni^{3+}O_2^{2-} = Ag^{2+}Ag^{1-}Ni^{3+}O_2^{2-}$, which guarantees the existence of subvalent states; and (iii) is satisfied by the existence of degenerate pseudo-spin states on the honeycomb lattice, which guarantees the lattice distorts by bifurcation, thus satisfying an analogue of Peierls/Jahn Teller theorem\cite{garcia1992dimerization, jahn1937stability}, which states that distortions that lift degeneracies in lattices are more stable.} Moreover, Ag bilayered hybrids with alternating monolayers and bilayers within the same material have been observed, which correspond to other subvalent states. For instance, $\rm Ag_3^{2/3+}Ni_2^{3+}O_4^{2-} = Ag^{2+}Ag^{1-}Ag^{1+}Ni_2^{3+}O_4^{2-}$ represents a saturation or hybrid effect by the mass-less $\rm Ag^{1+}$ fermion\cite{kanyolo2022advances},
\begin{multline}
    \rm Ag^{2+}Ag^{1-}Ni^{3+}O_2^{2-} + Ag^{1+}Ni^{3+}O_2^{2-}\\ \rightarrow \rm Ag^{2+}Ag^{1-}Ag^{1+}Ni_2^{3+}O_4^{2-}.
\end{multline}
Lastly, $\rm Ag^{1+}Ni^{3+}O_2^{2-}$ cannot be bilayered since $\rm Ag^{1+}$ is mass-less in the theory. 

\red{In the case of the material in Figure \ref{Fig_10} (introduced as ${\rm Ag}_2M_2\rm TeO_6$ with $M = \rm Co$), assuming $\rm Te^{6+}$, the under-saturated bilayered material is expected to be given by ${\rm Ag^{2+}_2Ag^{1-}_2Co}_2^{2+}{\rm Te}^{6+}{\rm O_6^{2-}}$ = $\rm Ag_4^{1/2+}Co_2^{2+}Te^{6+}O_6^{2-}$ or ${\rm Ag^{2+}_4Ag^{1-}_4Co}_2^{2+}{\rm Te}^{4+}{\rm O_6^{2-}}$ = $\rm Ag_8^{1/2+}Co_2^{2+}Te^{4+}O_6^{2-}$ ($\rm Ag$ coordination to $\rm O$ is assumed prismatic) with Ag sub-valency $+1/2$, consistent with the experimental observations ($\rm Co^{2+}, Te^{4+}, Te^{6+}$). Thus, we can consider the Te hybrid},
\begin{subequations}
\begin{multline}
    {\rm Ag^{2+}_2Ag^{1-}_2Co}_2^{2+}{\rm Te}^{6+}{\rm O_6^{2-} + Ag^{2+}_4Ag^{1-}_4Co}_2^{2+}{\rm Te}^{4+}{\rm O_6^{2-}}\\ \rightarrow {\rm Ag_6^{2+}Ag_6^{1-}Co_4^{2+}Te^{4+}Te^{6+}O_{12}^{2-}},
\end{multline}
\end{subequations}
which is bilayered, since ${\rm Ag_6^{2+}Ag_6^{1-}Co_4^{2+}Te^{4+}Te^{6+}O_{12}^{2-}} = {\rm 2Ag_6^{1/2+}Co_2^{2+}Te^{5+}O_6^{2-}}$ and hence $\rm Ag^{1/2+}$ is subvalent. However, the subvalency of Ag in the as reported $\rm Ag_2Ni_2TeO_6$ with bilayers has not yet been unequivocally ascertained experimentally.\cite{masese2023honeycomb}
\red{The discussion above is also valid for $M = \rm Ni$, as has been reported elsewhere}.\cite{masese2023honeycomb} Finally, the crystal structures of typical nickel-based honeycomb layered oxides showcasing their monolayered/bilayered/hybrid cationic lattices has been availed in Figure \ref{Fig_14}. The differing properties of their cations has been tabulated in Table \ref{Table_3}.

\subsection{Idealised model (fermions)}

\subsubsection{Non-relativistic treatment}


\red{Consider the primitive cell of the honeycomb lattice of cations given in Figure \ref{Fig_5} by the blue-shaded rhombus. Each primitive cell is comprised of two cationic sites as earlier discussed. We shall assign an index, $j = 1,2,3 \cdots k$ for each primitive cell on the honeycomb lattice. Figure \ref{Fig_10} (a) and (b), show that such a honeycomb lattice bifurcates into bilayers as seen in the STEM images (Figure \ref{Fig_10} (c) and (d)), with each primitive cell contributing a single cation from a specific site to form two triangular sub-lattices on top of each other, as shown in Figure \ref{Fig_13}. This implies that we can assign a wave function, $\varphi_{a}(\vec{r}_j)$ for each Ag atom, where $\vec{r}_j = (\vec{r}_{\uparrow} - \vec{r}_{\downarrow})_j$ is the displacement vector of the first cationic site in the primitive cell relative to the second.}

Performing the double rotation operation 
\red{twice} yields the identity matrix, $S^4 = I$, suggesting the honeycomb primitive cell 
\red{can be identified by} two eigenvalues, $\pm 1/2$, corresponding to two pseudo-spins defining a pair of triangular sub-lattices. Thus, 
\red{it is the two cationic sites with eigenvalues $\pm 1/2$ which bifurcate into bilayers}, 
\red{that is responsible for breaking scale invariance of the honeycomb lattice}. This is evident by realizing that re-scaling the basis, $\omega_1 \rightarrow \Omega\omega_1, \omega_2 \rightarrow \Omega\omega_2$, where $\Omega$ is the scaling factor, leaves $k = \omega_1/\omega_2 = 1/\bar{k}$ invariant in the monolayer, but is ill-defined for the bilayer due to bifurcation. 

\red{Proceeding}, it is prudent to consider the 2-spinor,
\begin{subequations}
\begin{align}
\varphi_j(\vec{r}_j)
= \begin{pmatrix}
\varphi_{\uparrow}(\vec{r}_j)\\\varphi_{\downarrow}(\vec{r}_j)
\end{pmatrix} \equiv \varphi_{\uparrow}(\vec{r}_j)|\uparrow \,\rangle + \varphi_{\downarrow}(\vec{r}_j)|\downarrow \,\rangle,\\
J_j^0 = |\varphi_j(\vec{r}_j)|^2 = |\varphi_{\uparrow}(\vec{r}_j)|^2 + |\varphi_{\downarrow}(\vec{r}_j)|^2,
\end{align}
\end{subequations}
where $J_j^0 = \rho_j(\vec{r}_j)g_j(\vec{r}_j)$ is \red{time component of the current density}, 
$g_j(\vec{r}_j)$ is the \red{pair-correlation function}, 
and $\rho_j(\vec{r}_j)$ the bulk density function of the unit cell at $j$,
\begin{subequations}
\begin{align}
    \frac{1}{2}\sigma_3|\uparrow \,\rangle = \frac{1}{2}S^{4n}|\uparrow \,\rangle,\\
    \frac{1}{2}\sigma_3|\downarrow \,\rangle = \frac{1}{2}S^{2(2n + 1)}|\downarrow \,\rangle
\end{align}
\end{subequations}
$n$ is a positive integer and,
\begin{subequations}
\begin{align}
\vec{\sigma} = (\sigma_1, \sigma_2, \sigma_3),\\
\sigma_1 = 
\begin{pmatrix}
0 & 1\\ 
1 & 0
\end{pmatrix},\,\,
\sigma_2 = 
\begin{pmatrix}
0 & -i\\ 
i & 0
\end{pmatrix},\,\,
\sigma_3 = 
\begin{pmatrix}
1 & 0\\ 
0 & -1
\end{pmatrix},
\end{align}
\end{subequations}
are the Pauli matrices. Thus, \red{we can impose, on each primitive cell, the Pauli Hamiltonian}\cite{bransden2003physics} with an RKKY interaction/Heisenberg term\cite{ruderman1954indirect, kasuya1956prog, yosida1957magnetic}, 
\begin{multline}\label{Pauli_eq}
    \mathcal{H}_j = \frac{1}{2m}\left (-i\vec{\nabla}_j - \vec{A}_j(\vec{r}_j) \right )^2 - \frac{1}{2m}\vec{B}_j(\vec{r}_j)\cdot\vec{\sigma}\\
    - J_{\rm RKKY}(\vec{r}_j - \vec{r}_{j'})\vec{\sigma}\cdot\vec{\sigma'},
\end{multline}
where 
$m$ is the effective mass at each cationic site, $\vec{\nabla}_j = \partial/\partial \vec{r}_j$, 
\begin{subequations}
\begin{align}
    \vec{B}_j(\vec{r}_j) = \vec{\nabla}\times\vec{A}_j(\vec{r}_j),\\
    \vec{A}_j(\vec{r}_j) = \frac{1}{m}(\vec{n}\times\vec{E}_j(\vec{r}_j)),
\end{align}
\end{subequations}
is the geometric curvature playing the role of pseudo-magnetic field and Berry connection playing the role of pseudo-gauge field respectively\cite{berry1984quantal, georgi2017tuning}, arising from 
\red{the conformal metric},
\begin{subequations}
\begin{align}
    dt_j^2 = g^j_{ab}(r_j)dr_j^adr_j^b = \exp(2\Phi_j(\vec{r}_j))(dx_j^2 + dy^2_j),
\end{align}
which is equivalent to the Liouville's equation\cite{zamolodchikov1996conformal}, 
\begin{align}\label{Liouville_j_eq}
    \nabla_j^2\Phi_j(\vec{r}_j) = -K_j(\vec{r}_j)\exp(2\Phi_j(\vec{r}_j)),
\end{align}
\end{subequations}
\red{with $\vec{E}_j(\vec{r}_j)$ the electric field, $K_j(\vec{r}_j)$ the Gaussian curvature, $\beta = 1/m$ acting as the effective cut-off thickness of the interlayer between the Te and Ni slabs, $g_{ab}(\vec{r}_j)$ the metric tensor}, 
\begin{align}
    \Phi_j(\vec{r}_j) = 4\pi\int d\vec{r}_j\cdot\vec{n}\times\vec{A}_j(\vec{r}_j),
\end{align}
and $\vec{n} = (0, 0, 1)$ the unit vector normal to the honeycomb lattice. 

The electric field satisfies Gauss' law,
\begin{align}\label{Gauss_eq2}
    \beta\vec{\nabla}_j\cdot\vec{E}_j(\vec{r}_j) = \rho_j(\vec{r}_j) g(\vec{r}_j) = \vec{n}\cdot\vec{B}_j(\vec{r}_j),
\end{align}
where $\rho_j(\vec{r}_j)$ is the 2D bulk number density and $g(\vec{r}_j)$ is the pair \red{pair correlation function} 
at each primitive cell. 
The \red{pair correlation function
is normalised as},
\begin{align}
    \int d^{\,2}r_j\,\vec{n}\cdot\vec{B}_j = \int d^{\,2}r_j \rho_j(\vec{r}_j) g_j(\vec{r}_j) = k - 1,
\end{align}
where $k = \omega_1/\omega_2$ is the number 
\red{cationic pairs} in the unit cell labeled by $j$. This implies that, for the primitive cell with $k = 1$, the integral vanishes. 
\red{Thus}, the Euler characteristic associated with the primitive cell corresponds to the geometric phase,
\begin{multline}
    \chi(\mathcal{A}_j) = \frac{1}{2\pi}\int d^{\,2}r_j\sqrt{\det(g^j_{ab})}K_j(\vec{r}_j)\\
    = -\frac{1}{2\pi}\int d^{\,2}r_j\nabla^2_j\Phi_j(\vec{r}_j) 
    = -\frac{1}{2\pi}\int d\vec{r}_j\cdot (\vec{n}\times\vec{\nabla}_j\Phi_j(\vec{r}_j))\\
    = -2\int d\vec{r}_j\cdot\vec{A}_j(\vec{r}_j)
    = -2\int d^{\,2}r_j\vec{n}\cdot\vec{B}_j(\vec{r}_j)\\
    = -2\beta\int d^{\,2}r_j\vec{\nabla}_j\cdot\vec{E}_j(\vec{r}_j)\\
    = -2\int d^{\,2}r_j \rho_j(\vec{r}_j) g_j(\vec{r}_j) = 2 - 2k = 0,
\end{multline}
\red{where $k = \omega_1/\omega_2 = 1$ is the genus of a 2-torus}, implying that,
\begin{align}\label{rho_g_K_eq}
    \rho_j(\vec{r}_j)g_j(\vec{r}_j) = -\frac{K_j(\vec{r}_j)}{4\pi}\exp(2\Phi_j(\vec{r}_j)).
\end{align}

\begin{figure*}
\begin{center}
\includegraphics[width=\textwidth,clip=true]{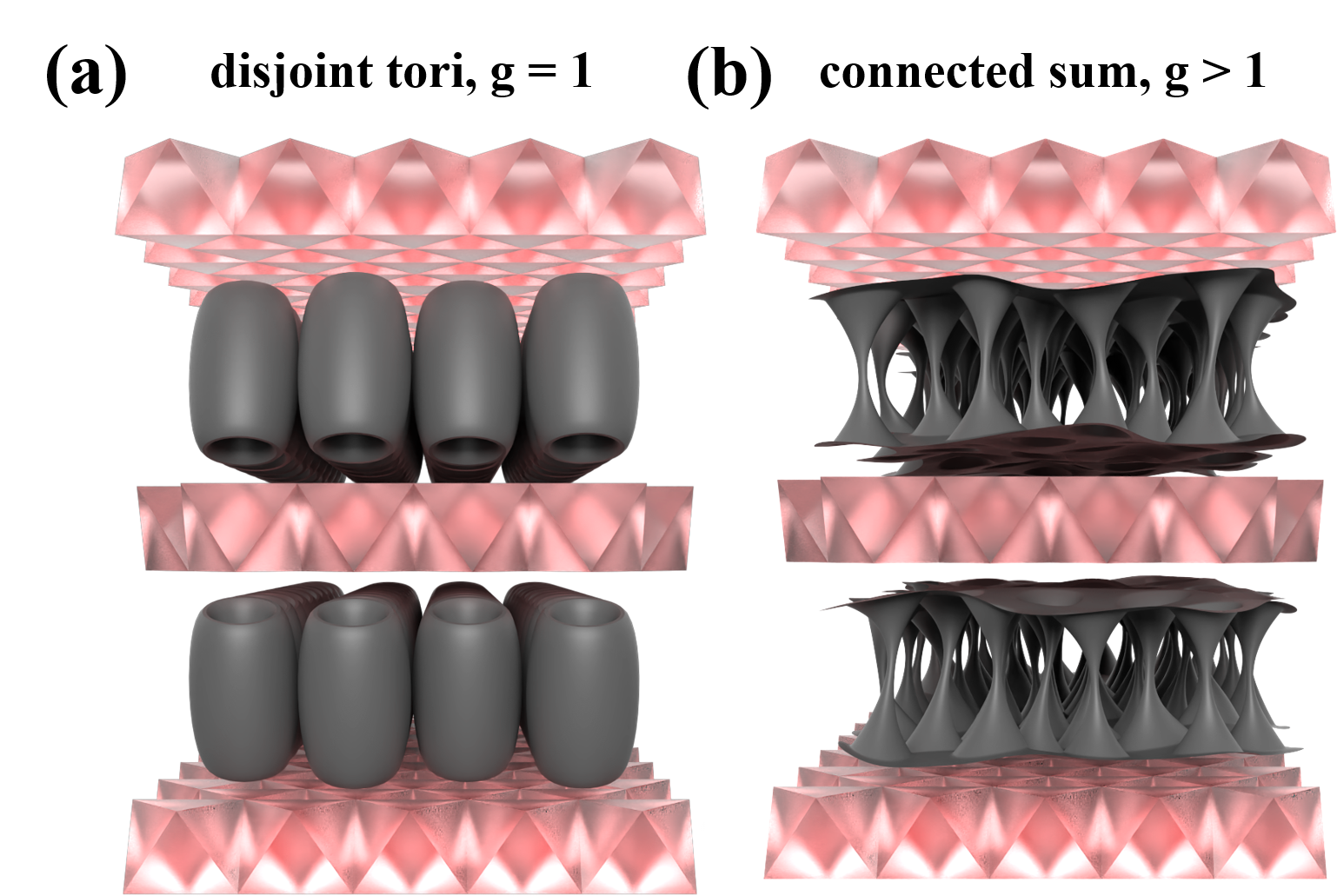}
\caption{
\red{(a) Illustration depicting primitive cells of the cationic lattice as disjoint tori ($g = 0$)} (b) The honeycomb lattice depicted as a connected sum of the tori in (a), which yields $g > 1$.}
\label{Fig_15}
\end{center}
\end{figure*}

Particularly, \red{we can define the primitive cell as a two-torus by identifying} 
the opposite sides of the primit cells with each other 
\red{following the illustration in Figure \ref{Fig_4}}.
\red{Consequently}, the entire honeycomb lattice can be treated as a connected sum of all the tori, $\mathcal{A}_j = T^2_j$ at each honeycomb primitive cell \red{with Euler characteristic}, 
$\chi(T^2_j) = 0$, as depicted in Figure \ref{Fig_15}(a), obtaining Figure \ref{Fig_15}(b). 
\red{The Euler characteristic of the emergent manfold $\mathcal{A} = T^2_1\#\cdots\#T^2_k$ thus becomes},
\begin{subequations}
\begin{align}
    \chi(\mathcal{A} = T^2_1\#\cdots\#T^2_k) = 2 - 2k,
\end{align}
where 
\red{used the property},
\begin{multline}
    \chi(\mathcal{M}_1\#\cdots\#\mathcal{M}_k)\\
    = \sum_{j = 1}^{k}\chi(\mathcal{M}_j) - (k - 1)\chi(S^2) = 2-2k,
\end{multline}
\end{subequations}
\red{of Euler characteristics of connected sums of manifolds, $\mathcal{M}_j$} and $S^2$ is the two-sphere with Euler characteristic, $\chi(S^2) = 2$. Thus, the number of cationic sites can be succinctly linked to the genus of the emergent manifold. 

\red{Moreover}, 
\red{making the choice},
\begin{subequations}\label{K_rho_eq}
\begin{align}
   \rho_j(\vec{r}_j) = -\frac{K_j(\vec{r}_j)}{4\pi}\exp(\Phi_j(r_j)),\\
    g_j(\vec{r}_j) = \exp(\Phi_j(\vec{r}_j)) = \exp(-\beta\phi_j(\vec{r}_j)),
\end{align}
where $\beta = 1/m$ and $\phi_j(\vec{r}_j) = -m\Phi_j(\vec{r}_j)$ is the 
\red{average work needed to bring the two pseudo-spin up and pseudo-spin down cations of effective mass, $m$ from infinite separation to a distance $r_j$ apart whilst assuming a cavity distribution function of order unity}\cite{hansen2013theory}, 
\red{we can set} $\rho_j(\vec{r}_j) = 0$ to obtain the solution, 
\begin{align}\label{conformal_eq}
    \Phi_j(\vec{r}_j) = \ln\left (\frac{C_j}{|(\vec{r}_{\uparrow} - \vec{r}_{\downarrow})_j|^{2\Delta}}  \right ),
\end{align}
\end{subequations}
in 
\red{2D with} $C_j$ and $\Delta$ 
constants. 
\red{This corresponds to $K_j = 0$, and hence $k = 1$, which is the primitive cell}. Proceeding, we can consider the Liouville action, in $d$ dimensions\cite{alvarez2013random}, 
\begin{subequations}
\begin{multline}\label{Liouville_action_eqq}
   I \propto \sum_j\int \frac{d^{d}x_j}{2}\sqrt{\det(\tilde{g}^j_{ab})}\left (\tilde{g}_j^{ab}\frac{\partial\Phi_j}{\partial x^{a}}\frac{\partial\Phi_j}{\partial x^{b}} - K_j\exp(2k\Phi_j) \right )\\
   + \sum_j\int \frac{d^{d}x_j}{2}\sqrt{\det(\tilde{g}^j_{ab})}Q(k)\tilde{R}_j\Phi_j,
\end{multline}
where $\tilde{g}^j_{ab}$ is a $d$-dimensional metric tensor at $j$, $\tilde{R}_j$ is the Ricci scalar associated with the metric tensor, $Q(k)$ is a parameter dependent on $k$, $d = 2, 3$ and,
\begin{align}\label{coordinates_eq}
    \vec{x}_j = (\vec{x}_{\uparrow} - \vec{x}_{\downarrow})_j = (\vec{r}_{\uparrow} - \vec{r}_{\downarrow}, z_{\uparrow} - z_{\downarrow})_j = (\vec{r}_j, z_j).
\end{align}
\end{subequations}
\red{Since the Ricci scalar vanishes ($\tilde{R}_j = 0$)} \red{w}hen the Euclidean metric is flat, $\tilde{g}^j_{ab} = \delta_{ab}$, 
the last term in eq. (\ref{Liouville_action_eqq}) vanishes and the Liouville action reduces to,
\begin{align}\label{Liouville_action_eqq2}
   I \propto \sum_j\int \frac{d^{d}x_j}{2}\left (\vec{\nabla}\Phi_j\cdot\vec{\nabla}\Phi_j - K_j\exp(2\Phi_j) \right ). 
\end{align}
Thus, for arbitrary $Q(k)$, eq. (\ref{Liouville_action_eqq2}) can be varied with respect to $\Phi_j(\vec{x})$ to yield eq. (\ref{Liouville_j_eq}) when $d = 2$ dimensions (corresponding to $z_{\uparrow} = z_{\downarrow}$ in eq. (\ref{coordinates_eq})) and $k = 1$. 

To appropriately define $\Delta$, 
\red{we make use of scale invariance}, 
\begin{align}
    \vec{x}_j \rightarrow \Omega\,\vec{x}_j,\\
    \Phi(\vec{x}_j) \rightarrow \Phi(\Omega\vec{x}_j) = \Omega^{-\Delta(d)}\Phi(\vec{x}_j),
\end{align}
when,
\begin{align}
    K_j = 0,\,\,\Delta(d) = \frac{1}{2}(d - 2),
\end{align}
where $\Omega$ is an arbitrary scaling factor. Thus, the correlation function can be set as the \red{pair correlation function}, 
\begin{align}\label{rad_eq}
    g_j(\vec{r}_j) = \frac{C_j}{|(\vec{x}_{\uparrow} - \vec{x}_{\downarrow})_j|^{2\Delta}} = \langle \Phi_j(\vec{x}_{\uparrow})\Phi_j(\vec{x}_{\downarrow})\rangle.
\end{align}

Moreover, by the Pauli hamiltonian in eq. (\ref{Pauli_eq}) with $\mathcal{H}_j = 0$ the ground state and $\vec{n}\cdot\vec{B}_j(\vec{r}_j) = -K_j\exp(2\Phi_j(\vec{r}_j)) = 0$, where the theory is scale invariant, we must have,
\begin{subequations}
\begin{align}
    \vec{p}_j(\vec{r}_j) = (\vec{p}_{\uparrow} + \vec{p}_{\downarrow})_j = \vec{A}_j(\vec{r}_j) = \frac{1}{m}(\vec{n}\times\vec{E}_j), 
\end{align}
which introduces the Chern-Simons current density\cite{dunne1999aspects}, 
\begin{align}\label{CS_eq}
    \vec{J}_j \propto \vec{p}_j = \vec{A}_j(\vec{r}_j) = \frac{1}{m}(\vec{n}\times\vec{E}_j).
\end{align}
\end{subequations}
On the other hand, when,
\begin{align}
    \vec{n}\cdot\vec{B}_j(\vec{r}_j) = -K_j\exp(2\Phi_j(\vec{r}_j)) \neq 0,
\end{align}
and the current density 
\red{takes the Chern-Simons form} given in eq. (\ref{CS_eq}), scale invariance and hence conformal symmetry appears to be broken since $K_j \propto \rho_j \neq 0$ and the particles are massive. Nonetheless, near the critical point where the phase transition occurs, the \red{pair correlation function} 
must take the form given in eq. (\ref{rad_eq}). However, this 
\red{is not evident} in $d = 2$ when $K_j \neq 0$ but is 
\red{manifest} in $d = 3$ when $\vec{z}_{\uparrow} \neq \vec{z}_{\downarrow}$ and,
\begin{align}
    \vec{r}_j \rightarrow \vec{x}_j = (\vec{r}_j, z_j),
\end{align}
where Liouville's equation given in eq. (\ref{Liouville_j_eq}) transforms into the Emden-Chandrasekhar equation\cite{kippenhahn1990stellar} with solution given by eq. (\ref{conformal_eq}) provided,
\begin{subequations}\label{conf_cond_eq}
\begin{align}
    K_j = K_0^j\Delta\exp\left (-f(\Delta)\Phi_j(\vec{x}_j)\right ),\\
     f(\Delta) = \Delta^{-1} - 2 = 0,
\end{align}
and the normalisation constant is given by, 
\begin{align}
    C_j^{-1/\Delta} = K_0^j/2.
\end{align}
\end{subequations}
We note that, the 
\red{Gaussian curvature}, $K_j$ given in eq. (\ref{Fokker_Planck_eq}) or eq. (\ref{K_curvature_eq}) differs from the form 
\red{herein} in eq. (\ref{conf_cond_eq}) since the theory is in $d = 3$ dimensions. 
\red{Evidently, we can retrieve} eq. (\ref{Fokker_Planck_eq}) 
\red{by} the transformations, $f(\Delta) \rightarrow \nu$, $\Phi_j \rightarrow \Phi_{\rm AC}$ and $z_j = (z_{\uparrow} - z_{\downarrow})_j = 0$, which transforms the 3D theory to the 2D conformal field theory discussed above. 
\red{This suggests that, at the critical point} where $z_j = 0$ and $K_j \neq 0$, 
\red{we ought to consider} conformal invariance of the entire parameter space of Liouville conformal field theory.\cite{zamolodchikov1996conformal} 

In particular, it is known that the field $V_j(k) = \exp(2k\Phi_j)$ or $V_j(\bar{k}) = \exp(2\bar{k}\Phi_j)$ is primary when the \red{scaling} dimension is given by, 
\begin{subequations}
\begin{align}
    \Delta = \frac{k}{2}(Q(k) - k) = \frac{\bar{k}}{2}(Q(\bar{k}) - \bar{k}),
\end{align}
whilst the marginal condition for conformal invariance of the primary field is $\Delta = 1/2$, where $\Delta$ appears in eq. (\ref{conf_cond_eq}) and $z_j \rightarrow 0$, which is equivalent to,
\begin{align}
    Q(k) = k + 1/k,\,\, Q(\overline{k}) = \overline{k} + 1/\overline{k},
\end{align}
where $k = 0, 1$, $\bar{k} = 0, 1$ is the number of 
\red{vacancies} in 
\red{each primitive} cell of the honeycomb lattice with basis vectors given in eq. (\ref{basis_vector_eq}). 
The central charge is given by, 
\begin{align}
    c = 1 + 6Q^2(k) = 1 + 6Q^2(\bar{k}).
\end{align}
\end{subequations}
\red{However, the theory is known to be unitary only for $c = 1$ and $c = \infty$ corresponding to $Q(k) = Q(\bar{k}) \rightarrow \infty$ ($k \rightarrow \infty$ or $\bar{k} \rightarrow \infty$) and $Q(k) = 0$, respectively. However, for $z_j \neq 0$, the theory is 3D, which breaks scale invariance due to, $K_j \neq 0$. This results in the bifurcation of the honeycomb lattice following the Pauli Hamiltonian with $\vec{B}_j(\vec{x}_j) \neq 0$ since $\vec{z}_{\uparrow} \neq \vec{z}_{\downarrow}$. Consequently, this bifurcation corresponding to a monolayer-bilayer phase transition further lowers the energy as depicted in Figure \ref{Fig_16}, leading to deviations $\delta E$ from the ground state},
\begin{multline}
    \frac{\delta E_j}{2} = \langle \varphi_{\uparrow}|\mathcal{H}_j(\vec{r}_j)|\varphi_{\uparrow}\rangle
    = \frac{1}{2m}\vec{n}\cdot\vec{B}_j(\vec{x}_j)\langle \varphi_{\uparrow}|\sigma_3|\varphi_{\uparrow} \rangle\\
    = +\frac{1}{2m}\rho_j(\vec{x}_j)g_j(\vec{R}_j) = -\langle \varphi_{\downarrow}|\mathcal{H}_j(\vec{r}_j)|\varphi_{\downarrow}\rangle \neq 0,
\end{multline}
Consequently, 
\red{the two layers must be separated by} the energy barrier,
\begin{subequations}\label{gap_eq}
\begin{multline}\label{delta E_eq}
    \delta E_j = \frac{\vec{n}\cdot\vec{B}_j}{2m}(\vec{x}_j)\left (\langle \,\uparrow|\sigma_3|\uparrow \,\rangle  - \langle \,\downarrow|\sigma_3|\downarrow \,\rangle \right )\\
    = \frac{1}{m}\rho_j(\vec{x}_j)g_j(\vec{x}_j)
    = -\frac{K_0^j\Delta}{4\pi m}\exp\left (2\Phi_j(\vec{x}_j)\right )\\
    = -\frac{\Delta/2\pi m}{|(\vec{x}_{\uparrow} - \vec{x}_{\downarrow})_j|^2},
\end{multline}
which, together with the RKKY/Heisenberg interaction term in eq. (\ref{Pauli_eq}) 
\red{are interpreted} as the argentophilic interaction between the Ag cations responsible for stabilizing the Ag bilayer.\cite{masese2023honeycomb} This interaction is attractive in nature, and thus loosely binds the opposite pseudo-spin Ag cations to each other, 
\red{interpreted as} a pseudo-boson analogous to a Cooper pair.\cite{tinkham2004introduction} 

\begin{figure*}
\begin{center}
\includegraphics[width=0.8\textwidth,clip=true]{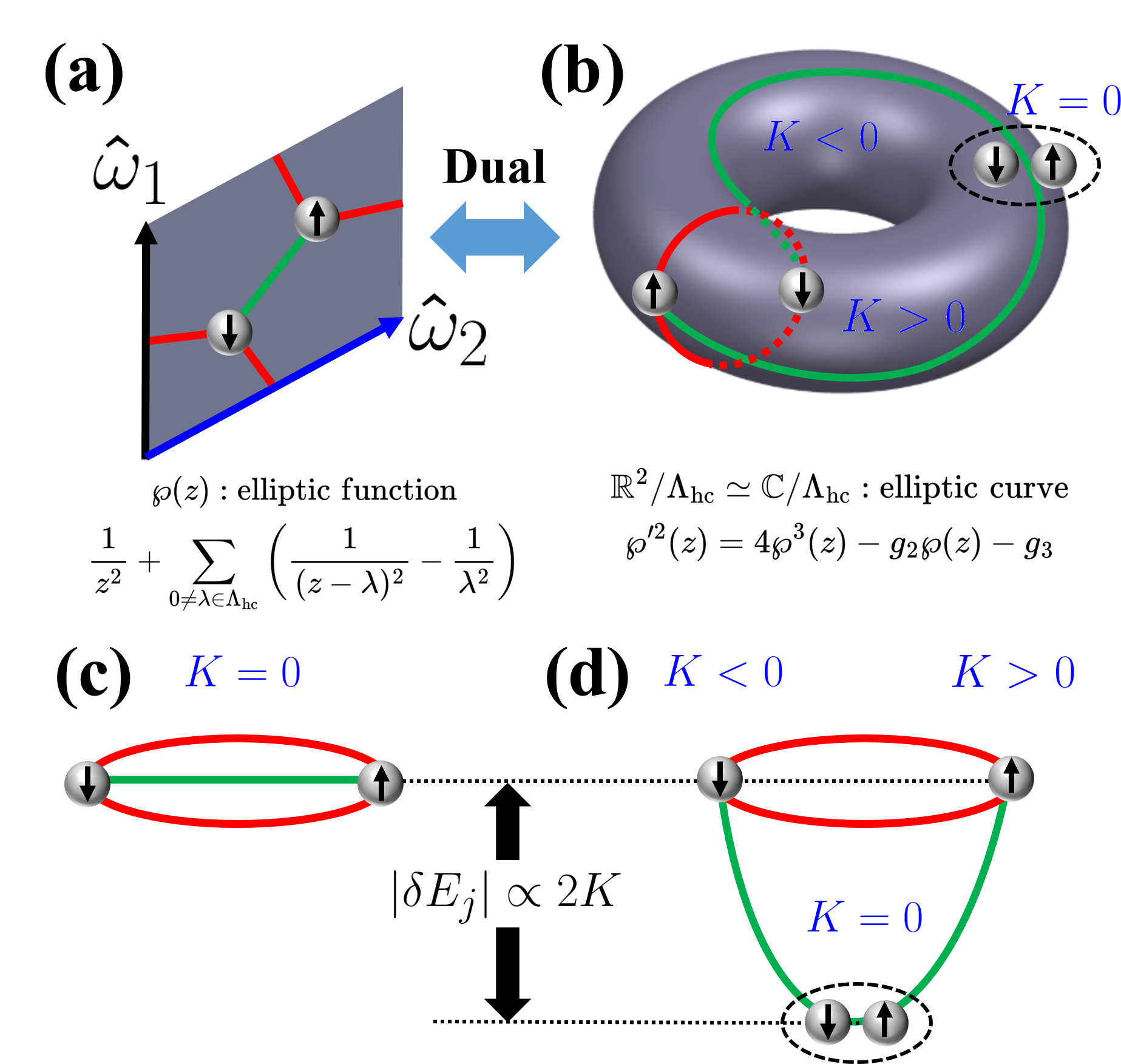}
\caption{Energy frustration (Pauli-exclusion principle) avoided by the dynamics exploiting the topology of a flat torus ($K = 0$) and two-torus ($K \neq 0$) respectively. Adjacent pseudo-spin up or down cations, indicated by black up or down arrows respectively, can be mapped onto each other along red and green lines, whereby the red honeycomb lattice lines connect adjacent opposite pseudo-spin cations which lie in adjacent primitive cells whereas the green honeycomb lattice lines connect adjacent opposite pseudo-spin cations which lie within the same primitive cell. (a) A primitive cell of the honeycomb lattice (given in Figure \ref{Fig_13}), showing a pair of pseudo-spin up and down cations. 
\red{The basis vectors projected to the upper half of the complex plane ($\omega_1, \omega_2 \in \mathbb{C}$) form} a hexagonal grid of basis vectors, 
$\lambda = n\omega_1 + m\omega_2$ ($n, m \in \mathbb{Z}$) with vertices at $z = \lambda$ corresponding to the Weierstrass elliptic function, $\wp(z)$.\cite{papanicolaou2021weierstrass} (b) The system exploiting the topology of the two-torus to avoid energy frustration. The two-torus is related to the flat torus in (a) by the transformations depicted in Figure \ref{Fig_4}. Points on the surface of the torus corresponds to points $z$ on the surface of the honeycomb primitive cell, which can be mapped (excluding the vertices, $\Lambda_{\rm hc}$) to points $(X, Y) = (\wp(z), \wp'(z))$ in a Weierstrass elliptic curve, $Y^3 = 4X^3 - g_2X - g_3$, where $g_2(\lambda)$ and $g_3(\lambda)$ are modular forms of weight $4$ and $6$ respectively comprising the partition function of the Liouville field in eq. (\ref{partition_Phi_eq}) given by the discriminant, $Z_{\rm \Phi}(b) = (2\pi)^{d/2}(g_2^3(b) - 27g_3^2(b))^{-d/24} \neq 0$ with $b = \omega_1/\omega_2$ and $d = 2$.\cite{abbott1991modular} The pseudo-spin down cation sits on the inside surface of the two-torus with negative curvature \red {($K < 0$)} whereas the pseudo-spin up cation sits on the outside surface with positive curvature \red {($K > 0$)} and paired cations with opposite pseudo-spins sits on the zero curvature region ($K = 0$). (c) A depiction of the red and green connections of opposite pseudo-spin cations of the entire honeycomb lattice as a flat torus ($K = 0$), which avoids energy frustration by mapping like pseudo-spin cations onto each other, leaving only a single red circle and a green line to represent the entire honeycomb lattice. The Gaussian curvature, $K$ vanishes everywhere along the green line. (d) The quasi-particle/excited states and ground state of the system with energy gap, $|\delta E| \propto K - (-K) = 2K$ (eq. (\ref{delta E_eq})). The $K = 0$ state persists along the green line only at the cross over between $K > 0$ and $K < 0$, where $K = 0$, which corresponds to a paired pseudo-fermionic state forming a pseudo-boson. Thus, the flat torus depiction in (c) is related to the two-torus by $\delta E_j \rightarrow 0$, which corresponds to the critical point.}
\label{Fig_16}
\end{center}
\end{figure*}

\red{\subsubsection{Anti-de Sitter space-time}}

To further analyse the 
\red{nature} of this attractive force between 
\red{like charges},  recall that within the idealised model previously discussed, the inverse temperature is given by $\beta = 1/m = 8\pi GM$, where $G \sim \ell_{\rm P}^2$ and $\ell_{\rm P}$ is the honeycomb lattice constant and $M$ is the total effective mass of the cations. Thus, using $\Delta (d = 3) = 1/2$, the 
\red{argentophilic interaction} satisfies the condition for the unit cell (eq. (\ref{AG_eq})),
\begin{align}\label{arg_eq}
    \left |\frac{\delta E_j}{M} \right | = \frac{2G}{|(\vec{x}_{\uparrow} - \vec{x}_{\downarrow})_j|^2} \equiv \frac{1}{\beta U_j(\vec{r})},
\end{align}
\end{subequations}
where $G$ \red{is the analogue of}
Newton's constant and $\langle U_j(\vec{r}) \rangle = M$ is the average potential energy of the cations. Moreover, since eq. (\ref{conformal_eq}) corresponds to eq. (\ref{Phi_eq2}) when $\Delta(d = 3) = 1/2$, 
\red{we set},
\begin{align}\label{KCG_eq}
    K_0^j = 1/G = 2/C_j^2
\end{align}
which 
\red{restricts eq. (\ref{arg_eq}) to} $|\delta E_j/M| = k = 1/\bar{k} = 1$ in the limit, $\vec{x}_j \rightarrow \vec{r}_j$. 
\red{In fact}, eq. (\ref{arg_eq}) is the analogue of eq. (\ref{AG_eq}), where $\nu = k = 1$ corresponds to the condition for the primitive cell. 
\red{C}onsidering the gravitation theory $\Delta(d = 3) = 1/2$ given by eq. (\ref{EFE_eq}), the metric in eq. (\ref{metric_ST_eq}) and the particle action correspond to the Newtonian limit, 
\begin{subequations}\label{grav_eq}
\begin{align}
    d\tau_j^2 \simeq \exp(-2\Phi_j(\vec{x}))dt_j^2 - d\vec{x}_j\cdot d\vec{x}_j,
\end{align}
and,
\begin{multline}
    S_j(\vec{x}_j) = -m\int d\tau_j\\
    = -m\int dt_j \sqrt{\exp(-2\Phi(\vec{x}_j)) - \frac{d\vec{x}_j}{dt_j}\cdot\frac{d\vec{x}_j}{dt_j}}\\ 
    \simeq \frac{m}{2}\int dt \left (\frac{d\vec{x}_j}{dt_j}\cdot\frac{d\vec{x}_j}{dt_j} + 1 - \exp(-2\Phi_j(\vec{x}_j))\right )\\
    = \frac{m}{2}\int dt \left (\frac{d\vec{x}_j}{dt_j}\cdot\frac{d\vec{x}_j}{dt_j} + 1 - \frac{\vec{x}_j\cdot\vec{x}_j}{2G}\right ), 
\end{multline}
\end{subequations}
respectively, where \red{we have used} eq. (\ref{conformal_eq}) and eq. (\ref{KCG_eq}) 
Introducing displaced coordinates, $\vec{y}_j = \vec{x}_j - \sqrt{2}\ell_{\rm P}\vec{n}$, where $d\vec{y}_j = d\vec{x}_j$, $\vec{n}\cdot\vec{n} = 1$, $\ell_{\rm P}^2 = G$ and $\vec{x}\cdot\vec{n} = 0$, 
eq. (\ref{grav_eq}) \red{is transformed} into,
\begin{subequations}\label{grav_eq2}
\begin{align}
    d\tau_j^2 \simeq \left (1 +  \frac{\vec{y}_j\cdot\vec{y}_j}{2G} \right )dt_j^2 - d\vec{y}_j\cdot d\vec{y}_j,
\end{align}
and,
\begin{align}
    S_j(\vec{y}_j) \simeq \frac{m}{2}\int dt \left (\frac{d\vec{y}_j}{dt_j}\cdot\frac{d\vec{y}_j}{dt_j} - \frac{\vec{y}_j\cdot\vec{y}_j}{2G}\right ). 
\end{align}
\end{subequations}
Thus, eq. (\ref{grav_eq2}) corresponds to the Newtonian limit,
\begin{subequations}
\begin{align}
    \nabla^2g_{00}(R_j) = -2\Lambda,\\
    R_j^2 = \vec{y}_j\cdot\vec{y}_j,
\end{align}
of the $1 + 3$ dimensional anti-de Sitter (AdS) space-time\cite{hubeny2015ads}, 
\begin{align}
    d\tau_j^2 = -g_{00}(R_j)dt_j^2 + g_{11}(R_j)dR_j^2 + R_j^2d\Omega_j^2,\\
    g_{00}(R_j) = \frac{1}{g_{11}(R_j)} = -\left (1 + \frac{R_j^2}{2G} \right ),
\end{align}
\end{subequations}
which 
\red{satisfies the} Einstein manifold $R_{\mu\nu} = \Lambda g_{\mu\nu}$ with a negative cosmological constant, 
$\Lambda = -3/2G = -3/2\ell_{\rm P}^2 < 0$ and $d\Omega^2$ the metric of the two-sphere. Finally, since, 
\begin{subequations}
\begin{align}
    \Phi_j(\vec{x}_j) = \ln \left (\frac{\vec{x}_j\cdot\vec{x}_j}{2G}\right ) = \ln k = \ln \bar{k} = 0,
\end{align}
corresponds to $k = 1/k = \bar{k} = 1/\bar{k} = 1$, the conformal field theory describing the primitive cell lives at the origin, $R_j^2 = 0$, 
\begin{align}
    |\vec{y}_j|^2 = |\vec{x}_j - \sqrt{2}\ell_{\rm P}\vec{n}|^2 = 0,
\end{align}
\end{subequations}
\red{corresponding to Minkowski space time}. 

\begin{figure*}
\begin{center}
\includegraphics[width=\textwidth,clip=true]{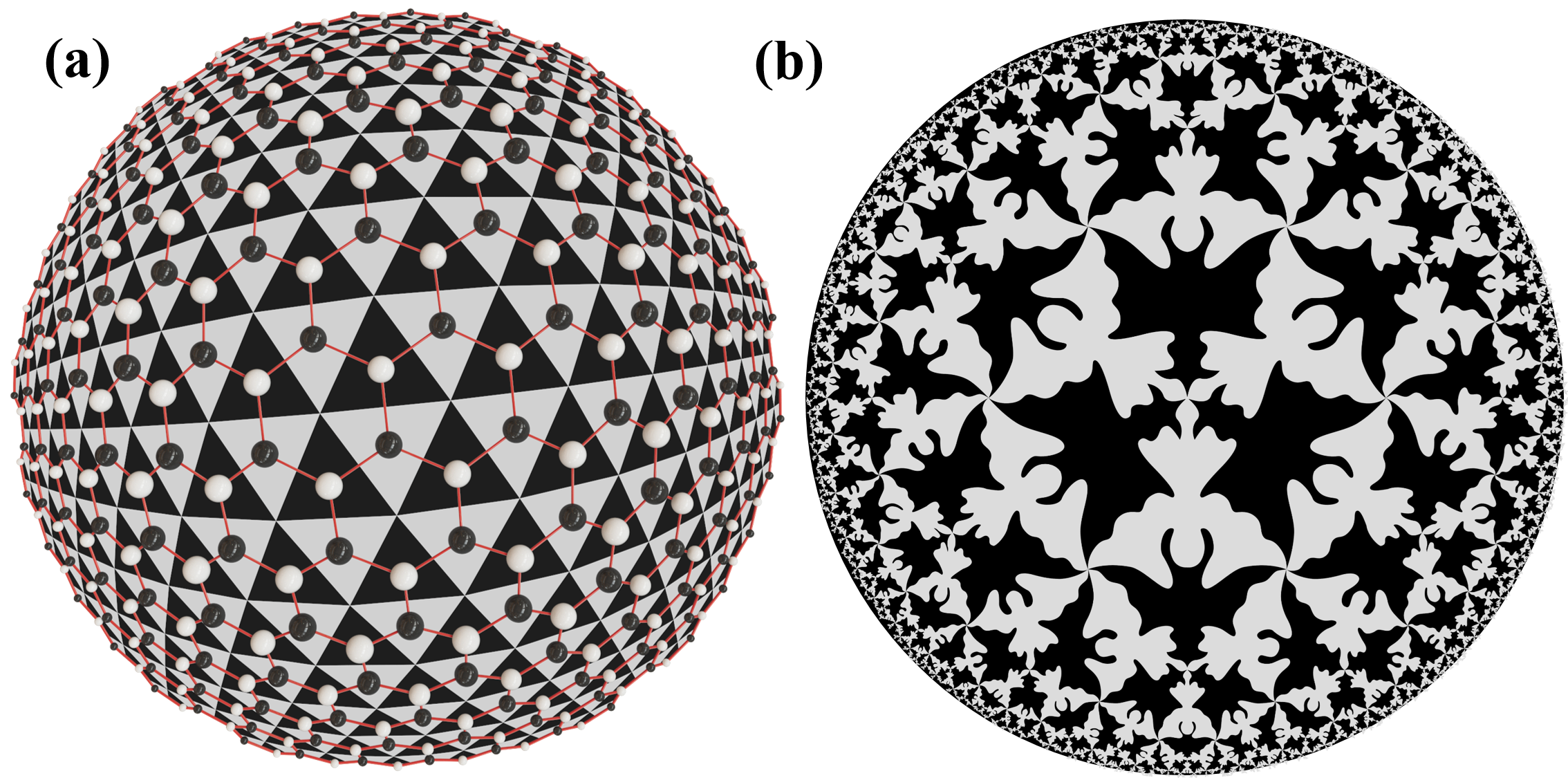}
\caption{
\red{Illustration of the regions occupied by cationic pseudo-spin states of the honeycomb lattice as angels and demons in 2D hyperbolic geometry. (a) The honeycomb lattice of cations depicted as the triangulation of the 2-sphere with genus $g = 0$ into a hexagonal simplex whose dual is the honeycomb lattice. (b) M. C. Escher's Circle Limit IV (Heaven and Hell) depicting the tessellation of the Poincar\'{e} disk (hyperbolic space) into regions with angels and demons.\cite{trott2004mathematica} The 2D-CFT in (a) can be mapped after bifurcation to Anti-de Sitter space --a 3D geometry analogous to (b)}}
\label{Fig_17}
\end{center}
\end{figure*}

This duality is analogous to AdS/CFT correspondence\cite{hubeny2015ads} whereby a conformal field theory lives at the boundary of AdS. Whilst such a boundary would correspond to the condition given in eq. (\ref{vz_eq}) in the idealised model equivalent to restricting cationic motion to the $x-y$ plane, ($dz_j/dt_j = \sqrt{g_{00}(r_j)} = 0$), in the case of bilayers, an exchange of cations between the triangular sub-lattices leads to $dz_j/dt_j \neq 0$. In fact, the velocity component in the $z$ direction becomes imaginary since,
\begin{subequations}
\begin{align}
    dz_j/dt_j = \sqrt{g_{00}(R_j)} = i\exp(-\Phi_j(R_j)).
\end{align}
\red{The exchange of cations can be understood as a result of quantum tunneling with traversal time $\tau_{\rm T} = \int_0^{\beta} dz_j m/|(p_z)_j| \sim 1/|(p_z)_j|$ measured by a Lamor's clock, where the wavevector, $(p_z)_j = mdz_j/dt_j$ of the wave function, $\varphi_j(z_j, \vec{r}_j) \propto \exp(i(p_z)_j z_j) = \exp(-|(p_z)_j|z_j)$ is imaginary.\cite{fevrier2018tunneling} Consequently, since $\Phi_j = 0$, the wave function penetrates a depth $\beta = 1/m$, the cut-off scale along the $z$ direction. Finally, recall that the unit cells with $k \neq 1$ can be mapped to the primitive cell with $k = 1$ via the generator, $T^k \in Q$ given in eq. (\ref{Tk_eq}). This corresponds to a mapping of $\Phi(r_j) = 0$ to $\Phi(R_j) \neq 0$, and hence represents mappings of the primitive cell to larger unit cells with $k \geq 1$. Consequently, the penetration depth becomes},
\begin{align}
    \beta \rightarrow \beta\exp(\Phi(R_j)) = \beta/\sqrt{-g_{00}(R_j)},
\end{align}
\end{subequations}
which corresponds to gravitational red-shift. A depiction of this duality as an illustration has been provided in Figure \ref{Fig_17}. 

\subsubsection{Relativistic treatment}

We shall consider the idealised model of cations captured by the field equations given in eq. (\ref{CFE_eq}).
\red{First}, 
\red{we introduce} the complex-valued function,
\begin{subequations}
\begin{align}
    f_{s,j} = 2\Delta\exp(\Phi_j/2)\exp(iS_s^j),
\end{align}
where $\Delta = (d - 2)/2$ is the \red{scaling} dimension,
$s = \pm 1/2$ labels the pseudo-spin (up, $\uparrow$ or down, $\downarrow$) of the cation with the action, $S_s^j$, where $j$ labels a particular primitive cell. 
Thus, we 
relate the two functions, 
$\Psi$ in eq. (\ref{CFE_eq}) and $f_{s,j}$, 
\red{via the following construction},
\begin{multline}\label{Construction_eq}
    \Psi = \sqrt{\frac{2\rho_0}{\beta}}\prod_{j = 0}^{k - 1}f_{1/2,j}f_{-1/2,j}\\
    = \sqrt{\frac{2\rho_0}{\beta}}(2\Delta)^{2k}\exp\left (\Delta\Phi \right )\exp(iS),
\end{multline}
\end{subequations}
\red{where}
\red{we have introduced the action, $S$ and field $\Phi$ by},
\begin{subequations}\label{S_Phi_eq}
\begin{align}
    S = \sum_{j = 0}^{k - 1}\sum_s S_s^j,\\
    \Phi = \frac{1}{\Delta}\sum_{j = 0}^{k - 1} \Phi_j,
\end{align}
\end{subequations}
with $S_s^0 = S_0/2$ a constant. 

Proceeding, we are interested in reproducing the Hamiltonian given by eq. (\ref{Pauli_eq}), as well as introducing a description of the bilayers as a phase transition with the \red{scaling} dimension as the order parameter. This will be achieved by 
\red{treating} $\Psi$ as the order parameter for the pairing cations. Such pairs of opposite spin particles are known to form by the Cooper pair mechanism in conventional (low-temperature) superconductors or some other exotic pairing mechanism in high-temperature superconductors\cite{tinkham2004introduction}, which creates a spin-zero bound state. The crucial progress here will be to formally relate $\Psi_{s,j}$ with 
\red{fermionic degrees of freedom}. 
\red{In particular}, we need to incorporate\red{, in eq. (\ref{Real_Imaginary_eq}),} 
the U($1$) Dirac current of the $j$-th pseudo-spin up or pseudo-spin down fermion given by $\bar{\psi}_s^j\gamma_{\mu}\psi_s^j$, 
where $\psi_s^j$ is a four-component spinor of the $j$-th cation with pseudo-spin $s$. The action 
\red{are} 
\red{defined} as,
\begin{subequations}\label{Psi_S_eq}
\begin{align}
    S_s^j = -m\int \left ( \frac{\bar{\psi}_s^j\gamma_{\mu}\psi_s^j}{|\Psi|^2} \right )dx^{\mu},\\
    |\Psi|^2 = \frac{2\rho_0}{\beta}(2\Delta)^{4k}\exp\left (2\Delta\Phi \right ),
\end{align}
\end{subequations}
where $\psi_{\uparrow}^j = (\varphi_{\uparrow}^{j}, 0, 0, 0)^{\rm T}$ or $\psi_{\downarrow}^j = (0, \varphi_{\downarrow}^{j}, 0, 0)^{\rm T}$ is the spinor 
\red{comprised of} a pseudo-spin up ($\uparrow$) or pseudo-spin down ($\downarrow$) wave function respectively given by $\varphi_s^{j}$, $\rm T$ is the transpose, $\gamma_{\mu}$ are the gamma matrices in curved space-time satisfying $\gamma_{\mu}\gamma_{\nu} + \gamma_{\nu}\gamma_{\mu} = 2g_{\mu\nu}$, $\bar{\psi}_s^j = (\psi_s^j)^{*\rm T}(\gamma^0)^{-1}$ is the Dirac adjoint spinor, $\gamma^{\mu} = e^{\mu}_{\,\,\bar{a}}\gamma^{\bar{a}}$ with $\gamma^{\bar{a}}$ the Dirac matrices in flat space-time satisfying  $\gamma_{\bar{a}}\gamma_{\bar{b}} + \gamma_{\bar{b}}\gamma_{\bar{a}} = 2\eta_{\bar{a}\bar{b}}$, and $e^{\bar{a}}_{\,\,\mu}, e^{\mu}_{\,\,\bar{a}}$ are tetrad fields satisfying $e^{\bar{a}}_{\mu}e_{\bar{a}\nu} = g_{\mu\nu}$ and $e^{\mu}_{\,\,\bar{a}}e_{\mu\bar{b}} = \eta_{\bar{a}\bar{b}}$ with $\eta_{\bar{a}\bar{b}}$ the Minkowski metric tensor \textit{i.e.} ${\rm diag} (\eta_{\bar{a}\bar{b}}) = (1, -1, -1, -1)$. 

For instance, \red{from} eq. (\ref{Construction_eq}), a pair of pseudo-fermions 
\red{at} a primitive cell, $j$ corresponds to $k = 1$,
\begin{subequations}\label{Def_Psi_eq}
\begin{align}
    \Psi = \sqrt{\frac{2\rho_0}{\beta}}(2\Delta)^2\exp\left (\Delta\Phi \right)\exp(iS_0),
\end{align}
where $S_0$ is a constant. Consequently, for $k$ pairs of fermions corresponding to a unit cell, the imaginary part of eq. (\ref{CFE_eq}) is normalised as,
\begin{multline}\label{norm_fermion_eq}
    \int_{\mathcal{V}} d^{\,3}x\,\nabla_{\mu}F^{\mu 0} = \beta\int_{\mathcal{V}} d^{\,3}x\, g^{0\mu}\frac{1}{2i} \left (\Psi^*\partial_{\mu}\Psi - \Psi\partial_{\mu}\Psi^* \right )\\
    = -\sum_s\sum_{j = 1}^{k - 1}\int_{\mathcal{V}} d^{\,3}x\,\bar{\psi}_s^j\gamma^{0}\psi_s^j\\ 
    = \sum_{j = 1}^{k - 1}\int_{\mathcal{V}} d^{\,3}x\,(\varphi_{\uparrow}^j)^*\varphi_{\uparrow}^j
    + \sum_{j = 1}^{k - 1}\int_{\mathcal{V}} d^{\,3}x\,(\varphi_{\downarrow}^j)^*\varphi_{\downarrow}^j\\
    = \sum_{j = 1}^{k - 1}\int_{\mathcal{V}} d^{\,3}x\,\bar{\psi}^j\gamma^{0}\psi^j = 2k - 2,
\end{multline}
\end{subequations}
where $\beta = 1/m$, we have introduced $\psi^j = (\varphi_{\uparrow}^j, \varphi_{\uparrow}^j, 0, 0)^{\rm T}$ with $(\gamma^0)^{-1}\gamma^0 = 1$ and assumed 
the wave functions $\varphi_{s}^j$ are appropriately normalised on the manifold $\mathcal{V}$,
\begin{subequations}\label{norm_fermion_eq2}
\begin{align}
    \int_{\mathcal{V}} d^{\,3}x\,(\varphi_s^j)^*\varphi_s^j = 1,\\
    \int_{\mathcal{V}} d^{\,3}x\,\bar{\psi}^j\gamma^{0}\psi^j = 2.
\end{align}
\end{subequations}

\red{Moreover, the normalisation in eq. (\ref{norm_fermion_eq})} is invariant under the exchange of pseudo-spin up ($\uparrow$) and pseudo-spin down ($\downarrow$) particle-particle pairs with pseudo-spin up ($\uparrow$) and pseudo-spin down ($\downarrow$) particle-anti-particle pairs. 
\red{In particular}, 
\red{the four-component spinor associated with a particular primitive cell takes the form},
\begin{subequations}
\begin{align}\label{psi_eq}
    \psi = (\varphi_{\uparrow}, \varphi_{\downarrow}, \tilde{\varphi}_{\uparrow}, \tilde{\varphi}_{\downarrow})^{\rm T},
\end{align}
where, $\tilde{\varphi}$ is the anti-particle wave function. 
\red{Thus, we have the corresponding compact notation},
\begin{align}
   \left (\frac{1}{2}\tau_0 + p\tau_3 \right )\bigotimes \left ( \frac{1}{2}\sigma_0 + s\sigma_3 \right )\psi = \psi_{p,s},
\end{align}
\end{subequations}
which represents a particle ($p = 1/2$) or anti-particle ($p = -1/2$) of spin-$\uparrow$ ($s = 1/2$) or spin-$\downarrow$ ($s = -1/2$), where $\sigma_0$ and $\tau_0$ are the $2\times2$ and $4\times4$ identity matrices respectively, $\sigma_3$ and $\tau_3$ are given by, 
\begin{subequations}
\begin{align}
\sigma_3 = 
\begin{bmatrix}
 1& 0\\ 
 0&-1 
\end{bmatrix},\,\,
\tau_3 = 
\begin{bmatrix}
 1&  0&  0& 0\\ 
 0&  1&  0& 0\\ 
 0&  0&  -1& 0\\ 
 0&  0&  0& -1
\end{bmatrix}.
\end{align}
\end{subequations}
and $\bigotimes$ is the tensor product. 
\red{For instance}, the spin-$\uparrow$ anti-particle, $\psi_{-\frac{1}{2},\frac{1}{2}} = (0, 0, \tilde{\varphi}_{\downarrow}, 0)$ corresponds to the matrix, 
\begin{align}
\left (\frac{1}{2}\tau_0 - \frac{1}{2}\tau_3 \right )\bigotimes \left ( \frac{1}{2}\sigma_0 + \frac{1}{2}\sigma_3 \right ) =   
    \begin{bmatrix}
 0&  0&  0& 0\\ 
 0&  0&  0& 0\\ 
 0&  0&  1& 0\\ 
 0&  0&  0& 0
\end{bmatrix},
\end{align}
acting on $\psi$ in eq. (\ref{psi_eq}). 
\red{Thus}, in eq. (\ref{Construction_eq}), 
\red{we employed} $\psi_{\frac{1}{2}, \frac{1}{2}}$, $\psi_{\frac{1}{2}, -\frac{1}{2}}$ pairs, where the condensate $\Psi$ 
\red{becomes charged}. 
Likewise, the $\psi_{-\frac{1}{2}, \frac{1}{2}}$, $\psi_{-\frac{1}{2}, -\frac{1}{2}}$ condensate has the opposite charge. However, the $\psi_{-\frac{1}{2}, \frac{1}{2}}$, $\psi_{\frac{1}{2}, -\frac{1}{2}}$ or the $\psi_{-\frac{1}{2}, -\frac{1}{2}}$, $\psi_{\frac{1}{2}, \frac{1}{2}}$ condensate would be neutral since particles and their anti-particles have opposite charge. 
Alternatively, a construction with Majorana fermions is possible, where $\psi_{\frac{1}{2}, \pm\frac{1}{2}} \propto \psi_{-\frac{1}{2}, \pm\frac{1}{2}}$, which renders the condensates neutral. 

\red{Proceeding}, we shall 
\red{consider} the particle-particle scenario, since we are interested in charged cations. Recall that, according to eq. (\ref{Killing_eq2}), the vector potential 
is set proportional to the four-velocity, $u^{\mu} = \beta A_{\mu}$. This suggests we can 
\red{identify} the super-current and quasi-particle current respectively as,
\begin{subequations}
\begin{align}
    J^{\mu}_{\rm S} = |\Psi|u^{\mu} = \beta|\Psi|^2A^{\mu},\\
    J_{\rm QP}^{\mu} = \sum_{j = 1}^{k - 1}\bar{\psi}^j\gamma^{\mu}\psi^j,
\end{align}
\end{subequations}
where, 
\begin{align}
    \psi^j = \psi^j_{\frac{1}{2},\frac{1}{2}} + \psi^j_{\frac{1}{2},-\frac{1}{2}},
\end{align}
is the Dirac spinor with two cations $\psi^j = (\varphi_{\uparrow}^j, \varphi_{\downarrow}^j, 0, 0)^{\rm T}$ of opposite spin in a unit cell. 
\red{The bosonic current is introduced in eq. (\ref{norm_fermion_eq}) by replacing the action $S_s^j$ by averages} 
for the paired pseudo-fermions,
\begin{subequations}\label{boson_eq}
\begin{align}
   S_s^j \rightarrow -\frac{m}{2}\int \frac{\langle \bar{\psi}^j\gamma_{\mu}\psi^j \rangle}{\sum_{l = 0}^{k - 1}\langle \bar{\psi^l}\psi^l\rangle}dx^{\mu},\\
   \langle \bar{\psi}^j\gamma^{\mu}\psi^j \rangle = u^{\mu}\langle \bar{\psi}^j\psi^j\rangle,
\end{align}
where $u^{\mu} = dx^{\mu}/d\tau$ is the center of mass four-velocity with $u^{\mu}u_{\mu} = -1$, $\langle \cdots \rangle$ is a quantum mechanical average 
\red{constrained by},
\begin{align}\label{constraint_eq}
    \sum_{j = 1}^{k - 1}\langle \bar{\psi^j}\psi^j\rangle = |\Psi|^2,
\end{align}
\end{subequations}
$|\Psi|^2$ is defined in eq. (\ref{Def_Psi_eq}) and $\psi^j$, $\bar{\psi}^j$ obey the Dirac equation in curved space time,
\begin{subequations}\label{Dirac_eq}
\begin{align}
    i\gamma^{\mu}(D_{\mu}\psi^j) = m\psi^j,\\
i(\tilde{D}_{\mu}\bar{\psi}^j)\gamma^{\mu} = m\bar{\psi}^j,
\end{align}
\end{subequations}
where $D_{\mu}\psi = (\partial_{\mu} - \frac{1}{4} \omega_{\mu}^{\bar{a}\bar{b}}\gamma_{\bar{a}}\gamma_{\bar{b}} - iA_{\mu})\psi$, $\tilde{D}_{\mu}\bar{\psi} = -\partial_{\mu}\bar{\psi} - \bar{\psi}(\frac{1}{4} \omega_{\mu}^{\bar{a}\bar{b}}\gamma_{\bar{a}}\gamma_{\bar{b}} + iA_{\mu})$, $\omega_{\mu}^{\bar{a}\bar{b}} = e^{\bar{a}}_{\,\,\alpha}\partial_{\mu}e^{\bar{b}\alpha} + \Gamma^{\beta}_{\,\,\mu\alpha}e^{\bar{a}}_{\,\,\beta}e^{\bar{b}\alpha}$ is the spin connection, $\Gamma^{\alpha}_{\,\,\mu\nu} = \frac{1}{2}g^{\alpha\beta}(\partial g_{\mu\beta}\partial x^{\nu} + \partial g_{\beta\nu}/\partial x^{\mu} - \partial g_{\mu\nu}/\partial x^{\beta})$ are the Christoffel symbols and $e_{\mu}^{\,\bar{a}}$ are the tetrad fields. In the non-relativistic limit, the Dirac equation yield the Pauli Hamiltonian given in eq. (\ref{Pauli_eq}).\cite{bransden2003physics} 

\red{On the other hand}
energy momentum tensor 
\red{appropriate} for the Einstein Field Equations given in eq. (\ref{EFE_eq}) is, 
\begin{multline}\label{Dirac_energy_eqq}
    T^{\mu\nu} = - \frac{1}{4mi}\left \langle \sum_{j = 1}^k \bar{\psi}^j\gamma^{\mu}D^{\nu}\psi^j + \bar{\psi}^j\gamma^{\nu}D^{\mu}\psi^j \right \rangle\\
    -\frac{1}{4mi}\left \langle \sum_{j = 1}^k (\tilde{D}^{\mu}\bar{\psi}^j)\gamma^{\nu}\psi^j + (\tilde{D}^{\nu}\bar{\psi}^j)\gamma^{\mu}\psi^j \right \rangle.
\end{multline}
Thus, the trace of eq. (\ref{EFE_eq}) yields the real part of eq. (\ref{CFE_eq}) given by $R = \beta|\Psi|^2$ as expected, where we have used eq. (\ref{constraint_eq}) to obtain the trace. Moreover, since $T^{\mu\nu}$ in eq. (\ref{Dirac_energy_eqq}) must satisfy eq. (\ref{killing_EFE_eq}), we must take the approximation, 
\begin{subequations}
\begin{align}
    D^{\mu}\psi^j \simeq -iA^{\mu}\psi^j,\\ 
    \tilde{D}^{\mu}\bar{\psi}^j \simeq iA^{\mu}\bar{\psi}^j,    
\end{align}
\end{subequations}
with $\beta A^{\mu} = u^{\mu}$ and $\beta = 1/m$, which yields, 
\begin{align}
    T^{\mu\nu} \simeq |\Psi|^2u^{\mu}u^{\nu},
\end{align}
where we have used eq. (\ref{constraint_eq}). Thus, 
\begin{subequations}\label{norm_fermion_eq3}
\begin{align}
    S = \sum_s\sum_{j = 0}^{k - 1}S_s^j = m\int d\tau - S_0,
\end{align}
and,
\begin{multline}
    \int_{\mathcal{V}} d^{\,3}x\,\nabla_{\mu}F^{\mu 0} = \beta\int_{\mathcal{V}} d^{\,3}x\, g^{0\mu}\frac{1}{2i} \left (\Psi^*\partial_{\mu}\Psi - \Psi\partial_{\mu}\Psi^* \right )\\
    = -\sum_{j = 1}^{k - 1}\int_{\mathcal{V}} d^{\,3}x\,|\Psi|^2u^0 = 2k - 2,
\end{multline}
\end{subequations}
\red{where} $\Psi$ is the Cooper pair order parameter and $\beta|\Psi|^2$ 
\red{is} understood as the total 2D bulk density of $k$ pairs of cations. The pairing is for each primitive cell whilst $u^0 = \exp(\Phi(\vec{x}))$ plays the role of the pair correlation function. This means we can have $u^{\mu} = \exp(\Phi(\vec{x}))\xi^{\mu}$, where $\xi^{\mu} = (1, \vec{0})$ is the time-like Killing vector. 

Recall that, in the idealised model, the presence of a time-like Killing vector and a $z$-like Killing vector renders the theory 2D. Thus, the \red{scaling} dimension becomes, $\Delta (d = 2) = 0$, the Cooper pair order parameter vanishes, $\Psi \rightarrow 0$ and the normalisation in eq. (\ref{norm_fermion_eq3}) yields $k = 1$. However, when there is no $z$-like Killing vector, $\Delta (d = 3) = 1/2$, and eq. (\ref{norm_fermion_eq3}) corresponds to the Gauss-Bonnet theorem at the critical point when $z \rightarrow 0$. 


\red{\section{Experimental searches for topological and high-temperature-induced phase transitions}}

\subsection{High-temperature induced phase transitions}
\red {Honeycomb layered oxides, particularly those entailing the chemical composition $\rm {\it A}_2{\it M}_2TeO_6$ ({\it A} = Li, Na, K; {\it M}= transition metals such as Ni, Zn or alkaline-earth metals suchlike Mg), have recently been gaining traction for \magenta {not only their interesting magnetic phenomena and high-voltage electrochemistry, but also their fast ionic conductivity} and fascinating phase transitions.} \cite{kanyolo2022advances, kanyolo2021honeycomb}

\begin{figure*}
\begin{center}
\includegraphics[width=0.7\textwidth,clip=true]{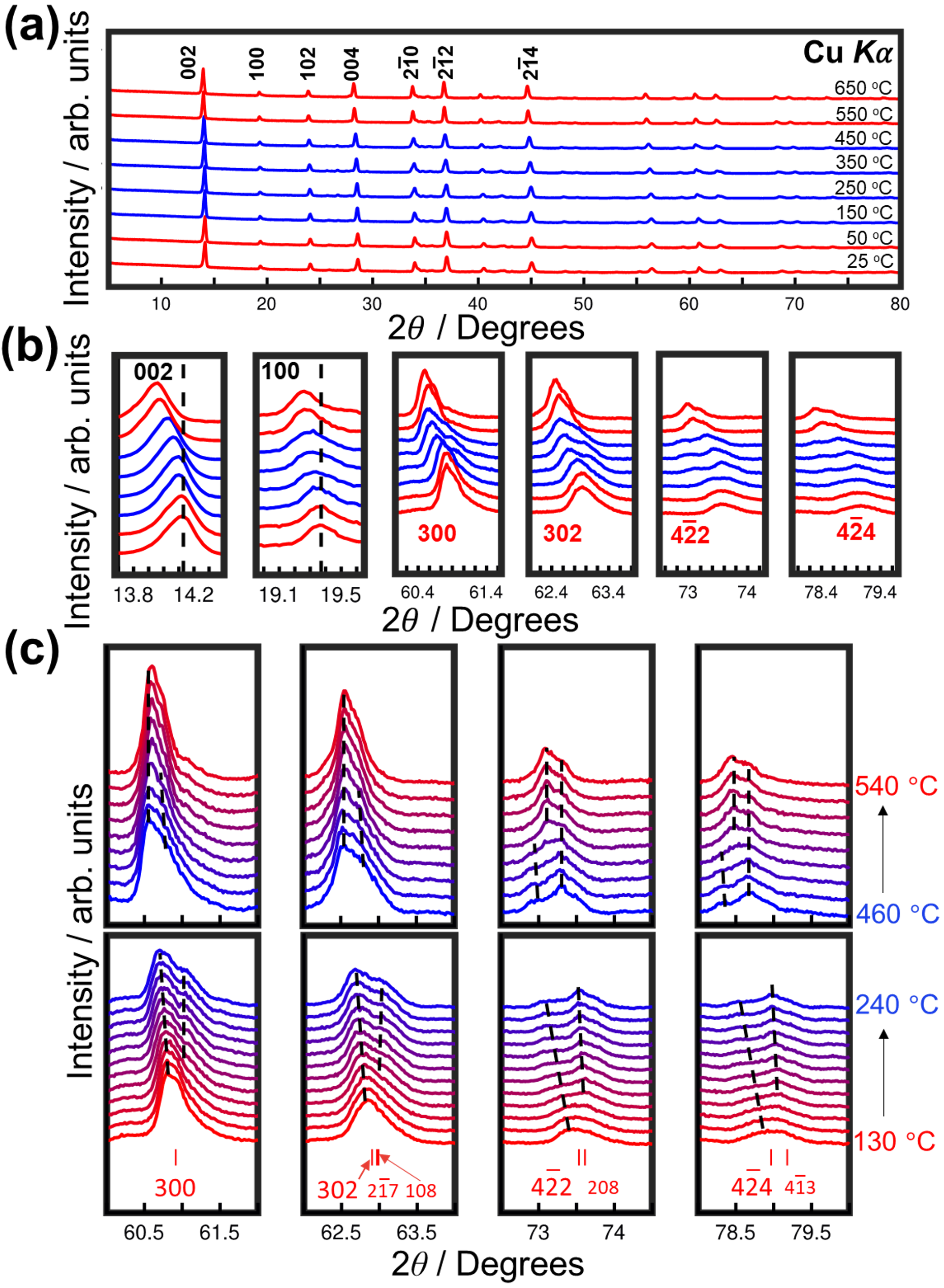}
\caption{High-temperature XRD measurements of honeycomb layered $\rm K_2Ni_2TeO_6$.\magenta {\cite {rizell2020high}} (a) Evolution of the XRD patterns of $\rm K_2Ni_2TeO_6$ 
\red {from 25 $^\circ$C to 650 $^\circ$C}. The room-temperature phase (indexed in a hexagonal lattice is shown in red, whilst a new intermediate monoclinic (pseudo-orthorhombic) phase is highlighted in blue. This intermediate phase further transforms back to the hexagonal lattice (red) with heating. (b) Shifts of the (002) and (100) Bragg peaks to lower diffraction angles with increase in temperature, indicating thermal lattice expansion along the {\it c}-axis and {\it a}-axis. Moreover, extinction of some Bragg diffraction peaks concomitant with the appearance of new ones is observed on heating, which attest to changes in the crystal symmetry. Notable is the appearance of (300), (302), (4-22) and (4-24) peaks that commences at around 130 $^\circ$C. These peaks further merge at above 550 $^\circ$C, hallmarking a reversion to the hexagonal lattice symmetry. (c) XRD patterns taken at temperature regimes where phase transition is noted. The lower panels show the XRD pattern evolution of selected Bragg peaks taken between 130 °C and 240 °C (at increments of 10 $^\circ$C), whereas the upper panels show XRD patterns taken between 460 °C and 540 °C.}
\label{Fig_18}
\end{center}
\end{figure*}

\red {Several of these materials show exemplary conductivities around 300$^\circ$C --- the temperature range where high-temperature energy storage systems like sodium-sulphur (Na-S) batteries operate.\cite{lu2010advanced} $\rm Na_2Ni_2TeO_6$ exhibits the highest ionic conductivity, reaching 10.1---10.8 S ${\rm m^{-1}}$ at 300$^\circ$C,\cite{evstigneeva2011new} demonstrating its potential as a fast ionic conductor. Structural changes occurring at high temperatures may lead to, for instance, enhanced ionic conductivity as marked by superionic phase transitions in several ionic solids.\cite{boyce1979superionic, hull2004superionics} As such, investigations on structural changes occurring at elevated temperatures have been performed on honeycomb layered tellurates such as $\rm Na_2Ni_2TeO_6$ and $\rm K_2Ni_2TeO_6$.\cite{zubayer2020}}

Figure \ref{Fig_18} shows the XRD patterns of $\rm K_2Ni_2TeO_6$ taken between 25$^\circ$C and 650$^\circ$C.\magenta {\cite {rizell2020high}} Two distinct phase transition regimes can be 
\red {observed}. 
\red {Splitting of multiple peaks is discerned upon heating to beyond 130$^\circ$C (Figure \ref{Fig_18}}) indicating a change in the lattice symmetry (from the initial hexagonal lattice to a monoclinic (pseudo-orthorhombic) lattice), as was further affirmed by ND measurements.\cite{zubayer2020} Although structural details of this new phase are beyond the reach of ND and XRD measurements, \magenta {{\it in situ}} TEM studies at elevated temperatures can be effective in elucidating the structural transitions.\magenta {\cite {wu2022situ}} Further heating of this intermediate phase leads to the merging of the peaks, as shown in Figure \ref{Fig_18}, finally reverting back to the initial hexagonal lattice. Upon cooldown, the XRD patterns of $\rm K_2Ni_2TeO_6$ were found to closely match those of the as-synthesised material before heating, confirming the reversibility of the observed structural changes. Although a majority of honeycomb layered oxides display reversible structural changes upon cooldown after heating process, honeycomb layered antimonates such as $\rm Li_{1.5}Na_{1.5}Ni_2SbO_6$ demonstrate irreversible structural changes upon cooldown.\cite{vallee2019}

Note also that shifts of Bragg diffraction peaks such as (002) and (100) towards lower diffraction angles with heating is discernible (Figure \ref{Fig_18}), indicating an overall thermal expansion of the $\rm K_2Ni_2TeO_6$ lattice with increment in temperature. Similar behaviour has also been observed in related honeycomb layered oxides such as $\rm Na_2Ni_2TeO_6$ and $\rm Na_3Ni_2SbO_6$.\cite{bera2020temperature, wang2019ordered} Anisotropic expansion was noted, with the expansion across the interlayers (interslabs) being a manifold higher than along the layers due to the weaker interlayer bonds of $\rm K_2Ni_2TeO_6$. A similar anisotropic expansion behaviour upon heating was also noted in related honeycomb layered tellurates such as $\rm Na_2Ni_2TeO_6$.\cite{bera2020temperature} 

\subsection{Electrochemically induced phase transitions/Jenga mechanism}
Apart from phase transitions observed at high temperatures, honeycomb layered oxides exhibit a variety of phase transitions when utilised as battery materials. Electrochemically extraction of the mobile alkali atoms during battery operation creates empty spaces (vacancies) that lead to variation in the layer stacking sequences as the material structure responds and adapts to the vacancies created. 
\red {Such structural response has been dubbed as the ‘{\it Jenga mechanism}’\cite {kanyolo2021honeycomb}, akin to what occurs in a game of {\it Jenga}: there is no deformation of the tower, since there are no empty spaces (vacancies) when the tower has all its pieces. However, the whole tower deforms every time you remove a piece of the Jenga blocks, and rearranges itself, without falling apart in the process, into a new structure. An analogous mechanism has been envisioned to occur in layered oxides during electrochemical extraction of cations from the lattice, with the entire structure rearranging to compensate for the vacancies created.\cite {kanyolo2021honeycomb}}


\red {Nevertheless, the {\it Jenga mechanism} does not explain why the cations would be organised into a honeycomb pattern, nor does it fully describe cationic transport in these layered materials. It is only when quantum mechanical concepts are used to correlate the thermodynamic quantities with geometric parameters that the cation diffusion occurring within the honeycomb layers is properly described.\cite {kanyolo2020idealised} The honeycomb pattern of cations has been concluded to be energetically favourable for these layered oxides in what is referred to in mathematics as Hale’s honeycomb conjecture. This conjecture states that the honeycomb pattern is the most efficient tiling of a floor that guarantees the unit tile covers the largest area with the least perimeter. It is this geometric concept that explains the thermodynamic considerations for cationic transport in honeycomb layered oxides.}

\subsection{Topological transitions}

\red{The topological features of the idealised model of cationic diffusion are captured by the Chern-Simons current}\cite{kanyolo2022cationic, dunne1999aspects}
\begin{subequations}
\begin{align}
    \vec{J}_{\rm AC} = \frac{k}{2\pi}\sigma\, (\vec{n}\times\vec{E}),
\end{align}
where $\sigma = 2\pi\beta D\rho = \mu\rho$ is the conductivity 
\red{exhibited by the primitive cell}, $\mu = 2\pi\beta D$ is the mobility (Einstein-Smoluchowski equation) and 
$k$ is the Chern-Simons level. 
\red{The current density does not correspond to the Hall current, but rather the spacial component of $J^{\mu}$ in eq. (\ref{Real_Imaginary_eq})} given by,
\begin{align}
    \vec{J} = \vec{n}\times\vec{J}_{\rm AC}.
\end{align}
\end{subequations}
\red{Thus, this implies that we should expect integer conductance spikes ($k \rightarrow k + n$, $n \in \mathbb{N}$) per 2D cationic lattice whenever sufficient activation energy is applied to extract a cation facilitating diffusion in the cathode}.

\begin{figure*}[!t]
\begin{center}
\includegraphics[width=2.0\columnwidth,clip=true]{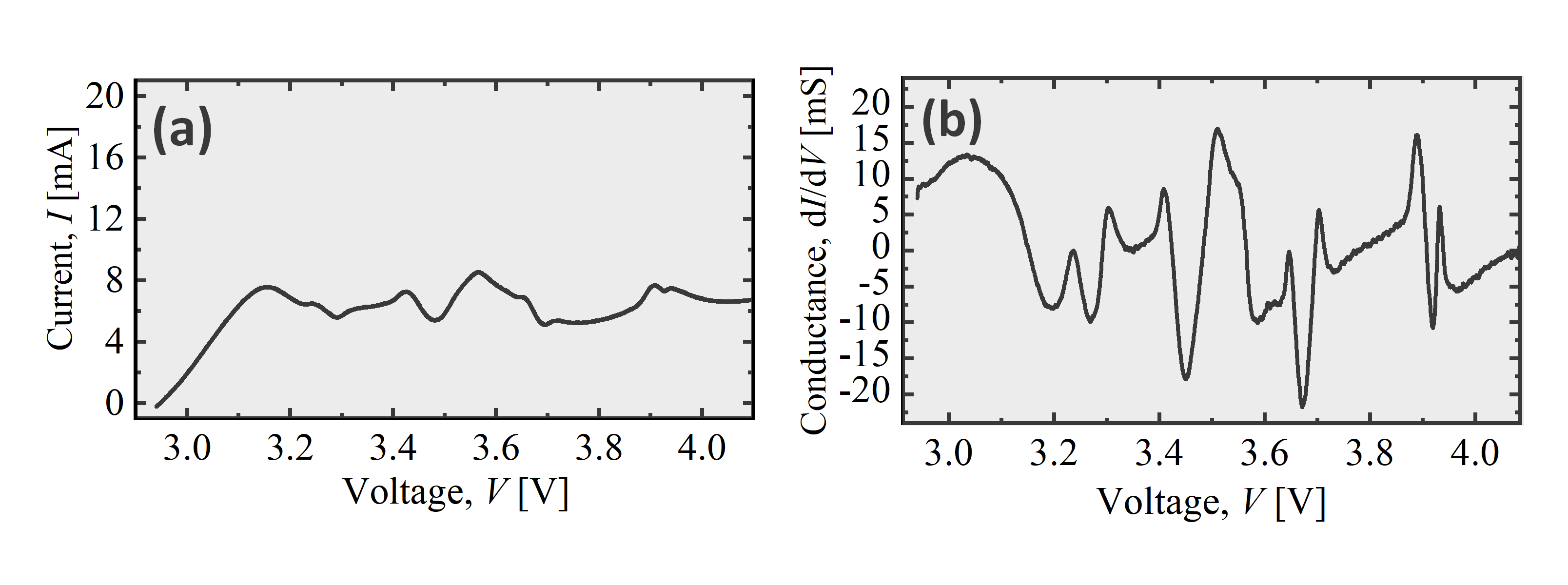}
\caption{
\red{(a) Current, $I$ -- Voltage, $V$ characteristics and (b) Conductance, $dI/dV$ --  Voltage, $V$ characteristics derived from cyclic voltammetry experiment of the $K$ cation extraction (charging) process with \ce{K2Ni2TeO6} as the cathode in a two-electrode setup. $K$ metal was used as the counter electrode and the scanning rate was set to 0.1 mVs$^{-1}$.\cite{bard2022electrochemical} The sharp conductance spikes occur evenly distributed at a rough interval of 0.1 V within the voltage interval, 3.2 V to 4.0 V}. Figure reproduced with permission.\cite{kanyolo2022cationic}}
\label{Fig_19}
\end{center}
\end{figure*}

\red{Unfortunately, no conductance experiments with a single 2D cationic lattice have been reported to date.} 
\red{Chemical exfoliation techniques can be used to isolate single slabs, making the prospects for such experiments within reach}.\cite{bae2021kinetic, yuan2022magnetic}
\red{Nonetheless, since the activation energy for K is relatively low\cite{matsubara2020magnetism}, $E_{\rm a}^{\rm K} \simeq 121$ meV, low resolution peaks, $k \rightarrow k + n$ with $n \in \mathbb{N} \gg 1$ have been reported\cite{kanyolo2022cationic} for $\rm K_2Ni_2TeO_6$ as shown in Figure \ref{Fig_19}}. 
\red{However, pre-existing cationic vacancies and/or high activation energies have a tendancy to disfavour the efficient extraction process during cycling, leading to a solitary broad current peak centred at} the high voltage regime $I$--$V$ cycling characteristics for $A_2$\ce{Ni2TeO6} (with $A = \rm Na, Li$), since the activation energy for Na and Li is vastly greater than that of K, \textit{i.e.} $E_{\rm a}^{\rm Li} > E_{\rm a}^{\rm Na} > E_{\rm a}^{\rm K}$.\cite{kanyolo2021honeycomb}

\subsection{Pressure-induced effects}

\red{In the idealised model,\cite{kanyolo2020idealised, kanyolo2022cationic, kanyolo2021honeycomb} one can define a `gravitational' potential $\Phi$, related to the probability density in eq. (\ref{probability_eq2}) as $\Phi(x,y) = \frac{1}{2}\ln(\mathcal{P}(x,y))$, where $\mathcal{P}(x,y) = p(x,y)/\Omega$ appears in the line element, $ds$ defined locally by the conformal metric,
\begin{align}
    ds^2 = \sum_{ab}g_{ab}dr^adr^b = \mathcal{P}(x,y)(dx^2 + dy^2),
\end{align}
for the $ab$(or $xy$)-surface, where $g_{ab}$ is the metric tensor (first fundamental form). This potential, representing the average effects of potential energies such as the Vashishta-Rahman potential\cite{sau2016influence, sau2015ion, sau2015role}, governs the dynamics of the cations on the surface},
\begin{align}
    \frac{d^2\vec{r}}{d\tau^2} = -\vec{\nabla}\Phi + \frac{1}{m}\vec{F},
\end{align}
where $\vec{F}$ is the force. It can be compared to the Langevin equation (eq. (\ref{Langevin_eq})) when $\vec{\nabla}\Phi = t^{-1}d\vec{r}/d\tau$, where $t$ is the mean free time. Consequently, it follows from the definition that $\Phi (x,y)$ satisfies Liouville’s equation\cite{du2012liouville}, 
\begin{subequations}
\begin{multline}\label{Liouville_eq3}
    4\pi d\rho(r) = \left ( \frac{\partial^2}{\partial x^2} + \frac{\partial^2}{\partial y^2} \right ) \Phi(x,y)\\
    = -K\exp(2\Phi(x,y)) = -K\frac{p(x,y)}{\Omega},
\end{multline}
where $\rho(r) = \rho_0 g(r)$ is the local number density related to the radial distribution function, $g(r)$ and the bulk number density, $\rho_0$. The number density acts as the source of the `gravitational’ potential, $\Phi(x,y)$ where,
\begin{align}\label{Gaussian_curvature_eq}
    K = -\frac{1}{2\mathcal{P}}\left ( \frac{\partial^2}{\partial x^2} + \frac{\partial^2}{\partial y^2} \right )\ln\left ( \mathcal{P} \right )
\end{align}
\end{subequations}
is the Gaussian curvature of the $xy$ surface.\cite{gray2006modern} We can set the `density of states' proportional to the Gaussian curvature, $\Omega \equiv -K/4\pi$, which gives the expression for the population density, $\rho(x,y,d) = p(x,y)/d$. In this case, the geometric relation between the Euler characteristic, $\chi = 2 - 2g$ of the surface of genus $g$ and the Gaussian curvature (known as the Gauss-Bonnet theorem)\cite{chavel2006riemannian} counts the number of cations, $k = h$ on the surface (neglecting the boundary terms), 
\begin{multline}\label{Gauss_Bonnet_eq11}
    k - 1 = \int \rho(r) dV\\
    = \int p(x,y) dxdy = -\frac{1}{4\pi}\int K\exp(2\Phi(x,y))dxdy\\
    = -\frac{1}{4\pi}\int K\sqrt{g_{ab}}dxdy = -\chi/2,
\end{multline}
and the description is more palatable for a large number of cations $k = h \gg 1$. Thus, since the number per unit area, $p(x,y)$ can be taken to be independent of the inter-layer distance between two slabs in a honeycomb layered oxide, we find that decreasing the inter-layer distance increases the energy density/number per unit volume, $\rho(x,y)$ as expected, which in turn accordingly raises the pressure, assuming the equation of state is given by the perfect fluid formula, $\rho = wP$ with, \cite{eq_of_state}
\begin{align}
    T^{\mu\nu} = (\rho + P)u^{\mu}u^{\nu} + Pg^{\mu\nu},
\end{align}
the stress-energy momentum tensor coupled to eq. (\ref{EFE_eq}) and $P$ the pressure. Finally, since the number density is given by $\rho(r) = \rho_0g(r)$, the Gauss-Bonnet theorem in eq. (\ref{Gauss_Bonnet_eq11}) corresponds to the well-known normalisation formula for the radial distribution function (ref. \cite{tuckerman2010} page 156).

In the case of the 
pressure effects pertinent to the increase of inter-layer distance at fixed cationic size could be experimentally explored through chemical means, via the equation of state for the layered material. For instance, the curvature\cite{du2012liouville, chavel2006riemannian, gray2006modern} of the $ab$ plane in these materials correlates with the number of cations/vacancies and/or the topology of the surface\cite{kanyolo2020idealised, kanyolo2022cationic, kanyolo2021honeycomb}, which in turn requires the curvature to be proportional to the energy density (or more succinctly the number per unit volume) of the diffusing cations. Typically, the number per unit volume, $\rho$ at thermodynamic equilibrium is related to pressure, $P$ by an equation of state. For a perfect fluid comprised of cations, the equation of state is given by $P= w\rho$,\cite{eq_of_state} where $w$ is a parameter characterizing the fluid \textit{e.g.} the ideal gas law formula $PV=nRT$ corresponds to $\rho = n/V$ and $w=RT$, where $n$ is the atomic number of gas molecules and $R$ is the gas constant. In materials, the parameter $w$ can also depend on the radial distribution function, $g(r)$.\cite{tuckerman2010} This avails an avenue to chemically change $\rho$ or the curvature and hence $P$ in the material.

\red{Effects of positive pressure on materials is known to modify the transition temperature in superconductors.\cite{snider2020room} However, the effects of pressure on honeycomb layered oxides has not yet been experimentally studied. Nonetheless, simulations with $\rm {\it A}_2Ni_2TeO_6$ ($A = \rm Li, Na, K$) have been carried out whereby the interlayer distance is fixed to that of K and ionic radius varied, or alternatively the ionic radius is fixed to that of K, and the interlayer distance varied.\cite{sau2022insights} The K system values are chosen since it possesses the widest inter-layer distance and hence the largest reported ionic radius amongst the experimentally reported conventional honeycomb layered nickel tellurates.\cite{kanyolo2021honeycomb, kanyolo2022advances}} 
\red{Thus, applying positive pressure to the material is simulated by reducing the interlayer distance to smaller values with the ionic radius fixed, whilst negative pressure corresponds to decreasing the ionic radius with a fixed interlayer distance. Consequently, the unpressurised material would correspond to the interlayer distance scaling linearly with the ionic radius, which has been observed experimentally and in simulations.}\cite{sau2022insights, kanyolo2020idealised, kanyolo2021honeycomb} 

\red{\section{Computational techniques}}

Computational modelling techniques avail \red{exclusive insight into the mechanisms dictating the physicochemical properties of materials, at the atomic level, and are thus invaluable tools in the design of materials.\cite{hempel2023dynamics, khanom2023first, chakrabarti2023density} The main advantage of computational modelling is nested in its ability to} support and complement experimental \red{data} by unveiling \red{both} fundamental atomic\red{-}scale \red{mechanisms and properties that are difficult to attain entirely} from experimental measurements. This paves the way for \red{judicious} materials design \red{through} the \red{methodical} optimisation of functionality in conjunction with experimental analyses. Computational modeling methods \red{suchlike molecular dynamics (MD) and first-principles density functional theory (DFT)} have been utilised to predict various atomic-scale properties of materials. In the following subsections:

\red{
\begin{enumerate}[(i)]
    \item We will present short overviews of the computational modelling methods;
    \item What functionalities can be garnered computationally using these methods; and,
    \item Their significance in the design of chalcogen- and pnictogen-based honeycomb layered oxides.
\end{enumerate}
}

\red{The details of both} MD and DFT \red{techniques} have been \red{outlined} elsewhere,\cite{parr1980density, cohen2012challenges, geerlings2003conceptual, burke2012perspective, koch2015chemist, cohen2008insights, sham1983density, kohn1996density, runge1984density, shuichi1991constant, andersen1980molecular} and hence we will only cover pertinent applications to the target honeycomb layered oxide materials here.

\subsection{Density functional theory (DFT)}

DFT is a potent and well-utilised quantum mechanical/electronic structure modelling method that utilises functionals to \red{compute} the ground state \red{energies} of a many-body system from its electron density. The density functionals (DFs) in DFT can be categorised as follows: \red{local density approximation (LDA), generalised gradient approximation (GGA),} fully non-local range-separated DFs, double-hybrid DFs, hybrid DFs, and meta-GGA. These categorisations have been discussed in details elsewhere.\cite{mardirossian2017} Perdew and contemporaries have framed such methodologies in analogy to what has been dubbed the ‘Jacob’s ladder’,\cite{zhang2021, perdew2001jacob} which offers a general prescription for the selection and design of DF approximations. 

\magenta {A plethora of DFT simulation codes exist, such as Quantum Espresso, \cite{giannozzi2009quantum} DMol3, \cite{delley2010} Wien2K, \cite{blaha2001wien2k} CASTEP, \cite{segall2002} CRYSTAL, \cite{erba2017} Gaussian, \cite{towler1996} CP2K, \cite{hutter2014cp2k} VASP, \cite{hafner2008} amongst others.} \red{The choice of the code to use is contingent on its capabilities (for instance, interatomic potential, periodicity, exchange-correlation functional, $etc.$), information required from the material and so on. One also has to consider: (i) how to describe the periodicity of the investigated system, and (ii) how to explicitly treat the basis sets (delocalised or localised) and electrons (using all electron codes or pseudopotentials).} Vienna {\it ab initio} simulation package (VASP) simulation code has mainly been used to assess various electronic structural aspects of chalcogen- and pnictogen-based honeycomb layered oxides. \cite{huang2020, tada2022implications, masese2021mixed, berthelot2021stacking, wang2019ordered} In what follows, we shall highlight the various physicochemical properties assess via DFT simulation of chalcogen- and pnictogen-based honeycomb layered oxides and cover any limitations associated with elucidating the various physicochemical aspects with DFT.

\subsubsection{Material stability / phase stability}

\red{In order to progress with further theoretical and experimental studies, it is crucial to determine the phase stability of compounds predicted computationally. In principle, a {\it thermodynamic convex hull} is typically established} through comparing the energies of \red{entire compounds within the chemical space.} The assessment of the phase stability of a material using DFT has nowadays been enabled materials databases such as the Automatic Flow (AFLOW), \cite{curtarolo2012} Open Quantum Materials Database (OQMD) \cite{saal2013} and Materials Project (MP) \cite{jain2013}. Structures are typically generated from structural databases such as the universal structure predictor (USPEX) code, \cite{glass2006} {\it ab initio} random structure searching (AIRSS), \cite{pickard2011} amongst others.


\red{DFT can be used to determine the {\it energy above the convex hull} of a material, which is a pertinent descriptor to predict the stability of the material, by computation of the phase equilibria and energy convex hull. This approach has been utilised to predict the stability of $\rm NaKNi_2TeO_6$, $\rm Na_2{\it M}_2TeO_6$ ($M =$ Zn, Mg) and $\rm Li_{1.5}Na_{1.5}Ni_2SbO_6$.\cite {berthelot2021stacking, huang2020, vallee2019} Assessment of the chemical phase diagram of a given material using DFT ground-state computations avails insights its phase stability.} For instance, the phase stabilities of $\rm Na_2{\it M}_2TeO_6$ ($M =$ Zn, Mg) were \red{accurately evaluated by computing both the quaternary (Na–{\it M}–Te–O) and ternary (Na–Te–O and Na–{\it M}–O) phase compositional diagrams. \cite{huang2020}}


\red{Phase diagram computations using DFT show honeycomb layered tellurates such as} $\rm Na_2Zn_2TeO_6$ and $\rm Na_2Mg_2TeO_6$ \red{to exhibit larger electrochemical stability windows compared to sulphides, enhancing electrolyte-electrode compatibility upon} prolonged cycling when utilised as solid electrolytes for rechargeable Na-ion batteries. $\rm Na_2Zn_2TeO_6$ and $\rm Na_2Mg_2TeO_6$ were found \red{to show intrinsically higher maximum kinetic voltage limits (3.57 and 3.82 V, respectively) and larger electrochemical stability windows (2.25–3.23 V and 1.74–3.15 V, respectively)} compared with the sulphides that are commonly envisaged as promising solid electrolytes.\cite{huang2020} Moreover, \red{phonon and phase diagram computations revealed the structural stabilities} of both $\rm Na_2Zn_2TeO_6$ and $\rm Na_2Mg_2TeO_6$. 


On another note, \red{DFT was employed to assess the structural stability of the mixed-alkali honeycomb layered $\rm NaKNi_2TeO_6$ based on computation of formation energies of various crystal structural configurations.} \cite{berthelot2021stacking} By virtue of further DFT analysis, $\rm Na_3Ni_2SbO_6$ ($\rm NaNi_{2/3}Sb_{1/3}O_2$) was found to adopt a symmetric honeycomb configuration of Ni atoms around Sb atoms (6 Ni atom-ring configuration ($\rm Ni _6$-rings)) as the most energy favourable structure that enhances both its air and thermal stability. \cite{wang2019ordered} Further, \red{DFT computations have been used to predict a plethora of new} stable honeycomb layered oxides such as $A_2\rm Ni_2TeO_6$ ($A =$ Cu, Au, Ag, Cs, Rb, H), along with predicting new polytypes of honeycomb layered tellurates such as $\rm Li_2Ni_2TeO_6$ and $\rm Na_2Ni_2TeO_6$.\cite{tada2022implications} DFT has also been utilised in predicting the interlayer distances of optimised structures attained upon metal substitution in \red{${\rm Na_2}M_2\rm TeO_6$ ($M=$ Pd, V, Ba, Sr, Ca, Co, Zn, Mg),}\cite{huang2020} availing insights how doping engineering on ${\rm Na_2}M_2{\rm TeO_6}$ can benefit in enhancing superior Na-ion diffusion.

\subsubsection{Operating voltage}

\red{Electrochemical performance of battery materials is evaluated using galvanostatic (dis)charge profiles, which show the evolution of the voltage against the capacity during repetitive (dis)charging (cycling). In other words, (dis)charge profiles correlate to various diffusion mechanisms and various ion storage at distinct voltages. The average equilibrium voltage of a compound, theoretically, corresponds to the free energy difference between the discharged and charged phases and can be evaluated by considering the Gibbs free energy difference between the discharged and charged phases, the Faraday constant, the valency of ions and the number of ions in the discharged and charged phases. Typically, entropy need to be taken into account when computing the free energy. However, since the zero-point energy difference between the discharged and charged phases, configurational entropy and vibrational entropy are trivial at room temperature, the free energy difference is roughly equivalent to the total energy difference (from where the voltage is calculated).}


\red{DFT simulations of the voltage (potential) profiles of honeycomb layered oxides have demonstrated good agreement with experimental data,} as has been done for $\rm K_2Ni_2TeO_6$, \red{$\rm Li_4FeSbO_6$} and $\rm Na_3Ni_2SbO_6$.\cite{masese2018rechargeable, wang2019ordered, jia2017} DFT revealed that the existence of ordered $\rm Ni _6$-rings -rings in $\rm Na_3Ni_2SbO_6$ lead to super-exchange interaction forming \red{degenerate electronic orbital states and symmetric atomic configurations, which not only significantly enhance both structural stability and air stability, but also raise the operating voltage.}\cite{wang2019ordered}

\subsubsection{Cation migration barriers}

\red{DFT can compute cation migration barriers, which can then be compared with the experimentally determined diffusion coefficients and migration barriers to understand cation diffusion of materials. The attained insights can be utilised to purposefully screen various material design strategies (such as crystal structure, dopant concentrations, composition, $etc.$) to optimise cation diffusion.}

\red{DFT simulations using the nudged elastic band (NEB) methodology have widely been employed to compute the metal migration or diffusion behaviour of materials.\cite{olsson2022} NEB method demands optimised end and start points (typically identical lattice sites set apart at some distance). Crystalline structural materials generally possess numerous ion migration pathways linking different or same ion sites, inducing anisotropic diffusion and distinct energy barriers. An {\it a priori} guess of the possible migration path is made between the end and start points. Thereafter, the given migration pathway is divided into several reaction coordinates or steps (NEB images) which are optimised based on the DFT formalism to determine the transition state or saddle point. The {\it diffusion barrier} ({\it migration energy barrier}) is thenceforth computed as the total energy difference between the start point and the saddle point. The {\it diffusion barrier} attained for a particular diffusion pathway using NEB can then be utilised to acquire the {\it ionic diffusion coefficients} via the Arrhenius equation. Theoretically, both the {\it diffusion barrier} and {\it ionic diffusion coefficients} can be traced for all feasible diffusion pathways by repeating the NEB calculations for various diffusion pathways.} 


\red{NEB calculations have greatly contributed to understanding diffusion mechanism, at the atomic-scale, of fast Na-ion conductors such as ${\rm Na_2}M_2{\rm TeO_6}$ ($M =$ Mg, Zn)\cite{huang2020} and K-ion conductors ($\rm K_2Mg_2TeO_6$)\cite{masese2018rechargeable}.} For example, NEB calculations conducted on $\rm Na_2Zn_2TeO_6$ and $\rm Na_2Mg_2TeO_6$ \red{revealed activation energies} of 204 meV and 261 meV, respectively, \cite{huang2020} in good agreement with the experimental values attained from electrochemical impedance spectroscopy \red {and nuclear magnetic resonance measurements}\cite{li2018new, evstigneeva2011new}. The low \red{Na-ion} activation energies were found to \red{promote facile and collective fluid-like Na-ion transport along preferred 2D {\it honeycomb diffusion pathways}.} 


NEB calculations, however, do have some limitations and caution needs to be taken when performing such calculations. Particularly for compounds with intricate structures, it is crucial to determine the main mobile carriers along with their diffusional pathways. Furthermore, {\it a priori} guesses of the mobile carriers and their diffusional pathways can be extremely demanding, for materials with very disordered sub-lattices of mobile ions.


Complementing NEB with molecular dynamics (MD) simulations can yield an accurate description of the diffusion pathways. Compared with NEB calculations, \red{MD simulations directly visualise the trajectory of the mobile ions along with their dynamics.} MD simulations, which shall be tackled in another subsection, can \red{therefore significantly} complement NEB \red{calculations to determine diffusion mechanisms, particularly for disordered materials. Hopping of the mobile ions is activated thermally when performing MD simulations, and possible ion migrations paths are traced owing to thermal activation. Thus, in contrast to NEB calculations, MD simulations can trace the diffusion trajectories of mobile ions without the necessity for preassigning the diffusional paths.}

\subsubsection{Defect formation}

Defects (such as cation vacancies, interstitials and stacking disorders) can \red{profoundly affect the stability, ion-storage mechanism and cation diffusion barriers} within the structure of crystalline materials. DFT simulations can \red{aid to determine the nature of stacking disorders (faults) that lead to broadening of the X-ray diffraction peaks of materials, and can also be employed to screen structural configurations that not only can lead to reversible cation storage or increased cation occlusion capacity,} but also provide insights into the design of materials with fast cationic diffusion (high ionic conductivity).


For instance, DFT modelling has been \red{employed to depict the site distribution of Na atoms and also affirm stacking disorder models of} $\rm Na_2Zn_2TeO_6$.\cite {li2020} Stacking faults entailing in-plane shifts \red{of Te atoms} along the \red{stacking direction of the layers} were found to be dominant in $\rm Na_2Zn_2TeO_6$. This was experimentally \red{validated} by first \red{designing} a supercell model of $\rm Na_2Zn_2TeO_6$ and \red{then} simulating \red{XRD} patterns at various \red{stacking faults concentrations, aiding to} identify \red{the best} structural model \red{that matches} the experimental \red{results}. An accurate structural model of $\rm Na_2Zn_2TeO_6$ was proposed based on the analyses of synchrotron XRD data and DFT calculations. \cite {li2020}

\subsubsection{Band structure and phonon calculations}

\red{In practice, band structures, density of states (DOS) and band gaps are useful descriptors to gauge the functionalities of various energy and electronic materials, wherein the desired properties can be tweaked using various band structure engineering methodologies.}


For instance, the band structures of $\rm Na_2Zn_2TeO_6$ and $\rm Na_2Mg_2TeO_6$ solid-state electrolytes were calculated using DFT \red{to evaluate their redox stabilities against the electrodes as well as their electronic insulating properties. Electronic insulation is a prerequisite for a potential solid-state electrolyte, in order to circumvent electronic transport across the electrolyte.} $\rm Na_2Zn_2TeO_6$ and $\rm Na_2Mg_2TeO_6$ reveal \red{similar spin-down and spin-up} band structures and are thus non-magnetic materials.\cite {huang2020} Considering that \red{the conduction band minimum (CBM) and the valence band maximum (VBM)} provides the upper bound of the \red{reduction and oxidation potentials} of the electrolyte, \red{the band} gaps of $\rm Na_2Zn_2TeO_6$ and $\rm Na_2Mg_2TeO_6$ were calculated using a reliable exchange-correlation functional to be 3.67 eV and 4.69 eV, respectively.\cite {huang2020}


The emission of $\rm O_2$ gas is a \red{typically problematic} for \red{a vast majority of} oxide electrolytes. The projected electronic orbitals of \red{$\rm O^{2-}$} anions were found to dominate the VBM of both $\rm Na_2Zn_2TeO_6$ and $\rm Na_2Mg_2TeO_6$, meaning that \red{$\rm O^{2-}$} anions are preferentially oxidised to $\rm O_2$ gas at high voltages.\cite {huang2020} Nevertheless, calculations showed higher maximum kinetic voltage limits for both $\rm Na_2Zn_2TeO_6$ and $\rm Na_2Mg_2TeO_6$ comparable with those of other oxides. Phonon dispersions were further calculated to assess the stability of $\rm Na_2Zn_2TeO_6$ and $\rm Na_2Mg_2TeO_6$ structural models that were optimised using DFT.\cite {huang2020} \red{Na-ion transport} was \red{shown to correlate} with the “switch-on/switch-off” vibrational phonon modes of transition metal oxide octahedra, \red{suggesting that adjusting} the interlayer spacing is pivotal to tailoring the Na-ion transport in $\rm Na_2Zn_2TeO_6$ and $\rm Na_2Mg_2TeO_6$ honeycomb layered oxide electrolytes.


\red{An accurate calculation of band gaps of honeycomb layered oxides demands development of better performing exchange-correlation functionals. For instance, the hybrid functional HSE06 was adopted in calculating the electronic band structures of $\rm Na_2Zn_2TeO_6$ and $\rm Na_2Mg_2TeO_6$.\cite {huang2020} This was based on the caveat that the PBE exchange-correlation functional initially used to perform the computations tends to miscalculate the Lowest Unoccupied Molecular Orbital (LUMO) – Highest Occupied Molecular Orbital (HOMO) gaps. In addition, functionals such as the new meta-GGA SCAN have been reported to effectively model the electronic structures of layered materials. \cite{chakraborty2018} The new meta-GGA SCAN functionals can account for both the localised states and van der Waal interactions (dispersion); thus, can be envisioned to be efficacious in the future modelling of this class of honeycomb layered oxides. Nevertheless, considerable testing will be paramount to assess both the computational cost and accuracy of using new exchange-correlation functionals as compared to those currently employed.}

\subsection{Molecular dynamics (MD)}

\red{Simulations based on MD} constitute the foundation of contemporary atomistic modeling, \red{by visualising the spacial distribution of mobile ions to reveal the ion dynamics of materials at the atomic-scale} in a given time frame ($i.e.$, a simulation time up to the nanosecond level, and a subfemtosecond time resolution). \red{MD simulations visualise the probability density of the mobile ions, thus divulging the diffusional pathways within the given structural framework of the material. Such information availed by MD simulations is not within the reach of NEB calculations.}


There exists a myriad of softwares to \red{conduct} MD \red{calculations}. \red{Amongst popular} and reliable programs \red{include} GROMACS, \cite{van2005gromacs} \red{LAMMPS,\cite{plimpton1995} GULP,\cite{gale1997gulp}} DL POLY,\cite{smith2006} or TINKER,\cite{ponder2004tinker} \red{{\it etc}}. Common codes \red{that are commercially available} include the Forcite module implemented in Accelrys Materials Studio, AMBER,\cite{case2008amber} CHARMM\cite{brooks2009charmm, jo2008charmm} or GROMOS\cite{scott1999gromos}. Simulation configurations or trajectories are visualised using softwares such as Rasmol,\cite{sayle1995rasmol} gOpenMol\cite{bergman1997visualization} or VMD\cite{humphrey1996vmd}, amongst others. The \red{selection of the code to use depends, for instance, on the code capabilities (interatomic potential, periodicity, $etc$.) and the information required from the given material.}


MD simulations \red{are} categorised \red{as follows}: (i) classical MD \red{(force field-based)} simulations and (ii) {\it ab initio} MD (AIMD) simulations (which is a combination of DFT and MD). {\it Classical MD simulations} \red{rely on atomistic forces (a.k.a force fields or semi-empirical interatomic potentials) to describe the time evolution of ion interactions within a material. Classical MD is reliant on knowing the precise location of atoms in real space and usually restricted by the interatomic potential choices available. Although classical MD cannot be utilised to simulate electronic structures, computations are by far faster than those performed by AIMD. This is due to the fact that AIMD simulations solve initially the electronic structure to attain interatomic forces, whilst the force calculation in classical MD is simplified. Therefore, classical MD can not only explore systems at longer simulation times, but also accommodate largescale systems.}


\red{Since classical MD cannot model the ion dynamics of new materials for which their interatomic potentials are necessarily not available, AIMD calculations can be performed for any new material with intricate chemical compositional space. As for AIMD calculations,} the interatomic potentials are \red{substituted with} forces \red{attained via DFT to describe the time evolution of interatomic interactions of a given system. Recent years have witnessed AIMD calculations garner traction in the study of }ion-conducting materials including pnictogen- and chalcogen-based honeycomb layered oxides.\cite {huang2020, sau2022ring, bianchini2019nonhexagonal} \red{Nevertheless, compared to both MD and DFT, AIMD calculations are computationally costly and are mostly confined to model systems at shorter time scales. For instance, the time scales of MD simulations are usually in the nanosecond range, whereas AIMD trajectories are in the picosecond range.}


\red{MD computations visualise, in real time, numerous ion migration scenarios. Thus, MD simulations avail diffusion mechanism information at the atomistic scale, information of which is unattainable with other methods. Crucial diffusional information (such as radial distribution functions, correlation function, spacial probability density of mobile ions, site occupancy, jumping rate, amongst others) can be quantified using MD simulations. Further, diffusion coefficients and conductivity values, which satisfy the Arrhenius relation, can be estimated using MD computations. To fully account for the varied migration modes of ions within a given material, prefactors and activation energies can be calculated based on the Arrhenius relation.}


\red{To determine the conductivity and diffusion coefficients of mobile ions, the diffusion scenarios of ions are usually analysed over the duration of the MD calculations. The mean-squared displacement of a given mobile ion over a stipulated time duration is determined in order to estimate the tracer diffusion coefficient. Moreover, the jump diffusion coefficient (usually regarded as diffusion coefficient) can be evaluated from mean-squared displacement of the center of mass of all mobile ions, which correlates with collective macroscopic migration of numerous ions. A pertinent diffusion descriptor in computation modelling techniques (such as MD and kinetic Monte Carlo simulations) is the Haven ratio, which can be estimated from the ratio between the tracer diffusion coefficient and the jump diffusion coefficient.}


\red{Using the Nernst-Einstein relation, the ionic conductivity values can be approximated. Nonetheless, owing to strong ion migration correlations in fast ion-conducting materials, ionic conductivity values tend to be underestimated when using the tracer diffusion coefficient. Multiple AIMD calculations at various temperatures are employed to ascertain the ionic conductivities $\sigma$ of a given material, which obey the Arrhenius equation: $\sigma = \sigma_0\exp(-E_{\rm a}/k_{\rm B}T)$. By fitting logarithmic plots of ($\sigma$) against $1/T$, the prefactor ($\sigma_0$) and the activation energy ($\rm E_a$) can be calculated. Ionic conductivity values at other temperatures can be estimated through extrapolation of the fitted Arrhenius plots, on the assumption that similar diffusion mechanisms occur at the temperatures extrapolated.}

Classical MD has been performed on $\rm Na_2{\it M}_2TeO_6$ ($M =$ Co, Zn, Mg, Ni) to explore the innate fast \red{Na-ion} diffusion \red{displayed by} these honeycomb layered oxides.\cite {sau2015role, sau2015ion, sau2016influence, sau2016ion, sau2014molecular} The Vashishta-Rahman interatomic potential is found to be effective in reproducing \red{a plethora of transport and structural} properties of ${\rm Na_2}M_2{\rm TeO_6}$ such as conductivity, and population of $\rm Na^{+}$ at different Na sites,\cite {sau2015ion} in excellent quantitative agreement with experiments.\cite {evstigneeva2011new} The Vashsishta-Rahman form of the inter-atomic potential has also been extended to reliably reproduce structural and transport properties of $\rm Li_2Ni_2TeO_6$ and $\rm K_2Ni_2TeO_6$.\cite {sau2022insights} $\rm Na_2Ni_2TeO_6$ has been reported, experimentally, to \red{display} the highest ionic conductivity (0.11 S cm$^{-1}$ at 573 K) in the series.\cite {evstigneeva2011new} Classical MD \red{calculations} further \red{show} that \red{the ionic conductivity of} $\rm Na_2Ni_2TeO_6$ can be further \red{enhanced} through \red{decreasing (by 20\%) the content of Na sandwiched between the transition metal slabs. \cite {sau2015role} Owing to strong ion-ion correlations, the Na-ion diffusion mechanism in $\rm Na_2Ni_2TeO_6$ is revealed to be highly cooperative.} Na occupies three crystallographically different sites within the lattice of \red{$\rm Na_2Ni_2TeO_6$} (denoted usually as Na1, Na2 and Na3).\cite {evstigneeva2011new} Based on the classical simulation of $\rm Na^{+}$ ion transport in $\rm Na_2Ni_2TeO_6$, $\rm Na^{+}$ ions migrate from Na1 to Na2 \red{crystallographic sites}, but with \red{low} contribution from Na3 \red{sites} owing to the potential energy of Na2 and Na1 sites being much lower than that of Na3.\cite {sau2015ion, sau2015role} It is worthy to mention that the \red{aforementioned description of the Na-ion diffusional mechanism in $\rm Na_2Ni_2TeO_6$ via experimental techniques had been elusive,} owing to the existence of numerous interstitial sites along with the fact that higher mobility Na-ions necessarily do not reside in equilibrium sites in experimental conditions. Nevertheless, a chiral circular pattern in the Na sub-lattice has been revealed, experimentally, using an inverse Fourier transform technique of both neutron and synchrotron X-ray powder diffraction.\cite {karna2017} More intriguing diffusional aspects of \red{$\rm Na_2Ni_2TeO_6$ at high temperatures 
\red{have} also been divulged using bond valence sum method of high-resolution neutron powder diffraction data.\cite {bera2020temperature}} Classical MD simulations \red{reveal} well-developed migration pathways of $\rm Na^{+}$ \red{linking Na sites in the repetitive sequence:} Na1–Na2–Na1–Na2–… emerging from calculated population profiles of $\rm Na_2Ni_2TeO_6$,\cite {sau2015ion, sau2015role} which were later experimentally confirmed\cite {bera2020temperature}.

AIMD simulations have been performed on $\rm Na_2Zn_2TeO_6$ and $\rm Na_2Mg_2TeO_6$,\cite{huang2020} in order to \red{divulge diffusional aspects such as Na-ion conductivity.} Mean-squared displacement values of Na-ion diffusion from 600 K to 900 K were attained and the calculated \red{ionic conductivities (at room temperature)} of $\rm Na_2Zn_2TeO_6$ and $\rm Na_2Mg_2TeO_6$ are 9.77 mS cm$^{-1}$ and 1.19 mS cm$^{-1}$,\cite{huang2020} which \red{are in good agreement with the experimental values attained from electrochemical impedance spectroscopy (EIS) and $^{23}$Na solid-state nuclear magnetic resonance (NMR) spectroscopy measurements. \cite{li2018new, evstigneeva2011new}} Complimenting AIMD simulation with NEB calculations, the \red{computed Na$^{+}$} diffusion coefficients of $\rm Na_2Mg_2TeO_6$ are \red{rather lower} than those of $\rm Na_2Zn_2TeO_6$ at the same temperature. The smaller activation energy of $\rm Na_2Zn_2TeO_6$ has been ascribed \red{to the wider interlayer (interslab)} spacing of $\rm Na_2Zn_2TeO_6$, which \red{enables facile diffusion of Na$^{+}$ and thus a} lower diffusion barrier.

Moreover, AIMD modelling has been performed on $\rm Na_2Zn_2TeO_6$, in another study,\cite{bianchini2019nonhexagonal} to elucidate the occupation of Na sites. $\rm Na_2Zn_2TeO_6$ was chosen over $\rm Na_2Mg_2TeO_6$ by virtue of its higher ionic conductivity and \red{more enhanced crystal purity reported from experiments. $\rm Na_2Zn_2TeO_6$ is found to possess a disordered sub-lattice of Na, whilst no dramatic changes can be discerned in the arrangement of $\rm TeO_6$ and $\rm ZnO_6$ octahedra. Na atoms are found to adopt myriad geometric configurations that are more energetically favourable than the typical honeycomb configuration, yet still retaining the prismatic coordination with oxygen atoms. Rhomboidal geometry of Na ions was revealed to be the most favourable configuration, consisting of ordered zigzag alignment of Na at non-equivalent crystallographic sites. In addition, high-pressure conditions were found to increasingly favour ordered arrangement of Na in triangular-pentagonal patterns. Moreover, annealing simulations of $\rm Na_2Zn_2TeO_6$ unveiled numerous geometric configurations with disordered Na sub-lattices that were energetically favourable. Altogether, AIMD results reveal Na atoms in $\rm Na_2Zn_2TeO_6$ to adopt myriad geometric configurations, attributable to the remarkable Na-ion mobility.\cite{bianchini2019nonhexagonal}}


AIMD simulation has also been done for $\rm Na_2LiFeTeO_6$,\cite{sau2022ring} a fast Na-ion conductor displaying a high ionic conductivity at \red{573 K ($i.e.$, 0.04 S cm $^{-1}$) }comparable to beta-alumina (a solid-electrolyte used in, for instance, sodium-sulphur (NaS) batteries). Based mainly on the analyses of the population density profiles and free-energy barriers, a distinct ring-like Na cationic diffusion was revealed. \cite{sau2022ring} Entropic contribution in Na-ion distribution along with the ring-like motion of Na-ions has been suggested to dictate the fast Na-ion transport in $\rm Na_2LiFeTeO_6$.


\red{The time-evolution information on atoms can be utilised to approximate the {\it van Hove correlation function} from MD simulations. In principle, the {\it van Hove correlation function} is splittable into: the distinct part correlation and self-part correlation. The distinct part correlation, in particular, shows} the time correlation of \red{mobility} of one or more adjacent ions. \red{AIMD simulations were employed to compute the} {\it van Hove functions} of $\rm Na_2Zn_2TeO_6$ and $\rm Na_2Mg_2TeO_6$,\cite{huang2020} indicating highly collective Na-ion diffusion with low diffusion barriers. The low activation energies \red{enable} a \red{facile} fluid-like Na-ion transport \red{along preferred 2D honeycomb diffusion pathways, manifested by collective Na-ion transport.} In addition, analysis of the {\it van Hove correlation function} calculated for $\rm Na_2LiFeTeO_6$ has lent support of the circular ring-like diffusion of Na-ions within the layers that is viewed to govern the\red{ir} innate fast Na-ion transport. \cite{sau2022ring}

\red{Extrapolation of the ionic conductivity values from the computed Arrhenius plots necessarily may not be physically valid or quantitatively precise, as many AIMD simulations are conducted using a small number of atoms at high temperatures (typically beyond 600 K). Moreover, better approximation of the error bounds and statistical variances from MD simulations is crucial, since the estimation of diffusion properties is contingent on the statistics of myriad diffusional events. In particular, the total number of diffusional events is very limited for AIMD simulations, which have a short time duration and a limited number of atoms.} Various procedures have been proposed to minimise errors when quantifying AIMD simulation results,\cite{he2018statistical} \red{which entail determining the statistical variances (contingent on the total number of effective ion hops) of diffusional properties during MD simulation. Thus, the quantitative assessment of diffusional properties of materials using AIMD simulations can only be applied at high temperatures and for materials with fast ion conduction. An AIMD simulation study performed to approximate the diffusion coefficients for a conventional fast ion conductor revealed a} standard deviation of 20-50\%.\cite{gao2020classical} \red{Thus, considering that the error bounds of diffusional results from AIMD simulations are non-trivial, a proper approximation of error bounds ought to be consistently performed.}


Diffusion and conductivity values of materials calculated using MD simulation may not necessarily match with experimental values. \red{Most of the computational techniques (along with AIMD simulations) mentioned herein, are mainly performed for bulk crystalline structures and are solely representative of their conductivity and diffusion in the bulk state. Conductivity of polycrystalline materials typically entails contribution from both the grain-boundary and bulk when measured using experimental techniques such as electrochemical impedance spectroscopy. Thus, the computed conductivity values can contravene with those experimentally measured, and can be exacerbated if any impurity phases or chemical compositional variations occur during the syntheses of honeycomb layered oxides.}

\red{\section{Muon spin rotation, relaxation and resonance}}


\red{Muon spin rotation, relaxation and resonance (also referred to as muon spin resonance spectroscopy and usually abbreviated as $\mu$SR)} is less popular than other spin spectroscopic techniques \red{suchlike electron spin resonance (ESR) and nuclear magnetic resonance (NMR)}. Nevertheless, $\mu$SR is a potent technique used to mainly probe the fundamental magnetic properties of a wide range of materials.\cite {schenck1995} $\mu$SR has also been applied to study superconductivity and interrogate solid-state ion diffusion of various functional \red{materials.\cite{mcclelland2020, maansson2013, blundell2004, yaouanc2011}}
	

Muons are electrically \red{charged elementary} particles. \red{Muons have a spin of 1/2 and possess an electric charge, akin to electrons. Although muons possess comparable properties as electrons, they are by far heavier than electrons. Muons consist of anti-muons ($\mu^+$) or muons ($\mu^-$) and cannot be broken down further into smaller particles.} The $\mu$SR technique involves implanting mainly spin-polarised positive muons ($\mu^+$) into a material. Owing to their short lifetime (around 2.2 $\mu$S), muons subsequently decay forming positrons. The muon spins respond to the local magnetic field \red{within} the crystal lattice \red{of a given} material \red{sample}, instantly emitting positrons \red{along the muon-spin direction. Muon polarisation following implantation inside the sample can be measured from the signature asymmetric time evolution of muon decay. Information on the sites muons reside within the crystal lattice can be obtained (alongside information relating to both the local dynamics and local structures), by monitoring the degree to which the muon spins are aligned along a particular direction (muon spin polarisation).} We shall refer readers to relevant bibliography on the fundamental physics of $\mu$SR technique.\cite {mcclelland2020, maansson2013, blundell2004, kanyolo2021honeycomb} 


In the following sub-sections, we discuss \red{the applications of $\mu$SR studies in revealing physicochemical properties of chalcogen- and pnictogen-based honeycomb layered oxides.}

\subsection{Probing magnetism in honeycomb layered oxides}


One advantage of $\mu$SR technique is that muons are sensitive \red{atomic probes of local magnetism} which can often detect effects that are too weak or elusive to be discerned by other methods. Therefore, $\mu$SR \red{can avail information to complement that attained via other techniques}, such as neutron diffraction. For instance, \red{neutron diffraction alongside $\mu$SR experiments have been utilised }to unveil the ground state spin dynamics of honeycomb layered $\rm Li_3Cu_2SbO_6$.\cite {bhattacharyya2021} Notably, dynamic $\rm Cu^{2+}$ (spin-half $d^9$ ions) spin fluctuations were found to persist even at very lower temperatures ($viz.$, 80 mK) devoid of static ordering of the spins. This is reminiscent of a quantum spin liquid ($i.e.$, exotic materials \red{with} potential applications in high-temperature conductivity, \red{quantum computing and data storage}), wherein the magnetic moments of such materials \red{behave similar to} a liquid and remain disordered even at \red{very low temperatures close to} absolute zero.

Complimenting with magnetic susceptibility measurements, $\mu^+$SR measurements have further been employed to elucidate the anti-ferromagnetic N\'{e}el transition of $\rm K_2Ni_2TeO_6$ centered at around 27 K. \cite{matsubara2020magnetism} Moreover, $\mu^+$SR oscillation signal that was observed at the N\'{e}el temperature was found to persist down to very low temperatures (2 K) suggesting commensurate spin ordering of $\rm Ni^{2+}$ in $\rm K_2Ni_2TeO_6$ down to 2 K. $\mu^+$SR measurements have also been performed on $\rm Na_2Ni_2TeO_6$ and $\rm Li_2Ni_2TeO_6$ to elucidate their magnetic spin ordering at low temperatures. \cite{zubayer2020}

\subsection{Investigating solid-state ion diffusion in honeycomb layered oxides}

\begin{figure*}
\begin{center}
\includegraphics[width=\textwidth,clip=true]{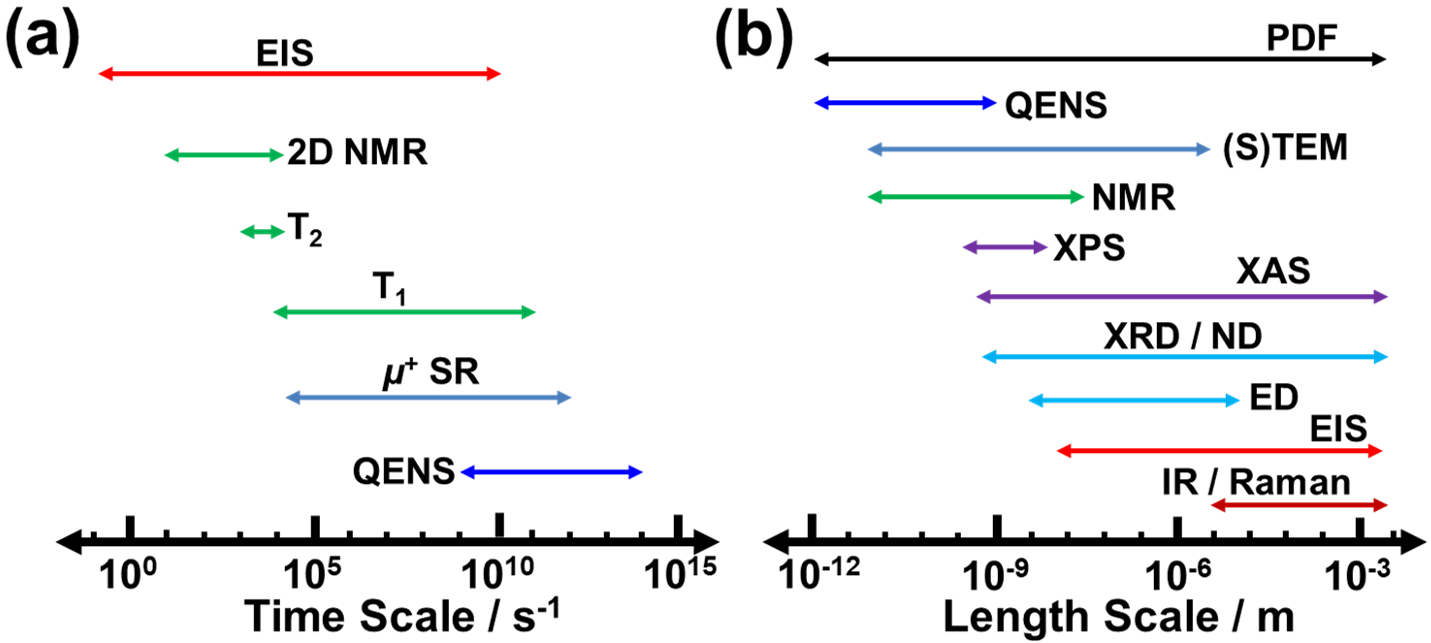}
\caption{Length and time scales that can be examined by various spectroscopic, imaging and diffraction techniques. The corresponding length or time scale for each technique is shown by the length of the arrow bars. For clarity, we hereafter provide the full naming of each abbreviated technique. 
\red {NMR (Nuclear Magnetic Resonance), EIS (Electrochemical Impedance Spectroscopy), $\rm T_2$ (spin-spin NMR relaxation (transverse NMR relaxation)), $\rm T_1$ (spin-lattice NMR relaxation (longitudinal NMR relaxation))}, $\mu^+$SR (muon spin resonance spectroscopy), QENS (Quasi-Elastic Neutron Scattering), PDF (Pair Distribution Function analysis), STEM (Scanning Transmission Electron Microscopy), XPS (X-ray photoelectron spectroscopy), XAS (X-ray Absorption Spectroscopy), XRD (X-ray Diffraction), ND (Neutron Diffraction), ED (Electron Diffraction), IR / Raman (Infrared and Raman Spectroscopy). Values for the length and time scales have been taken from ref. \citenum {gao2020classical}.}
\label{Fig_20}
\end{center}
\end{figure*}

\red{Each of the techniques used to investigate ion transport is responsive to specific time scale ranges. Figure \ref{Fig_20} shows the length and time scales that can be examined by various spectroscopic, imaging and diffraction techniques. Processes relating to macroscopic ion transport occur at comparatively longer length and time scales.\cite {gao2020classical} Macroscopic ion transport processes are typically probed using electrochemical impedance spectroscopy (EIS). Microscopic diffusion can be investigated employing techniques, suchlike} quasi-elastic neutron scattering (QENS), \red{NMR relaxometry, 2D NMR,} variable-temperature NMR and $\mu^+$SR. \red{NMR, QENS and $\mu^+$SR generally probe microscopic ion dynamics of materials.} 

\red{As aforementioned, $\mu^+$SR is a potent technique capable of probing ion diffusion in materials, owing to the sensitivity of implanted anti-muons ($\mu^+$) to the time evolution and local magnetism of spin polarisation. Thus, ion diffusion coefficients can be extracted through analysis of the perturbation induced on the embedded muons.} Muons \red{probe the mobility of ions} using the magnetic fields from the nuclei moving past them. \red{Although equivalent in magnitude to magnetic fields from paramagnetic moments, the magnetic fields emanating} from the nuclei are relatively small \red{when compared to magnetic fields arising from ordered electronic moments in ferromagnets or anti-ferromagnets.\cite {mcclelland2020muon} Therefore, materials with paramagnetic ions render the data analyses complex, akin to NMR. Nevertheless, the contribution of paramagnetic ions to muon spectra is rather distinct and hence easier to separate in the data analysis. Indeed, whether NMR is feasible in the material is a good indicator of the feasibility of $\mu^+$SR.} The nuclei of the ion of interest \red{ought to} have a magnetic moment with some significant abundance \red{to induce} changing magnetic fields measurable by muons.\cite {maansson2013} \red{Also, it is not feasible to probe ionic mobility in ordered phases of ferromagnets or anti-ferromagnets, but this does not pose a huge concern considering that few materials are magnetic at the temperature regimes where ion mobility is remarkable.}

Hereafter, we \red{briefly} highlight the advantages $\mu$SR technique have in comparison to other spectroscopic techniques utilised in studying ionic diffusion of materials: \cite{mcclelland2020muon, gao2020classical}
\red{
\begin{enumerate}[(i)]
    \item Although the presence of quadrupolar nuclei moments significantly affect NMR measurements, $\mu$SR measurements are not that affected by such nuclei moments and in principle can exploit them to garner further information on the muon stopping state and position;
    \item Distinct from neutron diffraction, muons are not scattered from the sample in $\mu$SR. The muons are instead implanted into the sample under investigation, analogous to NMR where nuclei probe the local environment. Thus, $\mu$SR measurement can avail insights into the innate bulk diffusion properties, since the muons implanted within the bulk material are not as prone to surface effects;
    \item Depending on the magnitude of the magnetic field at the muon site, $\mu$SR can measure magnetic fluctuation rates that bridge the gap between those detected by quasi-elastic neutron scattering (QENS) and NMR techniques;
    \item Whilst probing of ion dynamics of materials using conventional electrochemical impedance spectroscopy (EIS) is susceptible to external interferences (suchlike defects or grain boundaries), $\mu$SR can investigate ion dynamics inherent in crystalline lattice devoid of such extrinsic interferences;
    \item The time scale assessable by $\mu$SR technique allows to probe cationic diffusion on a time scale regime where a majority of consecutive short-range and long-range jumps of cations between interstitial sites arise;
    \item Compared with QENS that expends not only much longer data collection time but also larger sample mass for measurements, $\mu$SR measurement requires a much smaller sample mass.
\end{enumerate}
}

Hereafter, we will discuss how $\mu$SR technique has been used to probe microscopic ion transport of exemplar honeycomb layered oxides. 


Diffusion coefficients and activation energy of $\rm K^+$ in honeycomb layered $\rm K_2Ni_2TeO_6$ have been extracted using $\mu$SR measurements.\cite {matsubara2020magnetism} \red{Arrhenius analysis (performed by measuring the time evolution of the positron asymmetry as a function of temperature) revealed an activation energy of} 121(13) meV with $\rm K^+$ in $\rm K_2Ni_2TeO_6$ being mobile beyond 200 K. \red{Furthermore, complimenting $\mu$SR measurements with neutron diffraction,} the local self-diffusion coefficient of K-ion as a function of temperature could be estimated, yielding a room-temperature (300 K) diffusion coefficient of $0.13 \times 10^{-9} \rm cm^2s^{-1}$. This \red{estimated} value is one order of magnitude lower than that for archetypical layered oxide $\rm LiCoO_2$.\cite {sugiyama2009li} In the same vein, $\mu$SR measurements have also been \red{conducted} on $\rm Na_2Ni_2TeO_6$ and $\rm Li_2Ni_2TeO_6$, \cite {zubayer2020} providing a venue to assess ionic conductivity trend of honeycomb layered oxides encompassing mobile alkali ions.

It is worthy to mention that, \red{ionic diffusion values attained for a given material using multiple techniques necessarily do not yield the same value. There are a lot of variables and complexities involved, contingent on the scale of distance and time the particular measurement technique was undertaken and how the particular sample was prepared prior to measurements.} There exists a plethora of \red{theoretical and experimental methods via which the ionic diffusion properties ($e.g$., diffusion coefficient, activation energy, $etc$.) of materials} can be assessed for a comparison with the values obtained via $\mu$SR measurements. Examples include titration methods, including \red{galvanostatic and potentiostatic} intermittent titration techniques; spectroscopic techniques, suchlike EIS, NMR, Raman spectroscopies and secondary ion mass spectroscopies; neutron diffraction techniques such as QENS; electrochemical methods such as cyclic voltammetry; and computational techniques \red {suchlike} {\it ab initio} calculations and molecular dynamics simulations. \red{An arsenal of such techniques} avails the avenue for a holistic insight into \red{cation diffusion in} honeycomb layered oxides and their underlying mechanistics, where more than one technique is utilised. It will be interesting to see, in future publications, how the diffusional properties attained via the aforementioned techniques compare with those obtained using $\mu$SR measurements for $\rm K_2Ni_2TeO_6$ (as an example). Generally, \red{the activation energy values determined} by widely used spectroscopic techniques such as EIS can be expected to be much higher than that of $\mu^+$SR measurements.\cite{gao2020classical} This disparity arises from the fact that \red{EIS probes the ion transport both in grain and across grain boundaries (which can be impacted by the pellet densification process of the sample and so forth), whilst $\mu^+$SR measures ion transport within the grains (intragrain ion transport).}

\red{\section{Summary and Outlook}}

\red{\subsection{Cationic lattices 
}}

Herein, we first reviewed the important aspects of an idealised model of cationic diffusion in specific layered materials, whereby the number of cations or their vacancies is treated as the genus of an emergent 2D manifold without boundary\cite{kanyolo2020idealised, kanyolo2021honeycomb, kanyolo2021partition} whose partition function can be understood within the context of large $N$ theories.\cite{t1993planar, aharony2000large} Since cationic vacancies can be interpreted as topological defects of the manifold,\cite{kanyolo2020idealised} diffusion quantities have a dual geometric description, whereby the cationic vacancies, genus, Gaussian curvature and time-like Killing vector of a $1 + 3$ dimensional manifold are dual to the cations, cationic number, 2D charge density and the U($1$) gauge potential, respectively as summarised in Table \ref{Table_1}. Moreover, the 2D diffusion dynamics of the cations can be described by the field equations given in eq. (\ref{CFE_eq}), which constrain the trace of the Einstein Field Equations in eq. (\ref{EFE_eq}) with the inverse temperature given by $\beta = 8\pi GM$, where $M$ is the total effective mass of the cations equivalent to the average potential energy of the cations and $G$ is the mobility analogous to Newton's constant. Whilst eq. (\ref{CFE_eq}) contains a complex-Hermitian tensor ($K_{\mu\nu} = R_{\mu\nu} + iF_{\mu\nu}$), its structure differs from complex general relativity\cite{einstein1945generalization, einstein1948generalized} since the metric tensor and affine connection are real and torsion free. The 2D diffusion dynamics of the cations is retrieved from the $1 + 3$ dimensional theory of gravity by assuming the layers of the material are stacked along the $z$ coordinate. Thus, in addition to a $t$-like Killing vector, this introduces a $z$-like Killing vector, which guarantees a coordinate system where the 4D metric does not depend on $t$ and $z$.\cite{kanyolo2020idealised}

\red{In the case of bosonic lattices}, a Fermi level does not exist, implying that a particle-hole picture, where the particle and the vacancy carry separate pieces of information is precluded. Thus, the vacancies cannot be treated as holes, but an equivalent description for the dynamics of the cations carrying the same (thermodynamic) information. Consequently, a Bose-Einstein condensate of the cations\cite{kanyolo2020idealised} avails a prime avenue for an emergent geometric description of such vacancies as topological defects within a theory of diffusion on the honeycomb lattice in the context of emergent quantum geometry. This also implies that concepts such as Maxwell demon, linking information content to thermodynamical entropy of condensed matter systems may be relevant.\cite{maruyama2009colloquium} Conversely, describing the diffusion in layered materials comprising fermionic cations such as $^6\rm Li$ with this approach 
\red{poses a significant challenge}, requiring a more intricate description. 
Since their magnetic moment is readily traceable in nuclear magnetic resonance experiments, fermionic cations are typically introduced in meager amounts via doping techniques in order to improve resolution.\cite{pan20026li, lee20006li} Consequently, their overall effects on the diffusion properties are expected to be negligible in bosonic lattices. Nonetheless, if the vacancies are treated as holes it is expected that this particle-hole symmetry is rather befitting to cationic Majorana modes \textit{e.g.} with twist defects\cite{beenakker2013search, zheng2015demonstrating, bombin2010topological} which could be exploited to incorporate fermionic behaviour in the formalism.\cite{kanyolo2019berry} We have discussed another approach to dealing with fermionic lattices by incorporating a pairing mechanism, which transforms pairs of fermions into bosons\cite{tinkham2004introduction}, hence preserving the bosonic description.\cite{kanyolo2020idealised} Meanwhile, the honeycomb lattice can be shown to exhibit modular symmetries generated by $Q \in \rm SL_2(\mathbb{Z})/Z_2 \equiv PSL_2(\mathbb{Z})$, elements of the special linear group, $\rm SL_2(\mathbb{Z})/Z_2$ up to a sign in the cyclic group, $Z_2$\cite{kanyolo2022cationic}, 
\red{indicative of} a link between the theory of cations on the honeycomb lattice and conformal field theory.\cite{polchinski2005string} 

Additional considerations had to be incorporated in the case of fermionic lattices. In particular, it is well-known that the honeycomb lattice of graphene requires an additional degree of freedom to describe the orbital wave functions sitting in two different triangular sub-lattices, known as pseudo-spin.\cite{mecklenburg2011spin} Indeed this description is particularly useful for describing bilayered 
\green{materials}. Layered materials demonstrating a bilayer arrangement of metal atoms exist, a vast majority being Ag-based layered oxides and halides such as ${\rm Ag_2}M\rm O_2$ ($M$ = Co, Cr, Ni, Cu, Fe, Mn, Rh), $\rm Ag_2F$, $\rm Ag_6O_2$ (or equivalently as $\rm Ag_3O$), $\rm Ag_3Ni_2O_4$, and more recently ${\rm Ag_2}M_2\rm TeO_6$ (where $M$ = Ni, Mg, Co, Cu, Zn).\cite{allen2011electronic,schreyer2002synthesis,matsuda2012partially, ji2010orbital,yoshida2020static, yoshida2011novel, yoshida2008unique, yoshida2006spin, masese2023honeycomb, argay1966redetermination, beesk1981x, taniguchi2020butterfly} Despite having equal positive charges, Ag atoms in these compounds form idiosyncratic structural 
\green{materials} with cluster-like agglomerates of conspicuously short $\rm Ag^+ - Ag^+$ interatomic distances akin to those of elemental Ag metal, suggestive of unconventional weak attractive interactions between $d$-orbitals of monovalent Ag atoms ($d^{\rm 10}$-$d^{\rm 10}$ orbital interactions), what is referred to in literature as argentophilic interactions.\cite{jansen1980silberteilstrukturen} This postulation for the origin of weak attractive argentophilic interactions between Ag cations stems from diffuse reflectance spectroscopy measurements performed in a series of Ag-rich ternary oxides, which indicate a special electronic state of $\rm Ag^+$ in the ultraviolet-visible regime.\cite{kohler1985electrical} The unique structural features are accompanied by the formation of an empty orbital band of mainly Ag-$5s$ orbital character near the Fermi level, capable of accomodating additional electrons, which translates to a range of anomalous subvalent states in Ag cations.\cite{schreyer2002synthesis} In principle, subvalent Ag cations have been reported in Ag-rich oxide compositions such as $\rm Ag_5SiO_4$, $\rm Ag_5GeO_4$, $\rm Ag_5Pb_2O_6$, $\rm Ag_{13}OsO_6$, $\rm Ag_3O$, $\rm Ag_{16}B_4O_{10}$, and halides such as $\rm Ag_2F$ and the theoretically predicted $\rm Ag_6Cl_4$.\cite{derzsi2021ag, kovalevskiy2020uncommon, ahlert2003ag13oso6, jansen1992ag5geo4, jansen1990ag5pb2o6, argay1966redetermination, beesk1981x, bystrom1950crystal} Subvalency of Ag ($\rm Ag^{1/2+}$) in $\rm Ag_2NiO_2$ was demonstrated using X-ray absorption spectroscopy, resonant photoemission spectroscopy, magnetic susceptibility measurements and quantum chemical calculations.\cite{schreyer2002synthesis, wedig2006studies, yoshida2006spin, eguchi2010resonant, johannes2007formation} 
The underlying structural characteristics in such materials induces special physicochemical properties such as good metallic conductivity, as has been noted in $\rm Ag_2NiO_2$.\cite{yoshida2006spin} 

We considered the ground state of the theory with a single primitive cell ($k = 1$). The idealised model requires that the emergent manifold is a torus of genus $g = k = 1$, corresponding to two solutions given by a flat-torus with a vanishing Gaussian curvature and a two-torus with a finite Gaussian curvature. Thus, the theory lives on a torus, and is compatible with 2D Liouville conformal field theory. To see the relevance of the torus with respect to the pseudo-spins, recall that we argued that a finite Gaussian curvature breaks scale invariance and hence conformal symmetry. Within the honeycomb lattice shown in Figure \ref{Fig_13}, each pseudo-spin up (down) Ag cation within a primitive cell is bonded to three adjacent pseudo-spin down (up) Ag cations, where the two of the three pseudo-spin down (up) Ag cations lie on two different primitive cells adjacent to the primitive cell containing the pseudo-spin up (down) Ag cations. This ensures that there is no (geometric) spin frustration within the entire honeycomb lattice.\cite{toulouse1980frustration} However, given that each primitive cell is related to the others by translations along the basis vectors, the localised cations in adjacent primitive cells must occupy the same energy state. 

\begin{table*}
\caption{
Stacking sequences 
adopted by 
\red{select} silver-based bilayered honeycomb layered 
\green{materials}.\cite{kanyolo2022advances, kanyolo2021honeycomb} $\rm Tl_2MnTeO_6$ has been 
\red{included} for comparison.
}
\label{Table_4}
\begin{center}
\scalebox{1}{
\begin{tabular}{cc} 
\hline
\textbf{honeycomb layered framework} & \textbf{stacking sequence}\\
\hline\hline
$\rm Ag_2^{1/2+}F^{1-}$ & $U_{\rm F}V_{\rm Ag}W_{\rm Ag}U_{\rm F}$\\
$\rm Ag_6^{2/3+}O_2^{2-}$ & $U_{\rm O}V_{\rm Ag}W_{\rm Ag}U_{\rm O}$\\
$\rm Ag_2^{1/2+}\magenta {Ni^{3+}}O_2^{2-}$ & $U_{\rm O}V_{\rm Ni}W_{\rm O}U_{\rm Ag}V_{\rm Ag}W_{\rm O}U_{\rm Ni}V_{\rm O}W_{\rm Ag}U_{\rm Ag}V_{\rm O}W_{\rm Ni}U_{\rm O}V_{\rm Ag}W_{\rm Ag}U_{\rm O}V_{\rm Ni}U_{\rm O}$\\
$\rm Ag_6Ni_2TeO_6$ & $U_{\rm O}V_{\rm (Ni,Ni,Te)}W_{\rm O}W_{\rm Ag}V_{\rm Ag}V_{\rm O}U_{\rm (Ni,Ni,Te)}W_{\rm O}W_{\rm Ag}V_{\rm Ag}V_{\rm O}W_{\rm (Ni,Ni,Te)}U_{\rm O}$\\
$\rm Tl_2MnTeO_6$ & $U_{\rm O}V_{\rm (Mn, Te, -)}W_{\rm O}V_{\rm (-, -, Ag)}V_{\rm (Ag, - , -)}U_{\rm O}V_{\rm (Mn, Te, -)}W_{\rm O}$\\
\hline
\end{tabular}}
\end{center}
\end{table*}

In particular, since the interaction energy of any two adjacent Ag cations depends only on their relative distance in the $x - y$ plane (translation invariance), provided the cations are considered localised (\textit{i.e.}, $\vec{E} = (E_x, E_y, 0) = 0$ in Figure \ref{Fig_3} (a)), kinetic energy terms do not contribute to the energy suggesting that bonds of equal length imply that, considering only next neighbor interactions, all cation pairs occupy the same ground state. However, since the pseudo-spin of cations is assumed subject to the Pauli exclusion principle, this introduces energy frustration into the system, which precludes either translation invariance, localisation or both. Nonetheless, the topology of the system and hence a finite pseudo-magnetic field where $K \propto B_z \neq 0$ provides a recourse to treat the cations as pseudo-bosons avoiding the energy frustration, thus lifting the degeneracy by distortion/bifurcation. Thus, the system exploits the topology of a flat-torus (Gaussian curvature, $K = 0$) or the two-torus ($K \neq 0$) shown in Figure \ref{Fig_16}, where opposite sides of the primitive cell shown in Figure \ref{Fig_16} (a) are associated with each other, forming a flat-torus with vanishing Gaussian curvature ($K = 0$) or a two-Torus with a finite Gaussian curvature ($K \neq 0$) given in Figure \ref{Fig_16} (b). This maps the three pseudo-spin down (up) Ag cations to each other and hence identifies them as the same cation hence avoiding energy frustration. In addition, in the case of the two-torus ($K \neq 0$), opposite pseudo-spin pairs within a primitive cell experience an attractive interaction proportional to the finite Gaussian curvature, which acts as a pseudo-magnetic field\cite{georgi2017tuning} along the $z$ coordinate, leading to the energy gap and hence a bifurcation of the honeycomb lattice into two triangular sub-lattices with opposite pseudo-spins.

This monolayer-bilayer phase transition can be understood as the pairing of opposite pseudo-spin cations (Cooper pairs\cite{tinkham2004introduction}) within a given primitive cell as illustrated in Figure \ref{Fig_16} (c) and (d), leading to pseudo-spin zero bosons with an order parameter given by $|\Psi|^2 \propto K \propto \Delta = (d - 2)/2$, where $K \rightarrow 0$ or $(d \rightarrow 2)$ is the critical point of the phase transition. In particular, since the pseudo-bosons are not subject to Pauli exclusion, this mechanism avoids energy frustration. As a result, the system is gapped, with the energy difference between the two layers given by eq. (\ref{gap_eq}), corresponding to argentophilic interaction.\cite{masese2023honeycomb} Since the pseudo-magnetic field is proportional to the Gaussian curvature, which in turn is related to the cationic vacancy number density via the Gauss-Bonnet theorem, this critical phenomenon can be interpreted to correspond to the two-torus solution ($K \neq 0$). Alternatively, considering each triangular sub-lattice as a honeycomb lattice, after the bifurcation, each honeycomb sub-lattice consists of a vacancy and a cation such that the emergent manifold is of genus, $g = 0$. Thus, sufficient minimum activation energy of the order of the mass gap is needed to break the argentophilic bond, creating Ag quasi-particles and higher genus states during Ag de-intercalation processes. 

However, 
\red{the Ising model for the pseudo-spins considered appears to suggest that creating cationic vacancies in the honeycomb lattice ($-\Delta_{\nu} = 2\nu > 0$) is directly responsible for the bifurcation of the honeycomb lattice into its bipartite hexagonal sub-lattices}. Since vacancy creation 
\red{occurs discretely} costing activation energy proportional to the number of vacancies, $\nu$ whilst bifurcation corresponds to 
\red{a phase transition which spontaneous creates a vacancy and a cation in each unit cell}, the finite pseudo-magnetic field responsible for the two processes need to differ 
\red{quantitatively}. In fact, since a bifurcated lattice is the more stable structure, we should expect the activation energy (eq. (\ref{activation_E_eq})) $\mathcal{E} = 2\nu = -\Delta_{\nu} = -\chi(g) = 2g - 2 < 0$ to be negative, whilst for vacancy creation is positive, $\mathcal{E} > 0$. Since $\chi(g)$ is the Euler-Poincar\'{e} characteristic, this implies that for bifurcation, we have, $g = \nu + 1 = 0$ ($\chi(g = 0) = 2$, the Euler-Poincar\'{e} characteristic of the 2-sphere), 
\red{exploiting} the last remaining degree of freedom with $\nu = -1$, in order to spontaneously create a pseudo-magnetic field. 
\red{Consequently}, the 
\red{order parameter} corresponds to the scaling dimension, $\Delta(d) = (d - 2)/2 \equiv -2^{2g}\zeta(-2g)$, where $\zeta(s)$ is the analytic continuation of the Riemann zeta function ($d = 2, 3$).\cite{broughan2017equivalents, karatsuba2011riemann} 

In addition, we have availed \red{an equivalent} 
approach to the metallophilic interactions responsible for stabilising the bilayers, centred on the chemistry of group 11 elements.
\red{In particular, there are three coinage metal atom states, depending on the occupancy of the $nd$ and $(n + 1)s$ orbitals. Due to the odd number of electrons, the neutral atom is a fermion (as expected) with its spin state inherited from the spin of the valence electron. For coinage metal atoms ($A = \rm Cu, Ag, Au$), the $A^{1+}$ and $A^{1-}$ valence states are related by isospin rotation (SU($2$)) with the isospin given by $I = \mathcal{V}_A/2$ where $\mathcal{V}_A = 1+, 1-$ are the valence states, and $Y = 0$ is the electric charge of the neutral atom. Meanwhile, the $A^{2+}$ state is an isospin singlet with electric charge, $Y = \mathcal{V}_A = 2+$. Nonetheless, these three cation states $A^{2+}, A^{1-}$ and $A^{1+}$ must have an effective charge, $Q = +2, -1$ and $Q = +1$ respectively, obtained by the Gell-Mann–Nishijima formula and \red{are} treated as independent ions related to each other by $\rm SU(2)\times U(1)$, forming the basis for fractional valent (subvalent) states. Due to $sd$ hybridisation, all these three states are degenerate on the honeycomb lattice. Considering the case of $\rm Ag$, the degeneracy between $\rm Ag^{2+}$ and $\rm Ag^{1-}$ corresponds to right-handed and left-handed chirality of $\rm Ag$ fermions on the honeycomb lattice, treated as the pseudo-spin.\cite{masese2023honeycomb}} 

This picture explains the observed subvalent states of Ag in respective bilayered 
\green{materials} by SU($2$)$\times$U($1$) spontaneous symmetry breaking\cite{zee2010quantum}, leading to additional argentophilic bonds responsible for the bifurcation of the honeycomb lattice.\cite{kanyolo2022advances, masese2023honeycomb} 
\red{As a result}, introducing additional bonds that differ in length from the rest is expected to break scale invariance, 
\red{corresponding to a} monolayer-bilayer phase transition in a cationic lattice of fermions\cite{kanyolo2022cationic, kanyolo2022advances, masese2023honeycomb}, analogous to the Kekul\'{e}/Peierls distortion (2D) in \red{strained graphene}\cite{lee2011band, hou2007electron, ryu2009masses, chamon2000solitons, garcia1992dimerization, peierls1979surprises, peierls1955quantum} 
\red{expected} to generate Dirac masses for the ($1 + 2$)D pseudo-spin cations, $\rm Ag^{2+}$ and $\rm Ag^{1-}$. Thus, such materials have a prevant subvalent state of $1/2+$, obtained by $\rm Ag^{2+}Ag^{1-} = Ag^{1/2+}_2$. Some properties of Ag lattices in select honeycomb layered materials have been displayed in Table \ref{Table_3}. The FCC notation for select honeycomb layered 
\green{materials} exhibiting bilayers has been included in Table \ref{Table_4}.

Finally, apart from the aforementioned Ag-based compounds, subvalent compounds containing mixed atom clusters of Ag and Hg have been reported, \cite{weil2005hydrothermal} whilst metallophilic interactions in compounds entailing other coinage metal atoms such as gold and copper (aurophilic and cuprophilic interactions, respectively), have been envisaged.\cite{sculfort2011intramolecular, jansen2008chemistry} These fall beyond the scope of the present work. Moreover, thallophilic interactions \cite{childress2006thallophilic} can be envisaged in thallium-based layered oxides such as $\rm Tl_2MnTeO_6$\cite{nalbandyan2019preparation} (Stacking sequence written in FCC notation provided in Table \ref{Table_4}), which exhibits a bilayer arrangement of fermionic thallium (Tl) atoms. Since other materials such as $\rm Tl_2MnTeO_6$ share the aforementioned conditions with Ag-based systems, \textit{i.e.} (1) stable bonds between like charges of coinage metal atoms due to metallophilic interactions, (2) bilayers comprising 
a bifurcated bipartite honeycomb lattice; it is reasonable to expect the theoretical framework herein 
\red{also sheds light on the nature of} their monolayer-bilayer phase transition.

\subsection{Computational modeling}

\red {Computational modeling techniques (suchlike molecular dynamics (MD) and first-principles density functional theory (DFT)) have been employed to not only unveil fundamental atomic-scale mechanisms unattainable via experiments, but also predict various physicochemical properties of honeycomb layered materials. DFT has been employed to predict the stability of honeycomb layered oxides such as} $\rm NaKNi_2TeO_6$, $\rm Na_2{\it M}_2TeO_6$ ($M =$ Zn, Mg) and $\rm Li_{1.5}Na_{1.5}Ni_2SbO_6$,\cite {berthelot2021stacking, huang2020, vallee2019} by computation of the phase equilibria and energy convex hull. \red {At present, the computation of the phase equilibria and energy convex hull predominantly utilise DFT energies computed at 0 K. Considering the differences in the temperature-entropy (TS) and pressure-volume (PV) are usually minute in solid-state reactions, such DFT approximations are deemed reasonable. To evaluate the energetics of crystal structural configurations and finite-temperature properties (suchlike phase diagrams or voltage profiles) at elevated temperatures, Monte Carlo computations utilising the cluster expansion methodology can be employed to describe the configurational entropy in a given lattice model.} Furthermore, phonon calculations can be used to evaluate the vibrational contribution to the free energies, as has been done for ${\rm Na_2}M_2{\rm TeO_6}$ ($M =$ Mg, Zn).\cite{huang2020} 

\red {Phase stability computations using DFT face additional limitations. DFT energies of a given material influence the accuracy of phase stability computations. Caution therefore ought to be taken for materials for which the DFT energies may be inaccurately evaluated. In addition, typical DFT computations usually do not account for van der Waals interactions (or referred also as dispersion), which are common in layered materials. The adoption of DFT functionals (or correction terms) have been shown to faithfully reproduce the formation energies of layered materials, and their use ought to be considered in order to ameliorate both the phase stability predictions and the accuracy of DFT energies. Moreover, DFT energies of all materials in the pertinent compositional space are required to accurately evaluate the phase stability of a given system. False-positive predictions can occur when evaluating the stable phases of, for instance, a less-well studied or higher dimension material for which the low-energy compounds are not included or are unknown. Notwithstanding these challenges, the metastability as of a given system as computed using {\it energy above the convex hull} might not be adequate to determine, for example, the synthesisability of a compound–preferably efficacious for precluding} compounds with poor synthesisability.

Further, DFT computations are known to introduce intrinsic inaccuracies, since they demand the exchange-correlation energy to be approximated. Therefore, when performing DFT computations, one ought to be cognisant of such miscalculations. The exchange correlation functionals \red{ignore} van der Waals (vdW) interactions and long-range dispersion, \red{which are particularly critical in} modelling layered materials. \red{Various corrections can be introduced to account for long-range van der Waals interactions such as the utilisation of semi-empirical corrections ($e.g.$, DFT-D) and vdW functionals (such as vdW-DF and vdW-optPBE).\cite{chen2013interlayer, grimme2011effect, grimme2010consistent, grimme2007density, thinius2016reconstruction, becke2005density, johnson2006post, dal1996generalized, lee2012li, tsai2015} Semi-empirical corrections encompass long-range dispersion interactions, whilst vdW functionals directly uses the electron density to attain the dispersion interactions. The adoption of the aforementioned corrections has been shown to reproduce the formation energies, interlayer binding energies and structural parameters for layered materials (suchlike graphene and graphite), in good agreement with experimental data.} Moreover, DFT faces another limitation in predicting the structures of layered oxides with bilayer arrangement of \red{numismophilic} coinage metal atoms such as the recently reported ${\rm Ag}_2M_2{\rm TeO_6}$  \red{($M =$ Cu, Co, Ni, Mg, Zn)} manifesting Ag atoms bilayer domains. \cite{masese2023honeycomb, kanyolo2022advances} DFT computations often neglect metallophilic interactions which are known to be non-negligible especially for group 11 elements \red{(coinage metal atoms)} such as \red{Cu}, Ag \red{and} Au.

\red {DFT simulations using the nudged elastic band (NEB) methodology have widely been employed to compute alkali-ion migration or diffusion behaviour of honeycomb layered oxides. NEB computations can quantify the ion migration barriers and energy profiles for a selected diffusion path in a crystal structure. Nonetheless, {\it diffusion barriers} computed from NEB rely on the total energies derived from DFT, and typically do not consider the effect of pressure, for instance. Moreover, although {\it diffusion barriers} of mobile cations derived from DFT computations are found to coincide with experimental data (thus reliable), NEB calculations are challenging and expensive to perform, particularly for systems with many intricate diffusion pathways that demand an immense number of NEB images. To perform NEB computations, it is necessary to preassign diffusional pathways and mobile carriers (for instance, interstitialcy, vacancy and so forth) as input. A number of possible mobile carriers including their concentrations and formation energies ought to be assessed in order to explicitly establish the main mobile carrier for diffusion. Particularly for compounds with intricate structures, it is crucial to determine the main mobile carriers along with their diffusional pathways. Furthermore, {\it a priori} guesses of the mobile carriers and their diffusional pathways can be extremely demanding, for materials with very disordered sub-lattices of mobile ions.}


\red {Computational cost is a main drawback when performing both high-precision and large-scale DFT computations particularly for periodic and more intricate systems.\cite {ostrom2022} Although DFT computations generally entail the utilisation of so-called GGA functionals to model various structural aspects of materials, highly-accurate and robust computations are envisaged to adopt global hybrid functionals which can be tested to ensure that the DFT calculations are done within reasonable computational time and expense.\cite {csonka2010} Some of the merits of introducing global hybrid functionals in the modelling of the electronic properties (such as band structures and density of states (DOS)) of honeycomb layered oxides include: (i) reduced computational time and (ii) the better prediction of the band gap (as a consequence of improved estimation of the Kohn-Sham orbital energies).} It is also \red {widely-known} that GGA functionals underestimate the diffusion barriers \red{of materials} compared with \red{their} experimental results. Since each functional has its own pros and cons, a common trend is the utilisation of different functionals to obtain different properties. \red {Therefore, it is} crucial to understand the benefits and limitations of each approach in computational materials science.


\red{{\it Ab initio} molecular dynamics (AIMD) simulations have extensively been utilised in the study of new ion-conducting materials with novel chemistries and crystal structures. Whilst classical MD simulations are usually constrained by the force fields available, AIMD simulations are confined to fast ion-conducting materials at high temperatures and cannot be utilised to divulge diffusional events at any given scenario.\cite {he2018statistical} Further, AIMD simulations are confined to small-scale systems and are constrained to short time durations, since AIMD simulations are costly. In addition, extensive sampling of innumerable diffusional scenarios is imperative in order to guarantee the accuracy of AIMD simulations. Since inadequacy in the} number of ion hops \red{examined over the course of} the simulation can \red{cause inaccurate} diffusional properties, AIMD simulations \red{are therefore usually not suited to} study materials \red{possessing high activation energies and low ionic conductivities. On a similar note, AIMD simulations are generally conducted at elevated temperatures (usually beyond 600 K) in order to raise the number of diffusional scenarios scrutinised.}

\red{The approximated values attained from AIMD simulations are also susceptible to statistical errors and variances, apparently since the approximation of mean-squared displacement and diffusion coefficients entail both the statistical and stochastic averaging of diffusional scenarios during the simulation. If a number of underlying ion-hopping mechanisms can be ascertained,} Kinetic Monte Carlo \red{simulations are deemed alternative methods to surmount a few drawbacks of AIMD simulations. Nonetheless, disparities can still emerge whereby the diffusion the diffusion mechanism at a given temperature of interest completely differs from that at elevated temperatures. This is particularly the case for materials that undergo phase transitions at elevated temperatures, for instance, $\rm K_2Ni_2TeO_6$. \cite{zubayer2020}}

Another challenge lies in constructing \red{interatomic potentials} with the \red{precision} of top-level \red{first-principles computations.} Although Vashishta-Rahman potential \red{appears} to faithfully capture the \red{transport and structural} properties of this class of honeycomb layered oxides,\cite{sau2016influence, sau2015ion, sau2015role} more work needs to be done to develop \red{global inter-atomic potentials (force-fields) that are accurate, at a quantum-chemical level,} and which can \red{facilitate the convergence of} MD simulations with \red{entirely quantised nuclei and electrons.} Recent reports have shown \red{the possibility to use top-level first-principles computations to design interatomic potentials via introducing temporal and spacial physical symmetries into a machine-learning model.\cite {chmiela2018}}

\red{Insights from computation into the diffusion mechanisms of honeycomb layered oxides at longer time and length scales will be critical in the future research of their diffusion properties. This necessitates the advancement of computational methods with the versatility and accuracy of first-principles calculations yet cost-effective. }Such modelling methodologies \red{will aid to extend the time and scalability of first-principles MD simulations. Machine learning methods have currently been sought to improve the efficiency of computation.\cite{miwa2017, deng2019, li2017} Altogether, advancement in modelling methods will be critical for accelerated design of new honeycomb layered oxides with functionalities such as high ionic conductivity.}

\begin{figure*}
\begin{center}
\includegraphics[width=\textwidth,clip=true]{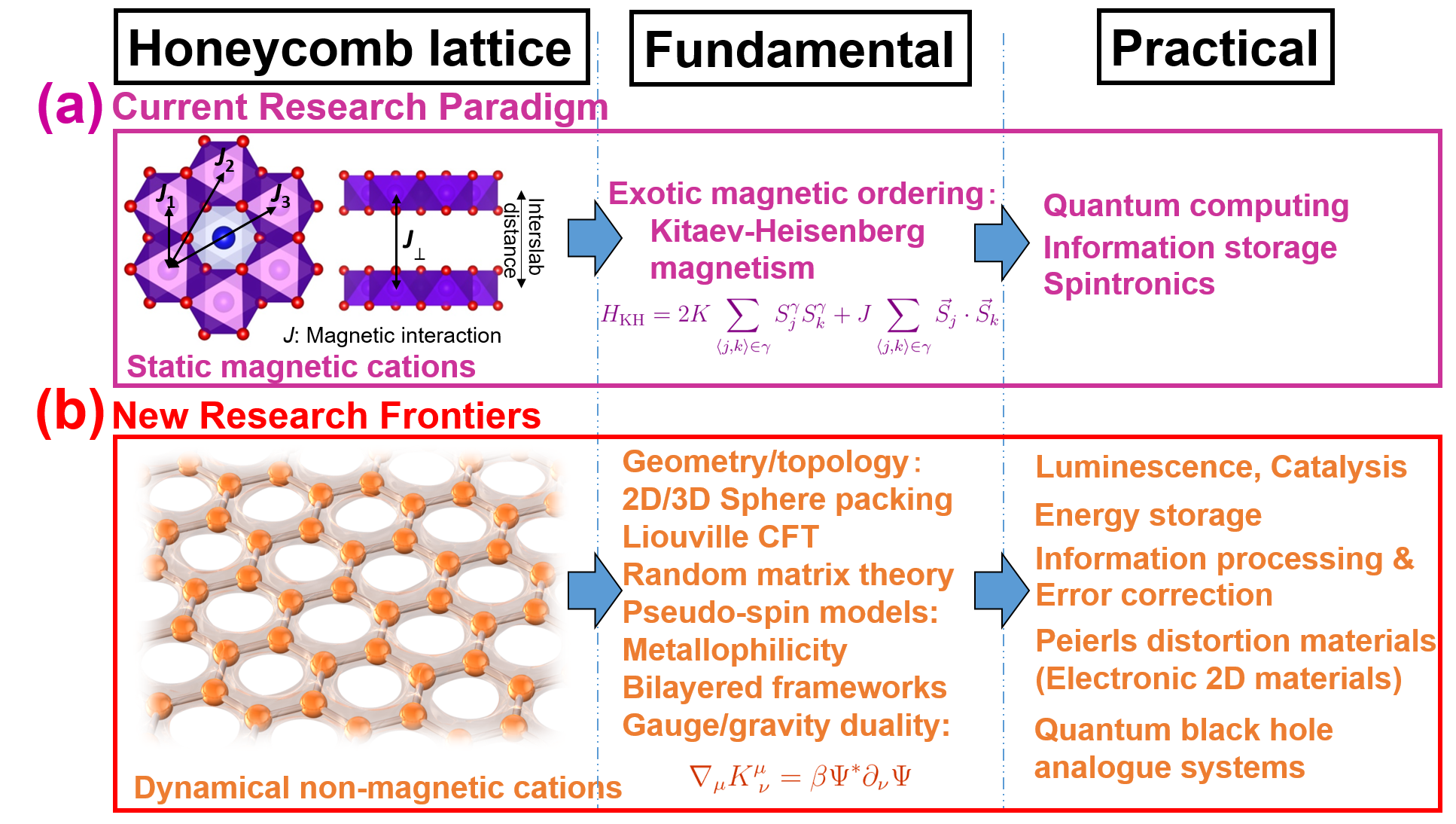}
\caption{Non-exhaustive examples of applications of honeycomb layered materials. (a) Current research on layered materials possessing static magnetic cations situated in a honeycomb lattice.\cite{kanyolo2021honeycomb} (b) New frontiers for layered materials with a honeycomb arrangement of mobile cations.\cite{kanyolo2020idealised, kanyolo2022advances, masese2023honeycomb, kanyolo2022cationic, kanyolo2021partition, kanyolo2022local, kanyolo2021reproducing, tada2022implications}}
\label{Fig_21}
\end{center}
\end{figure*}

\subsection{Solid-state ion diffusion}
\red {Macroscopic ion transport processes are typically probed using electrochemical impedance spectroscopy (EIS). Microscopic diffusion can be investigated employing techniques, suchlike} quasi-elastic neutron scattering (QENS), \red{nuclear magnetic resonance (NMR) relaxometry, 2D NMR,} variable-temperature NMR and $\mu^+$SR. \red{NMR, QENS and $\mu^+$SR generally probe microscopic ion dynamics of materials.} In particular, $\mu^+$SR is a potent technique \red {suited for} probing ion diffusion in materials, owing to the sensitivity of implanted anti-muons ($\mu^+$) to the time evolution and local magnetism of spin polarisation. 

Whilst layered materials with $\rm Li^+$ and $\rm Na^+$ \red{mobile ions} have \red{mainly been pursued},\cite {maansson2013} $\mu$SR has also been \red{utilised} to \red{probe $\rm K^+$ ion dynamics} in honeycomb layered $\rm K_2Ni_2TeO_6$.\cite {matsubara2020magnetism, kanyolo2021honeycomb, zubayer2020}  \red{Furthermore, muons have the capability to be utilised to probe the diffusion of other cations (suchlike Mg and Ca), provided that the target ionic species to be probed} have nuclear magnetic moment. \cite {mcclelland2020muon} \magenta {Recent reports on the synthesis of honeycomb layered $\rm BaNi_2TeO_6$,\cite {song2022influence} envisage the possibility to design honeycomb layered oxides encompassing alkaline-earth metal atoms (such as Mg and Ca) with targeted applications as high-voltage electrode materials for multivalent battery chemistries.\cite{orikasa2014high} Thus,} nuclei such as $\rm ^{25}Mg$ and $\rm ^{43}Ca$ are likely \red{to be enlisted in the domain of $\mu$SR.} $\rm ^{25}Mg$ muon spectroscopy works \red{albeit} with a \red{weak} signal, whereas signal arising \red{from} $\rm ^{43}Ca$ muon spectra is poor unless enriched. 

In addition, a number of honeycomb layered oxides encompassing Ag atoms exist (such as ${\rm Ag}_2M_2{\rm TeO_6}$ ($M =$ Mg, Ni, Co, Cu)) with projected applications as fast solid-state ionic conductors. \cite {masese2023honeycomb} However, Ag nuclei possess \red {very} abundant magnetic moments that are too small \red{to probe readily using} muons, making \red {$\rm ^{109 / 107}Ag$} muon spectroscopy measurements unwieldy. $\rm ^{1}H$ muon spectroscopy of honeycomb layered oxides encompassing hydrogen atoms is feasible, although one needs to separate \red{$\rm H^+$ and $\mu^+$ motion.} Moreover, \red{muonium can be formed when anti-muon ($\mu^+$) that is implanted in an insulating material captures an electron.} (For clarity to readers, muonium refers to an atom formed of an electron and a positive muon ($\mu^+$), analogous to hydrogen, with a very similar electronic structure (isoelectronic) but possessing only one-ninth of its mass). This process makes $\mu$SR a \red{potent technique to probe how hydrogen interacts with matter.}


\red {Although $\mu^+$SR can interrogate solid-state ion diffusion of various layered materials, caution is required when interpreting the attained data which at times can be ambiguous.  $\mu^+$ embedded in the sample is presumed to be static; however, there is a possibility of $\mu^+$ starting to diffuse above certain temperatures. Thus, the diffusive behaviour observed can emanate not only from intrinsic ion diffusion but also from $\mu^+$ diffusion. Negative muon ($\mu^-$) SR measurements have been proven to be effective in discriminating the diffusing species, \cite {sugiyama2020} although $\mu^-$SR spectrum demands extremely high counting statistics to achieve the same statistical precision as $\mu^+$SR. Nonetheless, a combined utilisation of $\mu^{\pm}$ SR \magenta {with nanoscale isotope imaging techniques (such as utilising vibrational spectroscopy in scanning transmission electron microscopy)\cite {senga2022}} will aid to obtain a holistic view of the solid-state cation diffusion in layered materials.}

\section*{Conclusion}

In this 
\red{treatise}, 
\red{we have highlighted recent advances in theoretical, experimental and computational models applicable to honeycomb layered 
\green{materials} that currently redefine the frontier of their research and applications as envisaged in Figure \ref{Fig_21}}. Owing to the intriguing concepts innate in honeycomb layered compounds that can accommodate various cations, we anticipate this work will be accessible not only to a wider community of experimentalists and theoreticians delving in Condensed Matter (materials science, solid-state (electro)chemistry, solid-state physics and solid-state ionics) and Electromagnetism (photonics, electromagnetic dynamics) but also in Mathematical Physics (number theory, topology, quasi-particle physics, modelling and simulation techniques). The discussed topics 
\red{will be relevant to established researchers and early career investigators alike within the aforementioned fields seeking to delve into new research avenues related to the science of honeycomb layered 
\green{materials}}. 

\section*{Acknowledgments}
The authors would like to acknowledge the financial support of TEPCO Memorial Foundation, Japan Society for the Promotion of Science (JSPS KAKENHI Grant Numbers
21K14730 and 23K04922) and 
Iketani Science and Technology Foundation. The authors also acknowledge fruitful discussions with D. Ntara during the cradle of the ideas herein. Both authors are grateful for the unwavering support from their family members (T. M.: Ishii Family, Sakaguchi Family and Masese Family; G. M. K.: Ngumbi Family). 

\bibliography{Theoretical}




\begin{appendix}

\red{\section{Detailed remarks on duality}}

The mathematical physics of lattices 
\red{has} played a pivotal role not only in 
\red{condensed matter theory} but also in other 
unexpected fields such as optimisation problems in mathematics and coding theory.\cite{hales2011revision, cohn2017sphere, viazovska2017sphere, zong2008sphere, cohn2009optimality, cohn2014sphere, afkhami2020high, hartman2019sphere, conway2013sphere} 
\red{The sphere packing problem is known to coincide} with certain conformal field theories (CFTs), provided quantities such as the scaling dimension and the central charge are 
identified with geometric quantities on a lattice. \red{Indeed}, gauge/gravity duality\cite{maldacena1999large, hubeny2015ads, ryu2006holographic, susskind1995world, bousso2002holographic} suggests such descriptions exist, \textit{albeit} the CFTs live on the boundary of the geometry with a suitable metric. 
\red{On the other hand}, layered materials with cations exhibit a myriad of characteristics reliant on their 2D hexagonal and honeycomb lattices. These characteristics are highly dependent on the exhibited symmetries, which in turn play a crucial role in determining the appropriate field theory describing the diffusion theory of cations.\cite{kanyolo2020idealised, kanyolo2022cationic, kanyolo2022advances} 

In the static regime where the cations are immobile (ground state), the problem of finding the appropriate lattice pattern has been \red{argued} to be equivalent to the 2D congruent sphere packing problem. On the other hand, the ground state for the diffusion theory corresponds to Liouville CFT with central charge, $c = 1$, 
\red{living} on the 
\red{two torus}.
\red{Since the ground state of the cations can be alternatively achieved by the minimisation of the electrostatic energy of cations, treated as a congruent sphere packing problem in $d = 2$, dimensions, this corresponds to finding linear programming bounds in mathematics, equivalent to the spinless modular bootstrap for CFTs under the algebra $U(1)^{\rm c}\times U(1)^{\rm c}$}.\cite{cohn2017sphere, cohn2009optimality, cohn2014sphere, afkhami2020high, hartman2019sphere} These realisations lead to the requirement that the underlying field theory must exhibit conformal symmetries. Indeed, we have shown that the appropriate conformal symmetries are manifest when the 2D field theory corresponds to Liouville CFT with central charge $c = d/2 = 1$, where $d = 2$ is the number of dimensions. In particular, the partition function, $\mathcal{Z}(q(b))$ ($q(b)$ is a complex variable (nome) analogue of the Boltzmann factor, and $b$ the analogue for temperature) of the hexagonal/honeycomb lattice exhibits the symmetries of the 2-torus (modular symmetries), whereby the \red{topological charge} 
corresponds to the number of cationic vacancies, related to the genus by the Poincar\'{e}-Hopf theorem.\cite{kanyolo2022cationic} Consequently, this avails a 
\red{quantitative} association (duality) between the theory of cations and their vacancies. 
\red{Mathematically}, quantities with a lattice description are related to geometric quantities by this duality 
\red{-- an observation with far-reaching implications}.

For instance, the cationic theory on the honeycomb lattice is shown to exhibit modular invariance as exemplified by invariance under $S$ and $T$, suitably defined. Since modular forms live on the lattice side of the duality, due to the modularity theorem\cite{darmon1999proof} 
elliptic curves 
must live on the 
\red{geometry} side of the duality. This has implications for other fields of research, \textit{e.g.} suggesting that error-correcting code optimisation problems (
\red{sphere packing problems}\cite{conway2013sphere}) in coding theory are dual to elliptic curve cryptography optimisation problems.\cite{koblitz1987elliptic} For the hexagonal lattices described herein, the Weierstrass elliptic function\cite{papanicolaou2021weierstrass},
\begin{subequations}\label{duality_eq}
\begin{multline}\label{Weierstrass_P_eq}
    \wp(z) = \frac{1}{z^2} + \sum_{0 \neq \lambda \in \Lambda_{\rm hc}} \left ( \frac{1}{(z - \lambda)^2} - \frac{1}{\lambda^2}\right)\\
    = \frac{1}{z^2} + \sum_{k = 1}^{\infty}(2k + 1)G_{2k + 2}z^{2k},
\end{multline}
suffices to capture this duality, where $\lambda = n\omega_1 + m\omega_2$ ($n, m \in \mathbb{Z}$), $\omega_1, \omega_2 \in \mathbb{C}$ are the basis vectors of the primitive cell projected to the upper half of the complex plane, $\mathbb{C}$ and,
\begin{multline}
    G_{2k}(b) = \sum_{(0, 0) \neq (n, m) \in \mathbb{Z}^2}\frac{1}{(m + nb)^{2k}}\\
    = (\gamma\tau + \delta)^{-2k}G(Q\cdot b)
\end{multline}
are the so-called Eisenstein series (modular forms of weight $2k \geq 4$, $k \in \mathbb{N}$) with $Q \in \rm PSL_2(\mathbb{Z})$ (eq. (\ref{modular_eq})). Thus, the hexagonal lattice projected to the complex plane is defined as the primitive cell given in Figure \ref{Fig_16}a up to the choice of the basis vectors. On the other hand, this defines the torus in Figure \ref{Fig_16}b as the Weierstrass elliptic curve, $Y^2 = 4X^3 - g_2X - g_3$ whereby $(X, Y) = (\wp(z), \wp'(z))$ are points on curve, and, 
\begin{align}
    \wp'(z) = \frac{d}{dz}\wp(z) = -2\sum_{\lambda \in \Lambda_{\rm hc}}\frac{1}{(z - \lambda)^3},
\end{align}
\end{subequations}
with $g_2 = 60G_4$, $g_3 = 140G_6$ modular forms of weight 4 and 6 respectively. In fact, these modular forms appear in the partition function of the Liouville field in eq. (\ref{partition_Phi_eq}) given by the discriminant, $Z_{\rm \Phi}(b) = (2\pi)^{d/2}(g_2^3(b) - 27g_3^2(b))^{-d/24} \neq 0$ with $b = \omega_1/\omega_2$ and $d = 2$ dimesions.\cite{abbott1991modular}

Whilst modular symmetries are generated by so-called $S$ ($\mathcal{Z}(b) = \mathcal{Z}(-1/b)$) and $T$ ($\mathcal{Z}(b) = \mathcal{Z}(b + 1)$) transformations, in the sphere packing problem optimised by the hexagonal lattice, the corresponding CFT is chiral, 
\red{implying the system is not $T$ invariant}. 
\red{Consider the partition function on the Riemann sphere which} scales as $\mathcal{Z}(b) \sim \cos(\pi b\mathcal{E}(b))$, where the energy $\mathcal{E}(b) = E_2(b)/12$ scales as the Eisenstein series of weight 2, $E_2(b) = 1 - 24q - 1/4\pi\Im(b) + O(q^2)$ (eq. (\ref{E2_eq}), $b \in \mathbb{N}$) with the Laurent series truncated at $O(q^2)$, the lack of $T$ invariance is an artifact of the lack of a modular form of weight 2 for the whole modular group holomorphic everywhere including at cusps.\cite{bump1998automorphic} Moreover, the choice above 
\red{has a pole} at $\Im(b) = 0$, which we associated with the infinity 
\red{discarded} by the zeta function regularisation. Physically the energy $\mathcal{E}(b) = 1/12 - 2 - \infty$ scales either as $E_2(b) - 3/\pi\Im(b)$ before regularisation or $E_2(b)$ after regularisation, 
\red{thus transforming between $S$ and $T$ modular invariance}. Consequently, both modular invariance under $S$ and $T$ invariance in $\mathcal{E}(b)$ 
\red{are physically realised}, \textit{albeit} non-simultaneously. However, due to the vacuum energy, $1/12$ originating from the cusp, $E_2(b = i\infty) = 1$, the partition function is not $T$ invariant, \textit{i.e.} $\mathcal{Z}(b + 1) \sim \cos(\pi b\mathcal{E}(b) + \pi/12) \neq \cos(\pi b\mathcal{E}(b))$. Recall that the factor of 1/12 is actually $c/12$, where $c = d/2 = 1$ is the central charge. Since we must have $c = 1$, this appears unavoidable. 

On the other hand, the complete modular invariance is achieved for other modular forms, $\mathcal{E}(b, \alpha) = -2C_{\alpha}(b)$, where $C_{\alpha}(b)$ instead is a family $\alpha \geq 0$ of cusp forms of weight 2\cite{koblitz2012introduction, lmfdb} ($C_{\alpha}(b = i\infty) = 0$) given by the series expansion, 
\begin{align}\label{cusp_form_eq}
    C_{\alpha}(b) = \sum_{n = 0}^{\infty}b_n(\alpha)q^n(b),
\end{align}
where $q(b) = \exp(2\pi ib)$ is the nome, $b_0(\alpha) = 0$, $b_1(\alpha) = 1$. 
\red{Thus, choosing the so-called cuspidal eigenforms of Hecke operators\cite{koblitz2012introduction}, $\hat{T}_{n \neq 0}$ satisfying},
\begin{align}\label{Hecke_eq}
    \hat{T}_nC_{\alpha}(b) = b_n(\alpha)C_{\alpha}(b), 
\end{align}
we shall define the first Betti number of the emergent manifold in the theory as the eigenvalue of the Hecke operator, $\hat{T}_1$, given by $b_1(\alpha)$. The other terms $\hat{T}_{n\geq 2}$ can be interpreted as boundary terms of the 2D manifold, \red{which must vanish}. Thus, 
the energy is calculated excluding terms of order ${\rm O}(q^2)$ and greater, and setting $b \in \mathbb{N}$ (like in the case of the Eisenstein series, $E_2$ discussed above). In this case, the truncation can thus be implemented by treating the nome $q$ as a Grassmann number, which introduces the nilpotent condition, $q^n = 0$ ($n \geq 2$). Consequently, 
\red{since} only the first Betti numbers of disjoint 2D manifolds without a boundary are additive under connected sums, we finally set $\mathcal{E}(b, \alpha = -1) = -2C_{\alpha = -1}(b) \equiv \mathcal{E}(b) = E_2/12 - 1/4\pi\Im(b)$, 
\red{and} write the full partition function as, 
\begin{multline}\label{full_partition_eq}
    \mathcal{Z}(b, d) = 2\sum_{g = 0}^{\infty}f_g(\lambda)\cos\left(\pi b\sum_{\alpha = -2\Delta(d)}^{g - 2\Delta^*(d)}\mathcal{E}(b, \alpha)\right)\\
    = 2\sum_{g = 0}^{\infty}f_g(\lambda)\cos\left(\pi b\chi(b, g, d)\right),
\end{multline}
where we have introduced $\Delta^*(d) = (1 - 2\Delta(d))/2$, 
$\Delta(d) = (d - 2)/2$ is the scaling dimension with $d = 2, 3$ and, 
\begin{multline}\label{chi_dash_eq}
    \chi(b, g, d) = \sum_{\alpha = -2\Delta (d)}^{g - 2\Delta^*(d)}\mathcal{E}(b, \alpha)\\
    = 2 - 2gq(b) - 2\Delta(d)/12 + {\rm O}(q^2(b)),
\end{multline}
with the sum over the particular modular forms labelled by $\alpha$ and $f_g(\lambda)$ is a function of a new variable $\lambda$ (to be defined). This reproduces $\chi(b \in \mathbb{N}, g, d = 3) = 2 - 2g - 1/12 = \chi(g) - 1/12$, where $-1/12$ can be interpreted geometrically as a boundary term introduced by the change in dimensions from 2D to 3D.   

Consequently, eq. (\ref{full_partition_eq}) 
\red{corresponds to} a modular invariant partition function for $d = 2$ as required (\textit{e.g.} sought after in \cite{kanyolo2022cationic} with $b = ik$, \textit{albeit} with a factor of $2\pi$ instead). Moreover, the $g = 0$ state in 3D is solely responsible for breaking modular invariance since $E_2(b)$ is not a cusp form, reproducing the boundary term (1/12). Physically, this is a manifestation of the relation given in eq. (\ref{delta_delta_eq}), where the Riemann zeta function at the 
\red{negative real line} is related to the scaling dimension, $\Delta(d)$ by, 
\begin{align}\label{E_char_average_eq}
    \langle \langle \chi(g) \rangle \rangle = -2^{2g}\zeta(-2g) = \Delta(d),
\end{align}
for $\Delta(d) = (d - 2)/2$ ($d = 2,3$) and $\chi(g) = 2 - 2g = -2\nu = \Delta_{\nu}$. 
Thus, \red{since} $\zeta(\Delta_g = 0) = -\Delta(d = 3) = -1/2$ \red{is non-vanishing, this} breaks modular invariance at genus, $g = 0$ by introducing a pseudo-magnetic field proportional to $\Delta(d)$ as expected, and hence the bifurcation of the honeycomb lattice (monolayer-bilayer phase transition).

\red{To motivate the involvement of the Riemann zeta function in the phase transition process, we can define $\Delta_g = -2g$ and take the order parameter/pseudo-magnetic field to be},
\begin{multline}\label{order_p_eq}
    |\varphi_{\Delta_g}(\vec{x})|^2 = \sum_{\vec{x}_i, \vec{x}_j \in \Lambda_d} \frac{-1}{|\vec{x}_i - \vec{x}_j|^{2\Delta_g}}\\
    = -2^{-\Delta_g}\sum_{n = 1}^{\infty}\frac{b_n}{n^{\Delta_g}} \equiv -2^{2g}L_{\Lambda_d}(-2g),
\end{multline}
where $|\vec{x}|^2/2 = |\vec{x}_i - \vec{x}_j|^2/2 = n \in \mathbb{N} > 0$ is the norm between lattice points $i$ and $j$, the sum is performed over the vectors $\vec{x}_i, \vec{x}_j \in \mathbb{R}^d$ with $\vec{x} = \vec{x}_i - \vec{x}_j$ re-scaled appropriately in order to be dimensionless, the constants $b_n$ are the number of vectors of norm $n$ in the lattice, obtained by performing the sum excluding the zero vector ($\vec{x} \neq 0$) and we have introduced a new function, $L_{\Lambda_d}(\Delta_g)$. Now, we can check that $\Delta_g$ is the scaling dimension for the order parameter since, 
\begin{align}\label{scaling_eq}
    |\varphi_{\Delta_g}(\Omega\vec{x})|^2 = \Omega^{-2\Delta_g}|\varphi_{\Delta_g}(\vec{x})|^2,
\end{align}
by eq. (\ref{order_p_eq}). 

Meanwhile, we can consider a lattice with the theta function (eq. (\ref{theta_eq})), 
\begin{align}
    \Theta_{\Lambda_d}(t) = \sum_{\vec{x} \in \Lambda_d}\exp(i\pi b(t)|\vec{x}|^2) = \sum_{n = 0}^{\infty}b_nq^n(b(t)), 
\end{align}
with $b(t) = it/\pi$. By the Poisson summation formula, the theta function is a modular function/form of weight $c = d/2 = 8n/2 = 4n$ with $n \in \mathbb{N} > 0$ whenever it is equivalent to the theta function of its dual lattice.\cite{chenevier2019automorphic} However, presently we 
\red{need not} require the series to be a modular form. Moreover,  
\red{to be consistent with eq. (\ref{order_p_eq}), we need to apply the condition $0 \neq \vec{x} \in \Lambda_d$}, which sets $b_0 = 0$. 
\red{Secondly}, we can check that the integral (Mellin transform\cite{weisstein2004mellin} normalised by the $\Gamma(s)$ function),
\begin{multline}\label{Mellin_transform_eq}
    \langle \Theta_{\Lambda_d} \rangle(s) \equiv \Gamma^{-1}(s)\times\left \{\mathcal{M}\Theta_{\Lambda_d}\right \}(s)\\
    = \Gamma^{-1}(s)\int_0^{\infty} \frac{dt}{t} t^s\sum_{n = 1}^{\infty}b_nq^n(b)\\
    \Gamma^{-1}(s)\sum_{n = 1}^{\infty}b_n\int_0^{\infty} \frac{dt}{t} t^s\exp(-2nt)\\
    = \Gamma^{-1}(s)2^{-s}\sum_{n = 1}^{\infty}\frac{b_n}{n^s}\int_0^{\infty}\frac{dt'}{t'} t'^s\exp(-t')\\
    = 2^{-s}L_{\Lambda_d}(s),
\end{multline}
is equivalent to the order parameter when $s = -2g$ and $\langle \Theta_{\Lambda_d} \rangle (-2g) + |\varphi_{\Delta_g}|^2 = 0$. In the integral, we have used $t' = 2nt$ and the Gamma function, 
\begin{align}
    \Gamma(s) = \int_0^{\infty} \frac{dt}{t} t^s\exp(-t).
\end{align}
Thus, the 
\red{equivalence} between eq. (\ref{Mellin_transform_eq}) and eq. (\ref{order_p_eq}) can be interpreted as the analogue of eq. (\ref{K_eq}). Indeed, Mellin transform has been proposed to relate CFT correlation functions to weakly coupled AdS gravity theories, in the spirit of AdS/CFT.\cite{fitzpatrick2011natural, fitzpatrick2012unitarity} However, any possible connection with this picture, while promising, will not be further explored herein.  

\red{Proceeding, it is clear that $b_{n \geq 1} = 1$ (equivalently, $L_{\Lambda_d}(-2g) = \zeta(-2g)$) corresponds to the phase transition discussed at $g = 0$, since the order parameter vanishes for $g \geq 1$, as expected, corresponding to the trivial zeros of the Riemann zeta function, $\zeta(s)$ after analytic continuation}.\cite{conrey2003riemann, karatsuba2011riemann, broughan2017equivalents} Moreover, by the modularity theorem\cite{darmon1999proof}, every 
\red{elliptic curve, $E$ over rational points, $\mathbb{Q}$, written as $E/\mathbb{Q}$ is related (up to isogenies) to a cusp form of weight 2 and level $N$, defined here as $C_{\nu}(b)$ (eq. (\ref{cusp_form_eq}))}, where $b_n(\nu)$ can be related to the number of solutions of the elliptic curve (mod, $n \in \rm prime\,\,number$).\cite{koblitz1987elliptic} 
Thus, for every elliptic curve, we can associate an order parameter taking the form given in eq. (\ref{order_p_eq}) using eq. (\ref{Mellin_transform_eq}), where $L_{\Lambda_d}(s) = L_E(s)$ is the L-function of the elliptic curve.\cite{koblitz1987elliptic} Since we are interested in the negative even numbers of $s$ associated with the Euler characteristic, $s = \chi(g) - 2$, the values of $L_E(s)$ must be calculated by analytic continuation of $L_{\rm E}(s)$ using their appropriate functional equation.\cite{koblitz1987elliptic,lmfdb}

Meanwhile, 
we indeed recover eq. (\ref{Large_N_cations_eq}) in the limit,
\begin{multline}\label{j_invariant_eq}
    \mathcal{Z}(b = ik, d) = \lim_{-ib \rightarrow \infty}2\sum_{g = 0}^{\infty}f_g(\lambda)\cos(\pi b\chi(b, g))\\
    = \lim_{k \rightarrow \infty}\sum_{g = 0}^{\infty}2f_g(\lambda)\cosh(\pi k\chi(k, g))\\
    = \sum_{g = 0}^{\infty}f_g(\lambda)\mathcal{N}^{\chi(g, d)}, 
\end{multline}
where 
$\chi(g, d) = 2 - 2g - 2\Delta(d)/12$ is the 
\red{Euler} characteristic and $\mathcal{N} = \exp(\pi k)$. 
Consequently, this takes the form of the free energy of a U($\mathcal{N}$) 
\red{large $\mathcal{N}$} theory with coupling constant, $g_{\rm c}$ satisfying $g_{\rm c}^2 = \lambda/\mathcal{N}$, for $\mathcal{N}(k) \rightarrow \infty$ at fixed $\lambda$.\cite{t1993planar, aharony2000large} Moreover, the coefficients $f_g(\lambda)$ should have their Taylor expansion in $\lambda$, different for every genus, $g$ of the Feynman diagram,
\begin{align}\label{feynman_no_eq}
    f_g(\lambda) = \sum_{n = 0}^{\infty}\frac{f_g^{(n)}(0)}{n!}\lambda^n = \sum_{n = 0}^{\infty}a_n(g)\lambda^n,
\end{align}
in order for eq. (\ref{j_invariant_eq}) to be equivalent to the sum of Feynman diagrams of the form, 
\begin{align}\label{feynman_diag_eq}
   \left (\frac{\mathcal{N}}{\lambda}\right )^{-E}\left(\frac{\mathcal{N}}{\lambda}\right)^{V} \mathcal{N}^F = \lambda^{E - V}\mathcal{N}^{F - E + V},
\end{align}
as is required by typical large $\mathcal{N}$ theories, where $E - V = n \in \mathbb{Z} \geq 0$ in eq. (\ref{feynman_no_eq}), $F$ is the number of loops, each contributing a factor of $\mathcal{N}$, $E$ is the number of propagators each contributing a factor of $\lambda/\mathcal{N}$ and $V$ the number of vertex interactions each contributing a factor of $\mathcal{N}/\lambda$, thus yielding the topology of the diagram as, $\chi(g) = F - E + V = 2 - 2g$. The Feynman diagrams serve as the triangulation of the emergent manifold. 
Thus, $a_n(g)$ in eq. (\ref{feynman_no_eq}) is the number of Feynman diagrams triangulating the manifold of topology $g$ with $E - V = n \in \mathbb{N} \geq 0$. 

\red{Consequently}, the limit above given by $k \rightarrow \infty$ is consistent with 
\red{$\mathcal{N}(k) = \exp(\pi k) \rightarrow \infty$ which yields the large $\mathcal{N}$ expansion},
\begin{align}
    \mathcal{Z}(\mathcal{N}) = f_0(\lambda)\mathcal{N}^{2 - 2\Delta(d)/12} + f_1(\lambda) + {\rm O}(\mathcal{N}^{-2}).
\end{align}
\red{Thus}, since $\mathcal{Z}(b, d)$ (with $d = 2$) is expected to be modular invariant, in this limit the coefficients $f_g(\lambda)$ must be generated by modular functions/forms ($\mathcal{Z}(b, d = 2)$ need not be holomorphic, but only invariant under $\rm PSL_2(\mathbb{Z})$), which fixes $\lambda$ for a particular theory, and $S$ invariance ($k \rightarrow 1/k$) links the 
\red{theory at high temperature to the same theory at low temperature}. At first glance, this duality appears perplexing, but nonetheless makes physical sense by recalling that at low temperature, the lattice is 
\red{comprised of cations with few vacancies whereas the converse (vacancies with few cations) applies for high temperature}. Consequently, $S$ transformation suggests the theory of cationic lattices described is equivalent/dual to a theory of vacancies. We could assume 
\red{a lattice of vacancies} and then perform the same arguments as before to arrive at eq. (\ref{j_invariant_eq}), thus treating vacancies no different from cations. However, this 
\red{would} not make physical sense especially 
\red{in situations where} no cations are present 
-- a location with a vacancy in empty space would appear no different from a location without one. 

To make progress, it is clear that we ought to introduce additional features 
\red{which distinguish the lattice of vacancies from empty space itself}. These features are treated as holes/handles in the emergent manifold, $\mathcal{A}$ whereby their numbers correspond to the genus of $\mathcal{A}$. Consequently, the duality is palatable and more importantly geometric in nature (as required) corresponding to the Liouville CFT (2D quantum gravity) discussed. However, the path integral given by, 
\begin{multline}\label{LCFT_eq}
    \mathcal{Z}_{\rm LCFT}(b = ik) = Z_{\Phi}Z_{g_{ab}}\\
    = \int_{\rm string} \mathcal{D}[\Phi]\exp\left (-\frac{k}{2}\int d^{\,2}r\,\vec{\nabla}\Phi\cdot\vec{\nabla}\Phi \right )\times\\
    \int \mathcal{D}[g_{ab}]\exp\left(-\frac{k}{4}\int_{\mathcal{A}}d^{\,2}r\sqrt{\det(g_{ab})}R_{\rm 2D}\right),
\end{multline}
with $R_{\rm 2D}$ the Ricci scalar, cannot be computed directly using $g_{ab}$, since the measure $\int \mathcal{D}[g_{ab}]$ is ill-defined. Meanwhile, the field $\Phi(\vec{x})$ is coupled to the metric, $g_{ab}(\vec{x}) = \exp(2\Phi(\vec{x}))\delta_{ab}$ in Liouville CFT. Nonetheless, by construction, we performed the path integral by a decoupling via the transformations, 
\begin{subequations}
\begin{align}
    \int \mathcal{D}[g_{ab}] \rightarrow \sum_{\vec{x} \in \Lambda_{2c}},\\
    \frac{1}{4\pi}\int d^{\,2}r\sqrt{\det(g_{ab})}R_{\rm 2D} = \chi(g) \rightarrow -|\vec{x}\,|^2,
\end{align}
\end{subequations}
which yields the sphere packing partition function given in eq. (\ref{sphere_packing_eq}), when $\Phi$ is treated as a bosonic string theory in 2D. In this case, $d = 2c$ corresponding to a central charge, $c = 1$. Unfortunately, since, $Z_{\Phi}(b + 1) = \exp(-i2\pi/12)\eta^{-1}(b) = \exp(i\pi/12)Z_{\Phi}(b)$, the sphere packing partition function is not modular invariant. This poses a challenge in associating $\mathcal{Z}(b, d = 2)$ in eq. (\ref{j_invariant_eq}) with a modular function/form.  

\red{Fortunately, since eq. (\ref{LCFT_eq}) is an effective theory of the idealised model, the discarded additional degrees of freedom can be exploited towards finding a modular invariant partition function}. Working in Euclidean signature (Wick rotation, $t = -it_{\rm E}$), the action for eq. (\ref{CFE_eq}) is given by the form\cite{kanyolo2021partition}, 
\begin{subequations}
\begin{multline}\label{S_ideal_eq1}
    S_{\rm Ideal.}(R = |\Psi|^2) = -\frac{4k}{8\pi}\int d^{\,4}x_{\rm E}\sqrt{\det(g_{\mu\nu}^{\rm E})}\,K^{\mu\nu}(K_{\mu\nu})^{*}\\
    \frac{k}{8\pi}\int d^{\,4}x_{\rm E}\sqrt{\det(g_{\mu\nu}^{\rm E})}\left ( 2(D^{\mu}\Psi)^{*}(D_{\mu}\Psi) + |\Psi|^4 \right )\\
    + \frac{k}{8\pi}\int d^{\,4}x\,\sqrt{-\det(g_{\mu\nu})} R^{\mu\nu\sigma\rho}R_{\mu\nu\sigma\rho},
\end{multline}
where, the superscript $\rm E$ stands for Euclidean signature, $g_{\mu\nu}^{\rm E}$ is the metric in Euclidean signature, $\int d^{\,4}x_{\rm E} = \int dt_{\rm E}\int d^{\,3}x$, $\mu = 8\pi G$, $R$ is the Ricci scalar, $R_{\mu\nu\sigma\rho}$ is the Riemann tensor, $R_{\mu\nu}$ is the Ricci tensor and $K_{\mu\nu} = R_{\mu\nu} + iF_{\mu\nu}/4$ is the complex-Hermitian tensor, $F_{\mu\nu} = \partial_{\mu}A_{\nu} - \partial_{\nu}A_{\mu}$ and $D_{\mu} = \partial_{\mu} - iA_{\mu}$. For later convenience, we have re-scaled $\Psi = \sqrt{\rho}\exp(iS)$ to $\Psi = \sqrt{R}\exp(iS)$ such that, $R = 8\pi GM\rho = 8\pi GM|\Psi|^2$ becomes instead, $R = |\Psi|^2$. This action is equivalent to, 
\begin{multline}\label{S_ideal_eq2}
    S_{\rm Ideal.} = -\frac{k}{8\pi}\int d^{\,4}x_{\rm E}\sqrt{\det(g_{\mu\nu}^{\rm E})}\,\,\frac{1}{4}F^{\mu\nu}F_{\mu\nu}\\
    \frac{k}{8\pi}\int d^{\,4}x_{\rm E}\sqrt{\det(g_{\mu\nu}^{\rm E})}\,\,2(D^{\mu}\Psi)^{*}(D_{\mu}\Psi)\\
    + \frac{k}{8\pi}\int d^{\,4}x_{\rm E}\sqrt{\det(g_{\mu\nu}^{\rm E})}\,G,
\end{multline}
\end{subequations}
where $G = R_{\mu\nu\sigma\rho}R^{\mu\nu\sigma\rho} - 4R_{\mu\nu}R^{\mu\nu} + R^2$ is the Gauss-Bonnet density satisfying,
\begin{align}
    \frac{1}{4\pi}\int_{\mathcal{M}}d^{\,4}x\sqrt{-\det(g_{\mu\nu})}\,G = 2\pi\chi(\mathcal{M}). 
\end{align}
The equations of motion for $A_{\mu}$ are, 
\begin{align}\label{use_F_eq}
    \nabla^{\mu}F_{\mu\nu} = 4\Im\left(\Psi^{*}D_{\nu}\Psi\right) = 4mRu_{\nu},
\end{align}
where we have used $\partial_{\mu}S = mu_{\mu} + A_{\mu}$. Thus, together with the Bianchi identity, $\nabla^{\mu}R_{\mu\nu} = \frac{1}{2}\partial_{\nu}R$, we can write, 
\begin{align}\label{Idealised_eq}
    \nabla^{\mu}K_{\mu\nu} = \Psi^{*}D_{\nu}\Psi = \beta\Psi^{'*}D_{\nu}\Psi',
\end{align}
where $\beta = 8\pi GM = 1/m$ and $\Psi' = \sqrt{\rho}\exp(iS)$ has been re-scaled by $R = \beta\rho$. This is the central equation for the idealised model\cite{kanyolo2022local, kanyolo2021partition, kanyolo2020idealised, kanyolo2021reproducing}, where the re-scaling with $R = \beta\rho$ corresponds to the real part of the equation.

We can justify this re-scaling by 
\red{integrating the first term of  eq. (\ref{S_ideal_eq2}) by parts} to yield, 
\begin{multline}
    -\frac{k}{8\pi}\int d^{\,4}x_{\rm E}\sqrt{\det(g_{\mu\nu}^{\rm E})}\,\,\frac{1}{4}F^{\mu\nu}F_{\mu\nu}\\
    = -\frac{k}{8\pi}\int d^{\,4}x_{\rm E}\sqrt{\det(g_{\mu\nu}^{\rm E})}\,\,\frac{1}{2}\nabla_{\mu}A_{\nu}F^{\mu\nu}\\
    = \frac{k}{8\pi}\int d^{\,4}x_{\rm E}\sqrt{\det(g_{\mu\nu}^{\rm E})}\,\,\frac{1}{2}A_{\nu}\nabla_{\mu}F^{\mu\nu}\\
    = \frac{k}{8\pi}\int d^{\,4}x_{\rm E}\sqrt{\det(g_{\mu\nu}^{\rm E})}\,\,2A_{\nu}u^{\nu}mR\\
    = -\frac{bm^2}{4\pi}\int d^{\,4}x_{\rm E}\sqrt{\det(g_{\mu\nu}^{\rm E})}\,\,R,
\end{multline}
where we have used eq. (\ref{use_F_eq}) and $u^{\mu}u_{\mu} = -1$ and then set $A_{\mu} = mu_{\mu}$ as in the idealised model. 
\red{Thus, in order to generate the Einstein-Hilbert action, we set the coupling constant to $km^2/4\pi = 1/16\pi G$}, finding the expression, $k = 1/4Gm^2 = (8\pi GM)^2/4G = 2\pi\beta M = 2\pi\nu$, where $\nu \equiv M/m$. 
Consequently, the action is finally transformed into, 
\begin{multline}\label{S_ideal_eq3}
    S_{\rm Ideal.} = \frac{k}{8\pi}\int d^{\,4}x_{\rm E}\sqrt{\det(g_{\mu\nu}^{\rm E})}\,\,2(D^{\mu}\Psi)^{*}(D_{\mu}\Psi)\\
    - \frac{1}{16\pi G}\int d^{\,4}x_{\rm E}\sqrt{\det(g_{\mu\nu}^{\rm E})}\,\,R + k\pi\chi(\mathcal{M}).
\end{multline}
Varying with respect to $g^{\mu\nu}_{\rm E}$ yields the Einstein Field Equations corresponding to eq. (\ref{EFE_eq}) where, the energy-momentum tensor is given by, 
\begin{multline}
    T_{\mu\nu} = \frac{\nu}{2M}(D_{\mu}\Psi)^{*}(D_{\nu}\Psi) + \frac{\nu}{2M}(D_{\nu}\Psi)^{*}(D_{\mu}\Psi)\\
    - \frac{\nu}{2M}(D^{\sigma}\Psi)^{*}(D_{\sigma}\Psi)g_{\mu\nu},
\end{multline}
where $\nu = k/2\pi = \beta M$. Setting $u^{\mu} = \xi^{\mu}\exp(\Phi)$ (eq. (\ref{u_eq})) introduces the field, $\Phi$. Thus, $\nabla^{\mu}\nabla^{\nu}F_{\mu\nu} = 0$ implies $\nabla^{\mu}(Ru^{\mu}) = 0$, which can be rearranged to obtain $R = R_0\exp(-\Phi)$ with $R_0$ a constant. Hence, $\Psi = \sqrt{R}\exp(iS) = \sqrt{R_0}\exp(i\phi)\exp(i\int A_{\mu}dx^{\mu})$, where $\phi = \int p_{\mu}dx^{\mu} + i\Phi/2 $ and $\phi^{*} = \int p_{\mu} dx^{\mu} - i\Phi/2$ with $S = \int (p_{\mu} + A_{\mu})dx^{\mu}$ with $p_{\mu} = mu_{\mu}$. Thus, taking the trace of eq. (\ref{EFE_eq}) yields,
\begin{multline}
    R = 8\pi G\nu(D^{\mu}\Psi)(D_{\mu}\Psi)\\
    = 8\pi G\nu|\Psi|^2g^{\mu\nu}(\partial_{\mu}\phi)^{*}(\partial_{\nu}\phi)\\
    = \beta^2|\Psi|^2g^{\mu\nu}(\partial_{\mu}\phi)^{*}(\partial_{\nu}\phi), 
\end{multline}
where we have used $\nu = \beta M$. Since $R = |\Psi|^2$, 
\red{one obtains},
\begin{multline}
    m^2 = \frac{1}{\beta^2} = g^{\mu\nu}(\partial_{\mu}\phi)^{*}(\partial_{\nu}\phi)\\
    = g^{\mu\nu}\left (-p_{\mu} - \frac{1}{2}i\partial_{\mu}\Phi \right)\left (-p_{\nu} + \frac{1}{2}i\partial_{\nu}\Phi \right )\\
    = p^{\mu}p_{\mu} + \frac{1}{2}g^{\mu\nu}(\partial_{\mu}\Phi)(\partial_{\nu}\Phi),
\end{multline}
which requires $p_{\mu}p^{\mu} = m^2$ and $g^{\mu\nu}(\partial_{\mu}\Phi)(\partial_{\nu}\Phi) = 0$ for consistency. Thus, $\Phi$ can be considered as the action for a mass-less particle $\Phi = \int P_{\mu}dx^{\mu}$ satisfying $P^{\mu}P_{\mu} = 0$. The two modes can be reconciled with the coinage metal atom states, \textit{e.g.} the fermions given by $\psi^{\rm T} = (\rm Ag^{2+}, Ag^{1-})$ and $\rm Ag^{1+}$ respectively 
\red{-- the mass of the former generated by SU($2$)$\times$U($1$) symmetry breaking responsible for bilayers, with $m = \beta^{-1} = 1/8\pi GM$ the transition temperature}.

In fact, 
imposing the time-like, $\xi^{\mu}$ and space-like, $n^{\mu}$ Killing vectors as before requires the Lie derivative of the Ricci scalar with respect to the Killing vectors to vanish, $\xi^{\mu}\partial_{\mu}R = n^{\mu}\partial_{\mu}R = 0$, and the metric tensor can be chosen 
\red{such that it is independent of these coordinates}. This suggests the Ricci scalar term (Einstein-Hilbert (EH) action) can further be simplified by considering the cut-off $\beta$ (inverse temperature) and $\pi/m = \pi\beta$ (Compton wavelength) along $t_{\rm E}$ and $z$ to find, 
\begin{multline}\label{EH_action_eq}
    S_{\rm EH} = -\frac{M}{2\beta}\int d^{\,4}x_{\rm E}\sqrt{\det(g_{\mu\nu}^{\rm E})}R\\
    = -\frac{M}{2\beta}\int_0^{\beta} dt_{\rm E}\int_0^{\pi/m} dz\int d^{\,2}r \sqrt{\det(g_{\mu\nu}^{\rm E}(r))}R(r)\\
    = -\frac{k}{4}\int_{\mathcal{A}} d^{\,2}r \sqrt{\det(g_{\mu\nu}^{\rm E}(r))}R(r) = \pi k\chi(\mathcal{A}),
\end{multline}
where 
$\mathcal{A}$ is a sub-manifold of $\mathcal{M}$. 
Thus. eq. (\ref{EH_action_eq}) reproduces the Gauss-Bonnet term in the Liouville CFT integral (eq. (\ref{LCFT_eq})), provided,
\begin{align}
    \sqrt{\det(g_{\mu\nu}^{\rm E}(r))}R(r) = -\sqrt{\det(g_{ab})}R_{\rm 2D}(r).
\end{align}
Alternatively, using $R = \beta\rho$, we have,
\begin{multline}\label{EH_action_eq2}
    S_{\rm EH}^{\rm E} = -\frac{M}{2\beta}\int d^{\,4}x_{\rm E}\sqrt{\det(g_{\mu\nu}^{\rm E})}\,R\\
    = -\frac{M}{2}\int d^{\,4}x_{\rm E}\sqrt{\det(g_{\mu\nu}^{\rm E})}\,\rho\\
    = -\frac{M}{2}\int_0^{\beta u^0(\sigma)} d\sigma\int d^{\,3}x\sqrt{\det(g_{\mu\nu}^{\rm E})}\,\rho u^0\\
    = 2\pi^2\beta M\chi(\mathcal{A}) = \pi k\chi(\mathcal{A}),
\end{multline}
where $u^0(x) = u^0(\sigma) = dt_{\rm E}(x(\sigma))/d\sigma$ and we have set the normalisation for the number of particles as, $\int d^{\,3}x\sqrt{\det(g_{\mu\nu}^{\rm E})}\,\rho u^0(x) = -4\pi^2\chi(\mathcal{A})/u^0(\sigma)$, 
\red{inspired by} eq. (\ref{dimensionless_eq}). Thus, the latter works even for metrics 
\red{that do not admit any Killing vectors}.

Finally, the final form of the action considered in Euclidean signature becomes, 
\begin{multline}\label{S_ideal_eq_final}
    S_{\rm Ideal.} = \frac{k}{8\pi}\int d^{\,4}x_{\rm E}\sqrt{\det(g_{\mu\nu}^{\rm E})}\,\,2(D^{\mu}\Psi)^{*}(D_{\mu}\Psi)\\
    + k\pi\chi(\mathcal{A}) + k\pi\chi(\mathcal{M}).
\end{multline}
Performing the path integral, 
\begin{align}
    \mathcal{Z}_{\rm Ideal.} = \int \mathcal{D}[g_{\mu\nu}^{\rm E}]\int_{\rm string}\mathcal{D}[\Psi^{*},\Psi]\exp(iS_{\rm Ideal.}),
\end{align}
by similar assumptions to eq. (\ref{LCFT_eq}), 
\begin{subequations}
\begin{align}
    \int \mathcal{D}[g_{\mu\nu}] \rightarrow \sum_{\vec{x} = (\vec{y}, \vec{z}) \in \Lambda_{2d}},\\
    \chi(\mathcal{A}) \rightarrow -|\vec{y}\,|^2,\,\,\chi(\mathcal{M}) \rightarrow -|\vec{z}\,|^2,
\end{align}
\end{subequations}
and replacing the two bosonic fields $\Psi^{*}$ and $\Psi$ with mass-less bosonic strings vibrating in $d = 4$ dimensions yields the sphere packing partition function\cite{hartman2019sphere}, 
\begin{align}\label{E8_Z_eq}
    \mathcal{Z}_{\rm Ideal.} = \sum_{\vec{x} \in \Lambda_{2d}}\frac{ \exp(ib\pi|\vec{x}\,|^2)}{\eta^{2d}(b)},
\end{align}
where the dimension, $d = c = 4$ now acts as a central charge of the CFT and $b = ik$. The optimised lattice for $\mathcal{Z}_{\rm Ideal,}$ \textit{e.g.} obtained via linear programming bounds is $\Lambda_{2d} = \rm E_8$\cite{hales2011revision, cohn2017sphere, viazovska2017sphere, zong2008sphere, cohn2009optimality, cohn2014sphere, afkhami2020high, hartman2019sphere}, which was also proved in Viazovska's seminal work by finding so-called \textit{magic functions}.\cite{viazovska2017sphere} The theta function for the $\rm E_8$ root lattice is the $E_4(b)$ modular form of weight 4, 
\begin{multline}
    \Theta_{\rm E_8}(b) = \sum_{\vec{x} \in E_8}\exp(i\pi b|\vec{x}\,|^2) = \sum_{n = 0}^{\infty}b_nq^n(b)\\
    = E_4(b) = 1 + 240\sum_{n = 1}^{\infty}\sigma_3(n)q^n(b) = G_4(b)/2\zeta(4),
\end{multline}
where $\zeta(4) = \pi^4/90$, $|\vec{x}|^2/2 = n \in \mathbb{Z} \geq 0$ and $\sigma_3(n)$ is the sum of the cubes of the divisors of integer, $n$. 

\red{Thus, whilst the idealised model is a $d = 4$ theory, the optimised lattice satisfies the $d^* = 2d = 8$ dimensional sphere packing problem as an artifact of introducing a theory in 4D with two copies of the Euler characteristic ($\chi(\mathcal{A})$, $\chi(\mathcal{M})$) and (two copies of) bosonic strings ($\Psi^{*}$, $\Psi$), \textit{albeit} in some sense describing the hexagonal lattice as required}. To make sense of this using our FCC notation, recall one can build a hexagonal lattice using three copies of hexagonal sub-lattices \textit{e.g.} $U = uvw$, $V = uvw$ or $W = uvw$, each lattice now satisfying the derived partition function given by $\mathcal{Z}_{\rm Ideal.}$ in eq. (\ref{E8_Z_eq}). Thus, the partition function for $U$, $V$ or $W$ becomes the product of three copies of the idealised model partition function (eq. (\ref{E8_Z_eq})), 
\begin{multline}\label{j_invariant_eq2}
    \mathcal{Z}(b) = \mathcal{Z}_{\rm Ideal.}^u(b)\mathcal{Z}_{\rm Ideal.}^v(b)\mathcal{Z}_{\rm Ideal.}^w(b)\\
    = \frac{E_4(b)^3}{q\prod_{n = 1}^{\infty}(1 - q^n(b))^{24}} = 1728\frac{g_2(b)^3}{g_2(b)^3 - 27g_3(b)^2}\\
    = j(b) = \sum_{\nu = -1}^{\infty}f_{\nu + 1}q^{\nu}(b) = 
    1/q + 744 + {\rm O}(q),
\end{multline}
equivalent to the $j$-invariant of the Weierstrass elliptic curve discussed earlier describing the hexagonal lattice in the upper half of the complex plane (under the transformation, $\mathbb{R}^2 \rightarrow \mathbb{C}$) where $g_2(b) = 60G_4(b)$, $g_3 = 140G_6(b)$ and $q(b) = \exp(i2\pi b)$. Comparing eq. (\ref{j_invariant_eq2}) to eq. (\ref{j_invariant_eq}), we can match their coefficients ($f_g(\lambda) = f_{\nu + 1}$) for $d = 2$ which 
\red{justifies the modular invariance of $\mathcal{Z}(b, d)$}.

Finally, it is worth investigating the significance of eq. (\ref{Idealised_eq}) or equivalently eq. (\ref{CFE_eq}) in the context of the scaling dimension. Provided the time-like Killing vector exists, $\xi^{\mu}\partial_{\mu}R = \partial_0R = 0$ and the time component of the real part of eq. (\ref{CFE_eq}) vanishes. Consequently, one can show that the conformal dimension 
\red{corresponds to} the Berry phase\cite{kanyolo2022local}, 
\begin{subequations}\label{Berry_eq}
\begin{multline}
    \left\langle \left \langle \int_0^{\beta}dt_{\rm E}\langle \Psi|i\partial_0|\Psi \rangle \right \rangle\right\rangle = \pi\left\langle \left \langle \Delta_{\nu} \right \rangle\right\rangle = \pi\Delta(d), 
\end{multline}
where $\Delta_{\nu} = -2\nu$, $\nu$ is the first Chern number, 
\begin{align}
    \langle \Psi|i\partial_0|\Psi \rangle = \int d^{\,3}x\sqrt{\det(g_{\mu\nu}^{\rm E})} \Psi^*i\partial_0\Psi,
\end{align}
\end{subequations}
and we have used eq. (\ref{E_char_average_eq}). This 
\red{validates} the duality in Table \ref{Table_1}, where $-2\nu = \chi(g)$ and the Chern number scales as the Euler characteristic of the 2D manifold. Taking the double average in eq. (\ref{E_char_average_eq}) maps the Euler characteristic to the dimension-dependent scaling dimension, $\Delta(d)$, which is relevant for the action of the Liouville/gravitational field $\Phi(\vec{x})$, appearing in eq. (\ref{Newt_gravity_S_eq}). Moreover, it is also clear that $|\varphi_{\langle \langle \Delta_g \rangle \rangle}|^2$ given in eq. (\ref{order_p_eq}) is the sum of correlation functions in $d = 3$ ($\Delta(d = 3) = 1/2$) for $K = 0$ ,
\begin{align}
    \left\langle \Phi(\vec{x}_i)\Phi(\vec{x}_j) \right\rangle = \frac{-1}{|\vec{x}_i - \vec{x}_j|^{2\Delta(d)}},
\end{align}
calculated using eq. (\ref{Newt_gravity_S_eq}) to yield, 
\begin{multline}\label{fail_CFT_eq}
    |\varphi_{\langle \langle \Delta_g \rangle \rangle}(\vec{x})|^2 = \sum_{\vec{x}_i, \vec{x}_j \in \Lambda_d} \frac{-1}{|\vec{x}_i - \vec{x}_j|^{2\langle \langle \Delta_g \rangle \rangle}}\\
    = \sum_{\vec{x}_i, \vec{x}_j \in \Lambda_d} \left\langle \Phi(\vec{x}_i)\Phi(\vec{x}_j) \right\rangle = -2^{-\Delta(d)}\zeta(\Delta(d)) \neq 0,
\end{multline}
which unfortunately 
\red{does not vanish} as it should (to guarantee the order parameter as defined above selects a valid conformal field theory). Here, $\langle \langle \Delta_g  \rangle \rangle = -2^{2g}\zeta(-2g) = \Delta(d)$, where the average now includes $\vec{x} = 0$, obtaining (a variant of) eq. (\ref{E_char_average_eq}). 

Nonetheless, \red{we note that} 
$|\vec{x}|^2/2 = \nu \in \mathbb{Z} \geq 0$ becomes the statement that surface area of given regions in 
\red{3D space} must be discrete -- 
\red{proportional to} the first Chern number. 
\red{This is reasonable since} theories treating gravity as emergent 
\red{assume the condition of discrete (2D) surface area corresponding to the information content within the 3D region (in the spirit of holographic principle)}.\cite{verlinde2011origin, maldacena1999large, hubeny2015ads, ryu2006holographic} Thus, the non-vanishing of the order parameter signals that a slight modification of eq. (\ref{fail_CFT_eq}) is warranted to guarantee a CFT exits. Indeed, the fully-fledged theory need not admit a time-like Killing vector, requiring the real part of eq. (\ref{CFE_eq}) (imaginary part of eq. (\ref{Berry_eq})) to be non-vanishing. In particular, when the system is $d + 1 = 4$ dimensional and $\Psi(x_{\rm E}) = \sqrt{\rho(x_{\rm E})}\exp(iS(x_{\rm E}))$, we obtain, 
\begin{multline}
   \left\langle \left \langle \int_0^{\beta}dt_{\rm E}\langle \Psi|i\partial_0|\Psi \rangle \right \rangle\right\rangle\\
   = -\left\langle \left \langle \int d^{\,3}x\sqrt{\det(g_{\mu\nu}^{\rm E})}\rho(x_{\rm E})u_0(x_{\rm E}) \right\rangle\right\rangle\\
    + \left\langle \left \langle\int d^{\,3}x\sqrt{\det(g_{\mu\nu}^{\rm E})}\frac{1}{2}i\left (\rho(\beta, \vec{x}) - \rho(0, \vec{x})\right )\right\rangle\right\rangle\\
   = \pi(\Delta(d) + i\gamma),
\end{multline}
where we have 
\red{introduced a} real variable ($\gamma \in \rm real$), 
\begin{align*}
    \gamma = \frac{1}{2\pi}\left\langle \left \langle\int d^{\,3}x\sqrt{\det(g_{\mu\nu}^{\rm E})}\left (\rho(\beta, \vec{x}) - \rho(0, \vec{x})\right )\right\rangle\right\rangle.
\end{align*}
Thus, when 
\red{the space-time manifold admits a time-like Killing vector}, $\gamma = 0$ thus, $\rho(\beta, \vec{x}) = \rho(0, \vec{x})$, suggesting that $\rho(t_{\rm E} + \beta, \vec{x}) = \rho(t_{\rm E}, \vec{x})$ is periodic in $\beta$ and takes the form of a thermal correlation function, 
\begin{multline}
    \rho(t_{\rm E} + \beta, \vec{x}) = \rho(t_{\rm E}, \vec{x})\\
    = \frac{{\rm Tr}\left(\exp(-\beta\mathcal{H})\exp(\mathcal{H}t_{\rm E})\mathcal{O}(\vec{x})\exp(-\mathcal{H}t_{\rm E})\right )}{{\rm Tr}\left(\exp(-\beta\mathcal{H})\right )},
\end{multline}
for some unknown operator, $\mathcal{O}(\vec{x})$ and Hamiltonian, $\mathcal{H}$ where $\rm Tr$ is the trace over the eigenstates of $\mathcal{H}$.\cite{kanyolo2022local} Thus, since introducing a time-like Killing vector yields a finite $\gamma \in \rm real$, $\rho$ is no longer periodic. Moreover, the appropriate correlation function for the Liouville field/Newtonian potential is, 
\begin{align}
    \langle \Phi(\vec{x}_i)\Phi(\vec{x}_j) \rangle = \frac{-1}{|\vec{x}_i - \vec{x}_j|^{2(\Delta(d) + i\gamma)}}, 
\end{align}
which yields for $d = 3$, 
\begin{multline}\label{succeed_CFT_eq}
    |\varphi_{\langle \langle \Delta_g \rangle \rangle + i\gamma}(\vec{x})|^2 = \sum_{\vec{x}_i, \vec{x}_j \in \Lambda_d} \frac{-1}{|\vec{x}_i - \vec{x}_j|^{2(\Delta(d) + i\gamma)}}\\
    = 
    -2^{-(\Delta(d) + i\gamma)}\zeta(\Delta(d) + i\gamma) = 0,
\end{multline}
which ought to vanish for specific real values of $\gamma$ to guarantee a CFT exists. 

It is clear that, since (i) $\Delta(d = 3) = 1/2$; (ii) all $\gamma$ values are real; and (iii) eq. (\ref{order_p_eq}) is generalised, \textit{i.e.} also incorporates L-functions; finding the appropriate thermal correlation function, $\rho(x_{\rm E})$ that reproduces all the non-trivial zeros solves the generalised Riemann hypothesis.\cite{conrey2003riemann} Moreover, since the isometry group of the CFT ($d = 2$) before bifurcation of the honeycomb lattice is $SO(d, 2)$ 
(includes Poincar{\,e} invariance, dilatations and special conformal transformations), eq. (\ref{succeed_CFT_eq}) -- the isometry group of AdS$_{d + 1}$\cite{matsuda1984note} -- suggests that there exists a duality between the trivial and non-trivial zeros in the spirit of AdS$_{d + 1}$/CFT$_d$, where the gravity theory is given by the idealised model in $d + 2$ dimensions admitting no time-like and $z$-like Killing vectors. Indeed, we have already argued that the bifurcation of the honeycomb lattice, which discards the $z$-like Killing vector, corresponds to $\rm AdS_{3 + 1}$ \textit{albeit} with a time-like Killing vector. Finally, that the imaginary part of the non-trivial zeros of the Riemann zeta function might correspond to a Hermitian operator (\textit{e.g.} Hamiltonian\cite{bender2017hamiltonian, bellissard2017comment, bender2017comment, wolf2020will}) of a condensed matter system; or location of points in a (quasi-)crystal is not a novel idea.\cite{dyson2009birds, wolf2020will, moxley2017schrodinger} Needless to state, proving the CFT satisfying eq. (\ref{succeed_CFT_eq}) exists using \red{our present understanding} herein (equivalent to proving the Riemann hypothesis) is entirely beyond the scope and intention of this treatise. 

\end{appendix}

\end{document}